%% file: paper.tex
\documentclass[12pt,prd, aps, showpacs, superscriptaddress,floatfix]{revtex4}
\usepackage{graphicx}
\usepackage{bm}
\usepackage{multirow}
\usepackage{axodraw}
\usepackage{graphicx}
\usepackage{epsfig}
\usepackage{psfrag}
\usepackage{soul}
\usepackage{multirow}
\usepackage{mathptmx}
\usepackage{mathrsfs}
\usepackage{amsmath, amssymb}

\usepackage{dcolumn}

\newcommand\riken{RIKEN-BNL Research Center, Brookhaven National
  Laboratory, Upton, NY 11973, USA}
\newcommand\bnl{Brookhaven National Laboratory, Upton, NY 11973, USA}
\newcommand\edinb{SUPA, School of Physics, The University of
  Edinburgh, Edinburgh EH9 3JZ, UK}
\newcommand\epcc{EPCC, School of Physics, The University of
  Edinburgh, Edinburgh EH9 3JZ, UK}
\newcommand\cu{Physics Department, Columbia University, New York,
  NY 10027, USA}

\newcommand\swansea{Department of Physics, Swansea University,
  Swansea SA2 8PP, UK}
\newcommand\kanaz{Institute for Theoretical Physics,  Kanazawa University, Kakuma, Kanazawa, 920-1192, Japan}
\newcommand\tokyo{Department of Physics, University of Tokyo,
  Tokyo 113-003, Japan}
\newcommand\uconn{Physics Department, University of Connecticut,
  Storrs, CT 06269-3046, USA}
\newcommand\soton{School of Physics and Astronomy, University of
  Southampton,  Southampton SO17 1BJ, UK}
\newcommand\kek{Institute of Particle and Nuclear Studies,
  KEK, Tsukuba, 305-0801, Japan}
\newcommand\sokendai{Physics Department, Sokendai Graduate U.\ Adv.\
Studies, Hayama, Kanagawa, 240-0193, Japan}

\newcounter{Outline}
\newcounter{Introduction}
\newcounter{ChPT}
\newcounter{SimDetail}
\newcounter{LatticePS}
\newcounter{LatticeBaryon}
\newcounter{SuTwo}
\newcounter{SuThree}
\newcounter{Bk}
\newcounter{Vector}
\newcounter{Conclusions}
\newcounter{Acknowledgments}
\newcounter{Appendix}
\newcounter{Tables}
\newcounter{Figures}
\input{inputs/user_def.tex}

\begin{document}
\bibliographystyle{apsrev}

\title{Physical Results from 2+1 Flavor Domain Wall QCD and SU(2)
Chiral Perturbation Theory}

\author{C.~Allton}\affiliation{\swansea}
\author{D.J.~Antonio}\affiliation{\edinb}
\author{Y.~Aoki}\affiliation{\riken}
\author{T.~Blum}\affiliation{\riken}\affiliation{\uconn}
\author{P.A.~Boyle}\affiliation{\edinb}
\author{N.H.~Christ}\affiliation{\cu}
\author{S.D.~Cohen\footnote{Present address:  Jefferson Lab,
12000 Jefferson Avenue, Newport News, VA 23606}}\affiliation{\cu}
\author{M.A.~Clark}\affiliation{Center for Computational Science,
3 Cummington Street, Boston University, MA 02215, USA}
\author{C.~Dawson\footnote{Present address:  Dept. of Physics,
University of Virginia, 382 McCormick Rd. Charlottesville,
VA 22904-4714}}\affiliation{\riken}
\author{M.A.~Donnellan}\affiliation{\soton}
\author{J.M.~Flynn}\affiliation{\soton}
\author{A.~Hart}\affiliation{\edinb}
\author{T.~Izubuchi}\affiliation{\riken}\affiliation{\kanaz}
\author{A.~J\"uttner\footnote{Present address:
Institut f\"ur Kernphysik, Johannes Gutenberg-Universit\"at Mainz,
Johann-Hoachim-Becher-Weg 45, D-55099 Mainz,
Germany}}\affiliation{\soton}
\author{C.~Jung}\affiliation{\bnl}
\author{A.D.~Kennedy}\affiliation{\edinb}
\author{R.D.~Kenway}\affiliation{\edinb}
\author{M.~Li}\affiliation{\cu}
\author{S.~Li}\affiliation{\cu}
\author{M.F.~Lin\footnote{Present address:  Center for Theoretical
Physics, Massachusetts Institute of Technology, 77 Massachusetts Ave.,
6-319, Cambridge, MA 02139}}\affiliation{\cu}
\author{R.D.~Mawhinney}\affiliation{\cu}
\author{C.M.~Maynard}\affiliation{\epcc}
\author{S.~Ohta}\affiliation{\kek}\affiliation{\sokendai}\affiliation{\riken}
\author{B.J.~Pendleton}\affiliation{\edinb}
\author{C.T.~Sachrajda}\affiliation{\soton}
\author{S.~Sasaki}\affiliation{\riken}\affiliation{\tokyo}
\author{E.E.~Scholz}\affiliation{\bnl}
\author{A.~Soni}\affiliation{\bnl}
\author{R.J.~Tweedie}\affiliation{\edinb}
\author{J.~Wennekers}\affiliation{\edinb}
\author{T.~Yamazaki}\affiliation{\uconn}
\author{J.M.~Zanotti}\affiliation{\edinb}
\collaboration{RBC and UKQCD Collaborations}
%
%
\noaffiliation{BNL-HET-08/5, CU-TP-1182, Edinburgh 2008/06, KEK-TH-1232, RBRC-730, SHEP-0812}


\date{April 2, 2008}
\maketitle

\centerline{ABSTRACT}

We have simulated QCD using 2+1 flavors of domain wall quarks on a
$(2.74 {\; \rm fm})^3$ volume with an inverse lattice scale of
$a^{-1} = 1.729(28)$ GeV.  The up and down (light) quarks are
degenerate in our calculations and we have used four values for the
ratio of light quark masses to the strange (heavy) quark mass in
our simulations: 0.217, 0.350, 0.617 and 0.884.  We have measured
pseudoscalar meson masses and decay constants, the kaon bag parameter
$B_K$ and vector meson couplings.  We have used SU(2) chiral
perturbation theory, which assumes only the up and down quark masses
are small, and SU(3) chiral perturbation theory to extrapolate to
the physical values for the light quark masses.  While next-to-leading
order formulae from both approaches fit our data for light quarks,
we find the higher order corrections for SU(3) very large, making
such fits unreliable.  We also find that SU(3) does not fit our
data when the quark masses are near the physical strange quark mass.
Thus, we rely on SU(2) chiral perturbation theory for accurate
results.  We use the masses of the $\Omega$ baryon, and the $\pi$
and $K$ mesons to set the lattice scale and determine the quark
masses.  We then find $f_\pi = 124.1(3.6)_{\rm stat}(6.9)_{\rm
syst}\,{\rm MeV}$, $f_K = 149.6(3.6)_{\rm stat}(6.3)_{\rm syst}\,{\rm
MeV}$ and $f_K/f_\pi = 1.205(0.018)_{\rm stat}(0.062)_{\rm syst}$.
Using non-perturbative renormalization to relate lattice regularized
quark masses to RI-MOM masses, and perturbation theory to relate
these to $\overline{\rm MS}$ we find $ m_{ud}^{\overline{\rm
MS}}(2\,{\rm GeV}) = 3.72(0.16)_{\rm stat}(0.33)_{\rm ren}(0.18)_{\rm
syst}\,{\rm MeV}$, $m_{s}^{\overline{\rm MS}}(2\,{\rm GeV}) =
107.3(4.4)_{\rm stat}(9.7)_{\rm ren}(4.9)_{\rm syst}\,{\rm MeV}$,
and $\widetilde{m}_{ud}:\widetilde{m}_s = 1:28.8(0.4)_{\rm
stat}(1.6)_{\rm syst}$.  For the kaon bag parameter, we find $
B_K^{\overline{\rm MS}}(2\,{\rm GeV}) = 0.524(0.010)_{\rm
stat}(0.013)_{\rm ren}(0.025)_{\rm syst}$.  Finally, for the ratios
of the couplings of the vector mesons to the vector and tensor
currents ($f_V$ and $f^T_V$ respectively) in the $\overline{\textrm{MS}}$
scheme at 2\,GeV we obtain: $f_\rho^T/f_\rho=0.687(27)$;
$f_{K^\ast}^T/f_{K^\ast}=0.712(12)$ and $f_\phi^T/f_\phi=0.750(8)$.

\ifnum\theOutline=1
\newpage
\centerline{\Large \bf BEGIN OUTLINE}
\input{text_sections/Outline.tex}
\centerline{\Large \bf END OUTLINE}
\newpage
\fi

\refstepcounter{section}
\setcounter{section}{0}


\newpage
\section{Introduction}
\label{sec:Introduction}

\ifnum\theIntroduction=1
\input{text_sections/Introduction.tex}
\fi

\section{Chiral Perturbation Theory}
\label{sec:ChPT}

\ifnum\theChPT=1
\input{text_sections/ChPT.tex}
\fi

\section{Simulation Details and Ensemble Properties}
\label{sec:SimDetail}

\ifnum\theSimDetail=1
\input{text_sections/SimDetail.tex}
\fi

\section{Lattice Results for Pseudoscalar Masses and Decay Constants}
\label{sec:LatticePS}

\ifnum\theLatticePS=1
\input{text_sections/LatticePS.tex}

\fi

\section{Results for the Omega Baryon Mass}
\label{sec:LatticeBaryon}

\ifnum\theLatticeBaryon=1
\input{text_sections/LatticeBaryon.tex}
\fi

\section{Lattice scale, quark masses and decay constants using SU(2)
ChPT}
\label{sec:Su2}

\ifnum\theSuTwo=1
\input{text_sections/Su2.tex}

\fi

\section{Fitting masses and decay constants to NLO SU(3) ChPT}
\label{sec:Su3}

\ifnum\theSuThree=1
\input{text_sections/Su3.tex}

\fi

\section{$B_K$}
\label{sec:Bk}

\ifnum\theBk=1
\input{text_sections/Bk.tex}

\fi

\section{Vector Meson Couplings}
\label{sec:Vector}

\ifnum\theVector=1
\input{text_sections/Vector.tex}

\fi

\section{Conclusions}
\label{sec:Conclusions}

\ifnum\theConclusions=1
\input{text_sections/Conclusions.tex}
\fi

\ifnum\theAcknowledgments=1
\section*{Acknowledgments}

The calculations reported here were done on the QCDOC computers
\cite{Boyle:2005qc,Boyle:2003mj,Boyle:2005fb} at Columbia University,
Edinburgh University, and at Brookhaven National Laboratory (BNL).
At BNL, the QCDOC computers of the RIKEN-BNL Research Center and
the USQCD Collaboration were used.  The software used includes:  the
CPS QCD codes {\tt http://qcdoc.phys.columbia.edu/chulwoo\_index.html},
supported in part by the USDOE SciDAC program; the
BAGEL {\tt http://www.ph.ed.ac.uk/\~{}paboyle/bagel/Bagel.html} assembler
kernel generator for many of the high-performance optimized kernels;
and the UKHadron codes.

The authors would like to acknowledge useful discussions with Heinrich
Leutwyler, J\"urg Gasser, Johan Bijnens and Steve Sharpe.

T.B. and T.Y. (University of Connecticut) were partially supported
by the U.S.\ DOE under contract DE-FG02-92ER40716.  N.C., S.C, M.L,
S.L, M.F.L., and R.M. (Columbia University) were partially supported
by the U.S.\ DOE under contract DE-FG02-92ER40699.
C.J., E.S and A.S. (BNL) were partially supported by the U.S.\ DOE
under contract DE-AC02-98CH10886.
S.S (University of Tokyo) was partially supported in part by a JSPS
Grant-In-Aid for Scientific Research (C) (No.19540265).
M.D., J.F., A.J., and C.S (University of Southampton) were partially
supported by UK STFC Grant PP/D000211/1 and by EU contract
MRTN-CT-2006-035482 (Flavianet).
The work of the Edinburgh authors was supported by PPARC grants
PP/D000238/1 and PP/C503154/1. The former directly supported CMM,
JW and JMZ. PAB acknowledges support from RCUK and AH acknowledges
support from the Royal Society. The Edinburgh QCDOC system was
funded by PPARC JIF grant PPA/J/S/1998/00756 and operated through
support from the Universities of Edinburgh, Southampton and Wales
Swansea, and from STFC grant PP/E006965/1.

Computations for this work were carried out in part on facilities
of the USQCD Collaboration, which are funded by the Office of Science
of the U.S. Department of Energy.  We thank RIKEN, BNL and the U.S.\
DOE for providing the facilities essential for the completion of
this work.

\fi

\appendix
\ifnum\theAppendix=1
\input{text_sections/Appendix.tex}
\fi

\bibliography{paper}


\ifnum\theTables=1
\newpage
\input{tab/SimDetail.tab}

\input{tab/LatticePS.tab}

\input{tab/LatticeBaryon.tab}
\input{tab/Su2.tab}

\input{tab/Su3.tab}
\input{tab/Bk.tab}

\input{tab/Vector.tab}
\fi


\ifnum\theFigures=1
\newpage
\input{fig/ChPT.fig}
\input{fig/SimDetail.fig}

\input{fig/LatticePS.fig}
\input{fig/LatticeBaryon.fig}
\input{fig/Su2.fig}

\input{fig/Su3.fig}

\input{fig/Bk.fig}

\input{fig/Vector.fig}

\fi

\end{document}

%% file: inputs/user_def.tex
\newcommand{\ba}{\begin{eqnarray}}
\newcommand{\ea}{\end{eqnarray}}
\newcommand{\bas}{\begin{eqnarray*}}
\newcommand{\eas}{\end{eqnarray*}}
\newcommand{\be}{\begin{equation}}
\newcommand{\ee}{\end{equation}}
\newcommand{\bes}{\begin{equation*}}
\newcommand{\ees}{\end{equation*}}
\newcommand{\bi}{\begin{itemize}}
\newcommand{\ei}{\end{itemize}}

\newcommand{\bcentre}{\begin{center}}
\newcommand{\ecentre}{\end{center}}

\newcommand{\tm}{\widetilde{m}}
 
\font\tenmsb=msbm10 scaled\magstep1
\font\sevenmsb=msbm7 scaled\magstep1
\font\fivemsb=msbm5 scaled\magstep1
\newfam\msbfam
\textfont\msbfam=\tenmsb
\scriptfont\msbfam=\sevenmsb
\scriptscriptfont\msbfam=\fivemsb

\newcommand{\cl}[1]{\mathcal{#1}}

\newcommand{\mev}{\rm MeV}

\newcommand{\mres}{m_{\rm res}}
\newcommand{\ov}{\overline}

\def\rmsub#1#2{#1_{\mbox{\tiny #2}}}	    

\def\ln{\mathop{\rm ln}}		    
\def\cosh{\mathop{\rm cosh}}		    
\def\tr{\mathop{\rm tr}}		    
\def\det{\mathop{\rm det}}		    

\def\su3{SU(3)}

\def\tD{\mbox{D}\kern-0.65em\raise0.15ex\hbox{/}\kern0.15em} 
\def\sD{\mbox{\scriptsize D}\kern-0.5em\raise0.15ex\hbox{\scriptsize/}}
\def\ssD{\mbox{\tiny D}\kern-0.42em\raise0.15ex\hbox{\tiny/}}

\def\dslash{\hbox{\(\partial\)}\kern-0.5em\raise0.15ex\hbox{/}} 

\def\su3{SU(3)}

\def\mres{\rmsub{m}{res}}

\newcommand{\bteq}[1]{\boldmath$#1$\unboldmath}

\def\nicefrac#1#2{\leavevmode\kern.1em\raise.5ex\hbox{\the\scriptfont0 #1}\kern-
.1em/\kern-.15em\lower.25ex\hbox{\the\scriptfont0 #2}}

\newcommand{\str}{\mathrm{str}}

\newcommand{\bthree}{B_0}       
\newcommand{\fthree}{f_0}
\newcommand{\lthree}[1]{L_{#1}^{(3)}}

\newcommand{\btwo}{B}           
\newcommand{\ftwo}{f}
\newcommand{\ltwo}[1]{L_{#1}^{(2)}}

\newcommand{\bagzero}{B_{\rm PS}^\chi}
\newcommand{\bagzeroSup}[2]{#2{B}_{\rm PS}^{\chi #1}}

\newcommand{\SU}{\mathrm{SU}}

\newcommand{\lqcd}{\Lambda_{\textrm{QCD}}}

\newcommand{\Sig}{\Sigma}
\newcommand{\Sigd}{\Sigma^{\dagger}}
\newcommand{\e}{{\rm e}}

%% file: text_sections/Outline.tex
\section{Introdution}
\begin{enumerate}
\item DWF is QCD at finite $a$ with continuum symmetries
\item ChPT needed to get to physical $m_l$
\item DWF has continuum ChPT - fewer parameters than other lattice types
\item Can use SU(2) and SU(3) ChPT with our data
\item NLO SU(2) ChPT provides accurate extrapolation to physical $m_l$.
\item Discuss accuracy of NLO SU(3) ChPT from our data as a function
of valence quark masses.
\item Use $m_\Omega$, $m_\pi$, and $m_K$ as inputs
\item Calculate $f_\pi$, $f_K$ and agree with experiment within 5\%
\item Quote quark masses in $\overline{\rm MS}$
\item Calculate $B_K$ with total error of 5\%
\item Calculate vector meson decay constants to ??
\item Accuracies quoted above are estimates - true values TBD
\end{enumerate}

\section{Chiral Perturbation Theory}
\begin{enumerate}
  \item Why DWF has a continuum-like ChPT - mres and O(a) ambiguities
  \item SU(2) and SU(3) both possible at finite lattice spacing
  \item Range of accuracy of NLO ChPT can be determined from lattice
  \item Put in SU(2) description of Flynn and Sachradja
  \item Continuum like form has fewer parameters
  \item Point to formula listed in appendix
\end{enumerate}

\section{Simulation Details and Ensemble Properties}
\begin{enumerate}
\item Specify RHMC and preconditioning choices - present algorithm
  simply and minmally
\item Ensemble Properties
  \begin{enumerate}
    \item Average acceptance of each ensemble
    \item Plaquette and $\ov{\psi}\psi$  evolution
    \item Autocorrelations for plaquette and  meson correlators
    \item Topology as measured by gluon fields
  \end{enumerate}
\end{enumerate}

\section{Lattice Results for Pseudoscalar Masses and Decay Constants}

\begin{enumerate}
\item Describe sources, sinks and masses in measurements
\item Fitting procedure: simultaneous fits, uncorrelated vs. correlated
\item Measured quantities in lattice units at input quark masses
\item $\mres$, $m_{\rm PS}$, $Z_A$, $f_{PS}$ 
\item Finite size effects in lattice quantities 
\begin{enumerate}
\item  Compare $f_\pi$ for $16^3$ and $24^3$ - we see difference
\item  Does not seem to agree with finite V ChPT (Meifeng's thesis)
\end{enumerate}
\end{enumerate}

\section{Lattice Results for the Omega Baryon Mass}
\begin{enumerate}
\item Describe sources, sinks and masses in measurements
\item Fitting procedure: simultaneous fits, uncorrelated vs. correlated
\item Measured quantities in lattice units at input quark masses
\end{enumerate}

\section{Lattice scale, quark masses and decay constants
using SU(2) ChPT}
\begin{enumerate}
\item Best method since $m_K$ large
\item Use partially quenched fits
\item $m_\pi$, $m_K$ and $m_\Omega$ are inputs
\item Show fits and explain - give SU(2) LEC's
\item Outputs are $f_\pi$, $f_K$, quark masses
\item NPR to get quark masses in $\overline{\rm MS}$
\item Currently envisioned to quote values with our unphysical
$m_s^{\rm dyn}$ in this section, and need to generate an error for
unphysical $m_s^{\rm dyn}$.
\item Cross check scale from $r_0$.
\item Paragraph on scale from relativistic action for chram quarks,
quote result from Min's lattice proceedings.
\end{enumerate}

\section{Fitting masses and decay constants to NLO SU(3) Chiral
Peturbation Theory}
\begin{enumerate}
\item $m_s^{dyn}$ large for us.  Are not adding partialy higher order
terms to address how well NLO works.
\item Fit with all valence masses small
\item Discuss range of accuracy of NLO ChPT
\item LEC's found agree well with Bijnens, also compare with MILC.
\item $f_0$ is smaller than Bijnens preferred value.  Also compare
to MILC
\end{enumerate}

\section{$B_K$}
\begin{enumerate}
\item  How measured
\item  Table of values
\item  Finite volume
\item SU(2) chiral extrapolation
\item SU(3) chiral extrapolation
\end{enumerate}

\section{Vector Meson Couplings and Parton Distribution Amplitudes}
\begin{enumerate}
\item Southampton work
\end{enumerate}

\section{Conclusions}

%% file: text_sections/Introduction.tex
%

Numerical simulations of Quantum Chromodynamics (QCD) conventionally
discretize four-dimensional space-time by introducing a lattice
scale, $a$, which yields a well-defined path integral formulation
appropriate for study via importance sampling techniques.  Measurements
of observables at non-zero values of $a$ can then be extrapolated
to $a = 0$ to produce continuum results and, if discretization
errors are small, the continuum extrapolation is better controlled.
An important aspect of simulating QCD is choosing a lattice
discretization which reduces the $a$ dependence of observables
\cite{Symanzik:1983dc}.  Here, one is helped by knowing that if an
$O(a^2)$ accurate gauge action is used, such as the Wilson action,
and a discretization of QCD preserves the continuum chiral symmetries
of massless QCD, then the lattice theory can only have errors
quadratic in $a$, in combinations such as $(a \Lambda_{\rm QCD})^2$,
and $(ma)^2$, where $m$ is a quark mass.  These errors are much
smaller than the $O(am)$ discretization errors which can occur if
the lattice theory breaks chiral symmetry.  Additionally, if chiral
symmetry is broken at non-zero lattice spacing, renormalization of
the simplest operators, while straightforward, is involved
\cite{Bochicchio:1985xa}. (For a recent review, see \cite{Sommer:2006sj}.)

Preserving chiral symmetry at non-zero lattice spacing has a large
impact on many of the more complicated observables, such as matrix
elements of operators, that one wants to determine from lattice
QCD.  For observables such as hadron masses, which do not require
any renormalization, controlling non-zero lattice spacing effects
in the discretized theory is helpful.  For observables requiring
renormalization, such as weak matrix elements in hadronic states,
the presence of chiral symmetry at non-zero lattice spacing can be
vital to the renormalization and mixing of the relevant operators
\cite{Blum:2001xb}.  The chiral symmetries control the allowed
operator mixings, so simplifications take place.  Without this
control, the number of operators and mixings amongst them can make
renormalization very difficult, if not practically impossible.
Chiral symmetry can now be preserved at non-zero lattice spacing
with a variety of formulations
\cite{Kaplan:1992bt,Shamir:1993zy,Furman:1995ky,Narayanan:1993wx,Neuberger:1997fp,Hasenfratz:1998ri}.
(It is important to point out that some of the benefits of chiral
symmetry can be achieved without its presence at non-zero lattice
spacing.  For example, twisted mass Wilson fermion discretizations
\cite{Frezzotti:2000nk} make judicious use of chiral transformations
to calculate quantities without linear dependence on $a$ and
continuum-like renormalization properties, without having full
chiral symmetry.)

In addition to taking the limit $a \to 0$, to achieve accurate
physical results lattice simulations must reach the large volume
($V \to \infty$) limit and also the limit of physical light quarks.
For discretizations which preserve chiral symmetry at non-zero $a$,
simulations with arbitrarily light quarks can be done before taking
the continuum limit -- without chiral symmetry, lattice artifacts
can alter the chiral limit at non-zero $a$ and make the limits
non-commuting.  Current computer power does not allow simulations
with physically light up and down quarks, so an extrapolation from
masses used in the simulations to physical light quark masses must
be done.  Chiral perturbation theory \cite{Gasser:1983yg,Gasser:1984gg}
provides a theoretical framework for these extrapolations and, for
lattice QCD discretizations which preserve chiral symmetry, the
chiral perturbation theory is very similar to the continuum theory,
since there are only a few lattice alterations to it.

Of long recognized importance, the preservation of the global chiral
symmetries in discretized QCD has been achieved by Kaplan's proposal
\cite{Kaplan:1992bt} of four-dimensional fermions resulting from a
defect in a five-dimensional theory.  Further developments led to
the domain wall fermions of Shamir and Furman
\cite{Shamir:1993zy,Furman:1995ky}, that we use here, the overlap
formulation of Neuberger and Narayanan
\cite{Narayanan:1993wx,Neuberger:1997fp} and the perfect
actions of Hasenfratz \cite{Hasenfratz:1998ri}.  To date, the
domain-wall formulation has proven to be the most numerically
feasible.  In this approach, one introduces a fifth dimension (which
we label by the index $s$ and which has extent $L_s$) and only
achieves exact chiral symmetry in the $L_s \to \infty$ limit.
However, for finite $L_s$, chiral symmetry breaking effects can be
made small enough to be easily controlled, as we will discuss in
subsequent sections.  The presence of the fifth dimension (we use
$L_s = 16$ in this work) increases the number of floating point
operations required in a calculation by a factor of $O(L_s)$ over
conventional QCD discretizations such as Wilson and staggered
fermions, which do not preserve continuum chiral symmetries.

The domain wall fermion formulation has been used extensively in
numerical simulations for about a decade.  The original works were
primarily in the quenched approximation, although some early work
did involve QCD with two light quark flavors \cite{Blum:1998ud}.
More extensive 2 flavor simulations were done \cite{Aoki:2004ht}
and, with recent improvements in algorithms and computers, 2+1
flavor QCD simulations have been completed
\cite{Antonio:2006px,Antonio:2007tr,Boyle:2007qe,Antonio:2007pb,Boyle:2007fn}.  These previous calculations demonstrated that
domain wall fermion QCD shows the expected consequences of having
a controlled approximation to the full chiral symmetries of continuum
QCD.  In particular, the following important features were observed.
\begin{enumerate} \item
A mild dependence on $a$ in the $a \to 0$ limit in the quenched
approximation.
\item For both quenched and unquenched QCD, the
residual chiral symmetry breaking at non-zero $a$, as measured by
the additive contribution to the quark mass $\mres$, is readily
made a small fraction of the input bare quark mass at practical
values for $L_s$.
\item The operator mixing problem for domain
wall fermion QCD is essentially the same as the continuum problem.
\end{enumerate}

These previous calculations were not able to fully exploit one of
the main benefits of domain wall fermion QCD -- the ability to probe
the chiral limit of the theory at non-zero lattice spacing -- and hence
make accurate contact with the physical quark mass region.   With
recent advances in computers and algorithms, we have made considerable
progress in simulating with light quarks and on large volumes.  In
this paper, we report on simulations of domain wall fermion QCD,
with two light degenerate quarks and a single flavor heavier quark
(a 2+1 flavor simulation) at a single inverse lattice spacing of
$a^{-1} = 1.729(28)$ GeV in a (2.74 fm)$^3$ spatial volume.  We
used four different light dynamical quark masses in our simulations
and the ratios of these light masses to the physical strange quark
mass is 0.217, 0.350, 0.617 and 0.884.  A single value for the heavy
quark mass was used in all our simulations and its ratio to the
physical strange quark mass is 1.150.  (An accurate value for the
physical strange quark mass was, of course, not known until after
our simulation was complete.)  For mesons made of degenerate light
quarks, the corresponding pseuduscalar meson masses are 331 MeV,
419 MeV, 557 MeV and 672 MeV.  We have also done measurements with
a variety of valence quark masses, with a ratio of the smallest
mass to the physical strange quark mass of 0.110, corresponding to
a pseudoscalar meson with a mass of 242 MeV.  In a previous paper
\cite{Antonio:2006px}, we have given results from simulations with
the same gauge coupling constant, but on a smaller volume, which
gives us some understanding of finite volume effects.

To extrapolate from our simulation quark masses to the physical
values, we use chiral perturbation theory (ChPT), which is an
expansion of low energy QCD observables in powers of the meson
masses and momenta over the pseudoscalar decay constant.  Within
the general framework of ChPT one can consider only the pions to
be light particles, yielding an SU(2)$_{\rm L}$ $\times$ SU(2)$_{\rm
R}$ ChPT (which we will call SU(2) ChPT) or one can also consider
the kaons as light, yielding an SU(3)$_{\rm L}$ $\times$ SU(3)$_{\rm
R}$ ChPT (which we will call SU(3) ChPT).  In Section \ref{sec:ChPT},
we discuss the DWF corrections to ChPT and develop SU(2) ChPT for the
kaon sector, which we will later use to fit our data.

In Section \ref{sec:SimDetail} we give details of our simulations,
including the Rational Hybrid Monte Carlo (RHMC) algorithm that we
use to generate our lattices.  Section \ref{sec:LatticePS} describes
the sources and sinks we use for our pseudoscalar observables, our
methods of determining desired quantities and results in lattice
units.  We will also use the mass of the $\Omega$ baryon, which has
no corrections from chiral logarithms, as part of our scale setting
and the details of the measurement of $m_\Omega$ are given in Section
\ref{sec:LatticeBaryon}.

In Section \ref{sec:Su2} we fit our lattice values for the masses
and decay constants of pseudoscalars to partially quenched SU(2)
ChPT at NLO.  We find our data is well described by the theoretical
formula from Section \ref{sec:ChPT}, provided the pions have masses
below about 420 MeV.  We use the fits to SU(2) ChPT as the most
accurate way to extrapolate our data to the chiral limit, since
SU(2) ChPT does not require the kaon mass to be small, but only
requires $m_\pi \ll m_K$.  Using the results for pseudoscalar masses
from our SU(2) ChPT fits and our lattice values for the $\Omega$
baryon mass, we fix the lattice scale and bare quark masses using
the known masses of the $\pi$, $K$ and $\Omega$.  We find that our
inverse lattice spacing is $a^{-1} = 1.729(28)$ GeV.  In a separate
work \cite{Aoki:2007xm}, we have used non-perturbative renormalization
to calculate the multiplicative renormalization factor needed to
relate our bare lattice quark masses to continuum $\overline{\rm
MS}$ masses.  We find
\begin{eqnarray}
m_{ud}^{\overline{\rm MS}}(2\,{\rm GeV}) &=& 3.72(0.16)_{\rm stat}(0.33)_{\rm ren}(0.18)_{\rm syst}\,{\rm MeV}\,,\\
m_{s}^{\overline{\rm MS}}(2\,{\rm GeV}) &=& 107.3(4.4)_{\rm stat}(9.7)_{\rm ren}(4.9)_{\rm syst}\,{\rm MeV}\,,\\
\widetilde{m}_{ud}:\widetilde{m}_s &=& 1:28.8(0.4)_{\rm stat}(1.6)_{\rm syst}\,.
\label{eq:qmass_intro}
\end{eqnarray}
where $(\cdots)_{\rm stat}$, $(\cdots)_{\rm ren}$ and $(\cdots)_{\rm
syst}$ show the statistical error, the error from renormalization
and the systematic error.  We assume the light quarks to be degenerate
in this work.  We now predict values for $f_\pi$ and $f_K$ and find
$ f_\pi = 124.1(3.6)_{\rm stat}(6.9)_{\rm syst}\,{\rm MeV}$ and $
f_K = 149.6(3.6)_{\rm stat}(6.3)_{\rm syst}\,{\rm MeV}$.  Our fits
to SU(2) ChPT also determine the low energy constants (LECs) for
pseudoscalar masses and decay constants in SU(2) ChPT.  Furthermore,
implications of our results to CKM matrix elements are discussed.

In Section \ref{sec:Su3} we fit our light pseudoscalar data to SU(3)
ChPT.  Here we also find that our data is well represented by SU(3)
ChPT at NLO, provided our pseudoscalars have masses below about 420
MeV.  The failure of NLO SU(3) ChPT to fit our data when pseudoscalar
masses are near the physical kaon mass rules out using NLO SU(3)
ChPT in this mass region.  With light masses, we determine values
for the SU(3) LECs which agree well with values determined by others.
However, we find a small value for the decay constant in the SU(3)
chiral limit, which we denote by $f_0$ (a complete description of
our notation is given in the Appendix).  Our fits give $f_0 = 94$
MeV, with conventions such that the physical value is $f_\pi = 131$
MeV, and this value is smaller than generally found phenomenologically,
which we discuss further in Section \ref{sec:Su3}.  Along with this
we find that the size of the NLO corrections to SU(3) ChPT, relative
to the leading order term, are in the range of 50\% or more.  This
makes the convergence of SU(3) ChPT for these quark masses unreliable.
Thus, although it represents our data well and the parameters we
find generally agree with others, we find the systematic errors in
SU(2) ChPT substantially smaller and use it as our most accurate
means of extrapolating our data to the chiral limit.

In section \ref{sec:Bk}, we discuss our determination of the kaon
bag parameter, $B_K$, which is needed to relate indirect CP violation
in the standard model to experimental measurements.  This section
expands upon the data and analysis presented in \cite{Antonio:2007pb}.
Here we also find extrapolations to the physical quark masses to
be under much better control with SU(2) than with SU(3) ChPT.  We
present our estimates of systematic errors, including finite size
effects.  We find $ B_K^{\overline{\rm MS}}(2\,{\rm GeV}) =
0.524(0.010)_{\rm stat}(0.013)_{\rm ren}(0.025)_{\rm syst}$.

In Section \ref{sec:Vector} we present results for the couplings
of light vector mesons to vector and tensor currents. The results
for the ratios of the couplings of the vector mesons to the vector
and tensor currents ($f_V$ and $f^T_V$ respectively) in the
$\overline{\textrm{MS}}$ scheme at 2\,GeV are:
 $f_\rho^T/f_\rho=0.687(27)$;
$f_{K^\ast}^T/f_{K^\ast}=0.712(12)$ and $f_\phi^T/f_\phi=0.750(8)$.

%% file: text_sections/ChPT.tex

\def\su#1#2{\SU(#1)_\mathrm{#2}}

\label{sec:chiral_limit}

In this section, we discuss chiral perturbation theory for domain
wall fermions in subsection \ref{subsect:dwf_chpt}.  We develop
SU(2) ChPT for kaons in \ref{subsect:kchpt}.  Our notation and
an extensive list of the formulae from SU(2) and SU(3) ChPT that
we use in this work are given in the Appendix.

\subsection{Chiral Perturbation Theory for Domain Wall Fermions}
\label{subsect:dwf_chpt}

Our simulations are done with domain wall fermions, which have
explicit chiral symmetry breaking effects at non-zero lattice spacing.
These effects are controlled by the extent of the lattice in the fifth
dimension, denoted by $L_s$, whose value is chosen to make
such terms small, consistent with current computer power.  The
small chiral symmetry breaking that remains can be measured and
its effects taken into account, as has been discussed extensively
in the literature.  A recent review of this topic is available in
\cite{Sharpe:2007yd}.

As  previously discussed in \cite{Blum:2001sr,Blum:2001xb},
for a theory with $N$ quarks these explicit symmetry breaking
effects can be easily included by introducing an $N \times N$ matrix
parameter $\Omega$ into the domain wall fermion action.  This
parameter connects four-dimensional planes at the mid-point of the
fifth dimension and is included into the action by adding
\begin{equation}
  S_{\Omega} = - \sum_x \left\{ \overline{\Psi}_{x,L_s/2-1} P_L 
    \left( \Omega^\dagger -1 \right)
    \Psi_{x,L_s/2} + \overline{\Psi}_{x,L_s/2} P_R 
    \left( \Omega -1 \right) \Psi_{x,L_s/2-1} \right\}
\label{eq:omega}
\end{equation}
to the conventional action for domain wall fermions \cite{Blum:2000kn}.
Here, $\Psi_{x,s}$ represents a five-dimensional fermion field with
four spin components and suppressed flavor indices.  We recover
the conventional domain wall fermion action when we set
$\Omega = 1$.

If we let $\Omega$ transform as
\begin{equation}
  \Omega \to V_R \, \Omega \, V_L^\dagger
\end{equation}
under SU$(N)_{\rm L} \times SU$(N)$_{\rm R}$, then the domain wall
fermion Dirac operator possesses exact chiral symmetry.  Thus we
can use $\Omega$ to track the explicit chiral symmetry breaking
from domain wall fermions, in both the Symanzik style effective
Lagrangian for domain wall fermion QCD and in Green functions.
However, we note that $\Omega$ itself is not a small quantity, since
a general Green function involving the five-dimensional fermion
fields does not have any approximate chiral symmetry.  However, for
the effective action and low momentum limit of Green functions
made from the four-dimensional fields at the boundaries of the fifth
dimension, each power of $\Omega$ that enters should come with a
suppression related to the ratio of amplitudes of the low-energy
fermion modes between the boundaries and the midpoint in the fifth
dimension.

Consider a non-zero $a$ effective Lagrangian description of QCD with
domain wall fermions at finite $L_s$.  The presence of the parameter
$\Omega$ implies that the terms containing fermions,
up to operators of dimension five, are
\begin{equation}
  \frac{Z_m m_f}{a} \overline{\psi}\psi +
 \frac{c_3}{a} \left\{ \overline{\psi} \Omega^\dagger P_R \psi
  + \overline{\psi} \Omega P_L \psi \right\}
  + ac_5 \left\{ \overline{\psi} \sigma_{\mu \nu}
   F_{\mu \nu} \Omega^\dagger P_R \psi
  + \overline{\psi} \sigma_{\mu \nu} F_{\mu \nu} \Omega P_L \psi
    \right\}\, .
\label{eq:symanzik_fermion_omega}
\end{equation}
Here $m_f$ is the dimensionless input bare quark mass in the domain
wall fermion formulation, and $c_3$ and $c_5$ are dimensionless
parameters that represent the mixing of left- and right-handed
quarks between the five-dimensional boundaries.  These parameters
are of $O(e^{-\alpha L_s})$ at weak coupling;  for coarse lattices
where there are localized dislocations in the gauge fields corresponding
to changes in the topology, they are generically $O (a_1 e^{-\alpha
L_s} + a_2)/L_s$, where $a_2 \ne 0$ is due to the density of localized
topological dislocations \cite{Antonio:2007tr}.

Setting $\Omega = 1$, we have
\begin{equation}
  \frac{Z_m m_f}{a} \overline{\psi}\psi +
 \frac{c_3}{a} \overline{\psi} \psi
  + ac_5 \overline{\psi} \sigma_{\mu \nu}
   F_{\mu \nu} \psi \, .
\label{eq:symanzik_fermion}
\end{equation}
The combination $Z_m m_f + c_3$ is the total (dimensionless) quark mass
and we choose $L_s$ to control the contribution of the second term,
by changing the size of $c_3$.  Eq.~(\ref{eq:symanzik_fermion}) is
identical to the result for Wilson fermions, except that the
coefficients $c_3$ and $c_5$ are expected to be small, $O(10^{-3})$,
for realistic domain wall fermion simulations, compared to being
$O(1)$ as for Wilson fermions.

We can now discuss the application of chiral perturbation theory
to our domain wall fermion simulations at a fixed lattice spacing.
Our discussion will be for SU(3), but the results are easily
generalized.  We start from the conventional QCD SU(3) chiral
Lagrangian in the continuum and make use of the presence of the
$\Omega$ spurion field to add all additional terms to it.  Initially
we power count only in $\Omega$ and defer, for the moment, the
additional question of power counting in $a$.  We choose $\Sig(x)
= {\e^{2i\phi(x)/\fthree}}$, where $\Sig$ transforms as: $ \Sig \to
V_L \Sig V_R^{\dagger}, V_L, V_R \in$ SU(3).  We define $\widehat{\chi}
= 2 \bthree \; \mbox{diag}(m_u, m_d, m_s)$, where $\bthree$ is one
of the low energy constants (LECs) that enters in chiral perturbation
theory.  To $O(p^4)$ the continuum Lagrangian is:
\ba
{\cal L} & = &
  \frac{\fthree^2}{8} \mbox{Tr}\lbrack \partial_{\mu} \Sig
    \partial^{\mu} \Sigd
  \rbrack+\frac{\fthree^2}{8} \mbox{Tr}\lbrack  \widehat{\chi}\Sig +
    (\widehat{\chi}\Sig)^\dagger \rbrack \nonumber \\
 & + & 
\lthree{1} \left\{\mbox{Tr}[\partial_{\mu}\Sig (\partial^{\mu}\Sig)^{\dagger}]
\right\}^2 
 + \lthree{2} \mbox{Tr} \left [\partial_{\mu} \Sig (\partial_{\nu}\Sig)^{\dagger}\right]
\mbox{Tr} \left [\partial^{\mu}\Sig (\partial^{\nu}\Sig)^{\dagger}\right]\nonumber\\
& + & \lthree{3} \mbox{Tr}\left[
\partial_{\mu}\Sig (\partial^{\mu}\Sig)^{\dagger}\partial_{\nu}\Sig (\partial^{\nu}\Sig)^{\dagger}
\right ]
 + \lthree{4} \mbox{Tr} \left [ \partial_{\mu}\Sig (\partial^{\mu}\Sig)^{\dagger} \right ] 
\mbox{Tr} \left( \widehat{\chi} \Sig+  (\widehat{\chi}\Sig)^{\dagger} \right )
\nonumber \\
& + & \lthree{5} \mbox{Tr} \left [\partial_{\mu}\Sig (\partial^{\mu}\Sig)^{\dagger}
(\widehat{\chi} \Sig+ (\widehat{\chi}\Sig)^{\dagger})\right]
+ \lthree{6} \left[ \mbox{Tr} \left ( \widehat{\chi} \Sig+ (
  \widehat{\chi}\Sig)^{\dagger} \right ) \right]^2
\nonumber \\
& + &  \lthree{7} \left[ \mbox{Tr} \left ( \widehat{\chi} \Sig - (
\widehat{\chi}\Sig)^{\dagger} \right )
\right]^2
+ \lthree{8} \mbox{Tr} \left ( \widehat{\chi}\Sig \widehat{\chi}\Sig
+ (\widehat{\chi} \Sig\widehat{\chi}\Sig)^{\dagger} \right )\label{cont_chpt_l} \, .
\label{eq:chpt_nlo}
\ea
For the domain wall fermion case, we can generate the new terms
that arise by starting from the Lagrangian in Eq.~(\ref{cont_chpt_l}).
Since $\Omega$ transforms as $\widehat{\chi}$ or $\Sigma^\dagger$
does, new terms can be created by substituting $\Omega$ for
$\widehat{\chi}$ and $\Sigma^\dagger$ in Eq.~(\ref{cont_chpt_l}),
remembering that derivatives acting on $\Omega$ vanish.  Since the
dominant contribution of explicit chiral symmetry breaking in
Eq.~(\ref{eq:symanzik_fermion}) is an additive contribution to the
quark mass, we power count $\Omega$ as $O(p^2)$.  Keeping terms of
$O(p^4)$ and using $D_i$, $\tilde{D}_i$ for the LEC's for these
terms, we have additional contributions to Eq.~(\ref{cont_chpt_l})
of
\ba
& & \frac{f^2}{8} D_0 \mbox{Tr} \lbrack \Omega \Sigma +
  (\Omega\Sigma)^\dagger \rbrack + 
  \frac{f^2}{8} \tilde{D}_0 \mbox{Tr} \lbrack \Omega \widehat{\chi}^\dagger +
  \Omega^\dagger \widehat{\chi} \rbrack  \nonumber \\
& &  D_4 \mbox{Tr} \left [ \partial_{\mu}\Sig (\partial^{\mu}\Sig)^{\dagger} \right ]
\mbox{Tr} \left( \Omega \Sig+ (\Omega\Sig)^{\dagger} \right )
+D_5 \mbox{Tr} \left[ \partial_{\mu}\Sig (\partial^{\mu}\Sig)^{\dagger}
(\Omega \Sig+ (\Omega\Sig)^{\dagger})\right]\nonumber \\
& & 
+ D_6  \left[ \mbox{Tr} \left ( \Omega \Sig + ( \Omega \Sig )^{\dagger} \right )
\right]  \left[ \mbox{Tr} \left ( \widehat{\chi} \Sig + (\widehat{\chi}\Sig)^{\dagger} \right )
\right] 
 + \tilde{D}_6 \left[ \mbox{Tr} \left ( \Omega \Sig+ (\Omega \Sig)^{\dagger} \right )
\right]^2 \nonumber \\
& & + D_7  \left[ \mbox{Tr} \left ( \Omega \Sig - (\Omega\Sig)^{\dagger} \right )
\right] \left[ \mbox{Tr} \left ( \widehat{\chi} \Sig - (\widehat{\chi}\Sig)^{\dagger} \right )
\right]  + \tilde{D}_7 \left[ \mbox{Tr} \left ( \Omega \Sig- (\Omega\Sig)^{\dagger} \right )
\right]^2 \nonumber \\
 & &+ D_8  \mbox{Tr} \left ( \widehat{\chi} \Sig \Omega\Sig
+ (\widehat{\chi} \Sig \Omega\Sig)^{\dagger} \right )
+ \tilde{D}_8\mbox{Tr} \left ( \Omega\Sig\Omega\Sig
+ (\Omega \Sig\Omega \Sig)^{\dagger} \right ) \, .
\label{eq:chpt_dwf_nlo}
\ea
Terms which involve two derivatives, a factor of $\widehat{\chi}$
and a factor of $\Omega$ will be $O(p^6)$ and have been neglected.
Note that we have kept the term involving $\tilde{D}_0$, even though
it does not involve any $\Sigma$ fields.  Such a term does play a
role in determining the value for the chiral condensate through the
variation of the partition function with respect to the quark mass.

We would seem to have many new low energy constants to determine with
domain wall fermions at non-zero $a$.  However, the form of the Symanzik
effective Lagrangian for DWF QCD shows that the leading order (in $a$)
chiral symmetry breaking effect is a universal shift in the quark mass,
{\em i.e.} $c_3/a$ multiplies the dimension three operator
$\overline{\psi}\psi$.  Thus, there is no
difference in ChPT between $m_f$ and $c_3$, and we
can rewrite the terms in Eq.~(\ref{eq:chpt_dwf_nlo}) in terms of the
original LEC's of QCD, with a shifted quark mass, plus higher order
corrections.  Letting
$\chi = 2\bthree [\mbox{diag}(m_u, m_d, m_s) + c_3
\mbox{diag}(1,1,1)]$ and looking at the $D_4$ term as an example,
we have
\ba
 D_4 \mbox{Tr} \left [ \partial_{\mu}\Sig (\partial^{\mu}\Sig)^{\dagger} \right ]
\mbox{Tr} \left( \Omega \Sig+ (\Omega\Sig)^{\dagger} \right )
& = &
 \lthree{4} \mbox{Tr} \left [ \partial_{\mu}\Sig (\partial^{\mu}\Sig)^{\dagger} \right ]
\mbox{Tr} \left( \chi \Sig +
  (\chi\Sig)^{\dagger} \right ) \nonumber \\
  & + &  
 O(a c_5) \lthree{4} \mbox{Tr} \left [ \partial_{\mu}\Sig (\partial^{\mu}\Sig)^{\dagger} \right ]
\mbox{Tr} \left( \Sig + \Sig^{\dagger} \right ) \, .
  \label{eq:d4_nlo}
\ea
The last term is $ O(a) \times O(c_5) \times
O(p^2)$.  It is customary to power count $O(a)$ and $O(p^2)$ terms
as the same size for unimproved Wilson fermions, and this same term
appears there \cite{Bar:2003mh}, except that
$c_5^{\rm Wilson}$ is $O(1)$.  While for Wilson fermions, this term must
be kept at NLO in the chiral Lagrangian, for domain wall fermions,
where $c_5$ is very small, it can be neglected.

Examining all the terms in Eq.~(\ref{eq:chpt_dwf_nlo}), we see that the
complete NLO chiral Lagrangian for domain wall fermions is given
by Eq.~(\ref{eq:chpt_nlo}), with $\widehat{\chi} \to \chi$, where
$\chi$ is proportional to the sum of the input bare quark mass, $m_f$,
and the additive quark mass contribution which comes from $c_3$.
Since we will be working to NLO order of ChPT in this work, domain
wall fermions at non-zero lattice spacing should be described by
the chiral Lagrangian given in Eq.~(\ref{eq:chpt_nlo}).  When we
fit our data for specific quantities to the chiral formula following
from Eq.~(\ref{eq:chpt_nlo}), the LEC's $L_i$ will differ at
$O(a^2)$ from their continuum values.  Since we work at a single
lattice spacing, we will not be able to correct for these deviations
from continuum QCD.

The size of the residual symmetry breaking terms represented by
$c_3$ and $c_5$ in Eq.~(\ref{eq:symanzik_fermion_omega}) is most
easily studied by examining the five-dimentional current ${\cal
A}_\mu^a$ which can be easily defined for domain wall fermions and
is exactly conserved in the limit $m_f \rightarrow 0$ and $L_s
\rightarrow \infty$ \cite{Furman:1995ky}.  The DWF equations of
motion imply that this current obeys the divergence condition
\begin{equation}
  \Delta_\mu {\cal A}^a_\mu(x) = 2m_f J^a_5(x) + 2 J^a_{5q}(x) \,  ,
  \label{eq:axial_cc_diverg}
\end{equation}
where $J^a_5$ is a pseudoscalar density made up of quark fields on
the boundary of the fifth dimension and $J^a_{5q}$ is a pseudoscalar
density containing quark fields at $L_s/2 -1$ and $L_s/2$.  While the
$J^a_5$ term is the result of the usual chiral non-invariance of the
mass term, the $J^a_{5q}$ operator is expected to have vanishing 
matrix elements at low energy as $L_s \rightarrow \infty$.  It 
represents the effects of residual, finite--$L_s$, chiral symmetry
breaking.  For low energy Green functions, the midpoint term in
Eq.~(\ref{eq:axial_cc_diverg}) can be expanded as
\begin{equation}
  J^a_{5q} \sim \mres J^a_5 -
  \frac{(Z_{\cal A} - 1 )}{2} \Delta_\mu {\cal A}^a_\mu 
       + c^\prime_5 O_5^a.
  \label{eq:midpt_exp}
\end{equation}
Here we have introduced the $m_f$-independent parameter $\mres$,
related to the constant $c_3$ in Eq.~(\ref{eq:symanzik_fermion_omega}).
We also have a new lattice operator, $O_5^a$, similar to the axial
transform of the $c_5$ term in Eq.~(\ref{eq:symanzik_fermion_omega}),
which is carefully subtracted so that its matrix elements are of
order $a^2$ smaller than those of the operator $J^a_5$ at long
distances.  Since the low energy matrix elements of the midpoint
operator on the left-hand side of Eq.~(\ref{eq:midpt_exp}) will be
suppressed by a factor $\exp(-\alpha L_s)$ we expect the quantities
$\mres$, $(Z_{\cal A} - 1 )$ and $c^\prime_5$ to all be of this
order (at least in the perturbative regime).

While an expansion such as that written in Eq.~(\ref{eq:midpt_exp})
is typically written in terms of the operators of the effective
theory, the present form of this equation is useful because it 
can be combined with Eq.~(\ref{eq:axial_cc_diverg}) to yield:
\begin{equation}
  Z_{\cal A} \Delta_\mu {\cal A}^a_\mu(x) = 2 (m_f + \mres) J^a_5(x) 
                    + c_5^\prime O_5^a\, .
  \label{eq:axial_cc_renorm_div} 
\end{equation}
As is well known from the classic analysis of the case of Wilson
fermions~\cite{Bochicchio:1985xa}, the conservation of the vector
current and the vector Ward identities imply that product $m_f
J^a_5(x)$ in Eq.~(\ref{eq:axial_cc_renorm_div}) approaches its
continuum counterpart without multiplicative renormalization.  This
equation then implies that the product $Z_{\cal A} {\cal A}^a_\mu(x)$
will reproduce the standard continuum axial current when evaluated
in low energy Greens functions.

The appearance of the $(Z_{\cal A} - 1 )$ term on the right-hand
side of Eq.~(\ref{eq:midpt_exp}) and the consequent renormalization
of the five-dimensional DWF axial current is an effect that was
not recognized in earlier RBC or RBC-UKQCD work.  Such a possibility
was raised by Steve Sharpe~\cite{Sharpe:2007yd}.  However,
in this paper, Sharpe argues that $(Z_{\cal A} - 1 )$ is expected
to be of order $\mres^2$, not the order $\mres$ scale suggested by
the above argument.  Sharpe's argument relies on the chiral character
of the operators $V_\mu \pm A_\mu$, defined in terms of fields on 
the boundaries of the fifth dimension.  The mixing
between these operators implied by $Z_{\cal A} \ne 1$ can arise only
if a fermion and an antifermion move from one wall from the other,
a situation suppressed as $\mres^2$.  We believe that this generally
very useful argument does not apply in the present case because the
right-- and left--handed currents under consideration each span
one--half of the entire five dimensions and are not localized on
the left and right walls.  For example, two fermions can propagate
from the midpoint operator $J_{5q}$ to the same wall with only a
suppression of $[\exp(-\alpha L_s/2)]^2 = \exp(-\alpha L_s)$.  Note,
this argument only applies to the perturbative piece.  For tunneling
caused by near-zero modes of the four-dimension Wilson operator,
two such modes are needed implying a suppression more like $\mres^2$.

Finally it is informative to consider the implications of this
more complete analysis on the conventional calculation of the
residual mass from a ratio of low energy pion to vacuum matrix
elements.  We measure a quantitiy commonly called $\mres^\prime(m_f)$
and given by
\begin{eqnarray}
  \mres^\prime(m_f) &=& \frac{ \langle 0 | J^a_{5q} | \pi \rangle}
          {\langle 0 | J^a_5 | \pi \rangle}  \label{eq:mres_prime1} \\
   &=& \frac{ \langle 0 |\Bigl(\mres J^a_5 
           - \frac{(Z_{\cal A} - 1 )}{2} \Delta_\mu {\cal A}^a_\mu 
           + c^\prime_5 O_5^a\Bigr)| \pi \rangle}{\langle 0 | J^a_5 | \pi \rangle}
 \label{eq:mres_prime2}
\end{eqnarray}
We can then use the lattice equations of motion to evaluate
the $ \Delta_\mu {\cal A}^a_\mu$ term so that Eq.~(\ref{eq:mres_prime2})
takes the form:
\begin{equation}
\mres^\prime(m_f)
  = \mres
  + (\frac{1}{Z_{\cal A}} - 1 )(m_f+\mres) 
  + \frac{c^\prime_5}{2} \left( 1 + \frac{1}{Z_{\cal A}} \right)
    \frac{\langle 0 | O_5^a| \pi \rangle}{\langle 0 | J^a_5 | \pi \rangle}
 \label{eq:mres_prime3}
\end{equation}
This equation governs the $m_f$ dependence of $\mres^\prime(m_f)$.
Since, by construction, the third term on the right-hand side is
of order $a^2$, we might be tempted to neglect it and use
Eq.~(\ref{eq:mres_prime3}) to relate the $m_f$ dependence of
$\mres^\prime(m_f)$ to the difference $(1/Z_{\cal A} - 1)$.  However,
we can distinguish the dependence of $\mres^\prime(m_f)$ on the
valence and sea masses, $m_f^{\rm val}$ and $m_f^{\rm sea}$ ($m_x$
and $m_l$ in the general notation of the rest of this paper).  We
see that there is valence quark mass dependence in both the $1/Z_{\cal
A} - 1$ term and in the term containing $c_5^\prime$, while the
sea quark mass dependence is only in the $c_5^\prime$ term.  As
will be shown in Section \ref{sec:latticeps_mres}, $\mres^\prime$
depends to a roughly equal degree on both $m_f^{\rm val}$ and
$m_f^{\rm sea}$.  In particular, we find that
$d\ln(\mres^\prime)/dm_f^{\rm val} \approx -4$ and
$d\ln(\mres^\prime)/dm_f^{\rm sea} \approx 6$.
This implies that the third term in
Eq.~(\ref{eq:mres_prime3}), which can depend on both
the valence and sea quark masses, must be of an equal size to the
second.
In fact, if we introduce physical dimensions, we expect
the ratio
$\langle 0 | O_5^a| \pi \rangle / \langle 0 | J^a_5 | \pi \rangle$
to be of $O((a\Lambda_{\rm QCD})^2)$.  The effect of the
quark masses on this ratio will be to replace one of the factors
of $\Lambda_{\rm QCD}$ by the dimensioned factor $m_f^{\rm val}/a$
or $m_f^{\rm sea}/a$.  Thus, the quark mass dependent terms coming
from the third term in Eq.~(\ref{eq:mres_prime3}) are suppressed
by only by $O(a\Lambda_{\rm QCD}$) relative to the second term,
which is apparently insufficient to permit them to be neglected,
a fact alreadly pointed out by Sharpe\cite{Sharpe:2007yd}.

The arguments above show that we expect $Z_{\cal A} - 1$ to be
$O(\mres)$.  If all of the unitary mass dependence of $\mres^\prime(m_f)$
came from the $1/Z_{\cal A} - 1$ term, this would give $Z_{\cal A}
- 1 = -0.003$.  Similarly, the valence mass dependence would give
$Z_{\cal A} - 1 = 0.01$.  At present, we do not have sufficient
data to measure $Z_{\cal A} - 1.$ without contamination from the
$c_5^\prime$ term.  Thus, for the remainder of this work, we take
$Z_{\cal A} = 1$ and we expect this to introduce an error of 1\%
or less in our axial current normalization.

We also work with a local, four-dimensional axial current $A^a_\mu$,
which we renormalize so that $Z_A A^a_\mu = {\cal A}^a_\mu$, so that
\begin{equation}
  Z_A \Delta_\mu A^a_\mu(x) = 2\tilde{m} J^a_5(x) \, .
  \label{eq:axial_lc_diverg}
\end{equation}


\subsection{$\su2L\times\su2R$ Chiral Perturbation Theory for
Kaon Physics}
\label{subsect:kchpt}

\subsubsection{Introduction}

An important goal of this paper is the comparison of the mass
dependence of our lattice results with both $\su3L\times \su3R$
and $\su2L\times \su2R$ chiral perturbation theory (ChPT) at NLO
({\em i.e.} at one-loop order). For compactness of notation, throughout the
remainder of this section we will simply refer to $\SU(3)$ and
$\SU(2)$ ChPT. We will see that our data for pseudoscalar masses and
decay constants only agrees with NLO $\SU(3)$ ChPT with good precision
when the quark masses are small. This will be quantified in detail in
Sec.~\ref{sec:Su3}, where we will see that NLO $\SU(3)$ ChPT provides a
good description of our data when the average, dimensionless,
input valence mass
satisfies $m_\textrm{avg}<0.01$ (all our results will be obtained
using data for valence $u$ and $d$ quark masses below this value). In
order to attempt to extend the range of agreement to heavier masses
(to include the strange quark for example), one possible approach is
to use NNLO (or even higher order) ChPT. This introduces many
additional low energy constants (LECs) and we find that we have
insufficient data to determine all of these with sufficient precision.

Instead of attempting to fit our data using NNLO (or even higher
order) ChPT, we propose to use $\SU(2)$ ChPT at NLO. For pion physics,
where it is possible to satisfy the condition, $m_{\textrm{avg}}<0.01$
for the valence quarks, both $\SU(2)$ and $\SU(3)$ ChPT provide a good
description of our data. Moreover, we will see in Secs.~\ref{sec:Su2}
and \ref{sec:Su3} that the $\SU(2)$ LECs obtained directly are
consistent with those obtained by ``converting'' the $\SU(3)$ LECs to
$\SU(2)$~\cite{Gasser:1984gg,Gasser:2007sg} (conversion formulae
appear in Appendix~\ref{sec:app:chPTconv}). For kaons on the other
hand, the presence of the valence strange quark means that we do not
satisfy the condition $m_{\textrm{avg}}<0.01$ and so we propose to use
$\SU(2)$ ChPT at NLO. The effects of the strange quark mass are now
absorbed into the LECs of the $\SU(2)$ ChPT. This eliminates the
errors due to neglected higher powers of
$\chi_s/\Lambda_\mathrm{CSB}^2$ present when using $\SU(3)$ ChPT, but
now one has to ensure that the $u$ and $d$ quark masses are
sufficiently small to be able to neglect higher order terms in
$\chi_{ud}/\chi_s$\,. Here $\Lambda_\mathrm{CSB} \sim 4 \pi f_\pi$ is
the chiral symmetry breaking scale, while $\chi_{ud}$ and $\chi_s$ are
tree level masses of pseudoscalars made of $ud$ and $s$ quarks
(see Appendix~\ref{sec:appendix:notation}).

In this section we derive the NLO formulae for the behavior of the
kaon's leptonic decay constant $f_K$, its mass $m_K$, and the bag
parameter $B_K$ as a function of the light-quark masses in $\SU(2)$
ChPT. At each stage, we start by presenting the arguments and
calculations in the unitary theory, with valence and sea quark masses
equal, and then proceed to the partially quenched theory. In the
partially quenched theory in general the valence and sea strange quark
masses, $m_y$ and $m_h$ respectively, are also different and the
$\SU(2)$ LECs depend on both these masses. For compactness of
notation, we write these LECs with a single argument $m_h$, but it
should be remembered that if $m_y\neq m_h$, then $m_h$ should be
replaced by the pair of variables $m_y,m_h$. We note that in this
subsection quark masses written without a tilde have their usual
interpretation as lagrangian mass parameters in continuum ChPT or
PQChPT (a change from the notation defined in
Appendix~\ref{sec:appendix:notation} and used elsewhere in this
paper).

We start by introducing our notation. In the unitary effective theory,
the pion matrix, quark-mass matrix and the kaon fields are written in
the form:
\begin{equation}
\phi=\begin{pmatrix}
\pi^0/\sqrt{2}&\pi^+\\\pi^-&-\pi^0/\sqrt{2}\end{pmatrix}\,,\quad
M=\begin{pmatrix}m_l&0\\0&m_l\end{pmatrix}\,,\quad
K=\begin{pmatrix}K^+\\K^0\end{pmatrix}\,.
\end{equation}
We work in the isospin limit in which the two light quarks are
degenerate with mass $m_l$. The pion matrices $\xi$ and $\Sigma$ are
defined in the standard way:
\begin{equation}
\xi=\exp(i\phi/f)\qquad\textrm{and}\qquad \Sigma=\xi^2\,,
\end{equation}
where the constant $f$ is the pion decay constant ($f_\pi$) at lowest
order in the chiral expansion (we use the convention in which
$f_\pi=132\,\mev$). Under global left- and right-handed
transformations, $L$ and $R$ respectively, these quantities transform
as follows:
\begin{equation}
\Sigma \to L \Sigma R^\dagger\,,\qquad \xi \to L \xi U^\dagger = U
\xi R^\dagger\qquad\textrm{and}\qquad K \to U K\,,
\end{equation}
where $U$ is a function of $L$, $R$ and the meson fields, but
reduces to a global vector transformation when $L=R$. The pion
Lagrangian at lowest order is
\begin{equation}
L_{\pi\pi}^{(2)} =
 \frac{f^2}8 \tr \partial_\mu\Sigma\partial^\mu\Sigma^\dagger
 + \frac{f^2 B}4 \tr(M^\dagger\Sigma + M \Sigma^\dagger)
\end{equation}
where $f$ and $B$ are the usual leading order LECs and, to this order,
$m_\pi^2 = 2 B m_l$.

The results for $m_\pi^2$ and $f_\pi$ at NLO in the chiral expansion
are well known~\cite{Gasser:1983yg} and are presented in
Eqs.~(\ref{eq:chPTsu2:mPi}) and~(\ref{eq:chPTsu2:fPi}) in the
appendix. We now turn our attention to kaon physics using ``Kaon ChPT"
(KChPT). The corresponding chiral Lagrangian has already been
introduced by Roessl~\cite{Roessl:1999iu} in order to study $\pi K$
scattering close to threshold. At leading order the interaction of
kaons with soft pions is described by the Lagrangian
\begin{equation}
\label{eq:lkpi1}
L_{\pi K}^{(1)} = D_\mu K^\dagger D^\mu K - M^2 K^\dagger K\,,
\end{equation}
where the covariant derivative $D_\mu$ is constructed using the vector
field $V_\mu$ so that $D_\mu K$ transforms like $K$ under chiral
transformations:
\begin{equation}
D_\mu K = \partial_\mu K + V_\mu K \to U D_\mu K\,.
\end{equation}
$V_\mu$ itself is constructed from the pion fields and transforms as
\begin{equation}\label{eq:Vdef}
V_\mu= \frac12 \left(\xi^\dagger\partial_\mu \xi +
                    \xi\partial_\mu\xi^\dagger
               \right)
      \to U V_\mu U^\dagger + U \partial_\mu U^\dagger\,.
\end{equation}
We refer the reader to Eq.~(11) of \cite{Roessl:1999iu} for the higher
order terms in the $K\pi$ Lagrangian (terms up to $O(m_l^4)$ are
explicitly listed). For completeness we also present the axial field:
\begin{equation}
\label{eq:Adef}
A_\mu = \frac i2 \left( \xi^\dagger\partial_\mu \xi -
                        \xi\partial_\mu\xi^\dagger
                 \right)
      \to U A_\mu U^\dagger\,.
\end{equation}
In the following, when constructing Feynman diagrams from the $K\pi$
Lagrangian and the effective-theory local operators, we expand the
vector and axial fields in terms of pion fields:
\begin{equation}
V_\mu = [\phi,\partial_\mu \phi]/2f^2 + \cdots\qquad\textrm{and}\qquad
A_\mu = -\partial_\mu \phi/f + \cdots.
\end{equation}

The use of Partially Quenched Chiral Perturbation Theory (PQChPT), in
which we vary the valence quark masses independently of those of the
sea quarks, gives us more scope to determine the low-energy constants
of the unitary chiral theory and in this paper we will profit from
this opportunity. We consider the case with two degenerate light sea
quarks, two degenerate light valence quarks and two `ghost' degenerate
light commuting quarks to cancel the loop effects of the valence
quarks. Thus we are using the graded symmetry
method~\cite{Bernard:1992mk,Bernard:1993sv}, based on Morel's
ghost-quark trick~\cite{Morel:1987xk}. See the lectures by
Sharpe~\cite{Sharpe:2006pu} for a recent review of PQChPT and
references to the original literature (as well as many other
applications of chiral perturbation theory to lattice QCD). We let the
sea quarks have mass $m_l$ while the valence and ghost quarks have
mass $m_x$. Thus we build an $\SU(4|2)_L \times \SU(4|2)_R$ effective
theory. We identify the quark flavors using the ordered list
\begin{equation}
u_v, d_v, u_s, d_s, \tilde u_v, \tilde d_v
\end{equation}
where subscripts $v$ and $s$ are for valence and sea respectively,
while the tilde denotes a ghost quark. We remind the reader that the
strange quark mass does not appear explicitly here; rather, the LECs
are functions of the strange quark mass.

The matrix of Goldstone fields is written in block form
\begin{equation}\label{eq:phipq}
\Phi=\begin{pmatrix}\phi & \eta_1\\\eta_2 &
\tilde\phi\end{pmatrix}
\end{equation}
where $\phi$ is a $4\times4$ block containing normal mesons,
$\eta_1=\eta_2^\dagger$ is a $4\times2$ block of normal-ghost mesons
and $\tilde\phi$ is a $2\times2$ block of ghost-ghost mesons. The QCD
pions are present in the central $2\times2$ block of $\Phi$. As
before, we work with the quantities $\xi$ and $\Sigma$
\begin{equation}
\xi = \exp(i\Phi/f) \qquad\mbox{and}\qquad \Sigma=\xi^2\,,
\end{equation}
in terms of which the leading order PQ chiral Lagrangian is
\begin{equation}
\mathcal L_\mathrm{PQ}^{(2)} =
  \frac{f^2}8\, \str (D_\mu\Sigma)^\dagger D^\mu\Sigma +
  \frac{f^2B}4\,
     \str (\mathcal M \Sigma^\dagger + \Sigma\mathcal{M}^\dagger)\,,
\end{equation}
where $\mathcal M$ is the mass matrix
\begin{equation}
\mathcal M =
 \mathrm{diag}(M_v,M_s,M_v),
\quad \mbox{with}\quad
M_v=\mathrm{diag}(m_x,m_x)
\quad\textrm{and}\quad
M_s=\mathrm{diag}(m_l,m_l).
\end{equation}
The quadratic terms in $\mathcal L_\mathrm{PQ}^{(2)}$ are
\begin{equation}\label{eq:Lsu2quadpq}
\begin{aligned}
\mathcal L_\mathrm{PQ, quad}^{(2)} &= \frac12
\str(\partial_\mu\Phi\partial^\mu\Phi) -
 B \str (\mathcal M\Phi^2)\\
 &= \frac12 \tr(\partial_\mu\phi\partial^\mu\phi
               +\partial_\mu\eta_1\partial^\mu\eta_2
               -\partial_\mu\eta_2\partial^\mu\eta_1
               -\partial_\mu\tilde\phi\partial^\mu\tilde\phi)\\
 &{}\quad -B\tr\left(
  (\phi\phi+\eta_1\eta_2)\begin{pmatrix}M_v&0\\0&M_s\end{pmatrix}
   - \eta_2 \eta_1 M_s -\tilde\phi\tilde\phi M_s\right)\,.
\end{aligned}
\end{equation}
The Lagrangian in Eq.~(\ref{eq:Lsu2quadpq}) leads to a number of
propagators which appear in the Feynman diagrams in the following
sections. We distinguish the propagators for the following mesons:
\begin{itemize}
\item normal `charged' (off-diagonal) mesons ($\sim q_1 \bar q_2$):
\begin{equation}
\frac i{p^2-m_{12}^2},\qquad
 m_{12}^2 = B(m_1+m_2) = (\chi_1+\chi_2)/2\,,
\end{equation}
where $m_{1,2}$ are valence or sea masses;
\item `charged' ghost mesons:
($\sim \tilde q_1 \bar{\tilde q}_2$)
\begin{equation}
\frac{-i}{p^2-m_{12}^2},\qquad
 m_{12}^2 = B(m_1+m_2) = (\chi_1+\chi_2)/2\,.
\end{equation}
Here $m_{1,2}$ are valence masses;
\item `charged' quark-ghost mesons
($\sim q_1 \bar{\tilde q}_2$ or $\sim \tilde q_1 \bar q_2$)
\begin{equation}
\frac i{p^2-m_{12}^2},\qquad
 m_{12}^2 = B(m_1+m_2) = (\chi_1+\chi_2)/2\,,
\end{equation}
where $m_{1,2}$ are a valence mass from the ghost and a valence or sea
mass from the quark;
\item `neutral' mesons from diagonal parts of
$\phi$ and $\tilde\phi$:
\begin{equation}
\langle \Phi_{ii} \Phi_{jj} \rangle =
 i\,\frac{\epsilon_i \delta_{ij}}{p^2-\chi_i}
 - \frac iN \frac{p^2-\chi_l}{(p^2-\chi_i)(p^2-\chi_j)}\,,
\end{equation}
for $N$ sea quarks which are all degenerate
with mass $m_l$, with $N=2$ here and 
\begin{equation}
\epsilon_i = \begin{cases}+1 &\mbox{valence or sea}\\
 -1 &\mbox{ghost}\end{cases}\,.
\end{equation}
Note that the neutral propagators are determined after
implementing the constraint $\str\,[\Phi]=0$, which has the effect
that neutral valence-valence meson propagators can have
contributions from neutral sea quark states
(see~\cite{Sharpe:2006pu} for example).
\end{itemize}
We now add a kaon matter field
\begin{equation}
K = \begin{pmatrix}K_v^+\\K_v^0\\K^+\\K^0\\\tilde K^+\\\tilde K^0
    \end{pmatrix}
  \sim \begin{pmatrix}\bar s u_v\\\bar s d_v\\\bar s u_s\\
       \bar s d_s\\\bar s \tilde u_v\\\bar s \tilde d_v\end{pmatrix}\,.
\end{equation}
The partially quenched $K\pi$ Lagrangian is the generalization of
Eq.~(\ref{eq:lkpi1}) written in terms of the partially quenched fields.

In the following three subsections we derive the mass dependence of
$f_K$, $m_K^2$ and $B_K$ respectively.

\subsubsection{Chiral Behavior of $f_K$}

The kaon decay constant, $f_K$, is defined by the matrix element of
the axial vector current:
\begin{equation}\label{eq:fkdef}
\langle 0 | \bar u \gamma_\mu \gamma_5 s | K^-(p)\rangle
 \equiv -i f_K p_\mu\,.
\end{equation}
We now need to match the QCD axial vector current to operators in
KChPT. We start by doing this for left and right handed currents and
then build the axial vector current. The left-handed QCD current for
kaon decay is
\begin{equation}\label{eq:lhcurrent}
\frac12\,\bar q \gamma_\mu(1-\gamma_5) s
\end{equation}
where $q=u$ or $d$. It is convenient to promote $q$ to be a
$2$-component vector with components $u$ and $d$ and to introduce a
constant $2$-component spurion vector $h$ in order to be able to
project $u$ or $d$ as required; specifically we write the left-handed
current as:
\begin{equation}\label{eq:lhcurrenth}
\frac12\bar q h \gamma_\mu(1-\gamma_5) s
\end{equation}
The current in Eq.~(\ref{eq:lhcurrenth}) would be invariant under
$\mathrm{SU}(2)_L$ transformations if $h$ transformed as $h \to L
h$.

We match to the effective theory by building quantities linear in
$h$ and with a single Lorentz index, which would also be invariant
if $h$ transformed as above. At lowest order, we identify two
terms:
\begin{equation}\label{eq:lheffective}
(D_\mu K)^\dagger \xi^\dagger h, \qquad\mbox{and}\qquad K^\dagger
A_\mu \xi^\dagger h.
\end{equation}
By using equations of motion, operators with more covariant
derivatives acting on the kaon field (which transform in the same way
under chiral transformations) can be reduced to ones of higher order
in the chiral expansion.

For a right-handed current we simply take the $h$-transformation
to be $h \to R h$ and obtain the two operators
\begin{equation}\label{eq:rheffective}
(D_\mu K)^\dagger \xi h, \qquad\mbox{and}\qquad K^\dagger A_\mu
\xi h\,,
\end{equation}
which transform in the same way as
$\bar{q}h\gamma_\mu(1+\gamma_5)s$\,.

Under a parity transformation, the quantities in the effective
theory transform as:
\begin{equation}
K \to -K,\qquad \xi \to \xi^\dagger,\qquad A_\mu \to -A_\mu\,.
\end{equation}
Noting that parity transforms the left-handed current into the
right-handed current and vice-versa, we deduce that the currents
are of the form:
\begin{align}
\label{eq:kchpt-lh-current} J^L_\mu &= -L_{A1}\, (D_\mu K)^\dagger
\xi^\dagger h
            + i L_{A2}\, K^\dagger A_\mu \xi^\dagger h \\
J^R_\mu &= L_{A1}\, (D_\mu K)^\dagger \xi h
            + i L_{A2}\, K^\dagger A_\mu \xi h\,,
\end{align}
where $L_{A1}$ and $L_{A2}$ are low-energy constants. To the extent
that the kaon is regarded as heavy, the $L_{A2}$ term is subleading
(and in any case does not contribute to $f_K$ at tree or one-loop
level). The axial vector current $J^R_\mu - J^L_\mu$ is therefore
\begin{equation}
\label{eq:axial-current} J^5_\mu = L_{A1}\, (D_\mu K)^\dagger
(\xi+\xi^\dagger) h
            + i L_{A2}\, K^\dagger A_\mu (\xi-\xi^\dagger) h\,.
\end{equation}
We have introduced a factor of $i$ in the $L_{A2}$ term to make the
low energy constants real. This can be seen by considering charge
conjugation $\mathcal{C}$. For the quark current (taking $h$ to be
real), we have
\begin{equation}\label{eq:axialc}
J_\mu^5 = \bar q h \gamma_\mu \gamma_5 s
 \stackrel{\mathcal{C}}\to \bar s \gamma_\mu \gamma_5 h^\mathrm{T} q
 = (J_\mu^5)^\dagger.
\end{equation}
In the effective theory, the charge conjugation transformations
are
\begin{equation}\label{eq:chiptc}
\xi \to \xi^\mathrm{T}, \quad A_\mu \to A_\mu^\mathrm{T}, \quad
V_\mu \to -V_\mu^\mathrm{T}
\end{equation}
together with
\begin{equation}\label{eq:chiptc2}
K_a \to K_a^\dagger, \quad (D_\mu K)_a = \partial_\mu K_a + (V_\mu
K)_a
 \to \partial_\mu K_a^\dagger - (K^\dagger V_\mu)_a
 = (D_\mu K)_a^\dagger,
\end{equation}
where $a$ is a flavor label. Using these transformations, the current
in the effective theory is transformed into its Hermitian conjugate
under charge conjugation provided $L_{A1}$ and $L_{A2}$ are both real.

The leading contribution to $f_K$ in the chiral expansion can be
readily deduced. The first term in the axial current has a
$\partial_\mu K$ factor with no pion fields, so that at tree level
$f_K$ is fixed by $L_{A1}$, specifically
\begin{equation}
f^{(K)}(m_h) = 2 L_{A1}\,,
\end{equation}
where $m_h$ is the mass of the strange quark (we are using the
notation defined in Appendix~\ref{sec:appendix:notation})\,.

The NLO contribution to $f_K$ is obtained from the tadpole diagram
in Fig.~\ref{fig:diagrams}(a). From the $K\pi\pi$ vertex in the
$L_{A1}$ term in the axial current we obtain the contribution
\begin{equation}
2 L_{A1} (-ip_\mu) \frac{-iC_F}{f^2}
 \int \frac{d^4 k}{(2\pi)^4} \frac1{k^2-m_\pi^2}\quad =\quad
2 L_{A1} (-ip_\mu)\frac{-iC_F}{(4\pi f)^2}
 m_\pi^2 \left(\log \frac{m_\pi^2}{\Lambda_\chi^2} +
 \textrm{constant}\right)\,,
\end{equation}
where $C_F=3/4$ is the eigenvalue of the quadratic Casimir operator
for $\SU(2)$ in the fundamental representation and $\Lambda_\chi$ is
an arbitrary renormalization scale. There is no contribution at
one-loop order from the term in the axial current proportional to
$L_{A2}$ and so we have shown that the chiral behavior of $f_K$ is of
the form:
\begin{equation}
f_K=f^{(K)}(m_h)\left\{ 1 + c(m_h)\,\frac{m_\pi^2}{f^2}-\frac{m_\pi^2}{(4\pi f)^2}\,\frac34\,\log \frac{m_\pi^2}{\Lambda_\chi^2}\right\}\,,
\label{eq:fknlo}\end{equation}
where $c(m_h)$ is a low-energy constant and $m_\pi^2=\chi_l$\,.

Before proceeding to discuss the chiral behavior of $f_K$ in
partially quenched $\SU(2)$ ChPT, we compare the above calculation
with that of $f_B$ in Heavy Meson ChPT~\cite{Sharpe:1995qp}. In that
case the NLO contribution has the corresponding contribution to
Eq.~(\ref{eq:fknlo}), but in addition it has a second contribution
from the self-energy diagram in Fig.~\ref{fig:diagrams}(b). To
discuss this for both $f_B$ and $f_K$ simultaneously, let $P$ and
$P^*$ be pseudoscalar and vector mesons containing a light quark
($u$ or $d$) and a heavier antiquark ({\em e.g.} $s$ or $b$).
Let the masses of the $P$ and $P^*$ mesons be $M$ and $M_*$
respectively. The interaction of pions with pseudoscalar and vector
mesons, $P$ and $P^*$, takes the form $M g\, \partial^\mu \pi P
P^*_\mu/f$, where the $1/f$ arises because there is one pion and the
$Mg$ is put in for compatibility with heavy meson chiral perturbation
theory (where the fields are usually normalized with an implicit
factor of $\sqrt M$) so that the coupling constant $g$ is
dimensionless. The diagram in Fig.~\ref{fig:diagrams}(b), for a $P$
meson with momentum $p$, is proportional to $(Mg/f)^2 I$ where $I$ is
the Feynman integral:
\begin{equation}
\label{eq:Iintegral} I = \int \frac{d^4 l}{(2\pi)^4}
 \frac{l^\mu l^\nu \big[g_{\mu\nu}-(p-l)_\mu(p-l)_\nu/M_*^2\big]}
      {(l^2-m_\pi^2)( (p-l)^2-M_*^2)}.
\end{equation}
We now compare the behavior of this integral with $m_\pi^2$ in the
two cases: i) the heavy quark limit in which $M\to\infty$ with
$M_*^2-M^2=O(\Lambda_{\textrm{QCD}}^2)$ which is an approximation for
$B$-physics and ii) $M$ and $M_*$ both of $O(\lqcd)$ and not
degenerate, a situation which is appropriate for kaon physics.

In the heavy quark limit, we write $p=Mv+k$, where $v$ is the meson's
four-velocity ($v^2=1$) and the leading term in $I$ in the $1/M$
expansion is:
\begin{equation}\label{eq:iasymphqet}
I\simeq\frac{1}{2M} \int \frac{d^4 l}{(2\pi)^4} \frac{l^\mu l^\nu
\big[g_{\mu\nu}-v_\mu v_\nu\big]}
      {(l^2-m^2)(v{\cdot}(k-l)-\Delta)}\,,
\end{equation}
where $\Delta=(M_*^2-M^2)/2M$ which is of $O(\lqcd^2/M)$ and hence
negligible compared to $m_\pi$ in the heavy quark limit. We stress
that in this analysis we take the heavy-quark limit before the chiral
limit, {\em i.e.} we keep $\lqcd/M\ll m_\pi/\lqcd$\,. By power counting we
see that the component of $I$ contributing to the wavefunction
renormalization can have a term proportional to
\begin{equation}
\frac{v{\cdot}k}{M} m_\pi^2\log\frac{m_\pi^2}{\Lambda_\chi^2}
\end{equation}
and so can lead to terms proportional to $(g^2/f^2) m_\pi^2 \ln
(m_\pi^2/\Lambda_\chi^2)$ in the behavior of $f_B$. An explicit evaluation
of the diagram confirms that such terms are indeed
present~\cite{Sharpe:1995qp}.

In contrast, there is no contribution of the form $m_\pi^2 \log
m_\pi^2$ in KChPT, {\em i.e.} when we take the chiral limit of small
$m_\pi^2$ while keeping $M$ and $M_*$ fixed, non-degenerate and neither
especially large or small. To see this we combine the denominators in
$I$ using Feynman parametrization and write
\begin{equation}
I=\int_0^1 d\alpha\int \frac{d^4 l}{(2\pi)^4}\frac{N(\alpha)}{D^2(\alpha)}\,,
\end{equation}
where $\alpha$ is the Feynman parameter,
\begin{equation}\label{eq:Ndef}
N(\alpha)=
 (l+\alpha p)^2 -
 \frac{\left((l+\alpha p)\cdot (l-(1-\alpha)p)\right)^2}{M_*^2}
\end{equation}
and
\begin{equation}
D(\alpha)=l^2+\alpha(1-\alpha)p^2-(1-\alpha)\,m_\pi^2-\alpha M_*^2\,.
\end{equation}
The question we are addressing is whether there is a contribution to
$I$ of the form $m_\pi^2\log m_\pi^2$. If there is such a term then we
can isolate it by differentiating with respect to $m_\pi^2$, setting
$m_\pi^2\to 0$ and searching for a logarithmic (infrared) divergence
in
\begin{equation}
\left.\frac{dI}{dm_\pi^2}\right|_{m_\pi^2=0}=-2\int_0^1 d\alpha\, (1-\alpha)\int \frac{d^4 l}{(2\pi)^4}\frac{N(\alpha)}{D^3(\alpha, m_\pi^2=0)}\,.
\end{equation}
We use dimensional regularization in $D=4+2\varepsilon$ dimensions.
The contribution from the first term in $N$ in Eq.~(\ref{eq:Ndef}) is
\begin{equation}
\label{eq:didmpisq}
\begin{aligned}
\left.\frac{dI_1}{dm_\pi^2}\right|_{m_\pi^2=0} &=
 -2\int_0^1 \!\!d\alpha\, (1{-}\alpha)
  \int \frac{d^D l}{(2\pi)^D}\,
  \frac{l^2+\alpha^2 p^2}{[l^2+\alpha(1{-}\alpha)p^2-\alpha M_*^2]^3}
  \\
  &=\frac{-i}{(4\pi)^{2+\varepsilon}} \int_0^1 \!\!d\alpha\,
     (1-\alpha)\alpha^\varepsilon[M_*^2-(1{-}\alpha) p^2]^\varepsilon
     \left\{(2{+}\varepsilon)\Gamma(-\varepsilon)-
      \frac{\alpha p^2}{M_*^2-(1{-}\alpha)p^2}\right\}\,.
\end{aligned}
\end{equation}
The contribution to the self-energy is the coefficient of $p^2$ at
$p^2=M^2$, which we can isolate by differentiating with respect to
$p^2$ and setting $p^2=M^2$\,. By inspection we can readily verify
that there are no infrared singular terms and hence no term
proportional to $m_\pi^2\,\log m_\pi^2$ in $I$\,. Note that $M_*>M$
and hence $M_*^2-(1-\alpha)p^2>0$ for $p^2=M^2$ throughout the
integration region in $\alpha$. A parallel argument shows that there
is also no contribution to $I$ of the form $m_\pi^2\log m_\pi^2$ from
the second term in $N(\alpha)$ on the right-hand side of
Eq.~(\ref{eq:Ndef})\,.

Note also that $I$ does contain a term proportional to $m_\pi^4\log
m_\pi^2$. In this case we seek an infra-red divergence after
differentiating $I$ twice with respect to $m_\pi^2$ and setting
$m_\pi^2\to 0$\,.  The power of
$l^2+\alpha(1-\alpha)p^2-\alpha M_*^2$ in the integrand is $-4$ as
compared to $-3$ in Eq.~(\ref{eq:didmpisq}). After performing the
$l$-integration, we find that for small $\alpha$ the integrand
$\sim \alpha^{-1+\varepsilon}$, which diverges in $4$ dimensions and
is a signature of the presence of a $\log m_\pi^2$ term ({\em i.e.} a
$m_\pi^4 \log m_\pi^2$ term in $I$).

There are also no chiral logarithms from the tadpole diagram in
Fig.~\ref{fig:diagrams}(c). The $KK\pi\pi$ term in $L_{\pi K}^{(1)}$
in Eq.~(\ref{eq:lkpi1}) is proportional to $K (\partial_\mu K)
[\phi,\partial^\mu\phi]$ and so the integrand is an odd function of
the pion's momentum and the integral vanishes.

Finally in this section we derive the behavior of $f_K$ in the
partially quenched case, {\em i.e.} in PQKChPT. The axial vector current is
of the form of that in Eq.~(\ref{eq:axial-current}), but uses the
PQ $K$ and $\xi$ fields. As before, the spurion $h$ is chosen to pull
out the appropriate flavor. The tree and quadratic terms from this
current are
\begin{equation}
J_\mu^{5a} = 2 L_{A1} \partial_\mu K_b ^\dagger
 \left( 1 - \frac{\Phi^2}{2f^2} \right)_{ba} + \cdots
\end{equation}
where we have dropped terms with a single derivative acting on a
Goldstone field.

To calculate the decay constant for a kaon containing a valence light
quark, we take the flavor index $a=1$ or $2$. Choosing, for
illustration, $a=1$ we find
\begin{equation}\label{eq:pqkaxialcurrent}
2 L_{A1} \partial_\mu K_1^\dagger
 \left[ 1 - \frac1{2f^2}(
  \eta_{1,11}\eta_{2,11} + \eta_{1,12}\eta_{2,21}
  + \phi_{11}^2 + \phi_{12}\phi_{21}+ \phi_{13}\phi_{31}
  + \phi_{14}\phi_{41}) \right]\,.
\end{equation}
There are also terms containing $\partial_\mu K_2^\dagger$, but
these have Goldstone field pairings which cannot be contracted to
give a one-loop tadpole contribution and thus are not listed
above. The one-loop tadpole contribution to $f_{K_v^+}$ is
\begin{equation}\label{eq:fkpq1loop}
-L_{A1} \frac1{f^2} \int \frac{d^4l}{(2\pi)^4}
  \left\{ -\frac{2i}{l^2-\chi_x}
  + \frac i{l^2-\chi_x}
  + \frac{2i}{l^2-(\chi_x+\chi_l)/2}
  + \frac i{l^2-\chi_x}
    - \frac12 \frac{l^2-\chi_l}{(l^2-\chi_x)^2} \right\}.
\end{equation}
The 5 terms in the integrand of Eq.~(\ref{eq:fkpq1loop}) have the
following origin:
\begin{enumerate}
\item 
The first term comes from the quark-ghost mesons in the tadpole loop.
These have valence-valence quark masses and the minus sign arises from
the closed loop of anticommuting fields.
\item
The second term has the $\phi_{12}\phi_{21}$ propagator with both
quarks having valence masses.
\item
The third term has the $\phi_{13}\phi_{31}+\phi_{14}\phi_{41}$
propagators with one quark having the valence-quark mass and the other
the sea-quark mass.
\item[4\&5.]
The final two terms have a neutral propagator in the tadpole diagram
from the $\phi_{11}^2$ term on the rhs of
Eq.~(\ref{eq:pqkaxialcurrent}). Both quarks have valence-quark masses.
\end{enumerate}
We observe, as expected, that the tadpole contribution with the
valence-ghost propagator cancels that with the valence-valence one.
Extracting the chiral logarithms from the loop integrals we arrive at
the expression:
\begin{equation}
\label{eq:fkpqlogs}
f_{xh} = 2 L_{A1} \left\{ 1 -
 \frac1{(4\pi f)^2} \left[
 \frac{\chi_l+\chi_x}2 \log \frac{\chi_l+\chi_x}{2\Lambda_\chi^2}
 + \frac{\chi_l-2\chi_x}4 \log \frac{\chi_x}{\Lambda_\chi^2} \right]
\right\} + \cdots\ .
\end{equation}
The analytic terms can be proportional to $\chi_l$ and $\chi_x$ and so
we obtain the final result for the mass behavior of a meson with a
light valence quark $x$ and a heavier (strange) valence quark $h$:
\begin{equation}\label{eq:fxspqfinal}
\begin{split}
f_{xh}=f^K(m_h)&\left\{
 1+\frac{\lambda_3(m_h)}{f^2}\chi_l+\frac{\lambda_4(m_h)}{f^2}\chi_x
 \right. \\
 &\left. -\frac1{(4\pi f)^2}\left[
    \frac{\chi_x+\chi_{l}}2
    \log\frac{\chi_x+\chi_{l}}{2\Lambda_\chi^2} +
    \frac{\chi_l-2\chi_{x}}4
    \log\frac{\chi_x}{\Lambda_\chi^2}
    \right]\right\}\,,
\end{split}
\end{equation}
where $\lambda_{3,4}$ are LEC's and we remind the reader that if the
valence and sea strange quark masses are different then the LECs
$f^{(K)}$ and $\lambda_{3,4}$ depend on both these masses.
Eq.\,(\ref{eq:fxspqfinal}) agrees with the corresponding calculation
of the chiral behavior of $f_B$~\cite{Sharpe:1995qp}, when the terms
proportional to the square of the $BB^*\pi$ coupling are neglected.
The right-hand side of Eq.~(\ref{eq:fxspqfinal}) reduces to
Eq.~(\ref{eq:fknlo}) in the unitary limit $\chi_x=\chi_l$\,.

\subsubsection{Chiral Behavior of $m_K^2$}

We now consider the chiral behavior of $m_K^2$, starting in the
unitary theory in which the valence and sea masses are equal. In
principle the chiral logarithms could come from the tadpole diagram in
Fig.~\ref{fig:diagrams}(c). However, just as for the wavefunction
renormalization, there is no such contribution at one-loop order. The
$KK\pi\pi$ term in $L_{\pi K}^{(1)}$ in Eq.~(\ref{eq:lkpi1}) is
proportional to $K (\partial_\mu K) [\phi,\partial^\mu\phi]$ and so
the integrand is an odd function of the pion's momentum and the
integral vanishes. This is also the case in the partially quenched
theory and so, at NLO in the chiral expansion, the mass-dependence of
$m_K^2$ comes from the analytical terms coming from the higher order
terms in the $K\pi$ Lagrangian. For the partially quenched theory at
NLO
\begin{equation}\label{eq:mksqpq}
m^2_{xh}= B^{(K)}(m_h)\tilde{m}_h\left\{1+\frac{\lambda_1(m_h)}{f^2}\chi_l+\frac{\lambda_2(m_h)}{f^2}\chi_x\right\}\,.
\end{equation}
In the unitary theory we have the natural simplification of only a
single low-energy constant.

\subsubsection{Chiral Behavior of $B_K$}

The non-perturbative strong interaction effects in neutral kaon mixing
are contained in the matrix elements of the QCD operator
\begin{equation}
\bar s_L \gamma_\mu d_L \, \bar s_L \gamma^\mu d_L\,,
\end{equation}
between $K^0$ and $\bar{K}^0$ states. This operator is part of a
multiplet transforming under $\su2L$ on the two down quarks and is
symmetric under the interchange of the two $d_L$'s. Hence it
transforms as a triplet under $\su2L$ and is a singlet under
$\su2R$ (the analogous situation for neutral $B$-meson mixing is
that the corresponding operator is part of a $6$ representation of
$\su3L$~\cite{Grinstein:1992qt}).

In our effective kaon theory the combination $\xi K$ transforms as
$\xi K \to L \xi K$ under $\mathrm{SU}(2)_L$, and we can use it to
build the leading $\Delta S=2$ operator,
\begin{equation}
\label{eq:ds2operator}
\mathcal O_{ab} = 2\beta (\xi K)_{\{a}(\xi K)_{b\}},
\end{equation}
which is symmetrised in the flavor indices $a$ and $b$ and has a LEC
$\beta$ (the analogous construction for neutral $B$-meson mixing in
$\su3L\times \su3R$ heavy meson chiral perturbation theory leads to an
operator $4\beta_Q (\xi P^{(Q)\dagger})_{\{a} (\xi P^{(\bar
Q)})_{b\}}$~\cite{Grinstein:1992qt}, where $P^{(Q)}$ and $P^{(\bar
Q)}$ destroy mesons containing heavy $Q$ and $\bar Q$ quarks
respectively). As in the discussion of the axial current, operators
with one or both of the kaon fields on the right hand side of
Eq.~(\ref{eq:ds2operator}) replaced by $D^nK$, where $D$ is the
covariant derivative (the Lorentz indices have been suppressed)
transform in the the same way as $\mathcal O_{ab}$. Again, by using
the equations of motion, the leading component of these operators can
be reduced to $\mathcal O_{ab}$\,.

In order to evaluate $B_K$ we will need the matrix element
$\langle \bar K^0 | \mathcal O_{22} | K^0\rangle$ and thus need to use
the $K^0 K^0$ piece of $\mathcal O_{22}$. At tree level we find
\begin{equation}
4\beta = \frac83 m_K^2 f_K^2 B_K.
\end{equation}
In order to evaluate the one-loop contributions to the matrix element
(see Fig.~\ref{fig:bkoneloop}) we expand $\mathcal O_{22}$ up to
second order in the pion fields:
\begin{equation}
\mathcal O_{ab} = 2\beta\, K_a K_b
 - \frac{2\beta}{f^2} \left\{
   (\phi K)_a (\phi K)_b + \frac12\left[
    (\phi^2 K)_a K_b + K_a (\phi^2 K)_b
   \right]\right\} + \cdots\,.
\end{equation}
We observe, by comparing to the expansion of the axial vector
current in Eq.~(\ref{eq:axial-current}), that the two terms
containing $\phi^2$ lead to relative one-loop corrections each of
which is the same as the relative one-loop correction for $f_K$. Since
we are calculating a matrix element proportional to $f_K^2 B_K$, these
two corrections will not affect $B_K$ and we need only calculate the
one-loop correction from the $(\phi K)_a(\phi K)_b$ term. Thus the
relevant component of $\mathcal O_{22}$ is the $K^0 K^0$ term in
\begin{equation}
-\frac{2\beta}{f^2}(\phi K)_2(\phi K)_2
 = -\frac{2\beta}{f^2} \frac{\pi^0\pi^0}2 K^0 K^0 + \cdots\ .
\end{equation}
Contracting the two pions into a loop leads to a contribution
\begin{equation}
-\frac{2\beta}{f^2}\int \frac{d^4k}{(2\pi)^4} \frac i{k^2-m_\pi^2}
 = -\frac{2\beta}{f^2} \frac1{(4\pi f)^2} m_\pi^2 \log
 \frac{m_\pi^2}{\Lambda_\chi^2} + \mbox{analytic terms}.
\end{equation}
Hence we find that
\begin{equation}
\label{eq:bkqcd}
B_K = B_K^\mathrm{tree} \left\{ 1 -
 \frac12 \frac1{(4\pi f)^2}m_\pi^2 \log\frac{m_\pi^2}{\Lambda_\chi^2}
 \right\} + \cdots\ .
\end{equation}
This agrees with the result for $B_B$ in Eq.~(3.8) of Sharpe and
Zhang~\cite{Sharpe:1995qp} when the $BB^*\pi$ coupling $g\to0$.

We now promote the above discussion to the partially quenched case.
The effective theory operator is still of the form of
Eq.~(\ref{eq:ds2operator}), but now the flavor labels take $6$ values
and $\xi$ is expanded in terms of $\Phi$ in Eq.~(\ref{eq:phipq}). This
echoes the discussion for partially-quenched neutral $B$-meson mixing
in Sharpe and Zhang~\cite{Sharpe:1995qp}, starting from the QCD
discussion in~\cite{Grinstein:1992qt}.

To obtain the mixing matrix element for neutral kaons with valence down
quarks, we look at the $K_2 K_2$ piece of the $\mathcal O_{22}$
operator, expanded up to second order in the Goldstone fields:
\begin{equation}
\mathcal O_{22} = 2\beta\, K_2 K_2
 - \frac{2\beta}{f^2} \left\{
   (\Phi K)_2 (\Phi K)_2 + \frac12\left[
    (\Phi^2 K)_2 K_2 + K_2 (\Phi^2 K)_2
   \right]\right\} + \cdots \,.
\end{equation}
As in the unquenched calculation above, the $(\Phi^2 K)_2 K_2$ terms
lead to chiral logarithms which are cancelled by those from $f_K^2$
when extracting $B_K$ from the $\Delta S=2$ matrix element. Hence the
pieces of the partially-quenched operator we need are:
\begin{equation}
2\beta \left( 1 - \frac1{f^2} \phi_{22}^2 \right) K_2 K_2\,.
\end{equation}
The $\phi_{22}$ field is a valence-valence neutral meson, so we find a
one-loop correction
\begin{equation}
-\frac{4\beta i}{f^2} \int \frac{d^4 l}{(2\pi)^4}
 \left\{ \frac1{l^2-\chi_x} -
 \frac12 \frac{l^2-\chi_{l}}{(l^2-\chi_x)^2}\right\}
 =
 -\frac{2\beta}{(4\pi f)^2} \chi_{l} \log\frac{\chi_x}{\Lambda_\chi^2} +
 \cdots
\end{equation}
For $B_K$ the result is:
\[
B_{xh} = B^{(K)}_{\rm PS}\left\{
 1 - \frac12 \frac1{(4\pi f)^2} \chi_{l} \log\frac{\chi_x}{\Lambda_\chi^2}
 \right\}+ \cdots
\]
This agrees with Eq.~(3.9) in Sharpe and Zhang~\cite{Sharpe:1995qp}
with $g\to 0$ and $N_f=2$. It also reduces to the QCD result above,
Eq.~(\ref{eq:bkqcd}), when $\chi_x = \chi_{l} = m_\pi^2$. Noting that
there are analytic terms proportional to both $\chi_l$ and $\chi_x$
the final result for the mass dependence of $B_K$ is that in
Eq.~(\ref{eq:chPTsu2K:BPS}) of the appendix
\begin{equation}
B_{xh} = B_\mathrm{PS}^{(K)}(m_h)
 \Bigg\{1+\frac{b_1(m_h)}{\ftwo^2}\chi_l
        + \frac{b_2(m_h)}{\ftwo^2}\chi_x
        - \frac{\chi_l}{32\pi^2\ftwo^2}
          \log\frac{\chi_x}{\Lambda_\chi^2}
\Bigg\}\,,
\end{equation}
where $b_{1,2}$ are LECs.

\subsubsection{Comments}
\label{sec:kchpt:comments}

We conclude this section with the observation that for all the
physical quantities considered in this section the chiral logarithms
in the $\SU(2)$ theory can be simply deduced from those in the
$\SU(3)$ theory. In all these cases the chiral behavior is of the
generic form:
\begin{equation}\label{eq:genericform}
O =
 O_\mathrm{LO}\,(1+\textrm{chiral logarithms} + \textrm{analytic terms})\,,
\end{equation}
where $O$ is the physical quantity (pseudoscalar mass-squared, decay
constant or bag parameter) and there is a single LEC at lowest order
($O_\mathrm{LO}$). The generic form in Eq.~(\ref{eq:genericform})
holds in both the $\SU(2)$ and $\SU(3)$ theories. As an example
consider the pseudoscalar decay constant $f_{xh}$, with a degenerate
heavy (strange) valence and sea quark with mass $m_h$ and partially
quenched up and down quarks. In the partially quenched $\SU(3)$ theory
at NLO we have from Eq.~(\ref{eq:chPTsu3:fK:valStrange}):
\begin{equation}
\begin{split}
f_{xh} = \fthree\Bigg\{1
 &-\frac1{8\pi^2\fthree^2}\bigg[
 \frac{\chi_x+\chi_{l}}4 \log\frac{\chi_x+\chi_{l}}{2\Lambda_\chi^2}
 +\frac{\chi_h+\chi_{l}}4 \log\frac{\chi_h+\chi_{l}}{2\Lambda_\chi^2} \\
 & \phantom{-\frac1{8\pi^2\fthree^2}\bigg[}
 +\frac{\chi_x+\chi_h}8 \log\frac{\chi_x+\chi_h}{2\Lambda_\chi^2}
 +\frac{\chi_h}4 \log\frac{\chi_h}{\Lambda_\chi^2} \bigg] \\
 &+\frac1{96\pi^2\fthree^2}\bigg[
 \Big( \chi_x-\chi_h-\chi_x\frac{\chi_{l}-\chi_x}{\chi_\eta-\chi_x}\Big)
   \log\frac{\chi_x}{\chi_h}\\
 & \phantom{+\frac1{96\pi^2\fthree^2}\bigg[}
 +\chi_\eta(\chi_h-\chi_x)\frac{\chi_\eta-\chi_{l}}{\chi_\eta-\chi_x}
    \Big(\frac{\log\frac{\chi_\eta}{\chi_x}} {\chi_x-\chi_\eta}
         -\frac{\log\frac{\chi_\eta}{\chi_h}}{\chi_h-\chi_\eta}
    \Big)
 \bigg] + \textrm{analytic terms}\Bigg\}.
\end{split}
\end{equation}
``Converting" to the $\SU(2)$ theory, we can expand in $m_x/m_h$ and
$m_{l}/m_h$ obtaining:
\begin{equation}\label{eq:fxhconversion}
\begin{split}
f_{xh}= \fthree \Bigg\{1
&-\frac1{8\pi^2\fthree^2} \frac{\chi_x+\chi_{l}}4
  \log\frac{\chi_x+\chi_{l}}{2\Lambda_\chi^2}
 +\frac1{64\pi^2\fthree^2}(2\chi_x-\chi_l)
  \log\frac{\chi_x}{\Lambda_\chi^2} \\
&+ \textrm{ analytic and higher order terms} \Bigg\}.
\end{split}
\end{equation}
The analytic terms now include contributions proportional to $\chi_h$
which do not vanish in the $\SU(2)$ chiral limit. They are absorbed
into the lowest order $\SU(2)$ LEC $f$, which now depends on $m_h$.
Whether or not the relation between $f$ and $f_0$ is well approximated
by one-loop $\SU(3)$ ChPT or whether higher order ($\chi_h^2$ and
higher powers) terms must be included depends on $m_h$ and on the
convergence of the series. In any case, if $f_{xh}$ is to satisfy the
generic form dictated by $\SU(2)$ ChPT given in
Eq.~(\ref{eq:genericform}) the chiral logs are fixed from
Eq.~(\ref{eq:fxhconversion}):
\begin{equation}\label{eq:fxhsu2}
\begin{split}
f_{xh}= \ftwo\Bigg\{1
&-\frac1{8\pi^2\ftwo^2}\frac{\chi_x+\chi_{l}}4
  \log\frac{\chi_x+\chi_{l}}{2\Lambda_\chi^2}
 +\frac1{64\pi^2\ftwo^2}(2\chi_x-\chi_l)
  \log\frac{\chi_x}{\Lambda_\chi^2} \\
&+ \textrm{ analytic and higher order terms} \Bigg\},
\end{split}
\end{equation}
which agrees with the result from the direct evaluation in the
$\SU(2)$ theory (see Eq.~(\ref{eq:fxspqfinal})). The same is true for
the other quantities being studied here. Of course, in general there
may be more than one operator in the effective theory with the same
quantum numbers as the QCD operator whose matrix element is to be
evaluated. In such cases there are more than one LEC at leading order,
Eq.~(\ref{eq:genericform}) does not apply and the simple arguments
presented here have to be generalized.

%% file: text_sections/SimDetail.tex
%
%

Following the work in \cite{Antonio:2006px,Allton:2007hx} we have
used the Iwasaki gauge action and the domain wall fermion action.
By producing smoother gauge fields at the lattice scale, for a fixed
lattice spacing in physical units, the Iwasaki action removes some
of the gauge field dislocations that contribute to the residual
chiral symmetry breaking for domain wall fermions at finite $L_s$.
While suppressing such dislocations improves residual chiral symmetry
breaking, it also suppresses topology change in the evolution
\cite{Antonio:2007tr}.  Since we want our ensembles to sample
topological sectors of the theory as well as possible, we have found
the Iwasaki gauge action to provide a reasonable balance between
these two, contradictory goals.

We generate ensembles with two degenerate light quarks, whose bare
input mass is given by $m_l$, and one heavy strange quark of mass
$m_h$.  We use the exact Rational Hybrid Monte Carlo (RHMC) algorithm
to generate ensembles.  During the course of this work, we made
improvements to the RHMC algorithm, as detailed in
\cite{Allton:2007hx}, yielding three variants of the RHMC algorithm,
0, I and II.  The original RHMC 0 algorithm
\cite{Clark:2006fx,Kennedy:1998cu} was used for the 2+1
flavor simulations in \cite{Antonio:2006px} and the RHMC I was used
for most of the simulations in \cite{Allton:2007hx}.  We also
compared the RHMC I and II algorithms in \cite{Allton:2007hx}.
We find the RHMC II algorithm to be the best version to date.  It
uses a single stochastic noise source to estimate ratios of
determinants, which reduces the size of the forces in the molecular
dynamics integration \cite{Aoki:2004ht}.  It produces ensembles
that change topology more rapidly than RHMC I, likely due to this
decrease in the fermionic force.  It uses a multiple time scale
Omelyan integrator and light quark preconditioning
\cite{Urbach:2005ji,Hasenbusch:2002ai}, which has the effect of
making the time spent solving the light quark Dirac equation less
than the time spent solving for heavier quarks.

We can briefly define the RHMC II algorithm.  Letting ${\cal D}(m_i)
= D_{\rm DWF}^\dagger(M_5, m_i) D_{\rm DWF}(M_5, m_i)$, where $D_{\rm
DWF}(M_5, m_i)$ is the domain wall fermion Dirac operator, $M_5$
is the domain wall height and $m_i$ is the mass of the quark we
wish to simulate, we write the fermionic contribution to the path
integral as
\begin{equation}
  \det\left\{\frac{{\cal D}(m_l)}{{\cal D}(1)}\right\} \;
  \det\left\{\frac{{\cal D}(m_h)}{{\cal D}(1)}\right\}^{1/2}
  =
  \det\left\{\frac{{\cal D}(m_l)}{{\cal D}(m_h)}\right\} \;
  \left[
    \det\left\{\frac{{\cal D}(m_h)}{{\cal D}(1)}\right\}^{1/2} 
  \right]^3 \, .
\label{eq:RHMCII}
\end{equation}
(Recall that for domain wall fermions, a regulator must be added to
remove the bulk infinity that would arise as $L_s \to \infty$.
The ${\cal D}(1)$ terms in Eq.~(\ref{eq:RHMCII}) are this regulator.)
Each ratio of determinants on the right-hand side of
Eq.~(\ref{eq:RHMCII}) is represented by a single stochastic estimator,
{\em i.e.} a single estimator is used for the ratio involving $m_l$
and $m_h$ and three estimators are used for the three, fractional
power determinants involving $m_h$ and 1.  The light quark ratio
is integrated on the coarsest time scale, using a conventional
Hybrid Monte Carlo algorithm.  The three 1/2 powers of ratios
involving the heavy dynamical quark are integrated on a 2 times
finer time scale using the RHMC algorithm.  For each integration
step of the RHMC and for each field, a single solution of the Dirac
equation is required with mass $m_h$ and two solutions with the
regulator mass of 1.  (The solutions are found with the conjugate
gradient algorithm.)  The regulator mass solutions take roughly
half as many conjugate gradient iterations in the RHMC as the $m_h$
solutions, but there are twice as many required.  Thus the time
spent solving the Dirac equation in the two cases is comparable.
Because of the three different stochastic estimators needed in the
RHMC, and the coarser time step for the HMC, the number of conjugate
gradient iterations in the RHMC are larger than in the HMC part of
the algorithm for the quark masses currently used.  This has made
simulating at lighter quark masses much less expensive.  The gauge
field is integrated on an even finer time scale than the RHMC.  For
a further review of the RHMC, see \cite{Clark:2006wq}.

As mentioned, we had already begun evolving configurations with the
RHMC 0 algorithm as improved versions were being developed.  Table
\ref{tab:evol} gives the molecular dynamics time when we changed
from one algorithm to the next, for each of our ensembles.  Since,
the RHMC II gives more decorrelated lattices than earlier versions,
all of the observables in this paper have been measured only on
configurations generated by the RHMC II algorithm.  The configurations
generated with the earlier algorithms are being used only for
equilibration.

As one indicator of how well the RHMC II algorithm is decorrelating
our ensembles, we have measured the global topological charge.  Of
course, for infinite volumes, global topological charge is not
relevant to local physics.  However, in finite volume knowing that
global topology is changing gives us evidence that there are local
topological fluctuations.  To calculate the topological charge the
configurations are first cooled by applying 30 updates with a
quenched, 5 loop improved gauge action with zero coupling strength,
$\beta=\infty$.  After cooling, the topological charge is measured
using a 5 loop improved gluonic topological charge operator, the
5Li method of \cite{deForcrand:1997sq} and the results are shown
in Fig.\ \ref{fig:topology}.  This figure shows that the RHMC
algorithm is sampling different topological sectors for each of our
ensembles and the histograms indicate, by their symmetry and shape,
that the topological landscape has been sampled reasonably well.

Figures \ref{fig:IntAutoCorr} and \ref{fig:IntAutoCorr20} show the
integrated autocorrelation time for the pion correlator at a
separation of 12 lattice spacings for the $m_l = 0.005$ ensemble.
Both show an integrated autocorrelation time for this quantity of
10 to 15 molecular dynamics time units.  Figure \ref{fig:plaq_m02}
shows the evolution of the plaquette for the $m_l = 0.02$ ensemble.
For this short-distance observable, equilibration took a few hundred
molecular dynamics time units.

%% file: text_sections/LatticePS.tex
%
%


\subsection{Measurements}
\label{sec:fitting} Since we are most interested in the light quark limit where
pion masses are close to the physical values, we put significant computational
effort into the measurements on the two lightest ensembles with $m_l =
0.005$ and $m_l = 0.01$. On these two ensembles, three separate measurements
have been done, which we shall refer to as Full-Partially-Quenched (FPQ),
Degenerate (DEG) and Unitary (UNI).  In the FPQ measurement we computed hadron
two-point correlators for all the combinations with valence quark masses $m_x$,
$m_y$ $\in$ \{0.001, 0.005, 0.01, 0.02, 0.03, 0.04\}. The DEG dataset consists
of hadron correlators with \emph{degenerate} quarks with the same list of
masses. In the UNI dataset only quark propagators with masses equal to the
light and strange sea quark masses were calculated, and the light-light,
light-strange and strange-strange meson correlators were then constructed. On
the $m_l = 0.02$ and $m_l = 0.03$ ensembles, only  the DEG and UNI calculations
were performed.  Details of the measurements, including the gauge
configurations used, separation of each measurement and the total number of
measurements, can be found in Table~\ref{tab:datasets}. For the $m_l = 0.005$
and $m_l = 0.01$ ensembles, we blocked the data so that each block contains
measurements from every 80 molecular-dynamics time units, while for the $m_l =
0.02$ and 0.03 ensembles the block size was chosen to be 40 time units to yield
a reasonable number of jackknife blocks to perform the analysis. 

In each of the FPQ measurements, we used Coulomb gauge fixed wall (W) sources of
size $24^3$, which were placed at $t = 5$ and $t = 59$. For each source, two
quark propagators were calculated, one with the periodic, and the other
anti-periodic, boundary condition in the temporal direction. The sum of these
two quark propagators (as a single quark propagator) was then used to construct
the meson correlators. The resulting cancellation of backward
propagating states has the effect of doubling the temporal extent
of the lattice, so over much of the lattice volume, there is no
excited state contribution to the hadron propagator.  We find this
works well with our Coulomb gauge fixed wall sources, which generally
have small statistical errors but are not tuned to remove excited
state contaminations.  The long plateaus and small statistical
errors allow us to work far enough from the source that excited
states are not a worry.

The DEG measurements used a Coulomb gauge fixed box source (B) of size $16^3$
which we have found to give the optimal early onset of the plateaus for
pseudoscalar mesons. The sources were placed at two timeslices, $t = 0$ and
$32$. 

In the UNI measurements, the propagators were calculated from Coulomb gauge
fixed hydrogen S-wavefunction (H) sources~\cite{Boyle:1999gx}, with radius
$r=3.5$, in lattice units, or gauge invariant gaussian (G) sources and sinks
with radius $r=4$.  Again the sources were placed at multiple timeslices for a
better sampling over the gauge fields. 

We also used two different interpolating operators for the pseudoscalar state,
namely, the pseudoscalar operator $P^a(x) = \ov{q}(x) \gamma_5 (\tau^a/2)
q(x)$, which we refer to using a short-hand notation $P$, and the axialvector
operator $A_\mu^a(x) = \ov{q}(x) \gamma_\mu\gamma_5(\tau^a/2) q(x)$, which we
refer to as $A$. Here $\tau^a$ is a flavor symmetry generator. Unless otherwise
specified, we only consider the time component of $A^a_\mu(x)$, {\em i.e.},
$A^a_4(x)$. Together with the different source/sink smearings described above,
this allows us to construct several different pseudoscalar meson correlators,
which are tabulated in Table~\ref{tab:corrlist}.  (Table~\ref{tab:Vcontract} has similar information for vector and tensor correlators, which
are discussed in detail in Section \ref{sec:Vector}.)
The notation in
Table~\ref{tab:corrlist} follows Ref.\cite{Antonio:2006px}, where quark
propagators are specified by the smearings applied in the source and sink. For
example, WL means a quark propagator calculated with a wall (W) source and a
local (L) sink. Meson correlators are then denoted by the the type of quark
propagators used in the contraction, {\em e.g.}, WL-WL refers to a meson correlator
with two WL quark propagators. In this case we may also use WL to refer to
the meson correlator unless ambiguities arise. 

In this paper we only consider meson correlators with zero momentum projection.
Our meson states are normalized such that the time dependence of the
correlators in the large $t$ limit can be expressed as
\begin{equation}
	{\cal C}_{O_{1}O_{2}}^{s_{1}s_{2}} (t) 
	= \frac{\langle 0 | O_{1}^{s_1} | \pi \rangle \langle \pi | O_{2}^{s_2}
| 0 \rangle }{2 m_{xy} V} \left [ e^{-m_{xy}t} +(-1)^p e^{-m_{xy}(T-t)} \right ],
\label{eq:corr-t}
\end{equation}
where the superscripts specify the smearings for the quark propagators, and the
subscripts specify the interpolating operators used. $m_{xy}$ is the ground-state
mass of a pseudoscalar meson composed of two valence quarks with masses $m_x$
and $m_y$.
 For convenience the amplitude of the correlator will be denoted  as
\begin{equation}
{\cal N}_{O_1 O_2}^{s_{1}s_{2}} \equiv \frac{\langle 0 |
O_1^{s_1} | \pi \rangle \langle \pi | O_2^{s_2} | 0 \rangle }{2 m_{xy} V}.
\end{equation}

\subsection{Residual Chiral Symmetry Breaking}
\label{sec:latticeps_mres}
As discussed in Section~\ref{subsect:dwf_chpt}, the finite size in
the fifth dimension of the domain wall fermion formulation in our
current simulations permits the mixing between the two light quark
states bound to the  boundary walls, resulting in a residual chiral
symmetry breaking~\cite{Furman:1995ky,Blum:1998ud,Antonio:2007tr}.
Close to the continuum limit, this effect can be quantified as a
residual mass term (denoted as $\mres$), which, up to $O(a^2)$,
behaves as a regular quark mass, making the total quark mass
effectively the sum of the bare input quark mass and the residual
mass. We determine the residual mass by computing the
ratio~\cite{Blum:1998ud,Antonio:2006px}:
\begin{equation}
	  R(t) = 
	  \frac{\langle \sum_{\vec{x}} J^a_{5q} (\vec{x}, t)
	    P^a(\vec{0},0) \rangle } 
	       {\langle \sum_{\vec{x}} P^a(\vec{x}, t) 
		 P^a(\vec{0},0) \rangle }, 
	       \label{eq:mres_ratio}
\end{equation}
where $J^a_{5q}$ is the ``mid-point" operator which mixes the quark
states from the left and right walls~\cite{Furman:1995ky,Blum:1998ud},
and $P^a$ is the pseudoscalar operator defined earlier.  For $t$
large enough that only pions contribute to the correlators
in Eq.~(\ref{eq:mres_ratio}), $R(t) \to \mres^\prime(m_f)$.
From the
DEG data sets, we obtained the residual mass for each pair of valence
quark masses by averaging the plateaus of $R(t)$ from $t = 10$ to
$t = 32$. Figure~\ref{fig:mres} shows the residual mass as a function
of the valence quark mass. We can see that the residual mass has a
linear dependence on the input quark masses, and the magnitude of
the slope is larger in the valence sector than the sea sector.  The
slopes have opposite signs, as was also observed in our previous
simulations on smaller volumes~\cite{Antonio:2006px, Allton:2007hx}.
In particular, $d\ln(\mres^\prime)/dm_f^{\rm val} = -4.1$ for the
ensemble with $m_l =0.005$ and $-4.5$ for the $m_l = 0.03$ ensemble.
We also find $d\ln(\mres^\prime)/dm_f^{\rm sea} = 6$ and, for
the dynamical quark mass dependence in the unitary case,
the result is $d \ln(\mres^\prime))/dm_l
= 1.1$.  In terms of the discussion at the end of Section
\ref{subsect:dwf_chpt}, it appears that the $Z_{\cal A} -1 $ and
dimension five terms are comparable in magnitude.  We define a
mass-independent residual mass for our simulations by evaluating
$\mres(m_l)$ at the $m_l = 0$ limit. The straight line in
Fig.~\ref{fig:mres} shows a linear fit to the four unitary points
with $m_x = m_y = m_l$. And the star is the resulting residual mass
at $m_l = 0$, which we determined to be
\begin{equation}
\mres = 0.00315(2). 
\end{equation}
It is worth noting that this result is in decent agreement with that obtained
from earlier simulations at the same gauge coupling but with a smaller lattice
volume~\cite{Allton:2007hx}, indicating that the finite volume effect for the
residual mass is small. 

\subsection{Pseudoscalar Meson Masses and Decay Constants}
\subsubsection{Results from the $m_l = 0.005$ and 0.01 ensembles}
To obtain the masses and decay constants for the pseudoscalar mesons from the
$m_l = 0.005$ and $m_l = 0.01$ ensembles, we focus on the FPQ data
sets, since the long plateaus of the FPQ data sets lessen the
uncertainty in choosing the fit ranges. In fact, including the UNI and DEG measurements in the simultaneous fits (described below) does not change significantly
either the central values or statistical errors of the final results. Thus here we only quote results from the FPQ data sets.

For each mass combination in the FPQ data sets, we performed a simultaneous fit
to the five correlators as specified in the second column of
Table~\ref{tab:corrlist}, to determine a common mass and a different amplitude
for each correlator. This approach has the advantage of averaging out the
systematic uncertainties due to different characteristics of the interpolating
operators ($P$ or $A$), and reducing the uncertainty of fit range choices. One
drawback is that, as the number of data points included in the fit ($\sim 220$)
is greater than the number of independent measurements we have in hand, the
covariance matrix is poorly resolved, and we lean on uncorrelated fits to
obtain results. The resulting pseudoscalar meson masses for all the mass
combinations of the $m_l = 0.005$ and 0.01 ensembles are given in
Table~\ref{tab:PSmass}. Note that the quoted $\chi^2$/dof's only give a
qualitative indication of how different fits compare to each other. They are
not meaningful quantitative indications of the goodness of the fits.

Besides a common mass for each valence mass combination, we also obtained an
amplitude for each correlator included in the simultaneous fit, namely  
\begin{equation}
{\cal N}_{AA}^{LW}, {\cal
  N}_{PP}^{LW}, {\cal N}_{AP}^{LW}, {\cal N}_{PP}^{WW} \ \ {\rm and} \ \ {\cal
  N}_{AP}^{WW}.
\label{eq:amps}
\end{equation}
Combinations of these amplitudes can be used to determine the pseudoscalar
decay constant, which we will describe in the following. 

The pseudoscalar decay constant can be determined by 
\begin{equation}
f_{xy} = \frac{Z_A |\langle \pi | A_4 | 0 \rangle|}{m_{xy}}, 
\label{eq:fpi-dfn}
\end{equation}
where $Z_A$ is the axial vector current renormalization constant which relates
the (partially) conserved axial vector current in the original five-dimensional
theory of domain wall fermions to the local four-dimensional axial vector
current~\cite{Blum:2001xb}. Alternatively, the axial Ward identity in
Eq.~(\ref{eq:axial_cc_diverg}) connects the divergence of the partially conserved axial vector current to the pseudoscalar density, allowing us to determine the pseudosclar decay constant through 
\begin{equation}
	f_{xy} = \frac{(\tm_x + \tm_y) |\langle \pi | P | 0 \rangle|}{m_{xy}^2},
	\label{eq:fpi-ps}
\end{equation}
where $\tm_{x,y} = m_{x,y} + \mres$. 

We determined $Z_A$ from the DEG data sets using
the improved ratio~\cite{Blum:2001xb} of $\langle {\cal A}_4(t) P(0) \rangle$
to $\langle A_4(t) P(0) \rangle$, where ${\cal A}_\mu(x)$ is the partially
conserved axial vector current for domain wall fermions~\cite{Furman:1995ky,Blum:2001xb}. Similar to the
residual mass, we computed the value of $Z_A$ at each unitary quark mass, and
extrapolated it to the chiral limit at $m_l = -\mres$, finding 
\begin{equation}
	Z_A = 0.7161(1), 
\end{equation}
which agrees very well with the result of our smaller-volume
simulations~\cite{Allton:2007hx}. 

There are in principle five methods~\cite{Lin:2007pt} to determine
the matrix element necessary for the $f_{xy}$ calculation in
Eq.~(\ref{eq:fpi-dfn}) or (\ref{eq:fpi-ps}).  These use jackknife
ratios of different amplitudes in Eq.~(\ref{eq:amps}).  We found
that different methods give statistically consistent results for
all quark masses except for the lightest or heaviest quark masses,
where systematic deviations for results obtained using
Eq.~(\ref{eq:fpi-dfn}) and Eq.~(\ref{eq:fpi-ps}) are observed.  The
methods where the pseudoscalar density is used make use of the
equation of motion for the conserved axial current and are more
indirect, since we have not systematically controlled higher order
renormalization effects here.  As was discussed in
Section~\ref{subsect:dwf_chpt}, the renormalization of the conserved
axial current is known to $O(\mres)$, and we know the renormalization
between the local and conserved current to high accuracy. Therefore,
we have chosen as our preferred result a determination of $f_{xy}$
using axial current matrix elements and postpone further investigation
of the differences we are seeing for future work.  We also find
that the statisical errors are the smallest when only the axial
vector current matrix elements are used in the analysis.  Thus, we
determine $f_{xy}$ by
\begin{equation} 
	f_{xy} = Z_A \sqrt{  \frac{2}{m_{xy}} \frac{{{\cal
	N}_{AP}^{LW}}^2}{{\cal N}_{PP}^{WW} } }. \label{eq:fpi-calc}
\end{equation} 
The final results for the $m_l = 0.005$ and 0.01 ensembles are shown in
Table~\ref{tab:fPS}. 

Comparing the results from the $m_l = 0.01$ ensemble with those obtained from a
smaller volume reported in Ref.\cite{Allton:2007hx}, we find that there is a
statistically significant 4\% difference at the unitary point of $m_x = m_y =
m_l = 0.01$, as shown in Fig.~\ref{fig:fpiFV}. As we shall see in
Section~\ref{subsec:SU2fit}, such a difference is about twice as large as
predicted by finite volume chiral perturbation theory. While it is possible
that ChPT still underestimates the finite volume effect for $f_{xy}$, we want
to point out that the $16^3\times32$ gauge configurations were generated using
RHMC I which could intrinsically induce long-range autocorrelations. Another difference is that there we determined $f_{xy}$ using correlators with much shorter plateaus
compared to the $24^3\times64$ ensembles discussed here, which means the
smaller-volume results may suffer from larger systematic errors. Although we
tried our best to estimate uncertainties on the physical quantities of the
$16^3\times32$ simulations, the errors might still be underestimated. With
these factors taken into account, the discrepancy between the lattice data and
the predictions of the finite volume ChPT may not be as significant as it
appears. 

\subsubsection{Results from the $m_l = 0.02$ and 0.03 ensembles}
For the two heavy ensembles with $m_l = 0.02$ and 0.03 we performed
uncorrelated simultaneous fits to four correlators from the DEG data sets:
$\mathcal{C}_{AA}^{LB}(t), \, \mathcal{C}_{PP}^{LB} (t),
\,\mathcal{C}_{AP}^{LB}(t)$ and $\mathcal{C}_{AP}^{BB}(t)$. The time slices
included in the fits are from $t = 10$ to $32$ for each correlator. The
pseudoscalar meson masses obtained this way are presented in
Table~\ref{tab:PSmass23}. The corresponding amplitudes of these correlators can
be used to determine the pseudoscalar meson decay constants by the following
relation: 
\begin{equation}
	f_{xy}   =  Z_A \sqrt{ \frac{2}{m_{xy}} \frac{{\cal N}_{AA}^{LB} \,
{\cal N}_{AP}^{LB}}{{\cal N}_{AP}^{BB}}}, \label{eq:fpi23} 
\end{equation}
the results of which are given in Table~\ref{tab:fpi23}.

%% file: text_sections/LatticeBaryon.tex
%
%

We have chosen to set the lattice scale using the $\Omega^-$ baryon mass because it is made of three strange valence quarks and so
does not have non-analytic light quark mass terms at NLO in chiral perturbation theory\cite{Toussaint:2004cj,Tiburzi:2004rh}. Thus, it is justified to linearly extrapolate in $m_l$ to obtain the physical value.

The interpolating field for the decuplet baryons is
\begin{eqnarray}
\chi_{\rm dec} &=& \varepsilon_{abc} [q^T_a C\gamma_{\mu} q_b] q_c,
\end{eqnarray}
which couples to both spin 3/2 and 1/2 states, the former being the ground state~\cite{Sasaki:2005ug}. For $q=s$, the strange quark, the operator destroys an $\Omega^-$ baryon.

The correlation function is constructed using the BL quark propagator
described in Section~\ref{sec:LatticePS} computed from sources set
at time slices 0 and 32 for every twentieth trajectory, starting
from a thermalized one. Results are averaged over these sources and
blocked in bins of size 80 trajectories to reduce auto-correlations.
The total number of measurements made varies between about 100 and
200, depending on the light sea quark mass (see Table~\ref{tab:omega
masses}).  To avail ourselves of all possible statistics, we average
over forward propagating particle states, and backward propagating
anti-particle states, as well as all three spatial directions for
the interpolating operator.

The average 
correlation function is fit using a jackknife procedure where the covariance matrix is estimated for each jackknife block. The minimum time slice used in the fit was varied in the range $4 \le t_{min}\le 10$; the maximum was fixed to 14. Though the mass values do not depend
sensitively on the choice of $t_{min}$,
we chose to take as central values the masses obtained from $t_{min}=8$ to minimize excited state contamination without losing control of the statistical error and because $\chi^2$ is acceptable for this $t_{min}$ value. The baryon mass values for valence quark masses $m_x=0.03$ and 0.04 are summarized in Table~\ref{tab:omega masses} and plotted in Fig.~\ref{fig:omega masses}.

Also shown in the figure is a linear extrapolation, for each
degenerate valence mass, to the physical value of the light sea
quark mass (solid lines).  The linear fit to the $m_x = 0.03$ data
has a $\chi^2$/d.o.f. of 1.2 and the $m_s=0.04$ data has a
$\chi^2$/d.o.f. of 1.4.  In the next section we use these values
to interpolate to the physical strange quark mass in order to set
the lattice scale using the $\Omega^-$ baryon mass.

A systematic error in the $\Omega^-$ mass, and hence in the
lattice spacing, results from the fixed unphysical value of the
strange sea quark mass in the simulation.  This effect can be
estimated from the slope of the baryon mass with respect to the
light sea quark mass (see Fig.~\ref{fig:omega masses}).
The effect, which is similar
for both valence quark masses, is about 2--3 \% for a (required)
shift in sea quark mass of about 0.008. Since this results from
varying both light quarks together, it should be divided by two for
the single strange quark in the simulation.  This is about a one
percent downward shift in the mass, which is smaller than the
statistical
errors, as shown in the next section, so we ignore it.

%% file: text_sections/Su2.tex
%
%
\subsection{$\SU(2)\times\SU(2)$ Chiral Fits}
\label{subsec:SU2fit}

In this subsection we present a brief overview of the techniques we use to perform the chiral extrapolations for pseudoscalar masses and decay constants. For our simulations with 2+1 flavors of dynamical fermions, the traditional  approach would be to use the SU(3) chiral perturbation theory to guide the extrapolations. However, as we shall see in Section~\ref{sec:SU3fit}, NLO SU(3) ChPT does not describe our data for the pseudoscalar mesons which contain a quark as heavy as the strange quark.
In order to extrapolate our results reliably to the physical quark masses (and in particular to the physical strange quark mass) we therefore use a different approach
and impose only SU(2) chiral symmetry in the light quark sector, as has been discussed in Section~\ref{sec:ChPT}.
We divide our discussion into two parts. In Section~\ref{subsubsec:pionsector} we discuss the chiral extrapolations in the pion sector and in Section~\ref{subsubsec:kaonsector} we extend the discussion to kaons. We present physical results for the quark masses and pseudoscalar decay constants using the SU(2) chiral fits in Section~\ref{sec:phys_result} below.

\subsubsection{The Pion Sector}\label{subsubsec:pionsector}

In the pion sector, the chiral dynamics of $\SU(2)_L\times \SU(2)_R$ symmetry is sufficient to
describe the physics we are interested in. The effects of the dynamical strange quark are
fully contained in the low-energy constants of $\SU(2)_L\times\SU(2)_R$ ChPT
~\cite{Gasser:1983yg}. There is a caveat. A posteriori we learn that
the dynamical strange quark mass used in our simulations is a little larger than the physical
one.  Ideally, we would need to
perform the simulations with several different dynamical strange quark
masses and interpolate the results to the physical strange quark mass. In the absence
of such simulations, the dynamical strange quark mass is fixed to
$m_h = 0.04$, and we need to be aware that the corresponding LECs will have a corresponding small systematic error (see\ Sec.~\ref{sec:sysErr:ms}).

The next-to-leading order partially quenched SU(2) formulae for the squared masses, $m_{xy}^2$, and decay
constants, $f_{xy}$, of the light pseudoscalar mesons composed of two
valence quarks with masses $m_x$ and $m_y$ can be derived
from \cite{Sharpe:2000bc}, as has been done, for instance, in \cite{Farchioni:2003bx}. For reference, we
summarize these formulae in Appendix \ref{sec:app:su2}. As discussed in Section~\ref{sec:ChPT}, for domain wall fermions the effective quark mass is the sum of the input quark parameters $m_{x,y,l,h}$ and the residual mass, which we denote as $\tm_{x,y,l,h} \equiv m_{x,y,l,h} + \mres$ (here $l$ and $h$ denote the light and heavy sea quarks respectively).
To next-to-leading order, this is the only correction
to the continuum chiral formulae. Since $f_{xy}$ and $m_{xy}^2$ share two unknown low energy
constants $B$ and $f$, we perform combined fits to both of
them, using the two ensembles with dynamical light quark masses of $m_l=0.005$
and $0.01$.  Such combined fits with valence quark masses $m_x,\, m_y = 0.001,0.005$, and 0.01 (which therefore satisfy the cut
$m_{\rm avg} \equiv(m_x+m_y)/2 \leq0.01$) are shown in Fig.~\ref{fig:SU2fits}.
The values of
the fit parameters are given in Table~\ref{tab:fitSU2}, where we quote the scale-dependent low energy constants
at two commonly used renormalization scales $\Lambda_\chi = 770$~MeV and 1.0~GeV. These partially quenched LECs (of $\SU(4|2)$ PQChPT) can be related to those in the physical unitary theory ($\SU(2)$), $l_3^r$ and $l_4^r$, by Eqs.~(\ref{eq:chPTsu2:LECrel:l3}) and (\ref{eq:chPTsu2:LECrel:l4}). In Table~\ref{tab:fitSU2} we also quote the values of $\bar{l}_3$ and $\bar{l}_4$, which are conventionally defined at the scale of the pion mass, see Eq.~(\ref{eq:chPTconv:barl}).

Using partially quenched NLO $\SU(2)$ chiral perturbation theory we are able to describe the data with the mass cut $m_{\rm avg} \equiv(m_x+m_y)/2 \leq m_{\rm cut}=0.01$. It is not possible however, to extend the range of masses significantly.  For example, the quality of the fit degrades significantly if instead we impose a mass cut of $m_{\textrm{avg}}\le 0.02$: the $\chi^2/{\rm d.o.f}$ increases from 0.3 ($m_{\rm cut}=0.01$) to 1.2 ($m_{\rm cut}=0.015$) and even 3.6 for $m_{\rm cut}=0.02$. (We only report the uncorrelated $\chi^2/{\rm d.o.f.}$ since we could not reliably determine the correlations, so these numbers most likely underestimate the ``true'' deviation from the fit.) This suggests that higher orders of chiral perturbation theory (NNLO or even more) are needed in this mass range. (See also the discussion in Sec.~\ref{sec:su2:sysErr:chPT} regarding the inclusion of analytic NNLO terms.)

\subsubsection{The Kaon Sector}\label{subsubsec:kaonsector}

In order to incorporate the valence strange quark mass in the chiral extrapolations, we need to extend the SU(2) ChPT to the kaon sector. As has been
discussed in Section~\ref{sec:ChPT}, this can be done by considering the kaon as a heavy meson under the assumption that $\tm_l \ll \tm_s$, in analogy
to the heavy-light meson (or heavy baryon) chiral perturbation theory~\cite{Sharpe:1995qp,
Booth:1994hx}. In this framework
the chiral expansions are ordered in powers of $\tm_l/\tm_h$ or equivalently $m_\pi^2/m_K^2$, as well as $m_\pi^2/\Lambda_\chi^2$.
We present the NLO chiral formulae
for $m_{xy}^2$ and $f_{xy}$, the masses and decay constants of pseudoscalar mesons composed of
a light valence quark with mass $m_x$ and a heavy valence quark with mass $m_y$
in Eqs.~(\ref{eq:fxspqfinal}) and (\ref{eq:mksqpq}) (see also App.~\ref{sec:app:su2kaon} for the complete set of formulae at NLO).

For kaons a new set of low-energy constants $f^{(K)}$, $B^{(K)}$ and $\lambda_{1,2,3,4}$
are introduced. These LECs depend on the valence and dynamical strange quark masses and we
have made this explicit by assigning a functional dependence, writing for instance, $f^{(K)}(m_h)$. Here $m_h$
represents both the masses of the heavy (strange) dynamical and valence quarks.
The parameters $B$ and $f$ are the usual LECs in the (pionic) SU(2) theory, and their values
are determined from the quantities in the pion sector, {\em i.e.}, their values are fixed to the ones given
in Table~\ref{tab:fitSU2}.

The strange quark mass dependence in the formulae for the meson masses and decay constants, Eqs.~(\ref{eq:fxspqfinal}) and (\ref{eq:mksqpq}) are contained in the low energy constants. Since it is not possible to tune the dynamical heavy quark mass $m_h$ perfectly to the mass $m_s$ of the physical strange quark, in principle we would wish to perform the calculations at several values of $m_h$ and interpolate the results to $m_s$. Since we only have results at a single value of $m_h$ ($m_h=0.04$) we rely as far as possible on using the behavour of our results with the valence strange quark masses, $m_y$, to extrapolate to the physical kaon.

Finally, we briefly summarise our fitting procedure. We start by fitting the mass dependence of
$m_{xy}^2$ and $f_{xy}$ using Eqs.(\ref{eq:fxspqfinal}) and (\ref{eq:mksqpq})
for $m_x \in [0.001, 0.01]$ and $m_y =  0.04$, with $B$ and $f$ fixed to the values in Table~\ref{tab:fitSU2}. Since these two formulae do not have any unknown parameters in common, the fits can be done independently.  We
then extrapolate to the physical value of the light quark mass to obtain $m_K(m_y = 0.04)$ and $f_K(m_y=0.04)$. (The physical light quark mass $m_{ud}$ is determined from the $\SU(2)$ pion sector fits, see \ref{sec:su2:detMassScale} for a detailed description.)
This step is shown in Fig.~\ref{fig:KSU2_ms0.04}, where the circles give the extrapolated values
$m_K(m_y=0.04)$ (right) and $f_K(m_y=0.04)$ (left).
Then similar fits are performed with $m_y = 0.03$, giving values for $m_K(m_y=0.03)$ and $f_K(m_y=0.03)$. Eventually,
we take the results at $m_y = 0.03$ and 0.04 and interpolate linearly between them.
As discussed in Sec.~\ref{sec:phys_result}, we define the physical strange quark mass from the interpolation of $m_{xy}^2$ to the physical value of $m_K^2$. Having determined the physical value of the strange quark mass, we
extract the kaon decay constant from the interpolation of $f_{xy}$ to this mass. The linear fits are shown in Fig.~\ref{fig:KSU2_interp}. In the same figure we also include the points obtained with $m_y=0.02$. Although at such a low mass the use of $\SU(2)$ chiral perturbation theory is questionable, we see that nevertheless the points lie very close to the line obtained from the interpolation between $m_y=0.04$ and $m_y=0.03$, suggesting that
the higher-order corrections
of $O(\tm_l/\tm_h)^2$ may be small.

\subsection{Lattice Scale, Physical Quark Masses and Decay Constants}
\label{sec:phys_result}
Having established the procedure for the chiral extrapolations, we now turn to obtaining the
physical results. In this subsection we start by explaining how the lattice scale and physical bare quark masses were determined. We then use the renormalization constant for the quark masses which was determined non-perturbatively in \cite{Aoki:2007xm} to obtain the physical quark masses in the $\overline{\rm MS}$ scheme at $\mu=2\,{\rm GeV}$.
Finally we present the results for $f_\pi$, $f_K$ and the CKM matrix element $|V_{us}|$ determined from our measured ratio $f_K/f_\pi$ together with experimental inputs.

\subsubsection{Determination of $m_{ud}$, $m_s$ and $a^{-1}$}
\label{sec:su2:detMassScale}

In order to determine the quark masses $m_{ud}$, $m_s$ and the lattice spacing $a$, we need to compare our lattice results for three quantities to their physical values. We take the pseudoscalar masses $m_\pi$ and $m_K$ as two of these. Natural choices for the third would be the $\rho$ mass or the Sommer scale. However, since the $\rho$ meson is a resonance with a finite width and there is an uncertainty of order 10\% in the value of the Sommer scale, we choose instead to use the mass of the $\Omega$ baryon. This is a
state composed of three valence
strange quarks. One advantage of using this baryon mass, is
that up to NLO in $\chi$PT it is free of logarithms containing the
light quark masses \cite{Toussaint:2004cj,Tiburzi:2004rh}.
Therefore the extrapolation of
the measured masses
to the light physical mass can be readily performed using a linear ansatz
without an uncertainty due to chiral logarithms.

The physical quark masses were obtained from the $\SU(2)\times\SU(2)$ fits described in
Section~\ref{subsec:SU2fit}. For the average light quark mass, $m_{ud} \equiv (m_u + m_d)/2$ ,
 we solved for a pion mass
of $m_\pi=135.0\,{\rm MeV}$, corresponding to the physical neutral pion mass,
while for the strange quark mass $m_s$
the fit to the kaon mass was solved at $m_K=495.7\,\rm{MeV}$, which is the
quadratically averaged neutral and charged kaon mass.

The determination of the three parameters is a coupled problem. The lattice scale is needed
to convert masses between lattice units and physical ones, whereas the quark masses are
needed for the extrapolation and interpolation in the light and strange quark
masses to obtain the mass of the $\Omega$ baryon.
We performed the determination iteratively, starting
with an initial ``guess" for the quark masses, fixing the lattice spacing by requiring that $m_\Omega$ takes its physical value, then using this value of $a$ to adjust the quark masses by requiring that $m_\pi$ and $m_K$ take their physical values and so on until no further
changes in the parameters were observed.
The final values for $1/a$,
$a$, $m_{ud}$, $m_s$ can be found in Table~\ref{tab:results}, with only the
statistical errors shown. (We will discuss the systematic errors in Sec.~\ref{sec:su2:sysErr}.)
We also find the quark mass ratio to be $\widetilde{m}_{ud}:\widetilde{m}_s\,=\,1:28.8(4)$.

An independent check of the lattice scale has been done from the spectrum of heavy-heavy
and heavy-strange quark states using the non-perturbatively determined
relativistic heavy quark action~\cite{Lin:2006ur,Christ:2006us}. The
preliminary analysis~\cite{Li:2007en} gave $1/a = 1.749(14)$ GeV, where
the error is statistical only. The agreement of this result with that obtained from the
$m_{\Omega^-}$ shown in Table~\ref{tab:results} suggests that the associated systematic errors are small.

\subsubsection{Quark Masses in $\ov{MS}$ Scheme}
\label{sec:su2:qmass}

In this section we present the results for the quark masses in the $\ov{MS}$ renormalization scheme.
The renormalization factor $Z_m=1/Z_S$ (where $S$ represents the scalar density) needed to convert the bare quark masses to the commonly used $\overline{\rm MS}$ scheme at a scale of 2 GeV has
been calculated with the same action on a  $16^3\times32\times16$ lattice
\cite{Aoki:2007xm}. (For details about this ensemble of
configurations, see\ \cite{Allton:2007hx}.) We first matched the bare
lattice operators to the RI-MOM scheme using the non-perturbative
Rome-Southampton technique \cite{Martinelli:1994ty}, and then performed a perturbative
matching to the $\overline{\rm MS}$ scheme.
The relation between the renormalized and bare quark masses is given by
\begin{equation}
m_{X}^{\overline{\rm MS}}(2\ {\rm GeV})\;=\;Z_m^{\overline{\rm MS}}(2\,{\rm GeV})\ a^{-1}\,(a\widetilde{m}_{X})\,,
\end{equation}
where $X=ud,\,s$ and for clarity we made explicit the factors of the lattice spacing $a$. In Ref.~\cite{Aoki:2007xm}
it was found that $Z_m^{\overline{\rm MS}}(2\,{\rm GeV})=1.656(48)(150)$, where
the first error is statistical, while the second is the combined systematic uncertainty due
to residual chiral symmetry breaking, the use of a three-loop matching factor between the RI-MOM and $\overline{\rm MS}$ schemes and mass effects.
With this renormalization constant, the results for the average light quark mass and the strange quark mass are determined to be
\begin{eqnarray}
m_{ud}^{\overline{\rm MS}}(2 \, {\rm GeV}) &=& 3.72(16)(33)\,{\rm MeV}, \\
m_s^{\overline{\rm MS}}(2 \, {\rm GeV})    &=& 107.3(4.4)(9.7)\,{\rm MeV},
\label{eq:qmass}
\end{eqnarray}
where the first error is the combined statistical error from the bare
lattice quark masses, the lattice spacing and $Z_m^{\overline{\rm MS}}$ and the second error is the
systematic error in $Z_m^{\overline{\rm MS}}$.

Related to the renormalization of the quark masses is the renormalization of the lowest order LEC $\btwo$, which is proportional to the quark condensate and therefore is renormalized by $Z_S=1/Z_m$, ensuring that the product of $\btwo$ and a quark mass does not depend on the renormalization scheme or scale. We have
\begin{eqnarray}
\btwo^{\overline{\rm MS}}(2\,{\rm GeV})&=&\left(Z_m^{\overline{\rm MS}}(2\,{\rm GeV})\right)^{-1}\,\cdot\,(1/a)\,\cdot\,a\btwo\\
 &=& 2.52(11)(23)\,{\rm GeV}\,,
\end{eqnarray}
where the first error is statistical and the second is the systematic uncertainty from renormalization. The renormalized value for the chiral condensate, $\Sigma=\ftwo^2\btwo/2$, reads
\begin{equation}
\Sigma^{\overline{\rm MS}}(2\,{\rm GeV})\;=\;\Big(255(8)(8)\,{\rm MeV}\Big)^3\,.
\end{equation}

\subsubsection{$f_\pi$, $f_K$ and $|V_{us}|$}
The chiral fits described in Section~\ref{subsec:SU2fit} also allow us to
determine the decay constants $f_\pi$ and $f_K$. Extrapolation
to the physical light quark mass $m_{ud}$ using the SU(2)$\times$SU(2) fits in
the pion sector gives $f_\pi=124.1(3.6)\,{\rm MeV}$, while interpolation  to
the physical strange quark mass $m_s$, as described in
Section~\ref{subsec:SU2fit}, gives $f_K=149.6(3.6)\,{\rm MeV}$. In both cases only the
statistical errors are given.  Systematic errors will be discussed in Section \ref{sec:su2:sysErr}.
These results can be compared to the physical values
\cite{Yao:2006px}, $f_\pi=$130.7(0.1)(0.36) and $f_K=$159.8(1.4)(0.44)~MeV.
For the ratio of the decay constants we find $f_K/f_\pi=1.205(18)$ compared to the experimental value
of 1.223(12).
See \footnote{In a recent publication \cite{Bernard:2007tk} an updated measurement of the CKM matrix element $V_{ud}=0.97418(26)$ \cite{Towner:2007np} from super-allowed nuclear $\beta$-decays has been used to determine $f_K/f_\pi=1.192(07)$ and $f_\pi=130.5(0.1)\,{\rm MeV}$ resulting in $f_K=155.6(0.9)\,{\rm MeV}$. (There $1=V_{ud}^2+V_{us}^2$ has been assumed, resulting in $V_{us}=0.2258(11)$.)} for possible implications of an updated result for the measured value of the CKM matrix element $V_{ud}$ \cite{Towner:2007np}.

An important application of the result for $f_K/f_\pi$ is the
determination of the CKM matrix element $|V_{us}|$, as has been pointed out in
\cite{Marciano:2004uf}. Using the experimentally determined branching ratios
$\Gamma(K\rightarrow\mu\nu(\gamma))$ and $\Gamma(\pi\rightarrow\mu\nu(\gamma))$, together with the
radiative electroweak corrections from \cite{Yao:2006px}, we obtain:
\begin{equation}
 |V_{us}|/|V_{ud}|=0.2292(034)_{\rm stat}(118)_{\rm syst}(005)_{\rm other},
\end{equation}
where the first two errors are statistical and systematic (as discussed in Sec.~\ref{sec:su2:sysErr} below), while the third error is the combined uncertainty from the measurement of the branching ratios and the radiative electroweak corrections \cite{Yao:2006px}.
Taking  $|V_{ud}|=0.97377(27)$
from super-allowed nuclear $\beta$-decays \cite{Yao:2006px}, we obtain
\begin{equation}\label{eq:vus}
|V_{us}|\;=\;0.2232(033)_{\rm stat}(115)_{\rm syst}(005)_{\rm other}
\end{equation}
(The third error now also contains the uncertainty in $|V_{ud}|$). This implies
\begin{equation}\label{eq:ckmunitarity1}
|V_{us}|^2+|V_{ud}|^2 = 0.9980(15)_{\rm stat}(51)_{\rm syst}(06)_{\rm other} \equiv 0.9980(54)_{\rm total}.
\end{equation}
Since $|V_{ub}|$ is negligible, our result implies that the unitarity relation is satisfied within the uncertainties, constraining the possible breaking of quark-lepton universality in models of new physics.
For completeness we now present the corresponding results obtained using the recently published value $|V_{ud}|=0.97418(26)$ \cite{Towner:2007np} (instead of the PDG value of $|V_{ud}|=0.97377(27)$ which was used in obtaining the results in Eq.~(\ref{eq:vus}) and Eq.~(\ref{eq:ckmunitarity1}) above):
\begin{eqnarray}
 |V_{us}| &=& 0.2232(033)_{\rm stat}(115)_{\rm syst}(005)_{\rm other}\,,\\
 |V_{us}|^2+|V_{ud}|^2 &=& 0.9989(15)_{\rm stat}(51)_{\rm syst}(06)_{\rm other} \equiv 0.9989(54)_{\rm total}\,.
\end{eqnarray}
We have also determined $V_{us}$ (with a smaller error) from the experimentally measured rates of semileptonic $K_{\ell 3}$ decays, by computing the form factor $f^+(q^2=0)$, where $q$ is the momentum transfer. The results (in particular $|V_{us}|=0.2249(14)$) and an outline of the calculation are presented in \cite{Boyle:2007qe}.

\subsection{\label{sec:su2:sysErr}Systematic Errors}

In this section we will discuss the systematic errors resulting from various sources: 1. chiral
extrapolations; 2. finite volume effects; 3. the unphysical dynamical strange
quark mass and 4. scaling errors.

\subsubsection{\label{sec:su2:sysErr:chPT}Errors arising from the chiral extrapolations}

As explained above, we obtain our results by using NLO chiral perturbation theory to fit the
measured values of the meson masses and decay constants in the range of valence quark masses satisfying $m_{\rm avg}\leq0.01$ and with $m_l=0.005$ and $0.01$. We do not have enough data to extend the analysis fully to NNLO where the chiral logarithms depend on the Gasser-Leutwyler coefficients and there are too many parameters to make the fits meaningful. In order to estimate the effects of the
(neglected) higher order terms in the chiral expansion we have therefore extended the range of quark masses to
$m_{\rm avg}\leq0.02$ but included only the analytic NNLO-terms in the fit functions. In the pion sector, the analytic terms at NNLO are quadratic in the $\chi$'s and symmetric under $\chi_x\leftrightarrow\chi_y$; there are four such terms:
\begin{equation}
 (\chi_x+\chi_y)^2\,,\;(\chi_x-\chi_y)^2\,,\;(\chi_x+\chi_y)\chi_l\quad\textrm{and}\quad \chi_l^2\,.
\end{equation}
We therefore have four new parameters for the behavour of the mass of the pseudoscalar mesons and four for the decay constants. Unfortunately, with only two values for the dynamical sea quark mass (which is proportional to $\chi_l$), we were not able to resolve the complete sea quark mass dependence up to NNLO in our fits. We therefore dropped the term proportional to $\chi_l^2$ and were then able to obtain stable fits with a good $\chi^2$/d.o.f. (about 0.1)\,. We stress that the NNLO analytic terms are necessary in order to obtain good fits; NLO chiral perturbation theory does not represent our data in the extended range of masses $m_{\rm avg}\leq0.02$. With the NNLO analytic terms included, not only do we obtain good fits to our data but the behavour of the chiral expansion is sensible, {\em i.e.} the relative importance of
successive terms in the series is as expected. We illustrate this point in Fig.~\ref{fit:Su2:nnloContr}, where, in the top two graphs, we plot the size of LO, NLO, and NNLO contributions (normalized such that the LO contribution equals one) for the masses-squared and decay constants of mesons with $m_x=m_y=m_l$ ({\em i.e.} for mesons in the unitary theory).
Up to a quark mass of 0.01 (the 3rd vertical dashed line in the plots, counting from the left) the NLO-contributions are much larger than the NNLO terms, which are negligible in this region.
As expected, the NNLO corrections become more important for larger quark masses and in particular, for masses close to 0.02 (rightmost vertical dashed line) they represent a significant fraction of the decay constant.
This study is reassuring, but, since we have neglected the chiral logarithms at NNLO terms, it is only a phenomenological estimate of the likely contributions from higher order terms in the chiral expansion.

As our final estimate of the errors due to the chiral extrapolation we take
\textit{twice} the difference between the results from our standard NLO $\SU(2)$-ChPT fit in the mass range $m_{\rm avg}\leq0.01$ and those from
including the analytic NNLO terms in the range $m_{\rm avg}\leq0.02$.
The factor of 2 was included, rather cautiously, because of the phenomenological nature of the NNLO analysis, and in particular because we were unable to study the sea-quark mass dependence.
The systematic error is reported in Table~\ref{tab:Su2:systErrSummary}.

The systematic error due to the extrapolation in the mass of the
light quark in the kaon quantities, $m_s$ and $f_K$, was also
estimated as twice the difference between our standard results and
those obtained using kaon $\SU(2)$ PQChPT in the mass range
$m_x\leq0.02$ with the values of $f$ and $B$ obtained from the NNLO
fit described above. The errors are also tabulated in
Table~\ref{tab:Su2:systErrSummary}. For this case we get a good
description of our data up to $m_x=0.02$ without the necessity of
adding NNLO analytic terms ($\chi_x^2$, $\chi_l^2$, and $\chi_x\chi_l$).
Therefore one expects those terms to have a negligible effect even
in the mass range $m_x\leq0.02$.  The lower panels in
Fig.~\ref{fit:Su2:nnloContr} show the LO and NLO contributions,
normalized so that the LO contribution equals one, to $m_{lh}^2$
and $f_{lh}$.  The horizontal axis is $\chi_l$, which is proportional
to the total light sea quark mass.  We see that the size of the NLO
contributions are at most 35\% for the range of masses where we have
data.

\subsubsection{\label{sec:su2:sysErr:FV}Finite volume effects}

In this section we estimate the errors due to the fact that the simulations are performed on a finite volume, $(aL)^3\,\approx\, (2.74\,{\rm fm})^3$. We do this by following the procedure proposed by Gasser and Leutwyler \cite{Gasser:1986vb,Gasser:1987ah,Gasser:1987zq} in which one compares results obtained
in ChPT using meson propagators in finite volume,
$G^{(L)}(x)$, to those obtained with the infinite-volume propagators, $G(x)$.
The finite-volume and infinite-volume propagators are related by
\begin{equation}
 G^{(L)}(x)\;=\;G(x)\:+\:\sum_{\vec{n}\neq0}G(x+L\vec{n})\,,
\end{equation}
with the 3-vector $\vec{n}\in\mathbb{Z}^3$.
This allows one to calculate the corrections in the PQChPT formulae for the meson masses and decay constants directly. We list the expressions in Appendix \ref{sec:appendix_fv}, where details of the numerical implementation can also be found. (See also refs. \cite{Bernard:2001yj,Aubin:2003mg,Aubin:2003uc} for the analogous discussion with staggered PQChPT.)
We now repeat the chiral fits using PQChPT with the finite-volume propagators and obtain a new set of fit parameters. For each physical quantity, we assign the difference between the results obtained using the finite-volume and infinite-volume formulae as the systematic error and tabulate these finite-volume errors in Table~\ref{tab:Su2:systErrSummary} for both the pion and the kaon sectors. In the latter case we find that the finite-volume corrections in $m_s$ and $f_K$ are actually negligible.
In Fig.~\ref{fig:Su2:fv} we plot the correction factor for the decay constants and squared masses, $(1-\Delta_{xy}^{L\,f})$ and $(1-\Delta_{xy}^{L\,m^2})$ (cf. App.~\ref{sec:appendix_fv} for their definition), respectively, for our two values of the dynamical light quark mass ($m_l=0.01$ and $0.005$).
Since the volume is reasonably large, these corrections are found to be below 1 percent, except for the very light valence masses.

We now compare our results for the finite-volume corrections obtained using ChPT (as described above) to the theoretical predictions by Colangelo et al.~\cite{Colangelo:2005gd}, who use a resummed L\"uscher formula.
The results in ref.~\cite{Colangelo:2005gd} are presented for the unitary theory ($m_x=m_y=m_l$ in our notation) and so we restrict the comparison to this case. We take the results in Tabs.\ 3 and 4 of \cite{Colangelo:2005gd} which we interpolate to the volume and pion masses used in our simulation. In Table \ref{tab:FVcompareCDH} we present the finite-volume corrections for the pion mass and decay constant estimated using 3 methods, SU(2) and SU(3) ChPT and that of Ref.~\cite{Colangelo:2005gd}. The quantities $R_m$ and $R_f$ are defined by
\begin{equation} R_m\;=\;\frac{m_{ll}(L) - m_{ll}(\infty)}{m_{ll}(\infty)}=\frac12\,\Delta_{ll}^{L\,m^2}\,,\;\;\;R_f\;=\;\frac{f_{ll}(L) - f_{ll}(\infty)}{f_{ll}(\infty)}=\Delta_{ll}^{L\,f}\,,\end{equation}
Within the uncertainties, we see a reasonable agreement between the estimates obtained with the different methods.

In an earlier publication~\cite{Allton:2007hx}, we presented results for the masses and decay constants obtained with the same action and coupling as in this paper, but on a smaller volume, about $(1.83\,{\rm fm})^3$ compared to $(2.74\,{\rm fm})^3$ used here.
(In~\cite{Allton:2007hx} a different method was used to extract the
lattice scale. Since the gauge action and coupling are identical, for consistency we quote here the volume obtained with the value for $a^{-1}$ used throughout this work.).

Specifically the lattice in~\cite{Allton:2007hx} has 16 points in each spatial direction as compared to 24 in this paper. We are therefore able to
compare the theoretical estimates of the finite-volume corrections described above, with the difference of our results on the two lattices with different volumes.
We compare the measured pion mass and decay constant at the common unitary point at $m_x=m_y=m_l=0.01$, $m_h=0.04$ for which the pion mass is $\approx 427\,{\rm MeV}$ (this is the smallest mass for which we have results on both lattices). On the smaller volume, both finite-volume corrected fits ($\SU(2)$ and $\SU(3)$) give results for the correction factors comparable to those interpolated from Colangelo et al..
To obtain these correction factors, we used the parameters obtained from the fits to the $24^3$ data and evaluated the finite-volume corrections for $L=16$ (instead of the original $L=24$). The $\SU(2)$ fits resulted in $R_m^{(16c)}=0.42(04)\%$ and $R_f^{(16c)}=1.66(14)\%$, where the superscript is meant to indicate that this is the finite-volume correction for a spatial volume of $L^3=16^3$. In Table~\ref{tab:FVcompare16c} we list the measured values of the meson masses and decay constants on the $24^3$ and $16^3$ lattices and compare their ratios to the prediction from finite volume ChPT and the resummed L\"uscher formula (interpolated from \cite{Colangelo:2005gd}). We obtain this prediction from the ratio of the finite-volume correction factor for the $16^3$ spatial volume and that for the $24^3$ spatial volume,
\begin{equation}\label{eq:su2:FVcomp}
\frac{m_\pi^{(16c)}}{m_\pi^{(24c)}}\:\overset{?}{=}\:\frac{1-R_m^{(16c)}}{1-R_m^{(24c)}}\,,\;\;\;
\frac{f_\pi^{(16c)}}{f_\pi^{(24c)}}\:\overset{?}{=}\:\frac{1-R_f^{(16c)}}{1-R_f^{(24c)}}\,.
\end{equation}
The superscripts in Eq.(\ref{eq:su2:FVcomp}) indicate the spatial volumes on which the results for $m_\pi$ and $f_\pi$ were obtained or for which the finite-volume corrections $R_m$ and $R_f$ were evaluated. The question mark reminds us that we are checking whether the measured values of the ratios are equal to the theoretical predictions for the finite-volume effects. Whereas (within the errors) the observed $(2.0\pm1.2)\%$ effect for the pion mass is somewhat better reproduced by the FV ChPT predictions than the $(3.7\pm1.1)\%$ for the decay constant, in both cases all three method tend to underestimate the observed FV effects. We make the following two comments.
First, the precision of FV ChPT at NLO and the resummed L\"uscher predictions for the finite-volume corrections are expected to improve with increasing volume. A possible reason for underestimating the difference between the results from the $16^3$ and $24^3$ lattices may be that the smaller volume is already borderline for the methods used here. Whereas the values for $m_\pi L$ would appear to be sufficient in both cases; on the $16^3$ lattice $m_\pi L\approx 4.0$ compared to $\approx 5.8$ on the $24^3$ lattice, it may be that the smaller lattice does satisfy the relation $L\,\gg\,(\sqrt{2} f_\pi)^{-1}\,\sim\,1\,{\rm fm}$ sufficiently. The second comment is that, as has already been pointed out in Sec.~\ref{sec:LatticePS}, the errors on the measured quantities on the smaller volume in general, and on the decay constants in particular, may be underestimated due to shorter plateaus. Furthermore, in this present paper we have obtained the decay constants using an improved ansatz (see also Sec.~\ref{sec:LatticePS}). For these reasons, we believe that the finite volume effects on the larger lattice are well described by our approach and give a reliable estimate for the systematic error.

\subsubsection{Effects of the unphysical dynamical strange quark mass}\label{sec:sysErr:ms}

In this subsection we estimate the systematic error due to the fact that the mass of the dynamical heavy quark ($m_h$) turns out, a posteriori, to be about 15\% larger that of the physical strange quark ($m_s$). To perform this estimate we exploit the fact that the LECs of SU(3) ChPT are independent of the quark masses, and use the NLO conversion formulae relating SU(3) and SU(2) LECs~\cite{Gasser:1984gg} given in  Eqs.~(\ref{eq:chPTconv:F}--\ref{eq:chPTconv:l4}), together with the results for the $\SU(3)$ LECs presented in Sec.~\ref{sec:SU3fit}. In this way we obtain the shifts in the SU(2) LECs:
\begin{eqnarray}
\btwo(m_s)-\btwo(m_h) &=& \bthree\,\bigg\{\frac{16}{\fthree^2} (2\lthree{6}-\lthree{4}) \Big(\chi_s-\chi_h\Big)\,-\,\frac1{72\pi^2\fthree^2}\Big(\chi_s\log\frac{2\chi_s}{3\Lambda_\chi^2}-\chi_h\log\frac{2\chi_h}{3\Lambda_\chi^2}\Big)\bigg\}\,,\nonumber\\\\
\ftwo(m_s)-\ftwo(m_h) &=& \fthree\,\bigg\{\frac{8}{\fthree^2} \lthree{4} \Big(\chi_s-\chi_h\Big)\,-\,\frac1{32\pi^2\fthree^2}\Big(\chi_s\log\frac{\chi_s}{2\Lambda_\chi^2}-\chi_h\log\frac{\chi_h}{2\Lambda_\chi^2}\Big)\bigg\}\,,\\
l^r_3(m_s)-l^r_3(m_h) &=& \frac1{576\pi^2}\log\frac{\chi_h}{\chi_s}\,,\\
l^r_4(m_s)-l^r_4(m_h) &=& \frac1{64\pi^2}\log\frac{\chi_h}{\chi_s}\,.
\end{eqnarray}
(Note: here one has to use $\bthree$ to determine $\chi_X$.) We use the shifted values of the SU(2) LECs to determine $m_{ud}$ and $f_\pi$ and again use the shift in those two quantities as the estimate of the corresponding systematic uncertainties. All these uncertainties are presented in
Table~\ref{tab:Su2:systErrSummary} and are small compared to the total errors.
(We did not quantify the systematic error on the separate $\ltwo{i}$, just on the $\bar{l}_{3,4}$ since these are the phenomenologically relevant observables.)

For the kaon sector the shift due to changing the dynamical heavy quark mass to its physical value can be obtained from Eqs.~(\ref{eq:chPTconv:FKaon}--\ref{eq:chPTconv:lam4}). We perform the conversion for fixed $m_y$ at $m_y=0.03$ and 0.04 separately and then in the same way as before linearly interpolate the heavy valence quark mass $m_y$ to $m_s$ to determine the shift in the kaon mass and the kaon decay constant. 
A shifted value for the physical strange quark mass is obtained by
solving the following system. Find a value of $m_s$ such that the
interpolation in $m_y$ to $m_y=m_s$ between $m^2_{ud\,y}(m_s,m_y)$
between $m_y=0.03$ and 0.04 gives the squared physical kaon mass.
Here $m^2_{ud\,y}(m_s,m_y)$ was obtained from the unitary extrapolation
in the light quark mass to $m_{ud}$ using shifted LECs.  The shifted
LECs are found by starting from
their values at the dynamical heavy quark mass used in the simulation
and changing to their values at the new value of $m_s$ (at fixed $m_y$).
By this method, we observe a 2\% shift in $\widetilde{m}_s$ and less than 0.5\% shift in $f_K$. Therefore, conservatively we will quote the systematic error from $m_h\neq m_s$ to be 2\% and 1\% for $\widetilde{m}_s$ and $f_K$, respectively.

\subsubsection{\label{sec:Su2:sysErr:scale}Scaling errors}

At present we only have data at a single value for the lattice spacing ($a=0.11\,{\rm fm}$) and so cannot perform a scaling study to extrapolate our results to the continuum limit. (We are currently generating an ensemble of configurations on a finer lattice and so will soon be able to study the discretization errors directly.)
We therefore take our central values and assign to them a systematic uncertainty of 4\% due to the missing continuum extrapolation. The 4\% is simply an estimate of $(a\Lambda_{\textrm{QCD}})^2$ with our value of $a^{-1}=1.73\,{\rm GeV}$, but we stress that it is only when we will have results at more than one value of the lattice spacing that we will be able to quantify this error reliably.

Table \ref{tab:Su2:systErrSummary} contains the 4\% estimate for scaling error for the measured quantities. Again, we did not calculate this uncertainty for the (phenomenologically uninteresting) LECs $\ltwo{i}$, but only for $\bar{l}_{3,4}$. Here one has to keep in mind, that a relative error affects the universal low energy scales
\begin{equation}
 \Lambda_{3,4}\;=\;\Lambda_\chi\cdot\exp\left(\frac{16\pi^2}{\gamma_{3,4}}l_{3,4}^r\right)\;=\;(139\,{\rm MeV})\cdot\exp({\bar{l}_{3,4}/2})\,,
\end{equation}
meaning that a 4\% uncertainty translates into an \textit{absolute} error of 0.08 for $\bar{l}_{3,4}$.

\subsection{Comparison of our results with other recent determinations.}

In this subsection we compare our results for the $\SU(2)$ LECs to those obtained in the continuum \cite{Bijnens:1998fm,Colangelo:2001df} and in other lattice simulations with either $N_f=2+1$ \cite{Aubin:2004fs,Bernard:2007ps} or $N_f=2$ \cite{DelDebbio:2006cn,Urbach:2007rt} dynamical fermions. We also compare the values we obtain for the quark masses with those from other recent simulations \cite{Blum:2007cy,Blossier:2007vv,Gockeler:2006jt,Bernard:2007ps,Ukita:2007cu,Ishikawa:2007nn}.

Our value for the decay constant in the $\SU(2)$ chiral limit, $f=114.8(4.1)_{\rm stat}(8.1)_{\rm syst}$ MeV, is consistent, within our uncertainties, with the phenomenological estimates of 122.3(0.5) \cite{Bijnens:1998fm} or 121.9(0.7) \cite{Colangelo:2003hf}. For the ratio $f_\pi/f$ we obtain 1.080(08) (with only the statistical error quoted, since the numerator and denominator are likely to be highly correlated), which agrees very well with the phenomenological results 1.069(04) \cite{Bijnens:1998fm} and 1.072(07) \cite{Colangelo:2003hf}. Previous lattice simulations give, e.g., $1.052(2)(^{+6}_{-3})$ \cite{Bernard:2007ps} or 1.075 \cite{Urbach:2007rt} for the ratio and for $\ftwo$ the results read $\ftwo=124.2\,{\rm MeV}$ \cite{Bernard:2007ps} and $121.6\,{\rm MeV}$ \cite{Urbach:2007rt}, respectively, where in Ref.\,\cite{Urbach:2007rt} no estimate of the error was provided.

From MILC's results for $\ftwo$ and the chiral condensate, $(\Sigma^{\overline{\rm MS}}(2\,{\rm GeV}))^{\nicefrac{1}{3}}\,=\,278(1)_{\rm stat}(5)_{\rm ren}(^{+2}_{-3})_{\rm syst}\,{\rm MeV}$ \cite{Bernard:2007ps} (compare to $255(8)_{\rm stat}(8)_{\rm ren}(13)_{\rm syst}\,{\rm MeV}$ obtained in this paper), one can via $\Sigma=f^2B/2$ deduce their result for the LO-LEC $\btwo^{\overline{\rm MS}}(2\,{\rm GeV})=2.79\,{\rm GeV}$ (no estimate for the error). Comparing this number with our value of $2.52(.11)_{\rm stat}(.23)_{\rm ren}(.12)_{\rm syst}\,{\rm GeV}$ shows approximately the same relative deviation as for the decay constant in the chiral limit, $\ftwo$.

In Table \ref{tab:LECs:su2} we compare the values for the NLO LECs $\bar{l}_{3,4}$ (we also include the  values obtained by converting the SU(3) LECs, see Sec.~\ref{sec:fit_comp}). Within the quoted uncertainties all these numbers agree, except the result for $\bar{l}_3$ as quoted in \cite{Leutwyler:2007ae}, which used the MILC $\SU(3)$-LECs as input obtained from a ``NLO plus analytic NNLO and beyond'' fit. A recent update of the MILC results finds a more consistent value for $\bar{l}_3$ obtained from a pure NLO fit. The two results in the table for $\bar{l}_4$ from the CERN Collaboration, where one was obtained by including NNLO analytic terms, not only agree with our result for the central value, but also suggest a comparable magnitude for the systematic error (as estimated from the difference between the NLO and partial NNLO fits).

Quark masses are computed in lattice simulations using a variety of actions. We end this section by making the observation that results obtained using non-perturbative renormalization (such as those in this paper) appear to be generally higher than those obtained by renormalizing the mass perturbatively (mostly using 2-loops). Whether or not this is significant or merely a coincidence is still to be investigated. We illustrate this point by tabulating recent results obtained using non-perturbative renormalization \cite{Blum:2007cy,Blossier:2007vv,Gockeler:2006jt} and perturbative renormalization \cite{Bernard:2007ps,Ukita:2007cu,Ishikawa:2007nn} in Table~\ref{tab:su2:quarkMasses}. Note that for the simulations in this table, both different fermion actions and different numbers of dynamical fermions were used.  The ratios $m_s/m_{ud}$ are also tabulated and agree better across the different computations.

%% file: text_sections/Su3.tex
%

In the previous section, we have fit our lattice data to NLO SU(2)
ChPT formulae and found good agreement.  Additionally, for quark
masses where we have data, we find that for $m_{ll}^2$, the NLO
terms are at most a 10\% correction, while for $f_{ll}$ they are
less than 30\%, as shown in Fig.\ \ref{fit:Su2:nnloContr}.  Our
estimates for NNLO effects in this range are appropriately small,
$< 5$\%, leading to the conclusion that SU(2) ChPT converges
reasonably well here.  This implies that our range of light quark
masses is small enough that NLO SU(2) ChPT provides more than just a
useful, smooth phenomenological function to fit our data to, rather
it represents the theoretically correct description of our data.

We now turn to fitting our data to SU(3) ChPT, with the following
two points in mind:  1) we want to determine how well our data is
fit by the NLO SU(3) ChPT formulae and 2) if the fits agree with
our data, how convincing is the convergence of the (now known) SU(3)
ChPT series for the quark masses where we have data.  We will see
that the answer to the first point is \textit{yes}, if we only
include observables involving light quarks, and is \textit{no} for
observables including the strange quark.  For the second point,
even in the light quark case, we find the convergence of the series
to be poor.

\subsection{$\SU(3)\times \SU(3)$ Chiral Fits}
\label{sec:SU3fit}

At NLO, the 2+1 flavor, partially quenched chiral
formulae~\cite{Sharpe:2000bc} for the squared masses, $m_{xy}^2$,
and decay constants, $f_{xy}$, of the light pseudoscalar mesons
composed of two non-degenerate quarks with masses $m_x$ and $m_y$
involve six unknown low energy constants (LECs), which we denote
as $B_0$, $f_0$ and $L^{(3)}_{4,5,6,8}$. (The complete formulae
used in our chiral fits are summarized in Appendix \ref{sec:app:su3}.)
As $f_{xy}$ and $m_{xy}^2$ share the unknown low energy constants
$B_0$ and $f_0$, we performed combined fits to both of them, using
the two ensembles with dynamical light quark masses of $m_l=0.005$
and $0.01$.  A reasonable $\chi^2/{\rm d.o.f.}$ of 0.7 (uncorrelated)
could only be achieved by imposing a cut in the average valence
quark mass of $m_{\rm avg}\equiv(m_x+m_y)/2 \leq 0.01$, corresponding
to partially quenched pion masses in the range of about 250 MeV to
420 MeV. The chiral fits with such a cut are shown in
Fig.~\ref{fig:SU3fits}, and the fit parameters are given in
Table~\ref{tab:fitSU3}, where, for convenience, we quote the
scale-dependent LECs at two commonly used chiral scales of
$\Lambda_\chi=770$ MeV and 1 GeV.  Only statistical errors are
quoted in this table, as discussed in more detail below. In these
fits, the valence quark masses are all in a region where NLO SU(3)
ChPT might be expected to be reasonably reliable (corresponding to
masses below 420 MeV for pseudoscalar mesons made of such valence
quarks), but we point out that the dynamical heavy quark mass, which
is approximately 15\% higher than the physical strange quark mass,
lies outside of our fit region ($m_{\rm avg} \leq 0.01$).  Even the
combination of our lightest valence quark mass of 0.001 with the
dynamical heavy quark mass leads to a pseudoscalar mass of $\approx
554$ MeV.

Extending the fit range to valence quarks satisfying $m_{\rm
avg}\equiv(m_x+m_y)/2 \leq 0.03$, or a partially quenched valence
pseudoscalar mass of up to 660 MeV, we find that the fit curves
miss almost all of the data points, resulting in a poor $\chi^2/{\rm
d.o.f}$ of 5.7.  This is shown in Fig.\ref{fig:SU3fits_bad}.  Similar
behavior is seen for a lower mass cut of $m_{\rm avg}\equiv(m_x+m_y)/2
\leq 0.02$, which allows the largest valence pseudoscalar masses
to be just above the kaon mass, and gives an uncorrelated $\chi^2/{\rm
d.o.f}$ of about 3.  From these fits, we conclude that we cannot
use SU(3) ChPT at NLO to obtain physical results for kaon observables,
such as $f_K$ or the physical strange quark mass from $m_K$, with
well-controlled interpolation or extrapolation errors.  For these
important quantities, we have used kaon SU(2) ChPT as described in
previous sections.

\subsection{\label{sec:su3:sysErr}Systematic Errors}

For the $\SU(3)$ PQChPT results, we have to consider the same sources
of systematic uncertainties (chiral extrapolation errors, finite
volume effects and scaling errors) as in the $\SU(2)$ case, except
that, in principle, the $\SU(3)$ approach allows one to directly
extrapolate the dynamical heavy quark mass to the value of the
physical strange quark mass.

For the chiral extrapolations, we first investigate the systematic
error from the convergence of NLO ChPT.  We have already seen that
we cannot use NLO SU(3) ChPT for valence quarks near the physical
strange quark mass, since the fits do not agree well with our data.
Focusing on the results from fits to the light quark region ($m_{\rm
avg}\equiv(m_x+m_y)/2 \leq 0.01$), where our data is well represented
by the NLO formula, in Fig.~\ref{fig:su3:NLO} we plot the LO and
NLO contributions to the pseudoscalar decay constant $f_{ll}$ in
the unitary case.  The values of $f_{ll}$ are plotted as functions
of the light quark mass parameter ($\chi_l\propto \widetilde{m}_l$)
and the heavy quark mass parameter ($\chi_h\propto \widetilde{m}_h$)
using our results from the $\SU(3)$ fit with the mass cut $m_{\rm
avg}\leq 0.01$. We see that the corrections are generally large,
even for relatively small light-quark masses. For example, the top
left-hand panel shows that for $m_h=0.04$ and for pseudoscalar
masses in the 250--400MeV range, i.e. in the region where we have
data, the NLO correction is as large as 70\% of the LO term.

In the preceding paragraph we found that, for the unitary case with
$m_\pi \approx 400$ MeV, the SU(3) ChPT expansion for the decay
constant is approximately of the form $\fthree(1 + 0.7 + O(p^4))$,
which leads one to expect the $O(p^4)$ term to be about $(0.7)^2
\approx 0.5$.  This would be a 30\% correction to the sum of LO
plus NLO terms.  However, the NLO fits agree with our data quite
well, certainly excluding corrections beyond the few percent level.
From this we conclude that the agreement between our data and the
SU(3) ChPT formula is not indicative of NLO SU(3) ChPT working well
for these quark masses, since this would imply anomalously small
$O(p^4)$ corrections.  It seems that the SU(3) ChPT formula are
serving as smooth fitting functions with sufficient flexibility in
the parameters to absorb the effects of higher order terms and to
match our data.  We would need more data to investigate this
hypothesis fully and, in particular, this data should come from
ensembles with more than the single $m_h$ we currently have. This
will be done in the future and if our hypothesis is correct then,
as discussed in the following section, part of the reason for the
small value of $\fthree$ is due to the fact that it absorbs some
of the NNLO corrections.  We now briefly discuss our limited attempts
to use our current data to obtain additional information but stress
that a complete analysis of the range of validity and precision of
SU(3) ChPT will have to wait until we have data at more values of
quark masses.

One can also try to extend the range of validity of $\chi$PT by
going from NLO to NNLO.  The complete continuum NNLO formulae are
available in the literature \cite{Bijnens:2006jv} and we have done
some preliminary fits of our data, augmented with results from an
ensemble with $m_l = 0.02$ \cite{Lin:THESIS}, to these formulae.
Many more LECs are needed at NNLO ($\lthree{i}$ for $i=0$ to 9 and
12 linear combinations of $K_i$) and it is not currently clear how
stable such fits will be.  We defer a discussion of these fits
pending completion of the ongoing analysis.

To further probe the convergence of the series, we have dropped
the additional logarithm terms that appear at NNLO and fitted to
the analytic terms, as we did for our SU(2) fits in Section
\ref{subsec:SU2fit}.  
The analytic terms for the meson masses and
decay constants for $\SU(3)$ are
\begin{equation}
 (\chi_x+\chi_y)^2\,,\;(\chi_x-\chi_y)^2\,,\;(\chi_x+\chi_y)\bar{\chi}
 \,,\;
 \bar{\chi}^2\,,\; \overline{\chi^2}=(2\chi_l^2+\chi_s^2)/3\,.
\end{equation}
With our present data, limited to a single value of $\chi_h$ and
to a small set of $\chi_l$'s, we were not able to include the last
two terms in our NNLO phenomenological fits. The fits led, {\em
e.g.}, to an approximate 10\% increase in the value of $f_0$ (for
this particular case a fit range of $m_{\rm avg}\leq 0.01$ has been
used) and to even more significant changes in some of the NLO LECs.
This supports our conjecture that $\SU(3)$ ChPT shows a slow
convergence and NNLO terms are indeed important. On the other hand,
since NNLO-terms are not negligible, taking into account only (some
of) the analytic NNLO terms and neglecting the logarithmic terms
is not sufficient to determine the chiral behaviour of observables
quantitatively. It is for this reason that we choose to use SU(2)
ChPT to determine our physical results and at this time we do not
quote any estimates for systematic errors for quantities from our
fits with $\SU(3)$ ChPT.

In spite of our reservations about the convergence of SU(3) ChPT,
we have used the corresponding formulae in finite volumes to estimate
finite volume effects (see Appendix~\ref{sec:appendix_fv}), as we
had done previously for the $\SU(2)$ case.  We find similar results
for the correction factors. Table \ref{tab:FVcompareCDH} also
contains the corrections obtained in the $\SU(3)$ case for the
unitary pions; for more details see the discussion in
Sec.~\ref{sec:su2:sysErr:FV}.

\subsection{\label{sec:fit_comp}Comparing $\SU(3)\times \SU(3)$ and
$\SU(2)\times\SU(2)$ Chiral Fits}

In this section, we compare the results of our SU(3) ChPT fits with
other results and also compare them with the results of our SU(2)
fits.  This serves to further probe the behavior of ChPT in the
region of quark masses we are studying.  We stress however, that
we believe that the convergence of the SU(3) series is relatively
poor and therefore at this stage any quantitative conclusions will
be limited.

To compare the SU(3) fit results with the previous ones obtained
in $\SU(2)$ PQChPT (Sec.~\ref{subsec:SU2fit}), we use
Eqs.~(\ref{eq:chPTconv:F})--(\ref{eq:chPTconv:l4}) (cf.\ also
\cite{Gasser:1984gg,Gasser:2007sg}) to match the three flavor ChPT
to the two flavor case at NLO.  The results for $\bthree$, $\fthree$
and the low energy constants $\bar{l}_{3,4}$ (for a definition of
the latter see Eq.~(\ref{eq:chPTconv:barl}) and \cite{Gasser:1984gg})
are shown in Table~\ref{tab:conv_SU3_SU2}.  Indeed, having a fixed
dynamical heavy quark mass in the $\SU(3)\times\SU(3)$ theory is
equivalent to the $\SU(2)\times\SU(2)$ theory, up to terms of
$O((m_l/m_s)^2)$.  The remarkable agreement between the LECs obtained
directly from the fit to the chiral behaviour using SU(2) ChPT and
those obtained by converting the SU(3) LECs using
Eqs.~(\ref{eq:chPTconv:F})--(\ref{eq:chPTconv:l4}) (see
Table~\ref{tab:conv_SU3_SU2}) is evidence that these terms of
$O((m_l/m_s)^2)$ are small. Although reassuring, we stress that
this in itself is insufficient to demonstrate fully the validity
of the SU(3) ChPT at NLO. For example, if the $m_s^2$ term in the
expansion for the decay constant were large, this would appear in
our NLO fits as a shift in $\fthree$ but the behaviour with $m_l$
would be equivalent to that in SU(2) ChPT (up to corrections of
$O((m_l/m_s)^2)$ of course). We suspect that this is the case and
use SU(2) ChPT to obtain physical results nevertheless, for the
remainder of this section, we take our SU(3) results at face value
and compare them to previous determinations of the LECs.

We first consider the ratio of the decay constants in the two and
three flavor chiral limit. From our fits to $\SU(2)$ and $\SU(3)$
ChPT we obtain $\ftwo/\fthree=1.229(59)$ (here we quote only the
statistical error), showing the influence of the strange quark
loops. An important observation is that the value of $f_0$, the
pion decay constant in the $\SU(3)$ chiral limit, is much smaller
than the measured pseudoscalar decay constant $f_{ll}$ at, say,
$m_l = 0.01$ and also much smaller than $f_\pi$.  This is due to
the large NLO correction shown in Fig.~\ref{fig:su3:NLO}. The small
value of $f_0$ may be a contributor to poor convergence, since the
chiral logarithms come in with a factor of $1/f_0^2$.  Of course,
one can rearrange the series and expand in $m_{xy}^2/f_\pi^2$
consistently to NLO and possibly improve the convergence.  We are
exploring these options, but if such a rearrangement, which only
effects the series at NNLO, markedly changes the convergence
properties, one is still lead to the conclusion that the series is
not well controlled.

If we compare our result for $\fthree$ to phenomenological estimates
by Bijnens \cite{Bijnens:2007yd}, our value of $f_0= 93.5(7.3)\,{\rm
MeV}$ (statistical error only) turns out to be substantially lower
than the preferred value from Bijnens, which is 124.0 MeV (NNLO,
alternative fits also published there give values of 114.7 (NLO),
99.6 (NNLO), and 113.7 (NNLO) MeV). His ratio $f_\pi/\fthree$ is
also different from ours (in contrast to the $\SU(2)$ case, where
for the ratio $f_\pi/f$ a better agreement has been achieved): we
obtain a value of 1.33(07) (statistical error only), whereas Bijnens'
preferred fit suggests $f_\pi/f_0=1.05$.  Interestingly, the $N_f=2+1$
lattice simulation by MILC \cite{Aubin:2004fs,Bernard:2007ps} also
observed a higher ratio of $f_\pi/\fthree=1.21(5)(^{+13}_{-3})$,
which translates into a central value for $f_0=106.0\,{\rm MeV}$
(using their value for $f_\pi=128.3(0.5)(^{+2.4}_{-3.5})$, we do
not quote an error for this number, since we do not know the
correlation between $f_\pi$ and $f_\pi/f_0$.)

For the ratio $\btwo/\bthree$ we obtain $\btwo/\bthree=1.03(05)$.
The very small deviation from 1 is perhaps surprising, and indicates
good agreement with the predictions of the large $N_c$ approximation
and only small Zweig rule violations in this case. A comparison of
the LO-LEC $\bthree$ is not directly possible, since this number
depends on the renormalization scheme (see Sec.~\ref{sec:su2:qmass}).
Instead, Bijnens quotes a number for the renormalization scheme
independent and dimensionless ratio $2B_0 m_{ud}/m_\pi^2$ of 0.736
(from the preferred NNLO-fit; alternative fits give: 0.991 (NLO),
1.129 (NNLO), 0.958 (NNLO)). From our $\SU(3)$ fit we obtain 0.995(41)
(statistical error only), which agrees better with the alternative
phenomenological fits than the preferred one.

The ratio of $\Sigma=f^2\,B/2$, the chiral condensate in the
two-flavor theory, and  $\Sigma_0=f_0^2\,B_0/2$ in the three-flavor
theory, can also be compared directly. We obtain
$\Sigma/\Sigma_0\,=\,1.55(21)$, whereas MILC quotes a value of
$1.52(17)(^{+38}_{-15})$. It should be noted however, that we obtain
slightly different values for $f/f_0$ and for $B/B_0$ than the MILC
collaboration.

In Table~\ref{tab:LECs:su3} we compare the NLO-LECs to phenomenological
NLO and NNLO fits \cite{Bijnens:2007yd} and the results of MILC's
$N_f=2+1$ dynamical lattice simulations \cite{Aubin:2004fs,Bernard:2007ps}
(all LECs in that table are quoted at the scale $\Lambda_\chi=770\,{\rm
MeV}$). In the fits by Bijnens, $\lthree{4}$ and $\lthree{6}$ were
set to zero (at $\Lambda_\chi=770\,{\rm MeV}\approx m_\rho$) as
motivated by Zweig-rule and large $N_c$ arguments. Whereas for the
latter our result agrees with this assumption within the statistical
uncertainty, for $\lthree{4}$ we observe some discrepancy (without
taking systematic errors into account).  Interestingly, our NLO
results for $\lthree{5}$ and $\lthree{8}$ agree very well with
Bijnens' numbers from the NNLO fit, but not for his NLO fit (for
which no systematic error is given however).  Within the reported
uncertainties, our NLO LECs agree nicely with the set of values
quoted by the MILC Collaboration.

In Fig.~\ref{fig:su2_su3_comp} we summarize some of the results
presented in this section.  Here we plot our measured values for
the decay constant, converted to physical units, versus the degenerate
valence pseudoscalar mass squared, along with the results of fits
to SU(2) and SU(3) ChPT.  The data for degenerate quarks is denoted
by filled symbols and for non-degenerate quarks, open symbols are
used.  The graph shows that we see small effects for non-degenerate
quarks.  The results of SU(2) and SU(3) partially quenched ChPT
fits to our data are shown, and both fits agree well with the data.
From the (now determined) fit functions, we plot the unitary SU(2)
chiral extrapolation.  Two of our data points lie on this curve,
but one sees that our measured decay constant for a pseudoscalar
mass of 420 MeV is about 30\% above the chiral limit value, $\ftwo$.
We also plot the SU(3) chiral limit curve.  For this curve, the
horizontal axis is the unitary meson mass in the theory where all
quarks are degenerate.  Therefore, none of our measured values must
lie on this line.  Figure~\ref{fig:su2_su3_comp} shows graphically
the large difference between $\fthree$ and our measurements.
This is graphical evidence of the poor convergence of NLO SU(3)
ChPT, with LECs as determined from our data.

In summary, we have found that we can fit our data well using NLO
SU(3) ChPT for average valence quark masses $< 0.01$, corresponding
to pseudoscalar masses below 420 MeV.  The SU(3) LECs we determine
are generally in reasonable agreement with continuum phenomenology
and other lattice results (note however, the small value of $\fthree$
which we find).  We see that the LECs from our SU(3) and SU(2) fits
agree well when a conversion is performed from SU(3) to SU(2).
However, since we find large corrections at NLO we would expect
significant ones also at NNLO. We are therefore not at all confident
that the systematic errors in the SU(3) LECs are currently under
control. In addition, the fits do not agree with our data when we
extend the NLO SU(3) ChPT theory to the kaon mass scale. It is
possible that with more data it will become feasible to perform
NNLO SU(3) fits and to control the systematic uncertainties with
sufficient precision. We will investigate this in the future but
we stress that whether or not this proves to be the case, we can
happily use SU(2) PQChPT to obtain predictions for physical results
(including quantities in kaon physics as explained in
section~\ref{subsect:kchpt}\,). This is what we do in this paper.

%% file: text_sections/Bk.tex
\subsection{Pseudoscalar Bag Parameter on the Lattice}

The kaon bag parameter is defined as the ratio of the neutral kaon mixing matrix element and its expectation value from the vacuum saturation approach,
\begin{equation}\label{eq:bk:kaonBag}
B_K\;=\;\frac{\langle\overline{K}^0\left|\mathcal{O}_{LL}^{\Delta S=2}\right|K^0\rangle}{\frac83f_K^2m_K^2}\,,
\end{equation}
where $m_K$ is the neutral kaon mass, $f_K$ the kaon decay constant, and 
\begin{equation}\label{eq:bk:4operator}
\mathcal{O}_{LL}^{\Delta S=2}\;=\;\Big(\bar{s}(1-\gamma_5)\gamma_\mu d\Big)\,\Big(\bar{s}(1-\gamma_5)\gamma_\mu d\Big)
\end{equation}
is the local, effective four quark operator, which couples to the left-handed quarks and induces a change in strangeness by $\Delta S=2$. In our simulations, we define the corresponding pseudoscalar bag parameter for a meson made from either valence or sea quarks with masses $m_x$ and $m_y$, which we measure by fitting the ratio
\begin{equation}\label{eq:bk:PSbag}
B_{xy}(t)\;=\;\frac{3}{8}\,\frac{\mathcal{C}_{P\mathcal{O}P}^{WLW}(t_{\rm src},t,t_{\rm snk})}{\mathcal{C}_{PA}^{WL}(t_{\rm src}, t)\mathcal{C}_{AP}^{LW}(t, t_{\rm snk})}
\end{equation}
to a constant $B_{xy}$ over some range in time $t$. Here, $t$ denotes the time coordinate at which we insert the four quark operator, $t_{\rm src}$ and $t_{\rm snk}$ are the temporal coordinates at which we insert Coulomb gauge fixed (spatial) walls of anti-kaon and kaon interpolating operators, respectively. The correlators using wall source and local sinks, which we construct from the WL-propagators obtained with quark masses $m_x$ and $m_y$, are defined as
\begin{eqnarray}
\mathcal{C}^{WLW}_{P\mathcal{O}P}(t_{\rm src},t,t_{\rm snk}) &=& \frac1{V}\,\sum_{y\in V}\,\langle\, q_w(t_{\rm src})P\bar{q}_w(t_{\rm src})\:\mathcal{O}_{LL}^{\Delta S=2}(y,t)\:q_w(t_{\rm snk})P\bar{q}_w(t_{\rm snk})\,\rangle\,,\\
\mathcal{C}^{WL}_{PA}(t_{\rm src},t) &=& \frac1{V}\,\sum_{y\in V}\,\langle\, q_w(t_{\rm src})P\bar{q}_w(t_{\rm src})\:q(y,t)A\bar{q}(y,t)\,\rangle\,,  \\
\mathcal{C}^{LW}_{AP}(t,t_{\rm snk}) &=& \frac1{V}\,\sum_{y\in V}\,\langle\, q(y,t)A\bar{q}(y,t)\:q_w(t_{\rm snk})P\bar{q}_w(t_{\rm snk})\,\rangle\,,
\end{eqnarray}
where $q_w(t)$ and $q(y,t)$ denote a Coulomb gauge fixed (spatial) wall smeared quark field and a local quark field, respectively. A summation over all spatial points for the local operators is performed, and a balancing volume factor $V=L^3$ included giving a spatial volume average that is statistically efficient. In addition, we average over propagators obtained with periodic and anti-periodic boundary conditions in the time direction, which results in a doubled time extent available for the plateau fit.

The values measured on the $24^3\times64$, $L_s=16$ lattices are given in Table~\ref{table:bk:values:24c}. There, the fit to the plateau was performed over the range $t\in[12,52]$ using an uncorrelated fit, since, due to the large time extent, correlated fits became unstable. Previously, our collaboration also obtained the pseudoscalar bag parameter on smaller $16^3\times32$, $L_s=16$ lattices with dynamical light quark masses $m_l\in\{0.01, 0.02, 0.03\}$ and $m_s=0.04$ at the same gauge coupling \cite{Allton:2007hx,Cohen:2006zza,Antonio:2007pb,CohenAntonio:Lat2007}. For comparison we quote those values in Table~\ref{table:bk:values:16c}. In this case the plateau was fitted over a range $t\in[12,22]$. 

We will have to extrapolate our measured values to the physical light and strange quark masses for which PQChPT in the $\SU(2)\times\SU(2)$ formulation will be used. These fits are discussed in the next part of this section; the result obtained for $B_K$ was already published in \cite{Antonio:2007pb}, but here we provide some more details on the chiral fitting procedure. In the remainder of this section, we employ $\SU(3)\times\SU(3)$ PQChPT to extract the bag parameter in the chiral limit and briefly discuss the data from the $16^3$ lattices. 

\subsection{$\SU(2)\times\SU(2)$ Chiral Fitting Procedure}

For the chiral extrapolation of $B_K$, we use $\SU(2)\times\SU(2)$ ChPT for the kaon sector, as introduced in Sec.~\ref{sec:chiral_limit} and applied to the kaon masses and decay constants in Sec.~\ref{subsec:SU2fit}. Specifically we use Eq.~(\ref{eq:chPTsu2K:BPS}) to extrapolate to the physical average light quark mass $m_{ud}$ determined in Sec.~\ref{sec:su2:detMassScale}. Here we consider the data from the $24^3$ lattices at values of $m_y=0.03$ and $0.04$ for the heavier quark mass approximating the strange quark mass. Of these two values only the latter describes a truly unquenched quark and we account for possible effects of a partially quenched strange quark treatment in our discussion of systematic errors. The fits using $\SU(2)\times\SU(2)$ ChPT for the kaon sector are shown in Fig.~\ref{fig:bK:SU2lightExtr}, whereas the fitted parameters are given in Table~\ref{table:bK:fitparamSU2}. The parameters $\btwo$ and $\ftwo$ have been fixed to their values obtained in the combined $\SU(2)\times\SU(2)$ ChPT fits for the pseudoscalar masses and decay constants, cf. Sec.~\ref{sec:Su2}. We applied a cut in the light valence quark mass of $m_x\leq0.01$. The low energy constant $b_1(m_y)$ multiplying the dynamical light quark mass in Eq.~(\ref{eq:chPTsu2K:BPS}) shows almost no dependence on the heavy quark mass, whereas $b_2(m_y)$ and $B_{\rm PS}^{(K)}(m_y)$ vary by 12\% and 4\%, respectively, when going from $m_y=0.03$ to $0.04$. Also the value of the kaon bag parameter extrapolated to the physical average light quark mass, $B_{ud\,y}$, increases by approximately 4\% due to this change in $m_y$.  

The interpolation to the physical value of the strange quark mass (obtained from the kaon mass, cf. Sec.\ref{sec:su2:detMassScale}) is done linearly as shown in Fig.~\ref{fig:bK:SU2strInt}. Included in the plot (and also in Table~\ref{table:bK:fitparamSU2}) is a point at $m_y=0.02$ for information only. It was not included in the linear fit and interpolation because this supposedly heavy quark mass is likely too light to be sufficiently separated from the simulated light quark masses to enable convergent effective treatment. However, the linear interpolation does not seem to deviate much even at this point, giving us confidence that non-linear effects in the strange quark mass region at least do not show up at the level of the numerical precision achieved. Finally, we quote for the unrenormalized physical value of the kaon bag parameter $B_K^{\rm lat}=0.565(10)$. We will now discuss the renormalization of $\mathcal{O}_{LL}^{\Delta S=2}$ and estimate all significant contributions to the systematic error.

\subsubsection{Renormalization}

The renormalization of $\mathcal{O}_{LL}^{\Delta S=2}$ has been treated in \cite{Aoki:2007xm} using the RI-MOM non-perturbative renormalization technique. Couplings to wrong chirality four quark operators are in principle admitted by the presence of a non-zero residual mass. It has been demonstrated in that paper that they are highly suppressed and small, using non-exceptional momenta to remove the spontaneous chiral symmetry breaking effects that might obscure this fact at the modest, accessible lattice momenta. Thus we only need to consider a simple, multiplicative renormalization. Here we quote the renormalization factors for the regularization independent (RI) and the modified minimal subtraction ($\overline{\rm MS}$) schemes both at $\mu=2\,{\rm GeV}$ as well as the renormalization group invariant (RGI) result:
\begin{eqnarray}
Z_{B_K}^{\rm RI}(2\,{\rm GeV})\,=\,0.910(05)_{\rm stat}(13)_{\rm syst}\,, & & B_K^{\rm RI}(2\,{\rm GeV})\,=\,0.514(10)_{\rm stat}(07)_{\rm ren}\,, \label{eq:bk:su2_RI}\\ 
Z_{B_K}^{\overline{\rm MS}}(2\,{\rm GeV})\,=\,0.928(05)_{\rm stat}(23)_{\rm syst}\,, & & B_K^{\overline{\rm MS}}(2\,{\rm GeV})\,=\,0.524(10)_{\rm stat}(13)_{\rm ren}\,, \label{eq:bk:su2_barMS}\\ 
Z_{B_K}^{\rm RGI} \,=\,1.275(10)_{\rm stat}(25)_{\rm syst}\,, & & \hat{B}_K\,=\,0.720(13)_{\rm stat}(14)_{\rm ren} \label{eq:bk:su2_RGI}\,, 
\end{eqnarray}
where the first error is the (combined) statistical error and the second error the systematic error from the renormalization. 

\subsubsection{\label{sec:bk:systErr}Systematic Errors}

\paragraph{Finite Volume Effects}

The spatial volume in our simulation is approximately $(2.7\,{\rm fm})^3$, therefore, from \cite{Becirevic:2003wk} it follows that the difference between the $B_{xy}$ measured on the finite lattice volume and the result in infinite volume is negligible for all points except the lightest ones (both valence quarks at a mass of 0.001), where the difference may be as large as 2\%. Excluding the point $m_l=0.01$, $m_x=m_y=0.001$ from the fit, the final result for $B_K^{\rm lat}$ remains almost unchanged. Comparing points with $m_l=0.01$ from the smaller volume simulation ($16^3$ lattice, $\approx(1.8\,{\rm fm})^3$ spatial volume) from Table~\ref{table:bk:values:16c} with the corresponding ones from the $24^3$ simulation (Table~\ref{table:bk:values:24c}), statistically marginal differences of up to 1\% are observed, see Sec.~\ref{sec:bk:16c} for a more detailed discussion. Conservatively, we adopt this as an estimate for finite volume effects affecting our final number for $B_K$. (Note, that for the lightest point in the $16^3$ ensemble ($m_x=m_y=m_l$) we have $m_\pi L\approx 4.0$, whereas for our lightest valence pion in both $24^3$ ensembles ($m_x=m_y=0.001$, $m_l=0.005$ or 0.01) we still have $m_\pi L \approx 3.4$, so it is reasonable to assume the finite volume effects in those points to be comparable to the observed effect from comparing $16^3$ and $24^3$ lattices.)

\paragraph{Scaling Effects}

Having only data at a single value for the lattice spacing does not allow to quantify the scaling effects affecting our data. The CP-PACS Collaboration calculated $B_K$ in a quenched calculation using the Iwasaki gauge action with Domain Wall valence quarks at two different lattice scales, namely $1/a = 1.81(4)$ and $2.81(6)\,{\rm GeV}$ (determined from the $\rho$-meson mass) keeping the physical volume approximately fixed \cite{AliKhan:2001wr}. Extrapolating their observed scaling violation to our coarser lattice spacing, we would expect a 3.5\% scaling effect. We assume a 4\% systematic error, which is in agreement with the scaling violations discussed in conjunction with the meson masses and decay constants in Sec.~\ref{sec:Su2:sysErr:scale}. 

\paragraph{Interpolation to the Physical Strange Quark Mass}

Currently only ensembles with a fixed value for the dynamical heavy quark mass are available. We estimate the effect of this 15\% too high sea quark mass from the measurements done at the $16^3$ lattices. There dynamical light quark masses of $m_l=0.02$ and $0.03$ have been simulated, which are closer to the physical value of the strange quark mass than the dynamical light quark masses available from the $24^3$ data. Comparing $B_{xy}$ for $m_x=0.01$, $m_y=0.04$ (lightest valence quark mass and dynamical strange quark mass) at the two aforementioned light sea quark masses, an increase of 3\% is observed (Table~\ref{table:bk:values:16c}). In that case, \textit{two} dynamical quark flavors were changed by $\Delta m=0.01$, whereas the \textit{single} strange quark mass only has to be changed by $0.0057$. Accordingly, the systematic error has to be scaled down by 2 (from number of flavors) times $\approx 1.8$ (from scaling $\Delta m$), resulting in an 1\% effect. (This is an exclusive sea quark effect and should not be confused with the 4\% difference discussed above, when changing the valence $m_y$ from 0.03 to 0.04 in the kaon itself.)

\paragraph{Extrapolation in the Light Quark Mass}

Including terms up to NLO in our chiral fit functions, we have to assign a systematic error to our extrapolation resulting from neglecting NNLO and higher order terms. The linear fit to the $16^3$ data gives a 6\% higher value for $B_K$, see Sec.~\ref{sec:bk:16c}. Assigning this difference to the (here included) NLO terms, the size of NNLO contribution can be estimated to be 2\% by scaling the observed 6\% difference by $\widetilde{m}_l/\widetilde{m}_s\approx0.4$ taken at the lightest value for quark mass in the linear fit, $m_l=0.01$, and the strange quark mass at the physical point $m_s$.  

\subsubsection{Final Result}

Combining the 1\% finite volume, 4\% scaling, 1\% heavy quark mass interpolation, and 2\% ChPT extrapolation systematic errors with the one from the NPR (except for $B_K^{\rm lat}$, of course), our final result in the different renormalization schemes considered reads:
\begin{eqnarray}
B_K^{\rm lat} &=& 0.565(10)_{\rm stat}(27)_{\rm syst}\,,\label{eq:bk:su2_lat:fin} \\
B_K^{\rm RI}(2\,{\rm GeV}) &=& 0.514(10)_{\rm stat}(07)_{\rm ren}(24)_{\rm syst} \;=\; 0.514(10)_{\rm stat}(25)_{\rm comb}\,, \label{eq:bk:su2_RI:fin}\\ 
B_K^{\overline{\rm MS}}(2\,{\rm GeV}) &=& 0.524(10)_{\rm stat}(13)_{\rm ren}(25)_{\rm syst} \;=\; 0.524(10)_{\rm stat}(28)_{\rm comb}\,, \label{eq:bk:su2_barMS:fin}\\ 
\hat{B}_K &=& 0.720(13)_{\rm stat}(14)_{\rm ren}(34)_{\rm syst} \;=\; 0.720(13)_{\rm stat}(37)_{\rm comb}\,, \label{eq:bk:su2_RGI:fin} 
\end{eqnarray}
where ``stat'' denotes the statistical error, ``syst'' the systematic error as discussed above and ``ren'' the error due to renormalization, ``comb'' is the combined systematic error from the latter two.

\subsection{Fits to $\SU(3)\times\SU(3)$ PQChPT}

Here we apply Eq.~(\ref{eq:chPTsu3:BPS:nondeg}) from $\SU(3)\times\SU(3)$ PQChPT to fit our data.
That means, we now also try to describe the dependence on the heavier (valence) quark mass by (PQ)ChPT.
As discussed in Sec.~\ref{sec:kchpt:comments}, if Eq.~(\ref{eq:chPTsu3:BPS:nondeg}) were applied for a fixed value of the heavier valence quark mass (also fixing the heavy sea quark mass), it would naturally revert to the kaon $\SU(2)\times\SU(2)$ form. If we were able to describe the valence mass dependence with the $\SU(3)\times\SU(3)$ PQChPT form up to the strange quark mass, this form would be applicable to a determination of $B_K^{\rm lat}$. However, it is also possible that this form can be used to describe only light valence masses $m_x$, $m_y$ to obtain the $\SU(3)\times\SU(3)$ low energy constant $\bagzero$ representing the pseudoscalar bag parameter in the three flavor theory in the limit of all three masses being zero, which is of phenomenological interest \cite{Bijnens:2006mr}. The only remnant terms involving a large mass in such a fit of the $\SU(3)\times\SU(3)$ PQChPT arise from the fixed dynamical heavy quark mass $m_h$, and we will estimate a systematic error from this.

We observe the same behaviour as for the meson masses and decay constants: a reliable fit including quark masses up to the strange quark mass is not possible with only terms of NLO included in the fit formula.
In Fig.~\ref{fig:bk:SU3fits} fits to the $24^3$ data are shown with two different ranges for the mass cut $m_{\rm avg}$. While the fit with  $m_{\rm avg}\leq0.01$ describes the data inside the fit range but badly fails at the heavier points, going to $m_{\rm avg}\leq0.02$ already fails to describe the data at the lowest masses. The fitted parameters are given in Table~\ref{table:bk:fitparamSU3}. Here we fixed $\bthree$ to its value obtained from the $\SU(3)\times\SU(3)$ fit for the meson masses and decay constants ($m_{\rm avg}\leq0.01$), cf. Sec.~\ref{sec:Su3}. Due to the failure to describe the data with one quark mass as heavy as the strange quark, a determination of $B_K^{\rm lat}$ is not meaningful using this ansatz. However, limiting the fit range to small masses, we will estimate the $\SU(3)\times\SU(3)$ LEC $\bagzero$.
In Fig.~\ref{fig:bK:cutDepSU3} the dependence of $\bagzero$ on the applied mass cut is shown. From that we conclude that at least within the statistical error the result is stable, therefore we quote (from a fit using $m_{\rm avg}\leq 0.01$):
\begin{eqnarray}
\bagzeroSup{\rm lat}{} &=& 0.266(26)_{\rm stat}\,,\label{eq:b0:su3_lat} \\
\bagzeroSup{\rm RI}{}(2\,{\rm GeV}) &=& 0.242(24)_{\rm stat}(03)_{\rm ren}\,, \label{eq:b0:su3_RI}\\ 
\bagzeroSup{\overline{\rm MS}}{}(2\,{\rm GeV}) &=& 0.247(24)_{\rm stat}(06)_{\rm ren}\,, \label{eq:b0:su3_barMS}\\ 
\bagzeroSup{}{\hat} &=& 0.339(33)_{\rm stat}(07)_{\rm ren}\,, \label{eq:b0:su3_RGI} 
\end{eqnarray}
using the same renormalization factors as in Eqs.~(\ref{eq:bk:su2_RI}--\ref{eq:bk:su2_RGI}) (first error statistical, second error systematics from renormalization). One has to keep in mind that this result was obtained by extrapolating $\widetilde{m}_h$ from its value used in the simulation to zero ({\em i.e.}, extrapolating $m_h\to-m_{\rm res}$). Considering Eq.~(\ref{eq:chPTsu3:BPS:nondeg}) which was used to fit the data, one sees that---except for the analytic terms in $I_{\rm disc}$ where it enters via $\chi_\eta$---the heavy quark mass only enters via $\bar\chi$, which multiplies the LEC $d$. By only using one dynamical heavy quark mass value, this LEC is exclusively determined by the dependence of $B_{xy}$ on the light dynamical quark mass, a very mild dependence as can be seen by comparing the columns for $m_l=0.005$ and $0.01$ in Table~\ref{table:bk:values:24c}. 
Two problems might arise. Firstly, the slope with respect to variations of $m_h$ may differ from the slope with respect to variations of $m_l$ due to higher order corrections. Secondly, with our fit procedure we might not be able to reliably determine this parameter. We assign the full size of this NLO estimate of the contribution,
\begin{equation}
 \left|\bagzero\,\frac{\chi_h}{3(4\pi \fthree)^2}\,d\right|\;=\;\left|\bagzero\,\frac{2\bthree\,\widetilde{m}_h}{48 \pi^2\,\fthree^2}\,d\right| 
\end{equation}
to be the systematic error of our chiral extrapolation in the heavy quark mass, which is a 10\% effect for the chosen cut-off, $m_{\rm cut}=0.01$. (Going to higher values in $m_{\rm cut}$ the contribution increases to as much as 28\% for $m_{\rm cut}=0.02$.) The NLO-contribution of the extrapolation in the light quark mass, which should be well under control by our fit procedure, we get from the difference of linearly extrapolating the measured $B_{xy}$ values at the dynamical points with $m_l=m_x=m_y=0.01$ and $0.005$ to $m_l=m_x=m_y=-m_{\rm res}$ ({\em i.e.} $\widetilde{m}_l=\widetilde{m}_x=\widetilde{m}_y=0$) and the chiral extrapolation to this point ($m_h=0.04$ is fixed in this procedure). By multiplying the resulting difference of 
\begin{equation}
 \Delta\;=\; B_{ll}^{\rm linear} (\widetilde{m}_l=0,\,m_h=0.04) \:-\: B_{ll}^{\chi{\rm PT}}(\widetilde{m}_l=0,\,m_h=0.04) \:\approx\: 0.154
\end{equation}
by $\chi_l/(4\pi\fthree)^2$, we determine the uncertainty due to neglected NNLO contribution to be 8\%.

For the systematic errors due to finite volume and scaling, the same applies as for the $\SU(2)\times\SU(2)$ analysis of $B_K$, so we assume them to be 1\% and 4\%, respectively. Eventually, with the 10\% strange and 8\% NNLO light quark mass extrapolation, the 1\% finite volume, 4\% scaling and NPR systematic error, we obtain (first error statistical, second combined systematic error):
\begin{eqnarray}
\bagzeroSup{\rm lat}{} &=& 0.266(26)_{\rm stat}(36)_{\rm comb}\,,\label{eq:b0:su3_lat_tot} \\
\bagzeroSup{\rm RI}{}(2\,{\rm GeV}) &=&  0.242(24)_{\rm stat}(03)_{\rm ren}(33)_{\rm syst} \;=\; 0.242(24)_{\rm stat}(33)_{\rm comb}\,, \label{eq:b0:su3_RI_tot}\\ 
\bagzeroSup{\overline{\rm MS}}{}(2\,{\rm GeV}) &=& 0.247(24)_{\rm stat}(06)_{\rm ren}(33)_{\rm syst} \;=\; 0.247(24)_{\rm stat}(34)_{\rm comb}\,, \label{eq:b0:su3_barMS_tot}\\ 
\bagzeroSup{}{\hat} &=& 0.339(33)_{\rm stat}(07)_{\rm ren}(46)_{\rm syst} \;=\; 0.339(33)_{\rm stat}(47)_{\rm comb}\,. \label{eq:b0:su3_RGI_tot} 
\end{eqnarray}
Our final number for $\bagzeroSup{}{\hat}$ is in agreement with the phenomenological estimates from (i) the large $N_c$ approximation, $\bagzeroSup{}{\hat}=0.38\pm0.15$ \cite{Bijnens:2006mr} ($\hat{B}_K^\chi$ in their notation), and (ii) $\bagzeroSup{}{\hat}=0.39\pm0.10$ \cite{Prades:1991sa} using the QCD-hadronic duality approach (which is close to the chiral limit value, see remark in \cite{Bijnens:2006mr}).

\subsection{\label{sec:bk:16c}Results from Simulations on a Smaller Volume}

As mentioned in the beginning of this section, our collaboration first performed a study of the pseudoscalar bag parameter at a smaller volume of $(1.8\,{\rm fm})^3$ with higher dynamical masses $m_l\in\{0.01, 0.02, 0.03\}$, see Table~\ref{table:bk:values:16c}. It turned out, that these masses were not light enough to probe the regime of ChPT, as was revealed by comparing a simple linear fit in the light quark mass with a fit to (PQ)ChPT. The extrapolations from these fits gave similar results within the numerical uncertainty, suggesting that the chiral logarithms occurring in NLO were not represented correctly, or---in other words---the influence of the linear terms in the PQ$\chi$PT formula was overestimated. Here, we just quote the result of extrapolating the three unitary points ($m_x=m_l\in\{0.01, 0.02, 0.03\}$, $m_y=m_h=0.04$) by a linear fit to the light quark mass at the physical point $m_{ud}$
\begin{equation}\label{eq:bklin:16c}
 B_{ud\,h}^{16^3,{\rm linear}} (m_h=0.04) \;=\;0.611(08)\,,
\end{equation}
which gives the value of the kaon bag parameter at a slightly too high strange quark mass. This has to be compared with $B_{ud\,y}=0.5789(97)$ at $m_y=0.04$ from Table~\ref{table:bK:fitparamSU2}, yielding a 6\% difference between these two approaches. (More details on the fits to the $16^3$ data have already been published in \cite{CohenAntonio:Lat2007}.) 

Also this data enables us to check for possible finite volume effects. The spatial volume increases by a factor $1.5^3=3.375$ when going to the $24^3$ lattices. Fig.~\ref{fig:bK:volume} shows a comparison of the measured $B_{xy}$ values at $m_l=0.01$ for the two different volumes, revealing only small changes in the range of valence quark masses $0.01\leq m_x,m_y \leq 0.04$, where data is available from both simulations, see also Table~\ref{table:bk:finiteVol}. The maximal deviation of 0.8\% is observed for the lightest points $(m_x,m_y)=(0.01,0.02)$ and $(0.02,0.02)$ and decreases down to 0.4\% for the heaviest point $(0.04,0.04$). (Accidently, $(0.01,0.01)$ shows almost no, {\em i.e.}, $\leq0.1\%$ deviation in the central value.)

%% file: text_sections/Vector.tex
%

In this section we discuss the couplings of the light vector
mesons $V$ to vector and tensor currents. These couplings
$f_V$ and $f_V^T$ are defined through the
matrix elements:
\begin{eqnarray}
\langle\,0\,|\,\bar{q}_2(0)\gamma^\mu
q_1(0)\,|\,V(p;\lambda)\,\rangle
&=&f_V\,m_V\,\varepsilon_\lambda^\mu\\
\langle\,0\,|\,\bar{q}_2(0)\sigma^{\mu\nu}
q_1(0)\,|\,V(p;\lambda)\,\rangle
&=&if_V^T(\mu)\,\left(\varepsilon_\lambda^\mu p^\nu-
\varepsilon_\lambda^\nu p^\mu\right)\,,
\end{eqnarray}
where $p$ and $\lambda$ are the momentum and polarization state of
the vector meson $V(p;\lambda)$ and $\varepsilon_\lambda$ is the
corresponding polarization vector. The tensor bilinear operator
$\bar{q}_2\sigma^{\mu\nu} q_1$ (and hence $f_V^T(\mu)$) depends on
the renormalization scheme and scale $\mu$. The final results will
be quoted in the $\overline{\textrm{MS}}$ scheme at $\mu=2$\,GeV. We have presented preliminary results
for the ratios $f_V^T/f_V$ in Ref.\,\cite{Donnellan:2007xr}.

\subsection{Experimental Determination of $f_V$}

The decay constants $f_V$ can be determined experimentally. For
the charged $\rho$ and $K^\ast$ mesons, one can use $\tau$ decays
to deduce $f_{\rho}$ and $f_{K^\ast}$ as illustrated by the
diagram in Fig.\ \ref{fig:pda_feynman},
where the curly line represents the $W$-boson. From the measured branching ratios one obtains the following
values for the decay constants~\cite{Becirevic:2003pn}:
\begin{eqnarray}
\textrm{Br}(\tau^-\to\rho^-\nu_\tau)=(25.0\pm
0.3)\%&\Rightarrow&f_{\rho^-}\,\simeq 208\,\textrm{MeV}\\
\textrm{Br}(\tau^-\to K^{\ast\,-}\nu_\tau)=(1.29\pm
0.03)\%&\Rightarrow&f_{K^{\ast\,-}}\!\!\simeq 217\,\textrm{MeV}\,.
\end{eqnarray}
One can also determine $f_{\rho^0}$ from the width of the decay of
the $\rho^0$ into $e^+e^-$ which gives $f_{\rho^0}=216(5)$\,MeV.
Similarly from the width of the decay $\phi\to e^+e^-$ one deduces
$f_\phi\simeq 233$\,MeV\,.

The couplings $f_V^T$ are not known directly from experiment but 
are used
as inputs in sum-rule calculations (see, for example, refs.~\cite{Ball:2006eu,Ball:2004ye})
and other phenomenological
applications to $B$-decays (see, for example, refs.~\cite{Beneke:2001at,Bosch:2001gv,Ali:2007sj,Beneke:2003zv}). 
Previous lattice results for the vector meson couplings are recalled below; determinations obtained using QCD sum-rules are nicely reviewed in \cite{Ball:2006eu}.
We now present our calculation and results for $f_V^T/f_V$, which
can then be combined with the experimental values of $f_V$ to
obtain $f_V^T$. For the $\phi$ we neglect the Zweig suppressed
disconnected contribution.

\subsection{Lattice Calculation of
$f_V^T/f_V$}\label{subsec:calculation_fv}

In order to determine $f_V^T/f_V$ it is sufficient to calculate
the following zero-momentum correlation functions for large values
of the Euclidean time $t$:
\begin{eqnarray}
C_{VV}^{s_1s_2}(t)&\equiv&\sum_{\vec{x},i}
\langle\,0\,|V_i^{s_1}(t,\vec{x})\,V_i^{s_2}(0)\,|\,0\,\rangle =
3f_V^{s_1}
f_V^{s_2}m_Ve^{-m_VT/2}\cosh\left(m_V\left(T/2-t\right)\right)\label{eq:CVV}\\
C_{TV}^{s_1s_2}(t)&\equiv&\sum_{\vec{x},i}
\langle\,0\,|T_{4i}^{s_1}(t,\vec{x})\,V_i^{s_2}(0)\,|\,0\,\rangle
=3f_V^{Ts_1}f_V^{s_2}
m_Ve^{-m_VT/2}\sinh\left(m_V\left(T/2-t\right)\right)\label{eq:CTV}\,,
\end{eqnarray}
where $V_i$ and $T_{4i}$ represent the vector and tensor
currents and $i=1,2,3$ is a spatial index.
$s_1$ and $s_2$ label the smearing at the sink and
at the source, respectively. From the
ratio
\begin{equation}\label{eqn:vec_ratio}
 R(t)=
 \frac{C_{TV}^{Ls_2}}{C_{VV}^{Ls_2}}=\frac{f_V^T}{f_V}\,
 \tanh(m_V(T/2-t))
\end{equation}
we readily obtain the ratio of
(bare) couplings.
\subsection{Results}\label{subsec:fvresults}

In Table \ref{tab:vector_masses} we summarize our results for the 
vector meson masses from fits
to (\ref{eq:CVV}) on the DEG data set (cf. Table \ref{tab:datasets}) and from 
fits to (\ref{eq:CVV}) and (\ref{eq:CTV}) on the 
UNI data set. In the latter case we average 
over various choices of the
source smearing function (cf. Table \ref{tab:Vcontract}) while always using
a point sink. 
We restrict our study to the unitary case in which the sea and valence
quarks have the same masses.
The bare strange quark mass is always fixed at 0.04. 
On the UNI data set, again averaging over the same 
choices for the  source and the sink, we also evaluate the ratios $f_V^T/f_V$. 
In each case Eq.~(\ref{eqn:vec_ratio}) exhibits
well pronounced plateaus which we fit to a constant.

In Table~\ref{tab:pda_frho} we present the bare values of $f_V^T/f_V$. 
It can be seen that the
measured results are obtained with excellent precision.
We have also
compared our results with those obtained on
a $16^3\times 32\times 16$ lattice for $m_l=$0.01, 0.02 and 0.03
\cite{Donnellan:2007xr}
(the properties of the ensembles on the $16^3$ lattice have been presented in
ref.~\cite{Allton:2007hx}). No significant finite volume effects were found.

>From Fig.~\ref{fig:pda_chiral} it can be seen that the dependence of
the bare $f_V^T/f_V$ on the masses of the light quarks is very
mild and so we restrict our chiral extrapolation to linear and quadratic functions in the quark mass as shown in the figure. For the ratio of bare couplings in the chiral limit we obtain:
\begin{equation}\frac{f_\rho^T}{f_\rho}=0.619(15)(18);\quad
\frac{f_{K^\ast}^T}{f_{K^\ast}}=0.6498(62)(60);\quad
\frac{f_{\phi}^T}{f_\phi}=0.6838(32)(22)\,,\label{eq:barefv}
\end{equation}
where the central value corresponds to the linear extrapolation
and the second error is the difference between the results from
the linear and quadratic extrapolations.

The bare results in Eq.\,(\ref{eq:barefv}) were obtained with the
notional strange quark mass of $m_h=0.04$ rather than the
physical value of $m_s = 0.0343$ (see Table\,\ref{tab:results}). The values of the ratios in
Eq.\,(\ref{eq:barefv}) are very similar for the $\rho,\ K^\ast$
and $\phi$ mesons and we correct for the change in $m_s$ by
linear interpolation in the valence quark mass
($m_h$ is fixed at 0.04). Thus, for example, for
the $K^\ast$ meson we interpolate between the $K^\ast$ and the $\rho$:
\begin{equation}
\frac{f_{K^\ast}^T}{f_{K^\ast}}(m_s=0.0343)=
\frac{f_{K^\ast}^T}{f_{K^\ast}}(m_h=0.04)+\frac{\Delta}{(0.04+m_{\textrm{res}})}\,
(0.0343-0.04)\,,
\end{equation}
where $\Delta=
f_{K^\ast}^T/f_{K^\ast}(m_h=0.04)-f_\rho^T/f_\rho$. After carrying out
a similar extrapolation for $f_\phi^T/f_\phi$ the corrected bare values
are then
\begin{equation}\frac{f_\rho^T}{f_\rho}=0.619(15)(18);\quad
\frac{f_{K^\ast}^T}{f_{K^\ast}}=0.6457(62)(60);\quad
\frac{f_{\phi}^T}{f_\phi}=0.6753(32)(22)\,.\label{eq:barefvcorr}
\end{equation}

We determine the renormalization constants non-perturbatively
using the Rome-Southampton method and run the results to
2\,GeV. The results for the renormalization constants of the tensor and (local) vector currents
were presented individually in ref.~\cite{Aoki:2007xm} and for the ratio we find $Z_T/Z_A=Z_T/Z_V=1.1101(92)\simeq1.11(1)$, where $Z_A$ is the renormalization constant of the local axial current.
The relation between the ratios of bare and renormalized matrix elements is then:
\begin{equation}
\frac{f_V^T(2\,\textrm{GeV})}{f_V}=\frac{Z_T(2\,\textrm{GeV}a)}{Z_V}
\,\frac{f_V^{T\,\textrm{bare}}(a)}{f^{\textrm{bare}}_V}=1.11(1)\,
\frac{f_V^{T\,\textrm{bare}}(a)}{f^{\textrm{bare}}_V}\,.\end{equation}
In the $\overline{\textrm{MS}}$ scheme with $\mu=2$\,GeV we
finally obtain:
\begin{equation}\frac{f_\rho^T}{f_\rho}=0.687(27);\quad
\frac{f_{K^\ast}^T}{f_{K^\ast}}=0.717(12);\quad
\frac{f_{\phi}^T}{f_\phi}=0.750(8)\,.\label{eq:finalpreliminary}
\end{equation}

These results can be compared with previous quenched lattice
results which we summarize in Table \ref{tab:f_V-previous}.
The QCDSF/UKQCD collaboration has also presented the result $f_\rho^T=168(3)$\,MeV using an $N_f=2$ $O(a)$ improved clover action with a range of lattice spacings
($0.07<a<0.11$\,fm)~\cite{Gockeler:2005mh}. Combining our result for
the ratio from Eq.\,(\ref{eq:finalpreliminary}) together with the
experimental value for $f_\rho$ we obtain a smaller value
$f_\rho^T=143(6)$\,MeV\,.

%% file: text_sections/Conclusions.tex
%
%

This paper reports the results of simulations of 2+1 flavor QCD
with domain wall fermions at a single lattice spacing, with a larger
volume than has been achieved previously.  The ensembles we have
generated have a length of $\approx 4,000$ molecular dynamics time
units, and we have seen a good distribution of global topological
charge, indicative of our algorithm's ability to sample phase space.
With our $(2.74 \; {\rm fm})^3$ spatial volume, and improvements
in the RHMC algorithm, our lightest pion made of dynamical quarks
has $m_\pi = 331$ MeV and our lightest valence pion has a mass of
$m_\pi = 242$ MeV.  For these two cases, we have $m_\pi L = 4.60$
and 3.36, respectively.

Because of the good chiral properties of domain wall fermions at
non-zero lattice spacing, the only correction to continuum chiral
perturbation theory at NLO order is the inclusion of
the residual mass, $\mres$, in the quark mass appearing in the ChPT
formula.  This means we have the same number of ChPT LECs to fit
as for continuum physics, although the LECs we determine can differ
from the continuum values by $O(a^2)$ terms.  The small number of
free parameters in our fits is a powerful advantage of working with
a fermion formulation with good chiral and flavor symmetries.  To
have good control over our chiral extrapolations, we have developed
SU(2) ChPT for kaon physics, which does not assume the kaon mass
is small.  We have fit our data to both SU(2) and SU(3) ChPT, at
NLO, and find both approaches give good fits to our data.  However,
for SU(3) ChPT, the NLO corrections are very large - as much as
50\% of the size of the LO term for the pseudoscalar decay constants
in the range of quark masses where we have data.  This poor convergence
of the series does not seem to be due to a single term, such as
having the dynamical heavy quark mass close to the physical strange
quark mass.  Our fits have used pseudoscalar particles with mass
$\lesssim 400$ MeV, where SU(3) ChPT might be expected to be more
accurate than we have found it to be.  We are investigating this
further, making use of the full continuum NNLO expressions of
\cite{Bijnens:2006jv}.

For our most reliable extrapolations to the physical values for the
light quark masses, we turn to SU(2) ChPT for kaons.  Here we have
found good fits, with the NLO corrections no larger than 35\% of
the LO terms for the light quark mass values we used in our
simulations.  This is still a somewhat large NLO contribution, but
makes the dropping of NNLO terms more sensible.  We have tried to
estimate our systematic errors by including possible analytic NNLO
terms in the fits.  For our central results, we use $m_\Omega$,
$m_\pi$ and $m_K$ to set the scale and determine $\widetilde{m}_{ud}$
and $\widetilde{m}_s$.  For $m_\Omega$, there are no light quark
chiral logarithms and for $\widetilde{m}_{ud}$ and $\widetilde{m}_s$
we use the SU(2) ChPT formula to perform the extrapolation.  We can
then predict the decay constants, $f_\pi$, $f_K$ and their ratio.
The fits also give values for the LECs of SU(2) ChPT.

We have estimated our systematics through several methods.  For the
systematic errors from our chiral fits, we have tried varying the
fit ranges and also including analytic NNLO terms, to see the effects
of these changes on our results.  Finite volume effects are also
estimated through ChPT, as are the errors due to the heavy quark
in our simulations being 15\% heavier than the physical value.  For
finite lattice spacing effects, we expect them to be $O(a \Lambda_{\rm
QCD})^2$ or about 4\%.  Scaling errors no larger than this have
been seen in the quenched case \cite{CP-PACS:BK}, and, since dynamical
quark loop effects change the scaling at a higher order in $\alpha_S$,
similar scaling is expected for the full QCD case studied here.
The size of this scaling error also agrees well with the deviation
of our central value of $f\pi$ from the experimental results, so
we have adopted a uniform 4\% $O(a^2)$ error for all of our results.
(Simulations are underway at a smaller lattice spacing, so that we
will soon have better control over the $a\to 0$ extrapolation.)

For $B_K$, which has already been published \cite{Antonio:2007pb},
we have given more details about our use of SU(2) ChPT
to extrapolate to the light quark limit.  We have also measured
the vector meson couplings for light vector mesons.

Our major results come from using SU(2) ChPT fits and
are summarized below:
\begin{eqnarray}
f&=&114.8(4.1)_{\rm stat}(8.1)_{\rm syst}\,{\rm MeV}\,,\\
B^{\overline{\rm MS}}(2\,{\rm GeV}) &=& 2.52(0.11)_{\rm stat}(0.23)_{\rm ren}(0.12)_{\rm syst}\,{\rm GeV}\,,\\
\Sigma^{\overline{\rm MS}}(2\,{\rm GeV}) &=& \Big(255(8)_{\rm stat}(8)_{\rm ren}(13)_{\rm syst}\,{\rm MeV}\Big)^3\,,\\
\bar{l}_3 &=& 3.13(0.33)_{\rm stat}(0.24)_{\rm syst}\,,\\
\bar{l}_4 &=& 4.43(0.14)_{\rm stat}(0.77)_{\rm syst}\,,\\
\Lambda_3 &=& 666(110)_{\rm stat}(80)_{\rm syst}\,{\rm MeV}\,,\\
\Lambda_4 &=& 1,274(92)_{\rm stat}(490)_{\rm syst}\,{\rm MeV}\,,\\
m_{ud}^{\overline{\rm MS}}(2\,{\rm GeV}) &=& 3.72(0.16)_{\rm stat}(0.33)_{\rm ren}(0.18)_{\rm syst}\,{\rm MeV}\,,\\
m_{s}^{\overline{\rm MS}}(2\,{\rm GeV}) &=& 107.3(4.4)_{\rm stat}(9.7)_{\rm ren}(4.9)_{\rm syst}\,{\rm MeV}\,,\\
\widetilde{m}_{ud}:\widetilde{m}_s &=& 1:28.8(0.4)_{\rm stat}(1.6)_{\rm syst}\,,\\
f_\pi &=& 124.1(3.6)_{\rm stat}(6.9)_{\rm syst}\,{\rm MeV}\,,\\
f_K &=& 149.6(3.6)_{\rm stat}(6.3)_{\rm syst}\,{\rm MeV}\,,\\
f_K/f_\pi &=& 1.205(0.018)_{\rm stat}(0.062)_{\rm syst}\,,\\
B_K^{\overline{\rm MS}}(2\,{\rm GeV}) &=& 0.524(0.010)_{\rm stat}(0.013)_{\rm ren}(0.025)_{\rm syst}\,.
\end{eqnarray}

These results have total errors for the decay constants of about
6\% and for the $\overline{\rm MS}$ quark masses of about 10\%.
The systematic error due to $O(a^2)$ errors will be reduced by the
lattices we are generating at a smaller lattice spacing ($1/a \approx
2.3$ GeV) and lighter quark masses on a similar physical volume.
The errors on the quark masses are largely due to renormalization
and should improve as better renormalization conditions are chosen
\cite{Aoki:2007xm}.

With the expected increases in computer power over the next few years,
we should be able to push to much lighter quark masses and
minimize our reliance on chiral perturbation theory in the extrapolation
to physical results.  This is possible for domain wall fermions, since
the chiral limit is decoupled from the continuum limit.  With lighter
quark masses, we should be able to check the convergence of the
chiral perturbation theory expansion and achieve better control over the
low energy constants.  Finally, the ensembles that we are generating
in these basic studies of the low energy properties of QCD are
useful for a wide variety of other measurements, including heavy
quark systems and hadronic weak interaction matrix elements.

%% file: text_sections/Appendix.tex
%
%

\section{\label{sec:appendix:notation}Notation, Conventions}

\subsection{Quark Masses}

Here we discuss our notation for quark masses.
In Section \ref{subsect:kchpt}, where we discuss Kaon ChPT, continuum
notation is used and $m$ represents the total quark mass.  Throughout
the rest of the paper, we need to distinguish between the input
bare quark mass in the domain wall fermion formulation and the total
quark mass, which includes the additive contribution to the quark 
mass from finite $L_s$.

Input (sometimes also called bare) quark masses for domain wall fermions
will be denoted by the symbol $m_X$, where the index $X\in\{l,\, h,\,x,\,y,\,ud,\,s\}$ distinguishes between different types of quarks:
\begin{itemize}
  \item $m_l$: dynamical light input quark mass (two degenerate flavors),
  \item $m_h$: dynamical heavy input quark mass (one flavor),
  \item $m_x$, $m_y$: valence input quark masses,
  \item $m_{ud}$: average light input quark mass at the physical point, {\em i.e.}, a meson made of two quarks of mass $m_{ud}$ acquires the experimentally measured mass of the neutral pion (this is related to the input quark masses of the up- and down-quarks via: $m_{ud}=(m_u+m_d)/2$),
  \item $m_s$: heavy input quark mass at the physical point (strange quark mass), {\em i.e.}, a meson made of one quark of mass $m_{ud}$ and one of mass $m_s$ acquires the quadratically averaged experimentally measured mass of the neutral kaons.
\end{itemize}
The total quark mass, which has the residual mass $m_{\rm res}$ added, will be denoted by $\widetilde{m}$,
\begin{equation}
 \widetilde{m}_X\;=\;m_X\:+\:m_{\rm res}\,.
\end{equation}
The total renormalized quark masses at the physical point carry a superscript indicating the renormalization scheme used. For example, physical masses in the $\overline{\rm MS}$ scheme renormalized at a scale $\mu=2\,{\rm GeV}$ read
\begin{eqnarray}
m_{ud}^{\overline{\rm MS}}(2\,{\rm GeV}) &=& Z_m^{\overline{\rm MS}}(2\,{\rm GeV})\cdot(m_{ud}+m_{\rm res})\;=\;Z_m^{\overline{\rm MS}}(2\,{\rm GeV})\,\widetilde{m}_{ud}\,,\\
m_s^{\overline{\rm MS}}(2\,{\rm GeV}) &=& Z_m^{\overline{\rm MS}}(2\,{\rm GeV})\cdot(m_s+m_{\rm res})\;=\;Z_m^{\overline{\rm MS}}(2\,{\rm GeV})\,\widetilde{m}_s\,,
\end{eqnarray}
where $Z_m^{\overline{\rm MS}}(2\,{\rm GeV})$ is the mass renormalization constant.

\subsection{Pseudoscalar Quantities}

The pseudoscalar quantities considered in this work (masses, decay constants, and bag parameters) will carry two indices, denoting the quark content. For example the mass of a pseudoscalar made from two valence quarks with mass $m_x$ and $m_y$ will be labeled $m_{xy}$, the decay constant of a meson with one valence quark and one dynamical heavy quark is $f_{xh}$ and the pseudoscalar bag parameter of a meson with one quark at the physical average light quark mass and one valence quark is $B_{ud\,y}$. Also we will use the obvious symbols for quantities at the physical point:
\begin{eqnarray*}
m_\pi\;=\;m_{ud\,ud}\,, && m_K\;=\;m_{ud\,s}\,,\\
f_\pi\;=\;f_{ud\,ud}\,, && f_K\;=\;f_{ud\,s}\,,\\
 && B_K\;=\;B_{ud\,s}\,.
\end{eqnarray*}

\subsection{Low energy constants in ChPT}

The following table lists the LECs we use in the PQChPT formulae:
\begin{center}
  \begin{tabular}{l|c|c}
    \hline\hline & LO & NLO \\\hline\hline
    PQ $\SU(3)\times\SU(3)$ & $\bthree$, $\fthree$, & $\lthree{4}$, $\lthree{5}$, $\lthree{6}$, $\lthree{8}$, \\
                            & $\bagzero$ &  $b$, $c$, $d$ \\\hline
    PQ $\SU(2)\times\SU(2)$ & $\btwo$, $\ftwo$        & $\ltwo{4}$, $\ltwo{5}$, $\ltwo{6}$, $\ltwo{8}$ \\\hline
    PQ $\SU(2)\times\SU(2)$ kaon & $B^{(K)}(m_h)$, $f^{(K)}(m_h)$,  & $\lambda_1(m_h)$, $\lambda_2(m_h)$, $\lambda_3(m_h)$, $\lambda_4(m_h)$,\\
                                 & $B_{\rm PS}^{(K)}(m_h)$ &  $b_1(m_h)$, $b_2(m_h)$ \\\hline\hline  
  \end{tabular}
  \end{center}
The renormalization scale dependent NLO LECs are defined at a scale $\Lambda_\chi$. For the decay constants, we use the normalization such that the experimentally measured value for the pion decay constant is $f_\pi\approx132\,{\rm MeV}\approx\sqrt{2}\cdot 93\,{\rm MeV}$.

Further, we will use the following abbreviations for (tree-level) masses in the formulae:
\begin{eqnarray}
 \chi_X &=& 2\bthree\,\widetilde{m}_X\,,\;\;\;\SU(3)\times\SU(3)\,,\\
 \chi_X &=& 2\btwo\,\widetilde{m}_X\,,\;\;\;\SU(2)\times\SU(2)\,,
\end{eqnarray}
where it should be clear from the context, whether the $\SU(3)\times\SU(3)$ or $\SU(2)\times\SU(2)$ expression is to be used. In addition, we define the average dynamical mass to be
\begin{equation}
 \overline{m} \;=\;\frac13\,(2m_l+m_h)\,,
\end{equation}
and the dynamical $\eta$-mass 
\begin{equation}
 m_\eta \;=\;\frac13\,(m_l+2m_h)\,. 
\end{equation}
(Note: this is not the tree-level $\eta$-mass at the physical point, which would be $(m_{ud}+2m_s)/3$.)


\section{PQChPT Fit Functions}


\subsection{\label{sec:app:su3}\bteq{\SU(3)\times\SU(3)}}

The following formulae hold for $N_f=2+1$ sea quark masses and two valence quarks ($\SU(5|3)$). The formulae for masses and decay constants were derived from \cite{Sharpe:2000bc}, while the ones for the bag parameter $B_{\rm PS}$ can be found in \cite{VandeWater:2005uq}.

\subsubsection{Squared Pseudoscalar Mass}

\paragraph{non-degenerate (\bteq{m_x \neq m_y})}

\begin{itemize}
 \item $m_x \neq m_\eta \neq m_y$:
\begin{eqnarray}\label{eq:chPTsu3:mPS:nondeg}
m_{xy}^2 &=& \frac{\chi_x+\chi_y}{2}\,\cdot\,\Bigg\{\,1\,+\,\frac{48}{\fthree^2}(2\lthree{6}-\lthree{4})\bar\chi\,+\,\frac8{\fthree^2}(2\lthree{8}-\lthree{5})(\chi_x+\chi_y) \nonumber\\
&& \:+\frac1{24\pi^2 \fthree^2}\bigg[ \frac{(\chi_x-\chi_l)(\chi_x-\chi_h)}{(\chi_x-\chi_y)(\chi_x-\chi_\eta)}\chi_x\log\frac{\chi_x}{\Lambda_\chi^2} \,+\, \frac{(\chi_y-\chi_l)(\chi_y-\chi_h)}{(\chi_y-\chi_x)(\chi_y-\chi_\eta)}\chi_y\log\frac{\chi_y}{\Lambda_\chi^2} \nonumber\\
&& \;+\frac{(\chi_\eta-\chi_l)(\chi_\eta-\chi_h)}{(\chi_\eta-\chi_x)(\chi_\eta-\chi_y)}\chi_\eta\log\frac{\chi_\eta}{\Lambda_\chi^2} \bigg] \Bigg\}
\end{eqnarray}
\item $m_x \to m_\eta \neq m_y$:
\begin{eqnarray}\label{eq:chPTsu3:mPS:nondegEta}
m_{xy}^2 &=& \frac{\chi_x+\chi_y}{2}\,\cdot\,\Bigg\{\,1\,+\,\frac{48}{\fthree^2}(2\lthree{6}-\lthree{4})\bar\chi\,+\,\frac8{\fthree^2}(2\lthree{8}-\lthree{5})(\chi_x+\chi_y) \nonumber\\
&& \:+ \frac1{24\pi^2 \fthree^2}\bigg[\frac{(\chi_y-\chi_l)(\chi_y-\chi_h)}{(\chi_y-\chi_x)^2}\chi_y\log\frac{\chi_y}{\Lambda_\chi^2} \nonumber\\
&& \;+ \frac{(\chi_x-\chi_y)\Big[(\chi_x-\chi_h)\chi_x+(\chi_x-\chi_l)(2\chi_x-\chi_h)\Big]-(\chi_x-\chi_l)(\chi_x-\chi_h)\chi_x}{(\chi_x-\chi_y)^2}\log\frac{\chi_x}{\Lambda_\chi^2} \nonumber \\
&& \;+ \frac{(\chi_x-\chi_l)(\chi_x-\chi_h)}{\chi_x-\chi_y}  \bigg] \Bigg\}
\end{eqnarray}
\end{itemize}
%
\paragraph{degenerate (\bteq{m_x = m_y})}

\begin{itemize}
\item $m_x = m_y \neq m_\eta$:
\begin{eqnarray}\label{eq:chPTsu3:mPS:deg}
m_{xx}^2 &=& \chi_x\,\cdot\,\Bigg\{\,1\,+\,\frac{48}{\fthree^2}(2\lthree{6}-\lthree{4})\bar\chi\,+\,\frac{16}{\fthree^2}(2\lthree{8}-\lthree{5})\chi_x \nonumber\\
&& \:+\frac1{24\pi^2 \fthree^2}\bigg[\Big(\frac{2\chi_x-\chi_l-\chi_h}{\chi_x-\chi_\eta} \,-\, \frac{(\chi_x-\chi_l)(\chi_x-\chi_h)}{(\chi_x-\chi_\eta)^2} \Big)\chi_x\log\frac{\chi_x}{\Lambda_\chi^2} \nonumber \\
&& \;+ \frac{(\chi_x-\chi_l)(\chi_x-\chi_h)}{\chi_x-\chi_\eta}\Big(1+\log\frac{\chi_x}{\Lambda_\chi^2}\Big) \,+\, \frac{(\chi_\eta-\chi_l)(\chi_\eta-\chi_h)}{(\chi_x-\chi_\eta)^2}\chi_\eta\log\frac{\chi_\eta}{\Lambda_\chi^2} \bigg] \Bigg\} \nonumber \\
\end{eqnarray}
\item $m_x=m_y \to m_\eta$:
\begin{eqnarray}\label{eq:chPTsu3:mPS:degEta}
m_{xx}^2 &=& \chi_x\,\cdot\,\Bigg\{\,1\,+\,\frac{48}{\fthree^2}(2\lthree{6}-\lthree{4})\bar\chi\,+\,\frac{16}{\fthree^2}(2\lthree{8}-\lthree{5})\chi_x \nonumber\\
&& \:+\frac1{24\pi^2 \fthree^2}\bigg[ \frac{\chi_l\chi_h}{2\chi_x} \,+\, \frac{5\chi_x-3\chi_l-3\chi_h}2 \,+\, (3\chi_x-\chi_l-\chi_h)\log\frac{\chi_x}{\Lambda_\chi^2}\bigg]\Bigg\}
\end{eqnarray}
\end{itemize}

\paragraph{Pion and Kaon masses \bteq{m_\pi^2} and \bteq{m_K^2}}

From Eq.~(\ref{eq:chPTsu3:mPS:deg}) we obtain for the degenerate meson mass in unquenched QCD ($m_x=m_l$)
\begin{eqnarray}\label{eq:chPTsu3:mPi}
m_{ll}^2&=&\chi_l\,\cdot\,\Bigg\{\,1\,+\,\frac{48}{\fthree^2}(2\lthree{6}-\lthree{4})\bar\chi\,+\,\frac{16}{\fthree^2}(2\lthree{8}-\lthree{5})\chi_l \nonumber\\
 &&\:+\frac{1}{24\pi^2\fthree^2}\bigg[\frac32 \chi_l\log\frac{\chi_l}{\Lambda_\chi^2}\,-\,\frac12\chi_\eta\log\frac{\chi_\eta}{\Lambda_\chi^2}\bigg] \,\Bigg\}\,,
\end{eqnarray}
and from Eq.~(\ref{eq:chPTsu3:mPS:nondeg}) for the kaon mass ($m_x=m_l$, $m_y=m_h$):
\begin{eqnarray}\label{eq:chPTsu3:mK}
m_{lh}^2&=&\frac{\chi_l+\chi_h}2\,\cdot\,\Bigg\{\,1+\,\frac{48}{\fthree^2}(2\lthree{6}-\lthree{4})\bar\chi\,+\,\frac{8}{\fthree^2}(2\lthree{8}-\lthree{5})(\chi_l+\chi_h) \nonumber\\
 &&\:+\frac{1}{24\pi^2\fthree^2}\, \chi_\eta\log\frac{\chi_\eta}{\Lambda_\chi^2} \,\Bigg\}.
\end{eqnarray}
So that for $m_x=m_l=m_{ud}$ and $m_y=m_h=m_s$ we get $m_\pi^2$ and $m_K^2$, respectively. These expressions agree with \cite{Gasser:1984gg}.

For a ``kaon'' made from light sea quark and valence quark ($m_x\neq m_l$), one obtains:\\
\begin{itemize}
\item $m_x \neq m_\eta$:
\begin{eqnarray}\label{eq:chPTsu3:mK:valLight}
m_{lx}^2&=&\frac{\chi_l+\chi_x}2\,\cdot\,\Bigg\{\,1+\,\frac{48}{\fthree^2}(2\lthree{6}-\lthree{4})\bar\chi\,+\,\frac{8}{\fthree^2}(2\lthree{8}-\lthree{5})(\chi_l+\chi_x) \nonumber\\
 &&\:+\frac{1}{24\pi^2\fthree^2}\bigg[ \frac{\chi_x-\chi_h}{\chi_x-\chi_\eta}\chi_x\log\frac{\chi_x}{\Lambda_\chi^2}\,+\,\frac{\chi_\eta-\chi_h}{\chi_\eta-\chi_x}\chi_\eta\log\frac{\chi_\eta}{\Lambda_\chi^2}\,\bigg]\Bigg\},
\end{eqnarray}
\item $m_x \to m_\eta$:
\begin{eqnarray}\label{eq:chPTsu3:mK:valLightEta}
m_{lx}^2&=&\frac{\chi_l+\chi_x}2\,\cdot\,\Bigg\{\,1+\,\frac{48}{\fthree^2}(2\lthree{6}-\lthree{4})\bar\chi\,+\,\frac{8}{\fthree^2}(2\lthree{8}-\lthree{5})(\chi_l+\chi_x) \nonumber\\
 &&\:+\frac{1}{24\pi^2\fthree^2}\bigg[ \chi_x - \chi_h\,+\,(2\chi_x-\chi_h)\log\frac{\chi_x}{\Lambda_\chi^2}\bigg]\Bigg\}.
\end{eqnarray}
\end{itemize}
Whereas for a ``kaon'' made from a valence quark ($m_x\neq m_h$) and a heavy sea quark, we have:\\
\begin{itemize}
\item $m_x \neq m_\eta$:
\begin{eqnarray}\label{eq:chPTsu3:mK:valStrange}
m_{xh}^2 &=& \frac{\chi_x+\chi_h}2\,\cdot\,\Bigg\{\,1+\,\frac{48}{\fthree^2}(2\lthree{6}-\lthree{4})\bar\chi\,+\,\frac{8}{\fthree^2}(2\lthree{8}-\lthree{5})(\chi_x+\chi_h) \nonumber\\
 &&\:+\frac{1}{24\pi^2\fthree^2}\bigg[ \frac{\chi_x-\chi_l}{\chi_x-\chi_\eta}\chi_x\log\frac{\chi_x}{\Lambda_\chi^2}\,+\,\frac{\chi_\eta-\chi_l}{\chi_\eta-\chi_x}\chi_\eta\log\frac{\chi_\eta}{\Lambda_\chi^2} \bigg]\Bigg\}\,,
\end{eqnarray}
\item $m_x \to m_\eta$
\begin{eqnarray}\label{eq:chPTsu3:mK:valStrangeEta}
m_{xh}^2 &=& \frac{\chi_x+\chi_h}2\,\cdot\,\Bigg\{\,1+\,\frac{48}{\fthree^2}(2\lthree{6}-\lthree{4})\bar\chi\,+\,\frac{8}{\fthree^2}(2\lthree{8}-\lthree{5})(\chi_x+\chi_h) \nonumber\\
 &&\:+\frac{1}{24\pi^2\fthree^2}\bigg[ (2\chi_x-\chi_l)\log\frac{\chi_x}{\Lambda_\chi^2}\,+\,\chi_x-\chi_l\bigg]\Bigg\}\,.
\end{eqnarray}
\end{itemize}

\subsubsection{Pseudoscalar Decay Constant}

\paragraph{non-degenerate (\bteq{m_x \neq m_y})}

\begin{itemize}
\item $m_x \neq m_\eta \neq m_y$:
\begin{eqnarray}\label{eq:chPTsu3:fPS:nondeg}
f_{xy} &=& \fthree\,\cdot\,\Bigg\{\,1\,+\,\frac{24}{\fthree^2}\lthree{4}\bar\chi \,+\,\frac4{\fthree^2}\lthree{5}(\chi_x+\chi_y) \nonumber\\
&& \:-\frac1{8\pi^2\fthree^2}\bigg[\frac{\chi_x+\chi_l}4 \log\frac{\chi_x+\chi_l}{2\Lambda_\chi^2} + \frac{\chi_y+\chi_l}4 \log\frac{\chi_y+\chi_l}{2\Lambda_\chi^2} \nonumber\\
&& \;+\frac{\chi_x+\chi_h}8 \log\frac{\chi_x+\chi_h}{2\Lambda_\chi^2} + \frac{\chi_y+\chi_h}8 \log\frac{\chi_y+\chi_h}{2\Lambda_\chi^2} \bigg] \nonumber \\
&& \:+\frac{1}{96\pi^2\fthree^2}\bigg[ \Big(\chi_x\frac{(\chi_l-\chi_x)(\chi_h-\chi_x)}{\chi_\eta-\chi_x}+\chi_y\frac{(\chi_l-\chi_y)(\chi_h-\chi_y)}{\chi_\eta-\chi_y}+(\chi_x-\chi_y)^2\Big)\frac{\log\frac{\chi_x}{\chi_y}}{\chi_x-\chi_y} \nonumber \\
&& \;+\chi_\eta(\chi_y-\chi_x)\frac{(\chi_\eta-\chi_l)(\chi_\eta-\chi_h)}{(\chi_\eta-\chi_x)(\chi_\eta-\chi_y)}\Big(\frac{\log\frac{\chi_\eta}{\chi_x}}{\chi_x-\chi_\eta}-\frac{\log\frac{\chi_\eta}{\chi_y}}{\chi_y-\chi_\eta}\Big) \nonumber \\
&& \;-\frac{(\chi_l-\chi_x)(\chi_h-\chi_x)}{\chi_\eta-\chi_x}\,-\,\frac{(\chi_l-\chi_y)(\chi_h-\chi_y)}{\chi_\eta-\chi_y} \bigg] \Bigg\}
\end{eqnarray}
\item $m_x \to m_\eta \neq m_y$:
\begin{eqnarray}\label{eq:chPTsu3:fPS:nondegEta}
f_{xy} &=& \fthree\,\cdot\,\Bigg\{\,1\,+\,\frac{24}{\fthree^2}\lthree{4}\bar\chi\,+\,\frac4{\fthree^2}\lthree{5}(\chi_x+\chi_y) \nonumber\\
&& \:-\frac1{8\pi^2\fthree^2}\bigg[\frac{\chi_x+\chi_l}4 \log\frac{\chi_x+\chi_l}{2\Lambda_\chi^2} + \frac{\chi_y+\chi_l}4 \log\frac{\chi_y+\chi_l}{2\Lambda_\chi^2} \nonumber\\
&& \;+\frac{\chi_x+\chi_h}8 \log\frac{\chi_x+\chi_h}{2\Lambda_\chi^2} + \frac{\chi_y+\chi_h}8 \log\frac{\chi_y+\chi_h}{2\Lambda_\chi^2} \bigg] \nonumber \\
&& \:+\frac{1}{96\pi^2\fthree^2}\bigg[ \frac{ \chi_x^3 + \chi_x^2(\chi_l+\chi_h-5\chi_y)+\chi_x\Big(5\chi_y(\chi_l+\chi_h)-5\chi_l\chi_h-2\chi_y^2\Big) -\chi_y\chi_l\chi_h }{2\chi_x(\chi_x-\chi_y)} \nonumber \\
&& \;+ \frac{\chi_x\Big(3\chi_y^2+\chi_l\chi_h-2\chi_y(\chi_l+\chi_h)\Big)+\chi_y\Big(2\chi_l\chi_h-\chi_y(\chi_l+\chi_h)\Big)}{(\chi_x-\chi_y)^2}\log\frac{\chi_x}{\chi_y} \bigg] \Bigg\}
\end{eqnarray}
\end{itemize}

\paragraph{degenerate (\bteq{m_x = m_y})}
\begin{eqnarray}\label{eq:chPTsu3:fPS:deg}
f_{xx} &=&  \fthree\,\cdot\,\Bigg\{\,1\,+\,\frac{24}{\fthree^2}\lthree{4}\bar\chi \,+\,\frac8{\fthree^2}\lthree{5}\chi_x \nonumber\\
&& \:-\frac1{16\pi^2\fthree^2}\bigg[(\chi_x+\chi_l)\log\frac{\chi_x+\chi_l}{2\Lambda_\chi^2}\,+\,\frac{\chi_x+\chi_h}2\log\frac{\chi_x+\chi_h}{2\Lambda_\chi^2} \bigg] \Bigg\}
\end{eqnarray}
%
\paragraph{Pion and Kaon decay constants \bteq{f_\pi} and \bteq{f_K}}
From Eq.~(\ref{eq:chPTsu3:fPS:deg}) we obtain for the pion decay constant in unquenched QCD ($m_x=m_l$)
\begin{eqnarray}\label{eq:chPTsu3:fPi}
f_{ll}&=&\fthree\,\cdot\,\Bigg\{\,1\,+\,\frac{24}{\fthree^2}\lthree{4}\bar\chi\,+\,\frac8{\fthree^2}\lthree{5}\chi_l \nonumber\\
 && \:-\frac{1}{16\pi^2\fthree^2}\bigg[\,\frac{\chi_l+\chi_h}2\log\frac{\chi_l+\chi_h}{2\Lambda_\chi^2}\,+\,2\chi_l\log\frac{\chi_l}{\Lambda_\chi^2} \,\bigg]\,\Bigg\}\,,
\end{eqnarray}
and from Eq.~(\ref{eq:chPTsu3:fPS:nondeg}) for the kaon decay constant ($m_x=m_l$, $m_y=m_h$):
\begin{eqnarray}\label{eq:chPTsu3:fK}
f_{lh}&=&\fthree\,\cdot\,\Bigg\{\,1\,+\,\frac{24}{\fthree^2}\lthree{4}\bar\chi\,+\,\frac4{\fthree^2}\lthree{5}(\chi_l+\chi_h) \nonumber\\
 && \:-\frac{3}{64\pi^2\fthree^2}\bigg[\,2\frac{\chi_l+\chi_h}2\log\frac{\chi_l+\chi_h}{2\Lambda_\chi^2}\,+\,\chi_l\log\frac{\chi_l}{\Lambda_\chi^2}\,+\,\chi_\eta\log\frac{\chi_\eta}{\Lambda_\chi^2} \,\bigg]\,\Bigg\}.\nonumber\\
\end{eqnarray}
Again, for $m_x=m_l=m_{ud}$ and $m_y=m_h=m_s$ we get $f_\pi$ and $f_K$, respectively, and these expressions agree with \cite{Gasser:1984gg}.

For a ``kaon'' made from a light sea quark and valence quark ($m_x\neq m_l$), one obtains:\\
\begin{itemize}
\item $m_x \neq m_\eta$:
\begin{eqnarray}\label{eq:chPTsu3:fK:valLight}
f_{lx}&=&  \fthree\,\cdot\,\Bigg\{\,1\,+\,\frac{24}{\fthree^2}\lthree{4}\bar\chi\,+\,\frac4{\fthree^2}\lthree{5}(\chi_l+\chi_x) \nonumber\\
 && \:+\frac1{8\pi^2\fthree^2}\bigg[\,-\frac38\chi_l\log\frac{\chi_l}{\Lambda_\chi^2}-\frac{\chi_x+\chi_l}4\log\frac{\chi_x+\chi_l}{2\Lambda_\chi^2} \nonumber\\
 && \;-\frac{\chi_l+\chi_h}{8}\log\frac{\chi_l+\chi_h}{2\Lambda_\chi^2}-\frac{\chi_x+\chi_h}{8}\log\frac{\chi_x+\chi_h}{2\Lambda_\chi^2} \nonumber\\
 && \; +\frac{(\chi_x-\chi_l)^2(\chi_\eta-\chi_h)}{12(\chi_\eta-\chi_x)^2(\chi_\eta-\chi_l)}\chi_\eta\log\frac{\chi_\eta}{\Lambda_\chi^2} \nonumber\\
 && \; +\frac{\chi_x^2(\chi_h-\chi_l-\chi_\eta)-\chi_\eta\chi_l(\chi_h-2\chi_x)}{12(\chi_\eta-\chi_x)^2}\log\frac{\chi_x}{\Lambda_\chi^2} \nonumber\\
 && \; -\frac{(\chi_l-\chi_x)(\chi_h-\chi_x)}{12(\chi_\eta-\chi_x)} \:\bigg]\,\Bigg\},
\end{eqnarray}
\item $m_x \to m_\eta$:
\begin{eqnarray}\label{eq:chPTsu3:fK:valLightEta}
f_{lx}&=& \fthree\,\cdot\,\Bigg\{\,1\,+\,\frac{24}{\fthree^2}\lthree{4}\bar\chi\,+\,\frac4{\fthree^2}\lthree{5}(\chi_l+\chi_x) \nonumber\\
 && \:+\frac1{8\pi^2\fthree^2}\bigg[\,-\frac38\chi_l\log\frac{\chi_l}{\Lambda_\chi^2}-\frac{\chi_x+\chi_l}4\log\frac{\chi_x+\chi_l}{2\Lambda_\chi^2} \nonumber\\
 && \;-\frac{\chi_l+\chi_h}{8}\log\frac{\chi_l+\chi_h}{2\Lambda_\chi^2}-\frac{\chi_x+\chi_h}{8}\log\frac{\chi_x+\chi_h}{2\Lambda_\chi^2} \nonumber\\
 && \;+\frac{\chi_l(\chi_h-3\chi_x)+\chi_x(\chi_h+\chi_x)}{24\chi_x}\,+\,\frac{\chi_l(\chi_h-\chi_l)}{12(\chi_l-\chi_x)}\log\frac{\chi_x}{\Lambda_\chi^2} \,\bigg]\,\Bigg\}.
\end{eqnarray}
\end{itemize}
Whereas for a ``kaon'' made from a valence quark ($m_x\neq m_h$) and a heavy sea quark, we have:\\
\begin{itemize}
\item $m_x \neq m_\eta$
\begin{eqnarray}\label{eq:chPTsu3:fK:valStrange}
f_{xh}&=& \fthree\,\cdot\,\Bigg\{\,1\,+\,\frac{24}{\fthree^2}\lthree{4}\bar\chi \,+\,\frac4{\fthree^2}\lthree{5}(\chi_x+\chi_h) \nonumber\\
&& \:-\frac1{8\pi^2\fthree^2}\bigg[\frac{\chi_x+\chi_l}4 \log\frac{\chi_x+\chi_l}{2\Lambda_\chi^2} + \frac{\chi_h+\chi_l}4 \log\frac{\chi_h+\chi_l}{2\Lambda_\chi^2} \nonumber\\
&& \;+\frac{\chi_x+\chi_h}8 \log\frac{\chi_x+\chi_h}{2\Lambda_\chi^2} + \frac{\chi_h}4 \log\frac{\chi_h}{\Lambda_\chi^2} \bigg] \nonumber \\
&& \:+\frac{1}{96\pi^2\fthree^2}\bigg[ \Big( \chi_x-\chi_h-\chi_x\frac{\chi_l-\chi_x}{\chi_\eta-\chi_x}\Big)\log\frac{\chi_x}{\chi_h} \nonumber\\
&& \;+\chi_\eta(\chi_h-\chi_x)\frac{\chi_\eta-\chi_l}{\chi_\eta-\chi_x}\Big(\frac{\log\frac{\chi_\eta}{\chi_x}}{\chi_x-\chi_\eta}-\frac{\log\frac{\chi_\eta}{\chi_h}}{\chi_h-\chi_\eta}\Big) \nonumber\\
&& \;-\frac{(\chi_l-\chi_x)(\chi_h-\chi_x)}{\chi_\eta-\chi_x}\bigg]\Bigg\}\,,
\end{eqnarray}
\item $m_x \to m_\eta$
\begin{eqnarray}\label{eq:chPTsu3:fK:valStrangeEta}
f_{xh}&=& \fthree\,\cdot\,\Bigg\{\,1\,+\,\frac{24}{\fthree^2}\lthree{4}\bar\chi \,+\,\frac4{\fthree^2}\lthree{5}(\chi_x+\chi_h) \nonumber\\
&& \:-\frac1{8\pi^2\fthree^2}\bigg[\frac{\chi_x+\chi_l}4 \log\frac{\chi_x+\chi_l}{2\Lambda_\chi^2} + \frac{\chi_h+\chi_l}4 \log\frac{\chi_h+\chi_l}{2\Lambda_\chi^2} \nonumber\\
&& \;+\frac{\chi_x+\chi_h}8 \log\frac{\chi_x+\chi_h}{2\Lambda_\chi^2} + \frac{\chi_h}4 \log\frac{\chi_h}{\Lambda_\chi^2} \bigg] \nonumber \\
&& \:+\frac{1}{96\pi^2\fthree^2}\bigg[\frac{\chi_x^3+\chi_x^2(\chi_l-4\chi_h)+3\chi_x\chi_h^2-\chi_h^2\chi_l}{2\chi_x(\chi_x-\chi_h)} \nonumber\\
&& \;+\chi_h\frac{\chi_x(\chi_h-\chi_l)+\chi_h(\chi_l-\chi_h)}{(\chi_x-\chi_h)^2}\log\frac{\chi_x}{\chi_h}\bigg]\Bigg\}\,.
\end{eqnarray}
\end{itemize}

\subsubsection{Scale-dependence of the NLO-LECs}
The NLO-LECs $\lthree{i}(\Lambda_\chi)$ at a renormalization scale $\Lambda_\chi$ can be converted to a different scale $\Lambda_\chi^\prime$ using \cite{Gasser:1984gg}
\begin{equation}\label{eq:chPTsu3:LECconv}
\lthree{i}(\Lambda_\chi^\prime)\;=\; \lthree{i}(\Lambda_\chi)\:+\:\frac{\Gamma_i}{16\pi^2}\log\frac{\Lambda_\chi}{\Lambda_\chi^\prime}\,,
\end{equation}
with
\begin{equation}\label{eq:ChPTsu3:LECconv:Gamma}
 \Gamma_4\:=\: \frac18\,,\;\;\; \Gamma_5\:=\: \frac38\,,\;\;\;
\Gamma_6\:=\: \frac{11}{144}\,, \;\;\; \Gamma_8\:=\: \frac5{48}\,.
\end{equation}
%

\subsubsection{Pseudoscalar Bag Parameter}
Here we follow the notation of \cite{VandeWater:2005uq} for the LEC's relevant for the pseudoscalar bag parameter and denote them by $b$, $c$, and $d$. Also here we only work in (PQ) $\SU(3)$ chiral perturbation theory.
\paragraph{non-degenerate (\bteq{m_x \neq m_y})}
\begin{eqnarray}
\label{eq:chPTsu3:BPS:nondeg}
B_{xy} &=& \bagzero\Bigg\{\,1\,+\,\frac{1}{24\pi^2\fthree^2}\bigg[\frac{I_{\rm conn}+I_{\rm disc}}{\chi_x+\chi_y} 
 +\frac{\chi_x+\chi_y}{4}b\,+\,\frac{(\chi_x-\chi_y)^2}{\chi_x+\chi_y}c\,+\,\frac32\bar\chi d\bigg]\Bigg\}\,,\\[15pt]
\label{eq:chPTsu3:BPS:Iconn}
I_{\rm conn} &=& \frac32(\chi_x+\chi_y)^2 \Big(-1-\log\frac{\chi_x+\chi_y}{2\Lambda_\chi^2}\Big) \nonumber\\
                 && - \frac32(3\chi_x+\chi_y)\chi_x\log\frac{\chi_x}{\Lambda_\chi^2} - \frac32(\chi_x+3\chi_y)\chi_y\log\frac{\chi_y}{\Lambda_\chi^2},\\[15pt]
\label{eq:chPTsu3:BPS:Idisc}
I_{\rm disc} &=& I_x + I_y + I_\eta\,,\\[15pt]
\label{eq:chPTsu3:BPS:Ix}
I_x &=& \frac{(3\chi_x+\chi_y)(\chi_l-\chi_x)(\chi_h-\chi_x)}{2(\chi_\eta-\chi_x)}\Big(-1-\log\frac{\chi_x}{\Lambda_\chi^2}\Big) \nonumber\\
&& -\,\bigg[ \frac{(3\chi_x+\chi_y)(\chi_l-\chi_x)(\chi_h-\chi_x)}{2(\chi_\eta-\chi_x)^2} 
  + \frac{(3\chi_x+\chi_y)(\chi_l-\chi_x)(\chi_h-\chi_x)}{(\chi_y-\chi_x)(\chi_\eta-\chi_x)} \nonumber \\
  && \;+ \frac{2(\chi_l-\chi_x)(\chi_h-\chi_x)-(3\chi_x+\chi_y)(\chi_h-\chi_x)-(3\chi_x+\chi_y)(\chi_l-\chi_x)}{2(\chi_\eta-\chi_x)}\bigg]\cdot \nonumber\\
  && \,\cdot\chi_x\log\frac{\chi_x}{\Lambda_\chi^2},\\[15pt]
\label{eq:chPTsu3:BPS:Iy}
I_y &=& I_x(x\leftrightarrow y),\\[15pt]
\label{eq:chPTsu3:BPS:Ieta}
I_\eta &=& \frac{(\chi_x-\chi_y)^2(\chi_x+\chi_y+2\chi_\eta)(\chi_l-\chi_\eta)(\chi_h-\chi_\eta)}{2(\chi_x-\chi_\eta)^2(\chi_y-\chi_\eta)^2}\chi_\eta\log\frac{\chi_\eta}{\Lambda_\chi^2}.
\end{eqnarray}

For $m_x \to m_\eta$, $m_x \neq m_y$, one obtains from the disconnected diagrams
\begin{eqnarray}
\label{eq:chPTsu3:BPS:IdiscEta}
    I_{\mathrm{disc}} & = & -\chi_l\frac{\chi_\eta^3+\chi_\eta^2(7\chi_h-16\chi_y)+\chi_\eta \chi_y(16\chi_h-9\chi_y)+\chi_h\chi_y^2}{4\chi_\eta (\chi_\eta-\chi_y)} \nonumber \\
    && -\frac{-7\chi_\eta^3-9\chi_h\chi_y^2+6\chi_y^3+\chi_\eta \chi_y(-16\chi_h + 7\chi_y)+\chi_\eta^2(\chi_h+18\chi_y)}{4(\chi_\eta-\chi_y)} \nonumber \\
    && + \frac{1}{2(\chi_\eta-\chi_y)^2} \log\frac{\chi_\eta}{\Lambda_\chi^2}\,\cdot \nonumber \\
    && \,\cdot\bigg[2\chi_\eta^4-8\chi_\eta^3\chi_y-\chi_y^2\Big(\chi_h\chi_y+\chi_l(\chi_y-4\chi_h)\Big) \nonumber \\
    && \; + \chi_\eta \chi_y\Big(\chi_l(7\chi_h-9\chi_y)+3\chi_y(\chi_y-3\chi_h)\Big) \nonumber\\
    && \; + \chi_\eta^2\Big(\chi_l(\chi_h-2\chi_y)+\chi_y(15\chi_y-2\chi_h)\Big)\bigg] \nonumber \\
    && + \frac{1}{2(\chi_\eta-\chi_y)^2}  \log\frac{\chi_y}{\Lambda_\chi^2}\,\cdot \nonumber \\
    && \, \cdot\bigg[\chi_y\Big(\chi_\eta^2(2\chi_h-3\chi_y)+\chi_\eta \chi_y(9\chi_h-11\chi_y)+\chi_y^2(\chi_h+2\chi_y)\Big) \nonumber \\
    && \; +\chi_l\Big(-\chi_\eta^2(\chi_h-2\chi_y)+\chi_y^2(\chi_y-4\chi_h)+\chi_\eta \chi_y(9\chi_y-7\chi_h)\Big)\bigg]
\end{eqnarray}
%
\paragraph{degenerate (\bteq{m_x = m_y})}

In this case $I_{\mathrm{disc}} \;=\;0$ and therefore
\begin{eqnarray}
\label{eq:chPTsu3:BPS:deg}
B_{xx} &=& \bagzero\Bigg\{\,1\,+\,\frac{1}{24\pi^2\fthree^2}\bigg[-9\chi_x\log\frac{\chi_x}{\Lambda_\chi^2}\,+\,\frac{\chi_x}{2}(b-6)\,+\,\frac32\bar\chi d\bigg]\Bigg\}\,,
\end{eqnarray}
%
\paragraph{Kaon bag parameter \bteq{B_K}} 
From Eq.~(\ref{eq:chPTsu3:BPS:nondeg}) one obtains the full (unquenched) QCD result ($m_x= m_l$, $m_y= m_h$)
\begin{eqnarray}\label{eq:chPTsu3:Bk}
B_{lh} &=& \bagzero\Bigg\{\,1\,-\,\frac1{16\pi^2\fthree^2}\bigg[\,(\chi_l+\chi_h)\Big(1+\log\frac{\chi_l+\chi_h}{2\Lambda_\chi^2}\Big)\,+\,\frac{3\chi_l+\chi_h}{2(\chi_l+\chi_h)}\chi_l\log\frac{\chi_l}{\Lambda_\chi^2} \nonumber\\
  &&\;+\frac{5\chi_l+7\chi_h}{2(\chi_l+\chi_h)}\chi_\eta\log\frac{\chi_\eta}{\Lambda_\chi^2}\,-\,b\frac{\chi_l+\chi_h}{6}\,-\,c\frac{2(\chi_l-\chi_h)^2}{3(\chi_l+\chi_h)}\,-\,d\bar{\chi}\,\bigg]\Bigg\}\,,
\end{eqnarray}
so that at the physical point ($m_l=m_{ud}$, $m_h=m_s$) we have

\begin{eqnarray}\label{eq:chPTsu3:BkPhysical}
  B_K &=& \bagzero\Bigg\{\,1\,-\,\frac{2}{(4\pi \fthree)^2}\bigg[\,M_K^2\Big(1+\log\frac{M_K^2}{\Lambda_\chi^2}\Big)\,+\,\frac{M_\pi^2(M_\pi^2+M_K^2)}{4M_K^2}\log\frac{M_\pi^2}{\Lambda_\chi^2}\nonumber\\
 &&\;+\frac{7M_K^2-M_\pi^2}{4M_K^2}M_\eta^2\log\frac{M_\eta^2}{\Lambda_\chi^2} \,-\,\frac{b}{6}M_K^2\,-\,\frac{2c}3\,\frac{(M_\pi^2-M_K^2)^2}{M_K^2}\,-\,\frac{d}{6}\Big(M_\pi^2+2M_K^2\Big)\,\bigg]\,\Bigg\}
\end{eqnarray}
Here we expressed everything in terms of meson masses: $M_\pi^2=\chi_{ud}$, $M_K^2=\frac12(\chi_{ud}+\chi_s)$, $M_\eta^2=\chi_\eta|_{m_l=m_{ud},m_h=m_s}$. (This agrees for the non-analytic terms with Eq.~(12) of \cite{Becirevic:2003wk}, analytic terms were not considered in that reference.)


\subsection{\label{sec:app:su2}\bteq{\SU(2)\times\SU(2)}}

The following formulae hold for $N_f=2$ (degenerate) sea quark masses (mass $m_l$) and two valence quarks. ($\SU(4|2)$). The formulae for masses and decay constants were derived from \cite{Sharpe:2000bc}, also cf.\ \cite{Sharpe:1997by}.

\subsubsection{Squared Pseudoscalar Mass}
%
\paragraph{non-degenerate}
\begin{eqnarray}\label{eq:chPTsu2:mPS:nondeg}
m_{xy}^2 &=& \frac{\chi_x+\chi_y}2\,\cdot\,\Bigg\{\,1\,+\,\frac{32}{\ftwo^2}(2\ltwo{6}-\ltwo{4})\chi_l\,+\,\frac{8}{\ftwo^2}(2\ltwo{8}-\ltwo{5})(\chi_x+\chi_y)\nonumber\\
 && \:+\frac1{16\pi^2\ftwo^2}\bigg[\,\frac{\chi_x-\chi_l}{\chi_x-\chi_y}\chi_x\log\frac{\chi_x}{\Lambda_\chi^2}+\frac{\chi_y-\chi_l}{\chi_y-\chi_x}\chi_y\log\frac{\chi_y}{\Lambda_\chi^2}\bigg]\,\Bigg\}
\end{eqnarray}
%
\paragraph{degenerate}
\begin{eqnarray}\label{eq:chPTsu2:mPS:deg}
m_{xx}^2 &=& \chi_x\,\cdot\,\Bigg\{\,1\,+\,\frac{32}{\ftwo^2}(2\ltwo{6}-\ltwo{4})\chi_l\,+\,\frac{16}{\ftwo^2}(2\ltwo{8}-\ltwo{5})\chi_x\nonumber\\
 && \:+\frac1{16\pi^2\ftwo^2}\bigg[\,\chi_x-\chi_l\,+\,(2\chi_x-\chi_l)\log\frac{\chi_x}{\Lambda_\chi^2} \bigg]\,\Bigg\}
\end{eqnarray}
%
\paragraph{Pion mass \bteq{m_\pi^2}}
From Eq.~(\ref{eq:chPTsu2:mPS:deg}) one obtains the degenerate meson mass in full (unquenched) QCD ($m_x=m_l$):
\begin{equation}\label{eq:chPTsu2:mPi}
m_{ll}^2 \:=\:  \chi_l\,\cdot\,\Bigg\{\,1\,+\,\frac{16}{\ftwo^2}\Big((2\ltwo{8}-\ltwo{5})+2(2\ltwo{6}-\ltwo{4})\Big)\chi_l\,+\,\frac1{16\pi^2\ftwo^2}\chi_l\log\frac{\chi_l}{\Lambda_\chi^2}\,\Bigg\}
\end{equation}
which gives for $m_x=m_l=m_{ud}$ the pion mass $m_\pi^2$ and agrees with \cite{Gasser:1983yg}. Furthermore, we can relate the LECs $\ltwo{4,5,6,8}$ from PQChPT to $l_3^r$ in ChPT:
\begin{equation}\label{eq:chPTsu2:LECrel:l3}
4\Big((2\ltwo{8}-\ltwo{5})+2(2\ltwo{6}-\ltwo{4})\Big) \;=\; l_3^r
\end{equation}
%
\subsubsection{Pseudoscalar Decay Constant}
%
\paragraph{non-degenerate}
\begin{eqnarray}\label{eq:chPTsu2:fPS:nondeg}
f_{xy} &=& \ftwo\,\cdot\,\Bigg\{\,1\,+\,\frac{16}{\ftwo^2}\ltwo{4}\chi_l\,+\,\frac{4}{\ftwo^2}\ltwo{5}(\chi_x+\chi_y)\nonumber\\
 && \:-\frac1{32\pi^2\ftwo^2}\bigg[(\chi_x+\chi_l)\log\frac{\chi_x+\chi_l}{2\Lambda_\chi^2}\,+\,(\chi_y+\chi_l)\log\frac{\chi_y+\chi_l}{2\Lambda^2}\bigg]\nonumber\\
 && \:+\frac1{64\pi^2\ftwo^2}\bigg[\,\chi_x+\chi_y-2\chi_l\,+\,\frac{2\chi_x\chi_y-\chi_l(\chi_x+\chi_y)}{\chi_y-\chi_x}\log{\frac{\chi_x}{\chi_y}}\bigg]\,\Bigg\}
\end{eqnarray}
%
\paragraph{degenerate}
\begin{equation}\label{eq:chPTsu2:fPS:deg}
f_{xx} \:=\: \ftwo\,\cdot\,\Bigg\{\,1\,+\,\frac{16}{\ftwo^2}\ltwo{4}\chi_l\,+\,\frac{8}{\ftwo^2}\ltwo{5}\chi_x\,-\,\frac{\chi_x+\chi_l}{16\pi^2\ftwo^2}\log\frac{\chi_x+\chi_l}{2\Lambda_\chi^2}\Bigg\}
\end{equation}
%
\paragraph{Pion decay constant \bteq{f_\pi}}
\begin{equation}\label{eq:chPTsu2:fPi}
f_{ll} \:=\: \ftwo\,\cdot\,\Bigg\{\,1\,+\,\frac{8}{\ftwo^2}(2\ltwo{4}+\ltwo{5})\chi_l\,-\,\frac{\chi_l}{8\pi^2\ftwo^2}\log\frac{\chi_l}{\Lambda_\chi^2}\,\Bigg\}.
\end{equation}
Again, for $m_x=m_l=m_{ud}$ this reproduces the pion decay constant $f_\pi$ and agrees with \cite{Gasser:1983yg}. Furthermore, we can relate the LECs $\ltwo{4,5}$ from PQChPT to $l_4^r$ in ChPT:
\begin{equation}\label{eq:chPTsu2:LECrel:l4}
4\Big(2\ltwo{4}+\ltwo{5}\Big)\;=\;l_4^r.
\end{equation}

\subsubsection{\label{sec:app:chPTconv}Conversion from \bteq{\SU(3)\times\SU(3)} to \bteq{\SU(2)\times\SU(2)}}

The conversion from the $\SU(3)$ LEC's to the $\SU(2)$ ones can be done (incl.\ terms up to NLO) following \cite{Gasser:1984gg} (see \cite{Gasser:2007sg} for NNLO, note the latter organizes the expansion in a slightly different way compared to the former).
\begin{eqnarray}
\label{eq:chPTconv:F}
\ftwo &=& \fthree \, \Bigg\{\,1\,-\,\frac{\chi_s}{32\pi^2 \fthree^2}\log\frac{\chi_s}{2\Lambda_\chi^2}\,+\,\frac{8}{\fthree^2}\lthree{4}\chi_s\,\Bigg\}\\
\label{eq:chPTconv:B}
\btwo &=& \bthree \, \Bigg\{\,1\,-\,\frac{\chi_s}{72\pi^2\fthree^2}\log\frac{2\chi_s}{3\Lambda_\chi^2}\,+\,\frac{16}{\fthree^2}\Big(2\lthree{6}-\lthree{4}\Big)\chi_s\,\Bigg\}\\
\label{eq:chPTconv:l3}
l_3^r &=& 4\Big(2\lthree{8}-\lthree{5}\Big)+8\Big(2\lthree{6}-\lthree{4}\Big)\,-\,\frac{1}{576\pi^2}\Big(1+\log\frac{2\chi_s}{3\Lambda_\chi^2}\Big)\\
\label{eq:chPTconv:l4}
l_4^r &=& 8\lthree{4}+4\lthree{5}\,-\,\frac{1}{64\pi^2}\Big(1+\log\frac{\chi_s}{2\Lambda^2_\chi}\Big)
\end{eqnarray}
Where one has to use $\bthree$ to evaluate the $\chi_X$ on the right hand side.
The relation of the $\SU(2)$ LECs in PQChPT, $\ltwo{i}$ to the unquenched $\SU(2)$ LECs $l_{3,4}^r$ is given by Eqs.~(\ref{eq:chPTsu2:LECrel:l3}, \ref{eq:chPTsu2:LECrel:l4}). To define them at the scale of the pion mass $m_\pi$, one has to use
\begin{eqnarray}
\label{eq:chPTconv:barl}
\bar{l}_i &=& \frac{32\pi^2}{\gamma_i}l_i^r\,-\,\log\frac{m_\pi^2}{\Lambda_\chi^2}\\
\label{eq:chPTconv:gamma34}
 && \gamma_3=-\frac12\,,\;\gamma_4=2\,.
\end{eqnarray}
%


\subsection{\label{sec:app:su2kaon}\bteq{\SU(2)\times\SU(2)} for the kaon sector}

Here, we choose to denote by $m_x$ always a light and by $m_y$ a heavier valence quark mass. The additional low-energy constants appearing here (LO: $B^{(K)}(m_h)$, $f^{(K)}(m_h)$, $B_{\rm PS}^{(K)}(m_h)$, NLO: $\lambda_{1,2,3,4}(m_h)$, $b_{1,2}(m_h)$) are in general dependent on the dynamical heavy quark mass $m_h$ and the valence heavy quark mass $m_y$. To simplify the notation the argument $m_h$ has to be viewed as placeholder for both of those. For more details see Sec.~\ref{sec:ChPT}. 

\subsubsection{Light-Heavy Squared Pseudoscalar Mass}
\begin{eqnarray}\label{eq:chPTsu2K:mPS}
m_{xy}^2 &=& B^{(K)}(m_h)\,\widetilde{m}_y\,\bigg\{\,1\,+\,\frac{\lambda_1(m_h)}{\ftwo^2}\chi_l\,+\,\frac{\lambda_2(m_h)}{\ftwo^2}\chi_x\bigg\}
\end{eqnarray} 
%

\subsubsection{Light-Heavy Pseudoscalar Decay Constant}
\begin{eqnarray}\label{eq:chPTsu2K:fPS}
f_{xy} &=& f^{(K)}(m_h)\Bigg\{\,1\,+\,\frac{\lambda_3(m_h)}{\ftwo^2}\chi_l\,+\,\frac{\lambda_4(m_h)}{\ftwo^2}\chi_x \nonumber\\
  && \;-\frac1{(4\pi\ftwo)^2}\bigg[\frac{\chi_x+\chi_l}2\log\frac{\chi_x+\chi_l}{2\Lambda_\chi^2}\,+\,\frac{\chi_l-2\chi_x}{4}\log\frac{\chi_x}{\Lambda_\chi^2}\bigg]\,\Bigg\}
\end{eqnarray}
%

\subsubsection{Light-Heavy Pseudoscalar Bag Parameter}
\begin{eqnarray}\label{eq:chPTsu2K:BPS}
B_{xy} &=& B_{\rm PS}^{(K)}(m_h)\Bigg\{\,1\,+\,\frac{b_1(m_h)}{\ftwo^2}\chi_l\,+\,\frac{b_2(m_h)}{\ftwo^2}\chi_x - \frac{\chi_l}{32\pi^2\ftwo^2}\log\frac{\chi_x}{\Lambda_\chi^2}\,\Bigg\}
\end{eqnarray}
%

\subsubsection{\label{sec:app:chPTconvKaon}Conversion from \bteq{\SU(3)\times\SU(3)} to kaon \bteq{\SU(2)\times\SU(2)}}

For the LECs appearing in the kaon mass and decay constant formulae (Eqs.~(\ref{eq:chPTsu2K:mPS}, \ref{eq:chPTsu2K:fPS})), we provide the relation to the $\SU(3)$-LECs as well:
\begin{eqnarray}
\label{eq:chPTconv:FKaon}
f^{(K)}(m_h,m_y) &=& \fthree\bigg\{1\,+\,\frac8{\fthree^2}\lthree{4}\chi_h+\frac{4}{\fthree^2}\lthree{5}\chi_y+\frac1{32\pi^2\fthree^2}\bigg[\frac{\chi_y-\chi_h}{3\chi_y-2\chi_h}\chi_y \nonumber\\
 && \:-\frac{\chi_h}2\log\frac{\chi_h}{2\Lambda_\chi^2}-\chi_y\log\frac{\chi_y}{2\Lambda_\chi^2}+\frac{\chi_h\chi_y^2}{(3\chi_y-2\chi_h)^2}\log\frac{\chi_y}{\Lambda_\chi^2} \nonumber\\
 && \:-\frac{\chi_h\chi_y^2}{(3\chi_y-2\chi_h)^2}\log\frac{2\chi_h}{3\Lambda_\chi^2}-\frac{\chi_h+\chi_y}{2}\log\frac{\chi_h+\chi_y}{2\Lambda_\chi^2}\bigg]\bigg\}\,,\\
\label{eq:chPTconv:BKaon}
B^{(K)}(m_h,m_y) &=& \bthree\bigg\{1\,+\,\frac{\chi_h}{\fthree^2}16(2\lthree{6}-\lthree{4})\,+\,\frac{\chi_y}{\fthree^2}8(2\lthree{8}-\lthree{5}) \nonumber\\
 &&\:+\,\frac{1}{72\pi^2\fthree^2}\bigg[9\frac{\chi_y-\chi_h}{3\chi_y-2\chi_h}\chi_y\log\frac{\chi_y}{\Lambda_\chi^2}\,+\,2\frac{\chi_h^2}{3\chi_y-2\chi_h}\log\frac{2\chi_h}{3\Lambda_\chi^2}\bigg]\bigg\}\,,\\
\label{eq:chPTconv:lam1}
\lambda_1(m_h,m_y) &=& 32(2\lthree{6}-\lthree{4})\,+\,\frac1{72\pi^2}\bigg[-\frac{18(\chi_y-\chi_h)^2}{(3\chi_y-2\chi_h)^2}\log\frac{\chi_y}{\Lambda_\chi^2} \nonumber\\
 &&\:+\,2\frac{5\chi_h-6\chi_y}{(3\chi_y-2\chi_h)^2}\chi_h\log\frac{2\chi_h}{3\Lambda_\chi^2}\,+\,\frac{\chi_h}{3\chi_y-2\chi_h}\bigg]\,,\\
\label{eq:chPTconv:lam2}
\lambda_2(m_h,m_y) &=& \frac{\fthree^2}{\chi_y}\,+\,8(2\lthree{8}-\lthree{5})\,+\,16\frac{\chi_h}{\chi_y}\Big(\lthree{4}-(2\lthree{6}-\lthree{4})\Big)\nonumber\\
 && +\frac1{4\pi^2}\bigg[ \frac{\chi_y-\chi_h}{2(3\chi_y-2\chi_h)}\log\frac{\chi_y}{\Lambda_\chi^2}-\frac{\chi_h}{4\chi_y}\log\frac{\chi_h}{2\Lambda_\chi^2}+\frac{\chi_h(3\chi_y-\chi_h)}{9\chi_y(3\chi_y-2\chi_h)}\log\frac{2\chi_h}{3\Lambda_\chi^2}\bigg]\,,\nonumber\\\\
\label{eq:chPTconv:lam3}
\lambda_3(m_h,m_y) &=& 16\lthree{4}+\frac1{16\pi^2}\bigg[-\frac12\log\frac{\chi_y}{2\Lambda_\chi^2}-\frac{4\chi_h^2-11\chi_h\chi_y+8\chi_y^2}{(3\chi_y-2\chi_h)^2} \nonumber\\
 && \:+\frac{15\chi_y-14\chi_h}{4(3\chi_y-2\chi_h)^3}\chi_y^2\log\frac{2\chi_h}{3\Lambda_\chi^2}+\frac{(3\chi_y-\chi_h)(\chi_y-\chi_h)(\chi_y-2\chi_h)}{(3\chi_y-2\chi_h)^3}\log\frac{\chi_y}{\Lambda_\chi^2}\bigg]\,,\nonumber\\\\
\label{eq:chPTconv:lam4}
\lambda_4(m_h,m_y) &=& 4\lthree{5} +\frac{1}{16\pi^2}\bigg[-\frac14\log\frac{\chi_h}{2\Lambda_\chi^2}+\frac{\chi_y-\chi_h}{3\chi_y-2\chi_h}\log\frac{\chi_y}{\Lambda_\chi^2}-\frac{\chi_y}{2(3\chi_y-2\chi_h)}\log\frac{2\chi_h}{3\Lambda_\chi^2}\bigg]\,.\nonumber\\
\end{eqnarray}

%
\section{\label{sec:appendix_fv}Finite Volume Correction for Pseudoscalar Masses and Decay Constants}

In the following we will provide the finite volume corrections for the meson decay constants and squared meson masses obtained in PQChPT for $N_f=2+1$ and $N_f=2$ sea quarks (cf. \cite{Gasser:1986vb,Gasser:1987ah,Gasser:1987zq,Bernard:2001yj,Aubin:2003mg,Aubin:2003uc}). The corrections are labeled $\Delta^{L\, f}_{xy}$ and $\Delta^{L\,m^2}_{xy}$, respectively, for decay constants and squared masses of mesons made of quarks with masses $m_x$ and $m_y$ in a finite (spatial) volume $L^3$. Labeling decay constants in finite volume $f^L_{xy}$ and those in infinite volume $f^{L\to\infty}_{xy}$ we have the following relations:
\begin{eqnarray}
f_{xy}^{L\to\infty}&=&f_{xy}^{L}\,\Big(\,1\,-\,\Delta^{L\,f}_{xy}\,\Big) \\
 &=& \fthree\,\Big(\,1\,+\,\chi{\rm PT}_{xy}^f\,\Big)\,, \\
f_{xy}^{L} &=& \fthree\,\Big(\,1\,+\,\chi{\rm PT}_{xy}^f\,\Big)\,\Big(\,1\,+\,\Delta^{L\,f}_{xy}\,\Big)\\
&=& \fthree\,\Big(\,1\,+\,\chi{\rm PT}_{xy}^f\,+\,\Delta^{L\,f}_{xy}\,\Big)
\end{eqnarray}
and similar for squared meson masses. Here we summarized the NLO contribution in infinite volume (PQ)ChPT (as given in the previous section) by $\chi{\rm PT}_{xy}^f$. (These equalities hold up to terms of NLO, higher order terms are neglected. Note further, that in case of $\SU(2)\times\SU(2)$ the $\fthree$ has to be replaced by $\ftwo$.

The Bessel functions of imaginary argument (modified Bessel functions of the $2^{\rm nd}$ kind) $K_n(x)$ enter the expressions for the finite volume corrections via
\begin{eqnarray}
\delta_1(x) &=& \frac4{x}\sum_{\vec{r}\neq0}\frac{K_1(rx)}r\,,\\
\delta_3(x) &=& 2\sum_{\vec{r}\neq0} K_0(rx)\,,\\
\delta_5(x) &=& -\sum_{\vec{r}\neq0} \frac{r}{x} K_1(rx)\,,
\end{eqnarray}
where the argument is typically $x=\sqrt{\chi_X}L$. For the numerical implementation, we made use of the multiplicities $m(n)$ and rewrite the sum as
\begin{equation}
 \sum_{\vec{r}\neq0}\,f(r)\;=\;\sum_{n>0}\,m(n)\,f(\sqrt{n})\,,
\end{equation}
where sum is evaluated until the relative change is less than the required precision $\epsilon$ or a maximum number $N$ for $n$ is reached. (Typically, we use $\epsilon=5\cdot10^{-4}$ and $N=100$, but checked that going to $\epsilon=5\cdot10^{-6}$, $N=1000$ does not change the result.) The multiplicities for $n\leq20$ can, {\em e.g.}, be found in \cite{Colangelo:2005gd}. Here we list $m(n)$ for $n=1,\ldots,100$:
\begin{eqnarray}
&& 6, 12, 8, 6, 24, 24, 0, 12, 30, 24, 24, 8, 24, 48, 0, 6, 48, 36, 24, 24,  \nonumber\\
&& 48, 24, 0, 24, 30, 72, 32, 0, 72, 48, 0, 12, 48, 48, 48, 30, 24, 72, 0, 24,  \nonumber\\
&& 96, 48, 24, 24, 72, 48, 0, 8, 54, 84, 48, 24, 72, 96, 0, 48, 48, 24, 72, 0,  \nonumber\\
&& 72, 96, 0, 6, 96, 96, 24, 48, 96, 48, 0, 36, 48, 120, 56, 24, 96, 48, 0, 24,  \nonumber\\
&& 102, 48, 72, 48, 48, 120, 0, 24, 144, 120, 48, 0, 48, 96, 0, 24, 48, 108, 72, 30\,. 
\end{eqnarray}
Note that we have the following relations involving the $\delta_i$:
\begin{eqnarray}
\frac{\rm d}{{\rm d}\chi}\,\chi\delta_1(\sqrt{\chi}L) &=& -\delta_3(\sqrt{\chi}L)\,,\\
\frac{\rm d}{{\rm d}\chi}\, \delta_3(\sqrt{\chi}L) &=& L^2\,\delta_5(\sqrt{\chi}L)\,.
\end{eqnarray}
Furthermore, by doing the substitutions in the finite volume correction part $\Delta^{L}_{xy}$
\begin{eqnarray}
\delta_1(\sqrt{\chi_X}L) &\rightarrow& \log\frac{\chi_X}{\Lambda_\chi^2}\,,\\
\delta_3(\sqrt{\chi_X}L) &\rightarrow& -\Big(1+\log\frac{\chi_X}{\Lambda_\chi^2}\Big)\,,\\
L^2\,\delta_5(\sqrt{\chi_X}L) &\rightarrow& -\frac1{\chi_X}\,,
\end{eqnarray}
one obtains at NLO the non-analytic part of the corresponding $\chi{\rm PT}_{xy}$ ({\em i.e.}\ without the analytic terms multiplying the LECs).

\subsection{SU(3)$\times$SU(3)}

\subsubsection{FV Correction for the Squared Pseudoscalar Mass}

\paragraph{non-degenerate ($m_x\neq m_y$)}
\begin{itemize}
\item  $m_x\neq m_\eta \neq m_y$
\begin{eqnarray}
\Delta_{xy}^{L\,m^2} &=& \frac1{24\pi^2\fthree^2}\bigg[ \frac{(\chi_\eta-\chi_l)(\chi_\eta-\chi_h)}{(\chi_\eta-\chi_x)(\chi_\eta-\chi_y)}\chi_\eta\delta_1(\sqrt{\chi_\eta}L) \nonumber\\
 && \;+\,\frac{(\chi_x-\chi_l)(\chi_x-\chi_h)}{(\chi_x-\chi_y)(\chi_x-\chi_\eta)}\chi_x\delta_1(\sqrt{\chi_x}L) 
 \,+\,\frac{(\chi_y-\chi_l)(\chi_y-\chi_h)}{(\chi_y-\chi_x)(\chi_y-\chi_\eta)}\chi_y\delta_1(\sqrt{\chi_y}L)\, \bigg]\nonumber\\
\end{eqnarray}
\item $m_x\to m_\eta$ ($m_y\neq m_\eta$)
\begin{eqnarray}
\Delta_{xy}^{L\,m^2} &=& \frac1{24\pi^2\fthree^2}\bigg[\,-\frac{(\chi_x-\chi_l)(\chi_x-\chi_h)}{\chi_x-\chi_y}\delta_3(\sqrt{\chi_x}L) \nonumber\\
 && \;+\frac{(2\chi_x-\chi_l-\chi_h)(\chi_x-\chi_y)-(\chi_x-\chi_l)(\chi_x-\chi_h)}{(\chi_x-\chi_y)^2}\chi_x\delta_1(\sqrt{\chi_x}L) \nonumber\\
 && \;+\frac{(\chi_y-\chi_l)(\chi_y-\chi_h)}{(\chi_x-\chi_y)^2}\chi_y\delta_1(\sqrt{\chi_y}L)\,\bigg]
\end{eqnarray}
\end{itemize}

\paragraph{degenerate}
\begin{itemize}
\item $m_x\neq m_\eta$
\begin{eqnarray}
\Delta_{xx}^{L\,m^2} &=& \frac1{24\pi^2\fthree^2}\bigg[ \frac{(\chi_\eta-\chi_l)(\chi_\eta-\chi_h)}{(\chi_\eta-\chi_x)^2}\chi_\eta\delta_1(\sqrt{\chi_\eta}L) \nonumber\\
 && \;+\,\frac{(\chi_x-\chi_\eta)(2\chi_x-\chi_h-\chi_l)-(\chi_x-\chi_l)(\chi_x-\chi_h)}{(\chi_x-\chi_\eta)^2}\chi_x\delta_1(\sqrt{\chi_x}L) \nonumber\\
 && \;-\,\frac{(\chi_x-\chi_l)(\chi_x-\chi_h)}{\chi_x-\chi_\eta}\delta_3(\sqrt{\chi_x}L) \,\bigg]
\end{eqnarray}
\item $m_x\to m_\eta$
\begin{eqnarray}
\Delta_{xx}^{L\,m^2} &=& \frac1{24\pi^2\fthree^2}\bigg[\,\chi_x\delta_1(\sqrt{\chi_x}L)\,-\,(2\chi_x-\chi_l-\chi_h)\delta_3(\sqrt{\chi_x}L)\nonumber\\
 && \;-\frac{(\chi_x-\chi_l)(\chi_x-\chi_h)}{2}L^2\delta_5(\sqrt{\chi_x}L)\,\bigg]
\end{eqnarray}
\end{itemize}

\subsubsection{FV Correction for the Pseudoscalar Decay Constant}

\paragraph{non-degenerate $m_x\neq m_y$}
\begin{itemize}
\item $m_x\neq m_\eta \neq m_y$
\begin{eqnarray}
\Delta_{xy}^{L\,f} &=& -\frac1{8\pi^2\fthree^2}\bigg[ \frac{\chi_x+\chi_l}4 \delta_1\left(\sqrt{\frac{\chi_x+\chi_l}2}L\right) \,+\,\frac{\chi_y+\chi_l}4 \delta_1\left(\sqrt{\frac{\chi_y+\chi_l}2}L\right) \nonumber \\
 && \;+\,\frac{\chi_x+\chi_h}8 \delta_1\left(\sqrt{\frac{\chi_x+\chi_h}2}L\right) \,+\,\frac{\chi_y+\chi_h}8 \delta_1\left(\sqrt{\frac{\chi_y+\chi_h}2}L\right) \bigg]\nonumber \\
 && +\frac1{96\pi^2 \fthree^2}\bigg[ -\frac{(\chi_l-\chi_x)(\chi_h-\chi_x)}{\chi_x-\chi_\eta}\delta_3(\sqrt{\chi_x}L)\,-\, \frac{(\chi_l-\chi_y)(\chi_h-\chi_y)}{\chi_y-\chi_\eta}\delta_3(\sqrt{\chi_y}L) \nonumber \\
 && \;+\,\Big(\frac{(\chi_l-\chi_\eta)(\chi_h-\chi_\eta)}{(\chi_x-\chi_\eta)^2}+\frac{(\chi_l-\chi_\eta)(\chi_h-\chi_\eta)}{(\chi_y-\chi_\eta)^2}-2\frac{(\chi_l-\chi_\eta)(\chi_h-\chi_\eta)}{(\chi_x-\chi_\eta)(\chi_y-\chi_\eta)}\Big)\,\chi_\eta\delta_1(\sqrt{\chi_\eta}L) \nonumber \\
 && \;+\,\Big(\frac{\chi_h+\chi_l-2\chi_x}{\chi_\eta-\chi_x}-\frac{(\chi_l-\chi_x)(\chi_h-\chi_x)}{(\chi_\eta-\chi_x)^2}-2\frac{(\chi_l-\chi_x)(\chi_h-\chi_x)}{(\chi_\eta-\chi_x)(\chi_y-\chi_x)}\Big)\,\chi_x\delta_1(\sqrt{\chi_x}L) \nonumber\\
 && \;+\,\Big(\frac{\chi_h+\chi_l-2\chi_y}{\chi_\eta-\chi_y}-\frac{(\chi_l-\chi_y)(\chi_h-\chi_y)}{(\chi_\eta-\chi_y)^2}-2\frac{(\chi_l-\chi_y)(\chi_h-\chi_y)}{(\chi_\eta-\chi_y)(\chi_x-\chi_y)}\Big)\,\chi_y\delta_1(\sqrt{\chi_y}L)\, \bigg]
\end{eqnarray}
\item $m_x\to m_\eta$, $m_y\neq m_\eta$
\begin{eqnarray}
\Delta_{xy}^{L\,f} &=& -\frac1{8\pi^2\fthree^2}\bigg[ \frac{\chi_x+\chi_l}4 \delta_1\left(\sqrt{\frac{\chi_x+\chi_l}2}L\right) \,+\,\frac{\chi_y+\chi_l}4 \delta_1\left(\sqrt{\frac{\chi_y+\chi_l}2}L\right) \nonumber \\
 && \;+\,\frac{\chi_x+\chi_h}8 \delta_1\left(\sqrt{\frac{\chi_x+\chi_h}2}L\right) \,+\,\frac{\chi_y+\chi_h}8 \delta_1\left(\sqrt{\frac{\chi_y+\chi_h}2}L\right) \bigg]\nonumber \\
 && +\frac1{96\pi^2 \fthree^2}\bigg[-\frac12(\chi_l-\chi_x)(\chi_h-\chi_x)L^2\delta_5(\sqrt{\chi_x}L) \,-\, \frac{(\chi_l-\chi_y)(\chi_h-\chi_y)}{\chi_y-\chi_x}\delta_3(\sqrt{\chi_y}L)  \nonumber \\
 && \;+\,\Big(1 - 2\frac{2\chi_x-\chi_l-\chi_h}{\chi_x-\chi_y} + 3\frac{(\chi_l-\chi_x)(\chi_h-\chi_x)}{(\chi_x-\chi_y)^2}\Big)\chi_x\delta_1(\sqrt{\chi_x}L)\nonumber\\
&& \:-\,\Big(2\chi_x-\chi_l-\chi_h - 2\frac{(\chi_l-\chi_x)(\chi_h-\chi_x)}{\chi_x-\chi_y}\Big)\delta_3(\sqrt{\chi_x}L)\nonumber\\
&& \;+\,\Big(\frac{\chi_h+\chi_l-2\chi_y}{\chi_x-\chi_y}-\frac{(\chi_l-\chi_y)(\chi_h-\chi_y)}{(\chi_x-\chi_y)^2}-2\frac{(\chi_l-\chi_y)(\chi_h-\chi_y)}{(\chi_x-\chi_y)^2}\Big)\,\chi_y\delta_1(\sqrt{\chi_y}L)\, \bigg] \nonumber \\
\end{eqnarray}
\end{itemize}

\paragraph{degenerate, $m_x=m_y$}
\begin{eqnarray}
\Delta_{xx}^{L\,f} &=& -\frac1{8\pi^2\fthree^2}\bigg[ \frac{\chi_x+\chi_l}2\,\delta_1\left(\sqrt{\frac{\chi_x+\chi_l}2}L\right)\,+\,\frac{\chi_x+\chi_h}4\,\delta_1\left(\sqrt{\frac{\chi_x+\chi_h}2}L\right)\,\bigg]
\end{eqnarray}
%

\subsection{SU(2)$\times$SU(2)}

\subsubsection{FV Correction for the Squared Pseudoscalar Mass}

\paragraph{non-degenerate ($m_x \neq m_y$)}
\begin{eqnarray}
\Delta_{xy}^{L\,m^2} &=& \frac1{16\pi^2\ftwo^2}\bigg[\frac{\chi_x-\chi_l}{\chi_x-\chi_y}\chi_x\delta_1(\sqrt{\chi_x}L)\,+\,\frac{\chi_y-\chi_l}{\chi_y-\chi_x}\chi_y\delta_1(\sqrt{\chi_y}L)\,\bigg]
\end{eqnarray}
\paragraph{degenerate ($m_x = m_y$)}
\begin{eqnarray}
\Delta_{xx}^{L\,m^2} &=& \frac1{16\pi^2\ftwo^2}\bigg[\chi_x\delta_1(\sqrt{\chi_x}L) \:-\: \Big(\chi_x-\chi_l\Big)\delta_3(\sqrt{\chi_x}L) \bigg]
\end{eqnarray}
%

\subsubsection{FV Correction for the Pseudoscalar Decay Constant}

\paragraph{non-degenerate ($m_x \neq m_y$)}
\begin{eqnarray}
\Delta_{xy}^{L\,f} &=& -\frac1{32\pi^2\ftwo^2}\bigg[\Big(\chi_x+\chi_l\Big)\delta_1\left(\sqrt{\frac{\chi_x+\chi_l}2}L\right)\,+\,\Big(\chi_y+\chi_l\Big)\delta_1\left(\sqrt{\frac{\chi_y+\chi_l}2}L\right)\bigg]\nonumber\\
 &&+\frac1{64\pi^2\ftwo^2}\bigg[\,-\Big(\chi_x-\chi_l\Big)\delta_3(\sqrt{\chi_x}L)\,-\,\Big(\chi_y-\chi_l\Big)\delta_3(\sqrt{\chi_y}L) \nonumber\\
 && \;+\frac{\chi_x+\chi_y-2\chi_l}{\chi_y-\chi_x}\chi_x\delta_1(\sqrt{\chi_x}L)\,+\,\frac{\chi_y+\chi_x-2\chi_l}{\chi_x-\chi_y}\chi_y\delta_1(\sqrt{\chi_y}L)\,\bigg]
\end{eqnarray}
\paragraph{degenerate ($m_x = m_y$)}
\begin{eqnarray}
\Delta_{xx}^{L\,f} &=& -\frac1{16\pi^2\ftwo^2}\Big(\chi_x+\chi_l\Big)\delta_1\Big(\sqrt{\frac{\chi_x+\chi_l}2}L\Big)
\end{eqnarray}

%% file: tab/SimDetail.tab
%
%

\clearpage

\begin{table}[htbp]
\caption{Improvements to the RHMC algorithm were implemented as the
ensemble generation progressed.  The table details the algorithms
used for each ensemble and the range, in molecular dynamics (MD)
time, where measurements were made.} \label{tab:evol}
\begin{tabular}{cccc}
$m_l$ & RHMC  & MD range & Measurement range\\
\hline
0.005 & II & 0-4435 & 900-4460\\
\hline
\multirow{3}{*}{0.01} & 0  & 0-549  \\
                      & I & 550-1454\\
                      & II   & 1455-5020 & 1460-5020\\
\hline
\multirow{3}{*}{0.02} & 0 & 0-1519\\
                      &I & 1520-1884\\
                      &II  & 1885-3610 & 1800-3560 \\
\hline
\multirow{3}{*}{0.03} & 0  &0-492\\
                      & I &493-929\\
                      & II  &930-3060 & 1260-3040 \\
\end{tabular}
\end{table}

%
%

\clearpage

\begin{table}[htbp]
\caption{Calculation parameters for the three separate datasets FPQ, DEG and UNI. The range and measurement
separation $\Delta$ are specified in molecular dynamics time units. $N_{\rm meas}$ is the number of 
measurements for each source position $t_{\rm src}$.  The total number of measurements is therefore 
$N_{\rm meas}\times N_{\rm src}$, where $N_{\rm src}$ is the number of
different values for $t_{\rm src}$.} \label{tab:datasets}
\begin{tabular*}{0.8\textwidth}{@{\extracolsep{\fill}}p{1in}|ccccc}
\hline
\hline
\centering{$m_l$} & Dataset & Range & $\Delta$ & $N_{\rm meas}$ & $t_{\rm src \; locations}$ \\
\hline
\centering{\multirow{3}{*}{0.005}} & FPQ & 900-4460 & 40 & 90 & 5,\,59 \\
                       & DEG & 900-4460 & 40 & 90 & 0,\,32 \\
                       & UNI & 900-4480 & 20 & 180& 0,\,32,\,16\\
\hline
\centering{\multirow{3}{*}{0.01}}  & FPQ & 1460-5020& 40 & 90 & 5,\,59 \\
                       & DEG & 1460-5020& 40 & 90 & 0,\,32 \\
                       & UNI & 800-3940 & 10 & 315& 0,\,32 \\
\hline
\centering{\multirow{2}{*}{0.02}}  & DEG & 1800-3560& 40 & 45 & 0,\,32 \\  
                       & UNI & 1800-3580& 20 & 90 & 0,\,32 \\
\hline
\centering{\multirow{2}{*}{0.03}}  & DEG & 1260-3020& 40 & 45 & 0,\,32 \\
                       & UNI & 1260-3040& 20 & 90 & 0,\,32 \\
\hline
\hline
\end{tabular*}
\end{table}

%
%

\begin{table}[htbp]
\caption{\label{tab:corrlist}Combinations of quark source and sink smearings used for correlation functions for
each of the datasets FPQ, DEG and UNI.  $A$ is the time component of the axial
current density operator and $P$ is the pseudoscalar density operator. XY-XY
denotes the contraction of two quark propagators with X-type smearing at source and
a Y-type smearing at sink. W=wall source, B=box source, H=hydrogen S-wave and L=local. }
\begin{tabular*}{0.65\textwidth}{@{\extracolsep{\fill}}|p{1in}|p{1in}|p{1in}|p{1in}p{0.01in}|}
\hline
 \centering{Operator Type} & \multicolumn{3}{c}{Dataset} &  \\ 
\hline
\centering{$\langle \cl{O}_{snk} \cl{O}_{src} \rangle$}  & \centering{FPQ} & \centering{DEG} & \centering{UNI}  & {\centering{}}\\
\hline
\centering{\multirow{2}{*}{$\langle A,A \rangle$ }}   & \centering{WL-WL}  & \centering{BL-BL}    & \centering{HL-HL}& {\centering{}}   \\
                       &      &          & \centering{LL-LL}   &{\centering{}}\\
\hline
\centering{\multirow{2}{*}{$\langle P,A \rangle $}}   &      &     & \centering{HL-HL}&{\centering{}}   \\
                       &      &     & \centering{LL-LL}   &{\centering{}}\\
\hline
\centering{\multirow{2}{*}{$\langle A,P \rangle$ }}   &\centering{WL-WL}  & \centering{BL-BL}  & \centering{HL-HL}& {\centering{}}        \\
                       &\centering{WW-WW}  &  \centering{BB-BB}    &    \centering{LL-LL}     &{\centering{}}\\
\hline
\centering{\multirow{2}{*}{$\langle P,P \rangle $}}   &\centering{WL-WL}  & \centering{BL-BL}    & \centering{HL-HL} &{\centering{}}\\
                       &\centering{WW-WW}  &     &      \centering{LL-LL}   &{\centering{}}\\
\hline
\end{tabular*}
\end{table}

%
%

\begin{table}[htbp]
\caption{The different source and sink contraction of the quark propagators for
the UNI dataset. 
$V=\gamma_i$ ($i=1,2,3$) is one of the  spatial components of the vector current
and $T=\sigma_{4i}$  are the calculated components of the tensor current.
XY-XY denotes contraction of two quark propagators with a 
X-type smearing at source
and a Y-type smearing at sink. H=hydrogen S-wave, G=Gaussian wave function (cf. \cite{Antonio:2006px}), 
and L=point. } \label{tab:Vcontract}
\begin{tabular*}{0.65\textwidth}{@{\extracolsep{\fill}}|p{1.0in}|p{1.5in}|p{1.5in}|}
\hline
 {Contraction} & \multicolumn{2}{c|}{ UNI Dataset} \\
\hline
\centering{$\langle \cl{O}_{snk} \cl{O}_{src} \rangle$}  & 
\multicolumn{1}{c|}{ m=0.005,\,0.02,\,0.03} & \multicolumn{1}{c|}{ m=0.01} \\
\hline&&\\[-8mm]
\centering{\multirow{3}{*}{$\langle V,V \rangle$}}
& \multicolumn{1}{c|}{ HL-HL}  & \multicolumn{1}{c|}{GL-GL}  \\
& \multicolumn{1}{c|}{ LL-HL}  & \multicolumn{1}{c|}{LL-GL}  \\
& \multicolumn{1}{c|}{ HL-LL}  & \multicolumn{1}{c|}{GL-LL}  \\
\hline
\centering{\multirow{3}{*}{
	$\langle T,V \rangle$
	}}   
& \multicolumn{1}{c|}{HL-HL} & \multicolumn{1}{c|}{GL-GL} \\
& \multicolumn{1}{c|}{LL-HL} & \multicolumn{1}{c|}{LL-GL} \\
& \multicolumn{1}{c|}{HL-LL} & \multicolumn{1}{c|}{GL-LL} \\
\hline
\end{tabular*}
\end{table}

%% file: tab/LatticePS.tab
%
%

\clearpage

\begin{table}[htbp]
\caption{Lattice values for the pseudoscalar meson mass and $\chi^2$/d.o.f.
calculated from an \emph{uncorrelated} simultaneous fit to five correlators of
the FPQ dataset on $m_l=0.005$ and $m_l=0.01$ ensembles.} \label{tab:PSmass}
\begin{tabular*}{0.6\textwidth}{@{\extracolsep{\fill}}cc|cccc}
\hline
\hline
       &         & \multicolumn{2}{c}{$m_l=0.005$} & \multicolumn{2}{c}{$m_l=0.01$}\\
$m_x$ & $m_y$  & $m_{xy}$ & $\chi^2$/d.o.f. & $m_{xy}$ & $\chi^2$/d.o.f. \\
\hline
0.001 & 0.001 & 0.1402(9) & 0.36(7) & 0.1432(10) & 0.18(6)\\
0.001 & 0.005 & 0.1680(8) & 0.38(8) & 0.1706(9) & 0.21(6)\\
0.005 & 0.005 & 0.1915(8) & 0.39(9) & 0.1938(8) & 0.25(6)\\
0.001 & 0.010 & 0.1971(8) & 0.38(8) & 0.1994(9) & 0.26(7)\\
0.005 & 0.010 & 0.2172(8) & 0.38(9) & 0.2194(8) & 0.25(7)\\
0.010 & 0.010 & 0.2400(7) & 0.36(9) & 0.2421(7) & 0.24(7)\\
0.001 & 0.020 & 0.2449(9) & 0.36(9) & 0.2473(9) & 0.30(8)\\
0.005 & 0.020 & 0.2611(8) & 0.34(9) & 0.2635(8) & 0.24(8)\\
0.010 & 0.020 & 0.2802(7) & 0.31(8) & 0.2825(7) & 0.22(7)\\
0.020 & 0.020 & 0.3155(7) & 0.26(8) & 0.3178(6) & 0.23(8)\\
0.001 & 0.030 & 0.2850(10) & 0.33(10) & 0.2877(10) & 0.33(10)\\
0.005 & 0.030 & 0.2990(8) & 0.32(9) & 0.3016(8) & 0.26(9)\\
0.010 & 0.030 & 0.3158(7) & 0.28(8) & 0.3183(7) & 0.24(8)\\
0.020 & 0.030 & 0.3477(6) & 0.23(7) & 0.3500(6) & 0.26(10)\\
0.030 & 0.030 & 0.3775(6) & 0.21(8) & 0.3797(6) & 0.29(11)\\
0.001 & 0.040 & 0.3204(11) & 0.34(13) & 0.3234(11) & 0.36(12)\\
0.005 & 0.040 & 0.3329(8) & 0.31(10) & 0.3358(8) & 0.31(11)\\
0.010 & 0.040 & 0.3482(7) & 0.26(8) & 0.3508(7) & 0.29(11)\\
0.020 & 0.040 & 0.3776(6) & 0.22(8) & 0.3800(6) & 0.30(12)\\
0.030 & 0.040 & 0.4055(6) & 0.21(8) & 0.4077(5) & 0.32(13)\\
0.040 & 0.040 & 0.4320(6) & 0.21(9) & 0.4341(5) & 0.34(13)\\
\hline
\hline
\end{tabular*}
\end{table}

\begin{table}[htbp]
\caption{Lattice values for the pseudoscalar decay constant calculated
using method described in Eq.~(\ref{eq:fpi-calc}).} 
\label{tab:fPS}
\begin{tabular*}{0.6\textwidth}{@{\extracolsep{\fill}}cc|cc}
\hline
\hline
$m_x$ & $m_y$  & $f_{xy}(m_l=0.005)$ & $f_{xy}(m_l=0.01)$  \\
\hline
0.001 & 0.001 & 0.0829(7) & 0.0860(7) \\
0.001 & 0.005 & 0.0846(7) & 0.0878(6) \\
0.005 & 0.005 & 0.0860(6) & 0.0891(6) \\
0.001 & 0.010 & 0.0867(7) & 0.0900(6) \\
0.005 & 0.010 & 0.0880(6) & 0.0910(6) \\
0.010 & 0.010 & 0.0898(6) & 0.0928(6) \\
0.001 & 0.020 & 0.0904(7) & 0.0928(7) \\
0.005 & 0.020 & 0.0915(6) & 0.0946(6) \\
0.010 & 0.020 & 0.0934(6) & 0.0962(6) \\
0.020 & 0.020 & 0.0969(6) & 0.0995(5) \\
0.001 & 0.030 & 0.0936(8) & 0.0972(8) \\
0.005 & 0.030 & 0.0947(6) & 0.0979(6) \\
0.010 & 0.030 & 0.0965(6) & 0.0994(6) \\
0.020 & 0.030 & 0.1001(6) & 0.1025(5) \\
0.030 & 0.030 & 0.1033(6) & 0.1056(5) \\
0.001 & 0.040 & 0.0965(8) & 0.1003(8) \\
0.005 & 0.040 & 0.0975(7) & 0.1008(7) \\
0.010 & 0.040 & 0.0993(6) & 0.1022(6) \\
0.020 & 0.040 & 0.1029(6) & 0.1053(5) \\
0.030 & 0.040 & 0.1062(6) & 0.1083(5) \\
0.040 & 0.040 & 0.1091(7) & 0.1111(5) \\
\hline
\hline
\end{tabular*}
\end{table}

\begin{table}[htbp]
\caption{Lattice values for the pseudoscalar meson mass and $\chi^2$/d.o.f.
calculated from an \emph{uncorrelated} simultaneous fit to four correlators of
the DEG dataset on $m_l=0.02$ and $m_l=0.03$ ensembles.} \label{tab:PSmass23}
\begin{tabular*}{0.6\textwidth}{@{\extracolsep{\fill}}c|cccc}
\hline
\hline
& \multicolumn{2}{c}{$m_l=0.02$} & \multicolumn{2}{c}{$m_l=0.03$}\\
$m_x$  & $m_{xx}$ & $\chi^2$/d.o.f. & $m_{xx}$ & $\chi^2$/d.o.f. \\
\hline
0.001 & 0.1493(11) & 0.7(4) & - & - \\
0.005 & 0.1993(9) & 0.5(3) & 0.2016(11) & 0.4(1)\\
0.010 & 0.2471(8) & 0.4(2) & 0.2500(10) & 0.3(1)\\
0.020 & 0.3227(7) & 0.3(2) & 0.3261(9) & 0.3(1)\\
0.030 & 0.3849(6) & 0.4(2) & 0.3884(9) & 0.2(1)\\
0.040 & 0.4396(6) & 0.4(2) & 0.4429(8) & 0.2(1)\\
\hline
\hline
\end{tabular*}
\end{table}

\begin{table}[htbp]
\caption{Lattice values for the pseudoscalar decay constant calculated
using method described in Eq.~(\ref{eq:fpi23}).} 
\label{tab:fpi23}
\begin{tabular*}{0.6\textwidth}{@{\extracolsep{\fill}}c|cc}
\hline
\hline
$m_x$ & $f_{xx}(m_l=0.02)$ & $f_{xx}(m_l=0.03)$  \\
\hline
0.001 & 0.0926(18) & - \\
0.005 & 0.0947(15) & 0.0961(14) \\
0.010 & 0.0969(13) & 0.0985(13) \\
0.020 & 0.1021(11) & 0.1038(12) \\
0.030 & 0.1077(10) & 0.1092(11) \\
0.040 & 0.1133(9)  & 0.1145(10) \\
\hline
\hline
\end{tabular*}
\end{table}

%% file: tab/LatticeBaryon.tab
\clearpage

\begin{table}[htdp]
\caption{Decuplet baryon masses for degenerate valence quarks. 
``traj." refers to the Monte Carlo trajectories used for measurements.  Numbers in ()'s give the total number of measurements, separated by 40 trajectories in each case. To reduce effects of auto-correlations, the measurements were block-averaged into bins of size 80 trajectories before fitting. }
\begin{center}
\begin{tabular}{ccccc}
\hline
$m_l$ & $m_{x}$ & traj. (\# meas) & $\chi^2$/d.o.f. & mass\\
\hline
0.005 & 0.03 &900-4460 (90), 920-4480 (90)  & 0.5(7) & 0.9647(65) \\
0.005 & 0.04 &  & 0.6(8) & 1.016(51)\\
0.01 & 0.03 & 1460-5020 (90), 1480-4040 (65) & 1.1(1.1) &0.9809(88)\\
0.01 & 0.04 &  & 1.2(1.2) & 1.033(60)\\
0.02 & 0.03 & 1600-3600 (50), 1900-3580 (43) & 1.7(1.6) & 1.030(11)\\
0.02 & 0.04 &  & 2.0(1.8) & 1.074(10)\\
0.03 & 0.03 & 1020-3060 (52), 1320-3040 (44) & 1.4(1.7) & 1.040(10)\\
0.03 & 0.04 & & 1.8(2.2) & 1.082(6)\\
\hline
\end{tabular}
\end{center}
\label{tab:omega masses}
\end{table}%

%% file: tab/Su2.tab
%
%
\clearpage

\begin{table}
\caption{\label{tab:fitSU2}Low energy constants obtained from $\SU(2)\times\SU(2)$ fits with a valence mass cut $m_{\rm avg}\leq0.01$. For convenience the LECs $\ltwo{i}$ are quoted in units of $10^{-4}$ at two different chiral scales, $\Lambda_\chi$ = 770 MeV and 1 GeV. The definitions of the low energy constants are given in the Appendix. (Only statistical errors are quoted.)} 
\begin{center}
%
\begin{tabular}{r*{4}{c}}
\hline\hline
 & $\btwo$ & $\ftwo$ & $\bar{l}_3$ & $\bar{l}_4$ \\\hline
 & 2.414(61) & 0.0665(21) & 3.13(.33) & 4.43(.14) \\\hline\hline
$\Lambda_\chi$ & $\ltwo{4}$ & $\ltwo{5}$ & $(2\ltwo{6}-\ltwo{4})$ & $(2\ltwo{8}-\ltwo{5})$ \\\hline
770 MeV & 3.3(1.3) & 9.30(.73) & 0.32(.62) & 0.50(.43) \\
1 GeV & 1.3(1.3) & 5.16(.73) & -0.71(.62) & 4.64(.43)\\
\hline\hline
\end{tabular}
\end{center}
\end{table}

\begin{table}
\caption{\label{tab:results}Lattice scale and unrenormalized quark masses in lattice units. Note $\tilde{m}_X \equiv m_X + \mres$.  Only the
statistical errors are given here.}
\begin{center}
\begin{tabular}{ccccccc}\hline\hline
  $a^{-1}$ [{\rm GeV}] & $a$ [{\rm fm}] & $m_{ud}$ & $\widetilde{m}_{ud}$ & $m_s$ & $\widetilde{m}_s$ & $\widetilde{m}_{ud}:\widetilde{m}_s$ \\\hline 
  1.729(28) & 0.1141(18)  & -0.001847(58) & 0.001300(58)  & 0.0343(16)   & 0.0375(16) & 1:28.8(4)\\
\hline\hline 
\end{tabular}
\end{center}
\end{table}

\begin{table}[ht]
\caption{Central value and statistical error (stat) for the fit parameters from $\SU(2)$ ChPT and systematic error estimates from four sources. These are: finite-volume effects (FV), lattice artefacts ($a^2$), the chiral extrapolation in the light quark masses (ChPT) and the unphysical value of the dynamical strange quark mass ($m_s$). For details we refer to Sec.~\ref{sec:su2:sysErr}. The LECs are quoted here only at a scale $\Lambda_\chi=1\,{\rm GeV}$, but the absolute value of the error is independent of $\Lambda_\chi$. The total systematic error is obtained by adding the separate errors in quadrature, neglecting any possible correlations between them.}
\label{tab:Su2:systErrSummary}
\begin{tabular}{ccc||cccc|c}
\hline\hline
 & & & \multicolumn{5}{c}{systematic errors} \\
 & value & stat & FV & $a^2$ & ChPT & $m_s$ & total  \\\hline 
$\btwo$ & 2.414 & (.061) & (.049) & (.097) & (.034) & (.017) & (.115) \\
$\ftwo$ & 0.0665 & (21) & (13)  & (27) & (36) & (01) & (47)\\
$(2\ltwo{6}-\ltwo{4}) \cdot 10^4$ & -0.7 & (0.6) & (0.1) & (--) & (1.8) & (--) & (1.8)\\ 
$(2\ltwo{8}-\ltwo{5}) \cdot 10^4$ & 4.64 & (.43) & (.95) & (--) & (3.20) & (--) & (3.34)\\
$\ltwo{4} \cdot 10^4$ & 1.3 & (1.3) & (0.3) & (--) & (1.8) & (--) & (1.8)\\
$\ltwo{5} \cdot 10^4$ & 5.16 & (.73) & (.16) & (--) & (8.20) & (--)&  (8.20)\\
$\bar{l}_3$ & 3.13 & (.33) & (.20) & (.08) & (.10) & (.02) & (.24)\\
$\bar{l}_4$ & 4.43 & (.14) & (.05) & (.08) & (.76) & (.04) & (.77)\\
$\widetilde{m}_{ud}$ & 0.001300 & (58) & (23) & (52) & (22) & (09) & (62)\\
$\widetilde{m}_s$ & 0.0375 & (16) & (00) & (15) & (04) & (08) & (17)\\
$\widetilde{m}_{ud}:\widetilde{m}_s$ & 1:28.8 & (0.4) & (0.5) & (1.2) & (0.6) & (0.6) & (1.6) \\  
$f_\pi [{\rm MeV}]$ & 124.1 & (3.6) & (1.9) & (5.0) & (4.4) & (0.2) & (6.9)\\
$f_K [{\rm MeV}]$ & 149.6 & (3.6) & (0.4) & (6.0) & (1.0) & (1.5) & (6.3)\\
$f_K/f_\pi$     & 1.205 & (18) & (14)& (48) & (34)& (12) & (62) \\
\hline\hline
\end{tabular}
\end{table}

\begin{table}[ht]
\caption{Finite volume correction factors obtained from the $\SU(2)$ and $\SU(3)$ PQChPT fits compared to the prediction by Colangelo, D\"urr, and Haefeli (CDH) \cite{Colangelo:2005gd} at the unitary pion masses of 331 MeV ($m_x=m_y=m_l=0.005$) and 419 MeV ($m_x=m_y=m_l=0.01$).} 
\label{tab:FVcompareCDH}
\begin{tabular}{c|*{3}{c}|*{3}{c}}
\hline\hline
 & \multicolumn{3}{c|}{$R_m [\%]$} & \multicolumn{3}{c}{$-R_f [\%]$} \\
$m_\pi [{\rm MeV}]$ & $\SU(2)$ & $\SU(3)$ & ${\rm CDH}$ & $\SU(2)$ & $\SU(3)$ & ${\rm CDH}$ \\\hline
331 & 0.09(.01) & 0.14(.04) & 0.13(.04) & 0.36(.03) & 0.56(.16) & 0.32(.00) \\
419 & 0.03(.00) & 0.04(.01) & 0.04(.01) & 0.10(.01) & 0.17(.05) & 0.09(.00) \\
\hline\hline
\end{tabular}
\end{table}

\begin{table}[ht]
\caption{The measured pion mass and decay constant at $m_x=m_y=m_l=0.01$, $m_h=0.04$ at two different volumes (for $16^3$ data see \cite{Allton:2007hx}). We compare directly the measured ratio to the predictions from $\SU(2)$ and $\SU(3)$ ChPT and the value interpolated from Colangelo, D\"urr, and Haefeli (CDH) \cite{Colangelo:2005gd}. For explanation see Eq.~(\ref{eq:su2:FVcomp}) and text.}
\label{tab:FVcompare16c}
\begin{tabular}{l|*{3}{c}|*{3}{c}}
\hline\hline
     & \multicolumn{3}{c|}{measured} & \multicolumn{3}{c}{$\frac{1-R^{(16c)}}{1-R^{(24c)}}$} \\
     & $24^3$ & $16^3$ & ratio &   $\SU(2)$ & $\SU(3)$ & CDH \\\hline
$m_\pi$ & 0.24211(75) & 0.247(3)  & 0.980(12) & 0.9961(04) & 0.9944(15) & 0.9942(22)\\
$f_\pi$ & 0.0928(6)   & 0.0895(7) & 1.037(11) & 1.0155(14) & 1.0252(70) & 1.0118(03)\\
\hline\hline
\end{tabular}
\end{table}

\begin{table}[ht]
\caption{Comparison of the $\SU(2)$ NLO-LEC $\bar{l}_3$ and $\bar{l}_4$ (defined at a scale of 139 MeV, cf. Eq.~(\ref{eq:chPTconv:barl})) with results from other $N_f=2+1$ and $N_f=2$ lattice simulations \cite{Bernard:2007ps,Urbach:2007rt,DelDebbio:2006cn} and a phenomenological estimate \cite{Colangelo:2001df}.}
\label{tab:LECs:su2}
\begin{tabular}{ccc|*{2}{c}}
\hline\hline 
 & $N_f$ & type & $\bar{l}_3$ & $\bar{l}_4$ \\\hline
this work, direct $\SU(2)$ fit & 2+1 & DWF & 3.13(.33)(.24) & 4.43(.14)(.77) \\
this work, conv. from $\SU(3)$ & 2+1 & DWF & 2.87(.28)(--) & 4.10(.05)(--) \\\hline
MILC\footnote{direct $\SU(2)$ fit, from \cite{Bernard:2007ps}} & 2+1 & stagg & 2.85(.07)(--) & -- \\
MILC\footnote{converted from $\SU(3)$, see \cite{Leutwyler:2007ae}} & 2+1 & stagg & 0.6(1.2) & 3.9(0.5) \\\hline
ETMC \cite{Urbach:2007rt} & 2 & TM-Wilson & 3.44(.08)(.35) & 4.61(.04)(.11) \\\hline
CERN \cite{DelDebbio:2006cn} & 2 & impr. Wilson & 3.0(0.5)(0.1) & 4.1(0.1)(--)\\
                             &   &             &               & 3.3(0.8)(--)\footnote{\begin{minipage}[t]{.9\textwidth}\linespread{.8}\footnotesize \begin{flushleft}First number obtained without additional NNLO-term, second number from fit including NNLO-term, see \cite{DelDebbio:2006cn} for details.\end{flushleft}\end{minipage}} \\\hline
phenom.\cite{Colangelo:2001df} &  &  & 2.9(2.4) & 4.4(0.2) \\ 
\hline\hline
\end{tabular}
\end{table}

\begin{table}[ht]
\caption{Comparison of light physical quark masses and their ratio obtained from recent lattice simulations using various types of fermion discretizations and different renormalization techniques (non-perturbative and perturbative). (In the cases where no error is given for the quark mass ratio, no error was quoted in the cited work. Since we do not know the correlations, we could not provide the error estimate here.) }
\label{tab:su2:quarkMasses}
\begin{tabular}{lcc >{$}c<{$} >{$}c<{$} >{$}c<{$}  }
\hline\hline
 & $N_f$ & type & m_{ud}^{\overline{\rm MS}}(2\,{\rm GeV}) [\rm MeV] & m_s^{\overline{\rm MS}}(2\,{\rm GeV}) [\rm MeV] & m_s:m_{ud} \\\hline
\multicolumn{6}{l}{using non-perturbative renormalization}\\
this work & 2+1 & DWF & 3.72(0.16)_{\rm stat}(0.33)_{\rm ren}(0.18)_{\rm syst} & 107.3(4.4)_{\rm stat}(9.7)_{\rm ren}(4.9)_{\rm syst} & 28.8(0.4)_{\rm stat}(1.6)_{\rm syst} \\
RBC \cite{Blum:2007cy} & 2 & DWF & 4.25(0.23)_{\rm stat}(0.26)_{\rm ren}  & 119.5(5.6)_{\rm stat}(7.4)_{\rm ren}  & 28.10(0.38)_{\rm stat} \\
ETMC \cite{Blossier:2007vv} & 2 & TM-Wilson & 3.85(0.12)_{\rm stat}(0.40)_{\rm syst} & 105(3)_{\rm stat}(9)_{\rm syst} & 27.3(0.3)_{\rm stat}(1.2)_{\rm syst} \\
QCDSF \cite{Gockeler:2006jt} & 2 & impr. Wilson & 4.08(0.23)_{\rm stat}(0.19)_{\rm syst}(0.23)_{\rm scale} & 111(6)_{\rm stat}(4)_{\rm syst}(6)_{\rm scale} & 27.2(3.2)   \\ \hline
\multicolumn{5}{l}{using perturbative renormalization}\\
MILC \cite{Bernard:2007ps} & 2+1 & stagg. & 3.2(0)_{\rm stat}(0.1)_{\rm ren}(0.2)_{\rm EM}(0)_{\rm cont} & 88(0)_{\rm stat}(3)_{\rm ren}(4)_{\rm EM}(0)_{\rm cont} & 27.2(0.1)_{\rm stat}(0.3)_{\rm syst} \\
PACS-CS \cite{Ukita:2007cu} & 2+1 & impr. Wilson &  2.3(1.1) & 69.1(2.5) & 30(?) \\
JLQCD \cite{Ishikawa:2007nn} & 2+1 & impr. Wilson & 3.54(^{+.64}_{-.35})_{\rm total} & 91.1(^{+14.6}_{-6.2})_{\rm total} & 25.7(?) \\
\hline\hline
\end{tabular}
\end{table}

%% file: tab/Su3.tab
%
%
\clearpage

\begin{table}
\caption{\label{tab:fitSU3}Fitted parameters from $\SU(3)\times\SU(3)$ fits with a valence mass cut $m_{\rm avg}\leq0.01$. For convenience the LECs $\lthree{i}$ are quoted in units of $10^{-4}$ at two different chiral scales, $\Lambda_\chi$ = 770 MeV and 1 GeV. The definitions of the low energy constants are given in the Appendix. (Only statistical errors are quoted here.)} 
\begin{center}
\begin{tabular}{r*{4}{c}}
\hline\hline
 & \multicolumn{2}{l}{$\bthree\:=\:2.35(.16)$} & \multicolumn{2}{l}{$\fthree\:=\:0.0541(40)$} \\\hline\hline
 $\Lambda_\chi$ & $\lthree{4}$ & $\lthree{5}$ & $(2\lthree{6}-\lthree{4})$ & $(2\lthree{8}-\lthree{5})$ \\\hline
770 MeV & 1.39(.80) & 8.72(.99) & -0.01(.42) & 2.43(.45) \\
1 GeV & -0.67(.80) & 2.51(.99) & -0.47(.42) & 5.19(.45) \\
\hline\hline
\end{tabular}
\end{center}
\end{table}

\clearpage
\begin{table}
\caption{\label{tab:conv_SU3_SU2} $\SU(2)\times\SU(2)$ low energy constants obtained directly from the $\SU(2)\times\SU(2)$ fits compared to those converted from the $\SU(3)\times\SU(3)$ chiral fits with a mass cut of $m_{\rm avg}\leq 0.01$.}
\begin{center}
\begin{tabular}{rl*{3}{c}c}
\hline
\hline 
           &                    &$\btwo$ & $\ftwo$ & $\bar{l}_3$ & $\bar{l}_4$ \\
\hline 
direct &$\SU(2)\times\SU(2)$    & 2.414(61) & 0.0665(21) & 3.13(.33) & 4.43(.14) \\
converted &$\SU(3)\times\SU(3)$ & 2.457(78) & 0.0661(18) & 2.87(.28) & 4.10(.05) \\
\hline\hline
\end{tabular} 
\end{center}
\end{table}

%
%
\begin{table}[ht]
\caption{Comparison of fitted $\SU(3)$ NLO-LECs $\lthree{i}$ at $\Lambda_\chi=770\,{\rm MeV}$ in units of $10^{-4}$ to phenomenologically obtained values by Bijnens (see Table 2 in \cite{Bijnens:2007yd}) and dynamical $N_f=2+1$ staggered lattice simulation (MILC), \cite{Aubin:2004fs} and \cite{Bernard:2007ps} (there, LECs were quoted at $\Lambda_\chi=m_\eta$, conversion was done according to Eq.~(\ref{eq:chPTsu3:LECconv}).) }
\label{tab:LECs:su3}
\begin{tabular}{c|*{6}{c}}
\hline\hline
              & $\lthree{4}$ & $\lthree{5}$ & $\lthree{6}$ & $\lthree{8}$ & $(2\lthree{8}-\lthree{5})$ & $(2\lthree{6}-\lthree{4})$ \\\hline
this work\footnote{\begin{minipage}[t]{.9\textwidth}\linespread{.8}\footnotesize \begin{flushleft}For reasons mentioned in Sec.\ref{sec:su3:sysErr}, we do not quote any systematic error for parameters obtained from the $\SU(3)$ fits.\end{flushleft}\end{minipage}}
                  &   1.4(0.8)(--) & 8.7(1.0)(--) & 0.7(0.6)(--) & 5.6(0.4)(--) & 2.4(0.4)(--) & 0.0(0.4)(--) \\\hline
Bijnens, NLO  &   $\equiv0$ & 14.6 & $\equiv0$ & 10.0 & 5.4 & $\equiv0$ \\
Bijnens, NNLO &   $\equiv0$ & 9.7(1.1) & $\equiv0$ & 6.0(1.8) & 2.3\footnote{\begin{minipage}[t]{.9\textwidth}\linespread{.8}\footnotesize \begin{flushleft}This value was derived from the quoted single values for $\lthree{5}$ and $\lthree{8}$; since we do not know the correlation between those two, we cannot provide the error estimate. \end{flushleft}\end{minipage}}  & $\equiv0$ \\\hline
MILC, 2007    &  1.3(3.0)$(^{+3.0}_{-1.0})$ & 13.9(2.0)$(^{+2.0}_{-1.0})$ & 2.4(2.0)$(^{+2.0}_{-1.0})$ & 7.8(1.0)(1.0) & 2.6(1.0)(1.0) & 3.4(1.0)$(^{+2.0}_{-3.0})$  \\
\hline\hline
\end{tabular}
\end{table}

%% file: tab/Bk.tab
%
%

\clearpage
\begin{table}
\caption{Pseudoscalar B-parameter $B_{xy}$ for pseudoscalar mesons with valence masses $m_x$, $m_y$ for two different values of the light sea quark mass $m_l$ ($m_h=0.04$) on $24^3\times64$ lattices, fit range 12--52 (uncorrelated fit)}
\label{table:bk:values:24c}
\begin{center}
\begin{tabular}{*{2}{D{.}{.}{3}}|*{2}{D{.}{.}{8}}}
\hline\hline
 & &\multicolumn{2}{c}{$B_{xy}$}\\
\multicolumn{1}{c}{$m_x$}&\multicolumn{1}{c|}{$m_y$}&\multicolumn{1}{c}{$m_l = 0.005$} & \multicolumn{1}{c}{$m_l = 0.01$} \\\hline
0.001 & 0.001 &       0.4691(83)&       0.4698(53) \\[2pt]
0.005 & 0.001 &       0.4910(66)&       0.4953(41) \\
      & 0.005 &       0.5079(46)&       0.5119(32) \\[2pt]
0.01 & 0.001  &       0.5136(57)&       0.5209(39) \\
     & 0.005  &       0.5267(37)&       0.5312(28) \\
     & 0.01   &       0.5421(29)&       0.5459(25) \\[2pt]
0.02 & 0.001  &       0.5494(52)&       0.5593(42) \\
     & 0.005  &       0.5588(31)&       0.5629(26) \\
     & 0.01   &       0.5697(25)&       0.5727(23) \\
     & 0.02   &       0.5911(23)&       0.5935(20) \\[2pt]
0.03 & 0.001  &       0.5777(55)&       0.5886(54) \\
     & 0.005  &       0.5850(32)&       0.5884(28) \\
     & 0.01   &       0.5933(26)&       0.5957(22) \\
     & 0.02   &       0.6105(23)&       0.6126(18) \\
     & 0.03   &       0.6267(22)&       0.6287(16) \\[2pt]
0.04 & 0.001  &       0.6012(64)&       0.6131(73) \\
     & 0.005  &       0.6070(38)&       0.6105(33) \\
     & 0.01   &       0.6137(30)&       0.6160(24) \\
     & 0.02   &       0.6280(23)&       0.6299(18) \\
     & 0.03   &       0.6418(21)&       0.6435(15) \\
     & 0.04   &       0.6550(19)&       0.6565(14) \\\hline\hline
\end{tabular}
\end{center}
\end{table}

\clearpage
\begin{table}
\caption{Pseudoscalar B-parameter $B_{xy}$ for pseudoscalar mesons with valence masses $m_x$, $m_y$ for three different values of the light sea quark mass $m_l$ ($m_h=0.04$) on $16^3\times32$ lattices, fit range 12--20 (uncorrelated fit)}
\label{table:bk:values:16c}
\begin{center}
\begin{tabular}{*{2}{D{.}{.}{2}}|*{3}{D{.}{.}{6}}}
\hline\hline
 & &\multicolumn{3}{c}{$B_{xy}$}\\
\multicolumn{1}{c}{$m_x$}&\multicolumn{1}{c|}{$m_y$}&\multicolumn{1}{c}{$m_l = 0.01$} & \multicolumn{1}{c}{$m_l = 0.02$} & \multicolumn{1}{c}{$m_l = 0.03$} \\\hline
0.01 & 0.01 & 0.546(8) & 0.539(8) & 0.527(7) \\[2pt]
0.02 & 0.01 & 0.577(6) & 0.569(6) & 0.556(6) \\
     & 0.02 & 0.598(5) & 0.589(5) & 0.580(5) \\[2pt]
0.03 & 0.01 & 0.600(6) & 0.594(6) & 0.579(5) \\
     & 0.02 & 0.617(4) & 0.609(5) & 0.600(4) \\
     & 0.03 & 0.633(4) & 0.626(4) & 0.618(3) \\[2pt]
0.04 & 0.01 & 0.620(5) & 0.616(6) & 0.599(5) \\
     & 0.02 & 0.633(4) & 0.627(4) & 0.618(4) \\
     & 0.03 & 0.647(4) & 0.641(4) & 0.634(3) \\
     & 0.04 & 0.659(3) & 0.655(3) & 0.648(3) \\[2pt]
0.05 & 0.01 & 0.636(5) & 0.634(6) & 0.616(4) \\
     & 0.02 & 0.648(4) & 0.643(4) & 0.634(3) \\
     & 0.03 & 0.660(3) & 0.655(3) & 0.648(3) \\
     & 0.04 & 0.671(3) & 0.667(3) & 0.661(3) \\
     & 0.05 & 0.682(3) & 0.679(3) & 0.673(3) \\\hline\hline
\end{tabular}
\end{center}
\end{table}

\clearpage
\begin{table}
\caption{Parameters and $\chi^2/\textrm{d.o.f.}$ for fits of $B_{xy}$ to Eq.~(\ref{eq:chPTsu2K:BPS}) as shown in Fig.~\ref{fig:bK:SU2lightExtr}. The parameters $\btwo=2.414(61)$ and $\ftwo=0.0665(21)$ were fixed from combined $\SU(2)\times\SU(2)$ meson masses and decay constants fits, Sec.~\ref{sec:Su2}; $\Lambda_\chi=1\,{\rm GeV}$.}
\label{table:bK:fitparamSU2}
\begin{center}
\begin{tabular}{D{.}{.}{2} D{.}{.}{7} D{.}{.}{5} D{.}{.}{6} c D{.}{.}{8}}
\hline\hline
\multicolumn{1}{c}{$\;\;m_y\;\;$} & \multicolumn{1}{c}{$\;\;B_{\rm PS}^{(K)}(m_y)\;\;$} & \multicolumn{1}{c}{$\;\;b_1(m_y)/\ftwo^2\;\;$}  & \multicolumn{1}{c}{$\;\;b_2(m_y)/\ftwo^2\;\;$} &\multicolumn{1}{c}{$\;\;\chi^2/\textrm{d.o.f}\;\;$} &\multicolumn{1}{c}{$\;\;B_{ud\,y}\;\;$}  \\\hline
0.02  & 0.513(10) & -5.2(1.6) & 8.18(.65) & 0.27 & 0.5249(86) \\
0.03  & 0.544(10) & -5.4(1.5) & 6.85(.65) & 0.20 & 0.5550(87) \\
0.04  & 0.568(11) & -5.4(1.7) & 6.05(.71) & 0.14 & 0.5789(97) \\
\hline\hline
\end{tabular}
\end{center}
\end{table}

\clearpage
\begin{table}
\caption{Fit parameters for fits of $B_{xy}$ to Eq.~(\ref{eq:chPTsu3:BPS:nondeg}) as shown in Fig.~\ref{fig:bk:SU3fits}. Values are quoted for different cuts $m_{\rm cut}$ in the average valence mass value. The paraemters $\bthree=2.35(.16)$ and $\fthree=0.0541(40)$ were fixed from combined $\SU(3)\times\SU(3)$ meson masses and decay constants fits, Sec.~\ref{sec:Su3}; $\Lambda_\chi=1\,{\rm GeV}$.}
\label{table:bk:fitparamSU3}
\begin{center}
\begin{tabular}{D{.}{.}{4} D{.}{.}{7} D{P}{.}{5}  *{2}{D{.}{.}{6}} c D{.}{.}{7}}
\hline\hline
\multicolumn{1}{c}{$\;\;m_{\rm cut}\;\;$} & \multicolumn{1}{c}{$\;\;\bagzero\;\;$} & \multicolumn{1}{c}{$\;\;b\;\;$} & \multicolumn{1}{c}{$\;\;c\;\;$} & \multicolumn{1}{c}{$\;\;d\;\;$} & \multicolumn{1}{c}{$\;\;\chi^2/{\rm d.o.f.}\;\;$} & \multicolumn{1}{c}{$\;\;B_K=B_{ud\,s}\;\;$} \\
\hline
0.01   & 0.266(26) & -4P5(1.6) & -0.20(.18) & 0.68(.60) &  1.8 & 0.484(15) \\
0.0125 & 0.249(24) & -2P5(1.5) &  0.87(.13) & 0.96(.61) &  2.7 & 0.556(12) \\
0.015  & 0.232(22) &  0P2(1.5) &  0.43(.13) & 1.25(.59) &  5.4 & 0.546(11) \\
0.0175 & 0.223(22) &  1P0(1.5) &  0.94(.12) & 1.49(.61) &  6.1 & 0.570(12) \\
0.02   & 0.205(19) &  4P2(1.4) &  0.40(.10) & 1.89(.58) & 10.2 & 0.553(10) \\
\hline\hline
\end{tabular}
\end{center}
\end{table}

\clearpage
\begin{table}
\caption{Finite volume comparison for the pseudoscalar bag parameter measured on $16^3$ ($\approx(1.8\,{\rm fm})^3$) and $24^3$ ($\approx(2.7\,{\rm fm})^3$) for dynamical quark masses $m_l=0.01$, $m_h=0.04$.}
\label{table:bk:finiteVol}
\begin{center}
\begin{tabular}{ *2{D{.}{.}{2}} |  D{.}{.}{6} D{.}{.}{8} c }
\hline\hline
 \multicolumn{1}{c}{$\;\;m_x\;\;$} &
 \multicolumn{1}{c|}{$\;\;m_y\;\;$} & 
 \multicolumn{1}{c}{$\;\;B_{xy}^{(16^3)}\;\;$} & 
 \multicolumn{1}{c}{$\;\;B_{xy}^{(24^3)}\;\;$} & 
 \multicolumn{1}{c}{$\;\;\left|B_{xy}^{(16^3)}-B_{xy}^{(24^3)}\right|/B_{xy}^{(24^3)}\;\;$} \\[2pt]
\hline
0.01 & 0.01 & 0.546(8) & 0.5459(25) & 0.0(1.5) \% \\[2pt]
0.02 & 0.01 & 0.577(6) & 0.5727(23) & 0.8(1.1)\% \\
     & 0.02 & 0.598(5) & 0.5935(20) & 0.8(0.9)\% \\[2pt]
0.03 & 0.01 & 0.600(6) & 0.5957(22) & 0.7(1.1)\% \\
     & 0.02 & 0.617(4) & 0.6126(18) & 0.7(0.7)\% \\
     & 0.03 & 0.633(4) & 0.6287(16) & 0.7(0.7)\% \\[2pt]
0.04 & 0.01 & 0.620(5) & 0.6160(24) & 0.6(0.9)\% \\
     & 0.02 & 0.633(4) & 0.6299(18) & 0.5(0.7)\% \\
     & 0.03 & 0.647(4) & 0.6435(15) & 0.5(0.7)\% \\
     & 0.04 & 0.659(3) & 0.6565(14) & 0.4(0.5)\% \\
\hline\hline
\end{tabular}
\end{center}
\end{table}

%% file: tab/Vector.tab
%
%

\clearpage

\begin{table}
 \begin{center}
 \caption{Results for the measured vector meson masses (DEG and UNI data sets). }
 \label{tab:vector_masses}
 \begin{tabular}{llllll}
 \hline\\[-8mm]
 \multicolumn{1}{c}{$m_l^{\rm sea}$} & 
 \multicolumn{1}{c}{$m_x$} & 
 \multicolumn{1}{c}{$m_y$} & 
 \multicolumn{1}{c}{DEG} & 
 \multicolumn{1}{c}{UNI} \\
 \hline\\[-8mm]
 {0.005} &0.005&0.005  &0.5053(58)&0.507(19) \\[-2mm]
	&0.005 &0.04   & 	  &0.5591(75)\\[-2mm]
	&0.04&0.04   &0.6227(19)&0.6183(48)\\
 \hline\\[-8mm]
 {0.01} &0.01 &0.01   &0.5288(45)&0.527(17) \\[-2mm]
        &0.01 &0.04   &	  &0.5887(89)\\[-2mm]
        &0.04 &0.04   &0.6295(18)&0.6319(48)\\
 \hline\\[-8mm]
 {0.02} &0.02 &0.02   &0.5789(55)&0.584(16) \\[-2mm]
        &0.02 &0.04   & 	  &0.612(13) \\[-2mm]
        &0.04 &0.04   &0.6453(33)&0.6453(84)\\
 \hline\\[-8mm]
 {0.03} &0.03 &0.03   &0.6317(33)&0.6239(76)\\[-2mm]
        &0.03 &0.04   &	  &0.6447(83)\\[-2mm]
        &0.04 &0.04   &0.6622(27)&0.6609(69)\\
 \hline\\[-8mm]
 \end{tabular}
 \end{center}
\end{table}

\begin{table}
\begin{center}
\caption{Results for the measured of couplings $f_V^T/f_V$.
\label{tab:pda_frho}}
\begin{tabular}{lcccc@{\hspace{1mm}}|@{\hspace{1mm}}c@{\hspace{1mm}}||cc}
\hline&&&&&&\\[-8mm]
$am_q$&0.03&0.02&0.01&0.005&physical&$\chi^2/$d.o.f\\[1mm]
\hline&&&&&&\\[-7mm]
$f_\rho^T/f_\rho$&0.6806(55)&0.6646(73)&0.645(13)&0.624(21)&0.619(15)&0.17\\[.5mm]
\hline&&&&&\\[-7mm]
$f_{K^\ast}^T/f_{K^\ast}$&0.6893(55)&0.6781(56)&0.6667(43)&0.6570(68)&0.6498(62)&0.11\\[.5mm]
\hline&&&&&&\\[-7mm]
$f_\phi^T/f_\phi$&0.6947(36)&0.6933(38)&0.6881(22)&0.6866(32)&0.6838(33)&0.10\\[.5mm]
\hline\\
\end{tabular}\end{center}\end{table}

\begin{table}
\caption{Previous quenched lattice results for the ratio of tensor
and vector decay constants.}
\label{tab:f_V-previous}
\begin{center}\begin{tabular}{c|c|c|c}
\hline&&&\\[-6mm]
Reference&$\frac{f_\rho^T(2\,\textrm{GeV})}{f_\rho}$&
$\frac{f_{K^\ast}^T(2\,\textrm{GeV})}{f_{K^\ast}}$&
$\frac{f_\phi^T(2\,\textrm{GeV})}{f_\phi}$\\[.5mm]
\hline Becirevic et
al.~\cite{Becirevic:2003pn}&$0.720(24)(^{+16}_{-00})$&
$0.739(17)(^{+3}_{-0})$&$0.759(9)(0)$\\
Braun et al.~\cite{Braun:2003jg}&$0.742(14)$&$-$&$0.780(8)$\\
\hline
\end{tabular}\end{center}
\end{table}

%% file: fig/SimDetail.fig
%
%
%

\clearpage
\begin{figure}[htbp]
\centering
\includegraphics[width=0.8\textwidth]{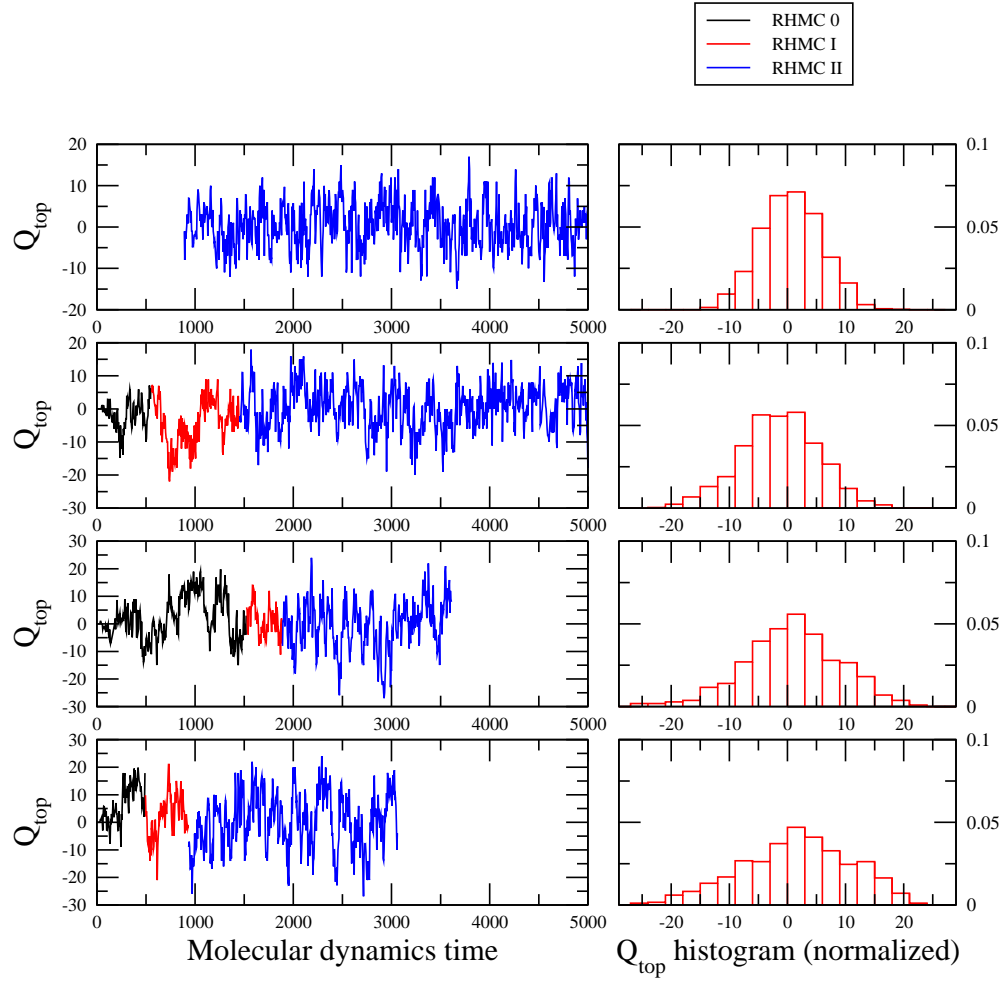}
\caption{The left graphs show topological charge, as measured on
each ensemble every 5th unit of molecular dynamics time.   The
right panels show normalised histograms of toplogical charge.
Notice the width of the histograms decreases as the light quark
mass is reduced.  For both the left and right graphs, the
light dynamical quark mass is $m_l=0.005,\,0.01,\,0.02,\,0.03$,
from top to bottom.}
\label{fig:topology}
\end{figure}

%
%
%

\clearpage
\begin{figure}[htbp]
\centering
\includegraphics[width=0.8\textwidth]{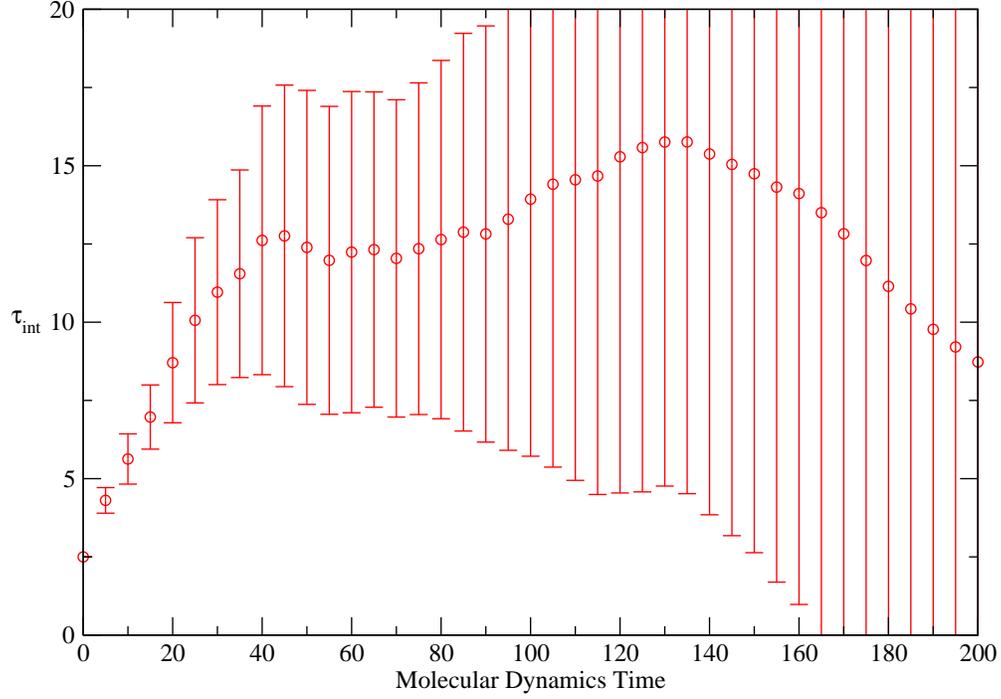}
\caption{Integrated autocorrelation time for 12th timeslice of the
unitary degenerate pseudoscalar correlator with local sources and
sinks, $\langle P^{LL},P^{LL} \rangle$, measured on the $m_l=0.005$
ensemble.  Measurements were done every fifth trajectory in the
range 900-1530 and a bin factor of two was used in the analysis.}
\label{fig:IntAutoCorr}
\end{figure}

%
%
%

\clearpage
\begin{figure}[htbp]
\centering
\includegraphics[width=0.8\textwidth]{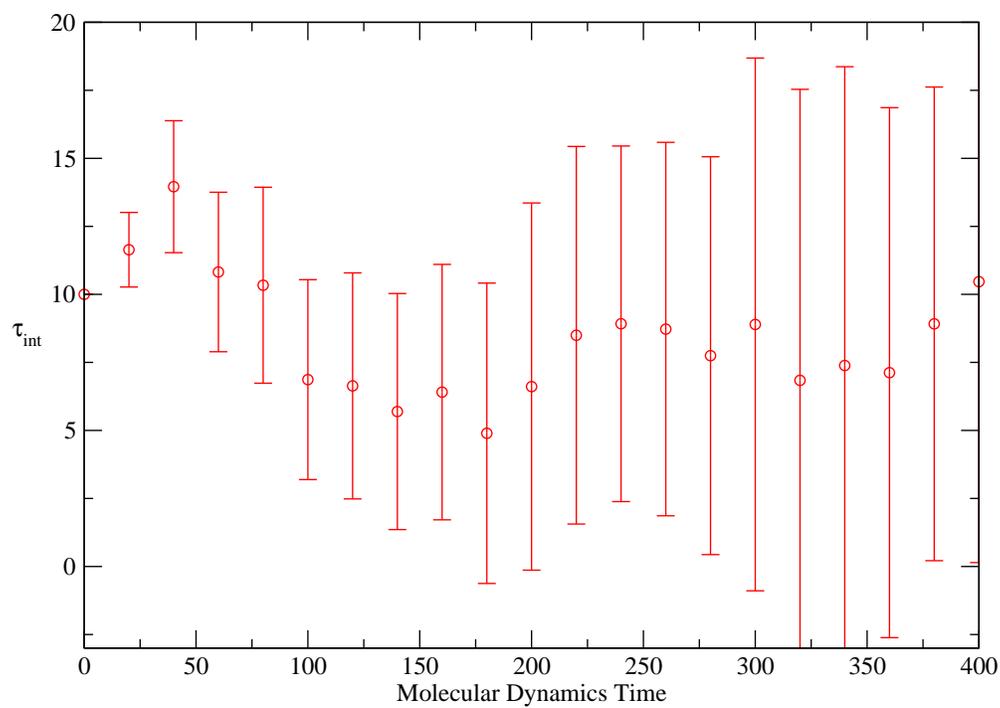}
\caption{Integrated autocorrelation time for 12th timeslice of the
unitary degenerate pseudoscalar correlator with a local sink and
a Gaussian source, $\langle P^{LL},P^{HH} \rangle$, measured on the
$m_l=0.005$
ensemble.  Measurements were done every twentieth trajectory in the
range 500-4500 and a bin factor of four was used in the analysis.}
\label{fig:IntAutoCorr20}
\end{figure}

%
%
%

\clearpage
\begin{figure}[t]
\centering
\includegraphics[width=4in]{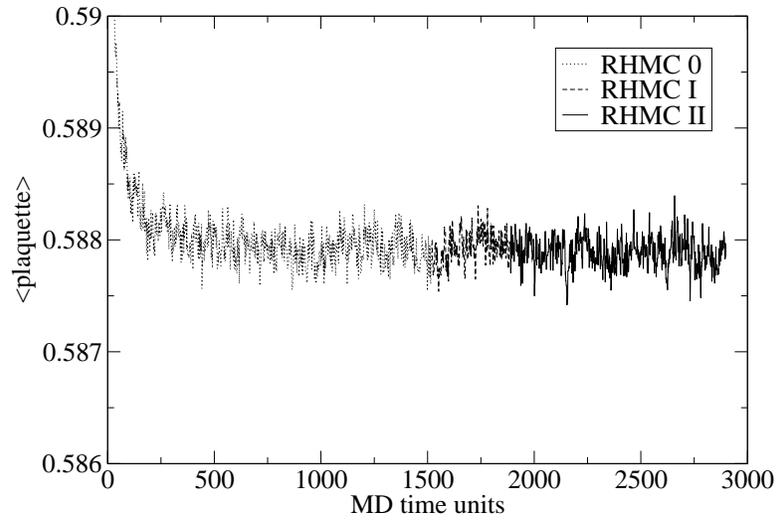}
\caption{Evolution history of the plaquette for the ensemble $m_l=0.02$}
\label{fig:plaq_m02}
\end{figure}

%% file: fig/LatticePS.fig
\begin{figure}[htbp]
\begin{center}
\includegraphics[width=0.8\textwidth]{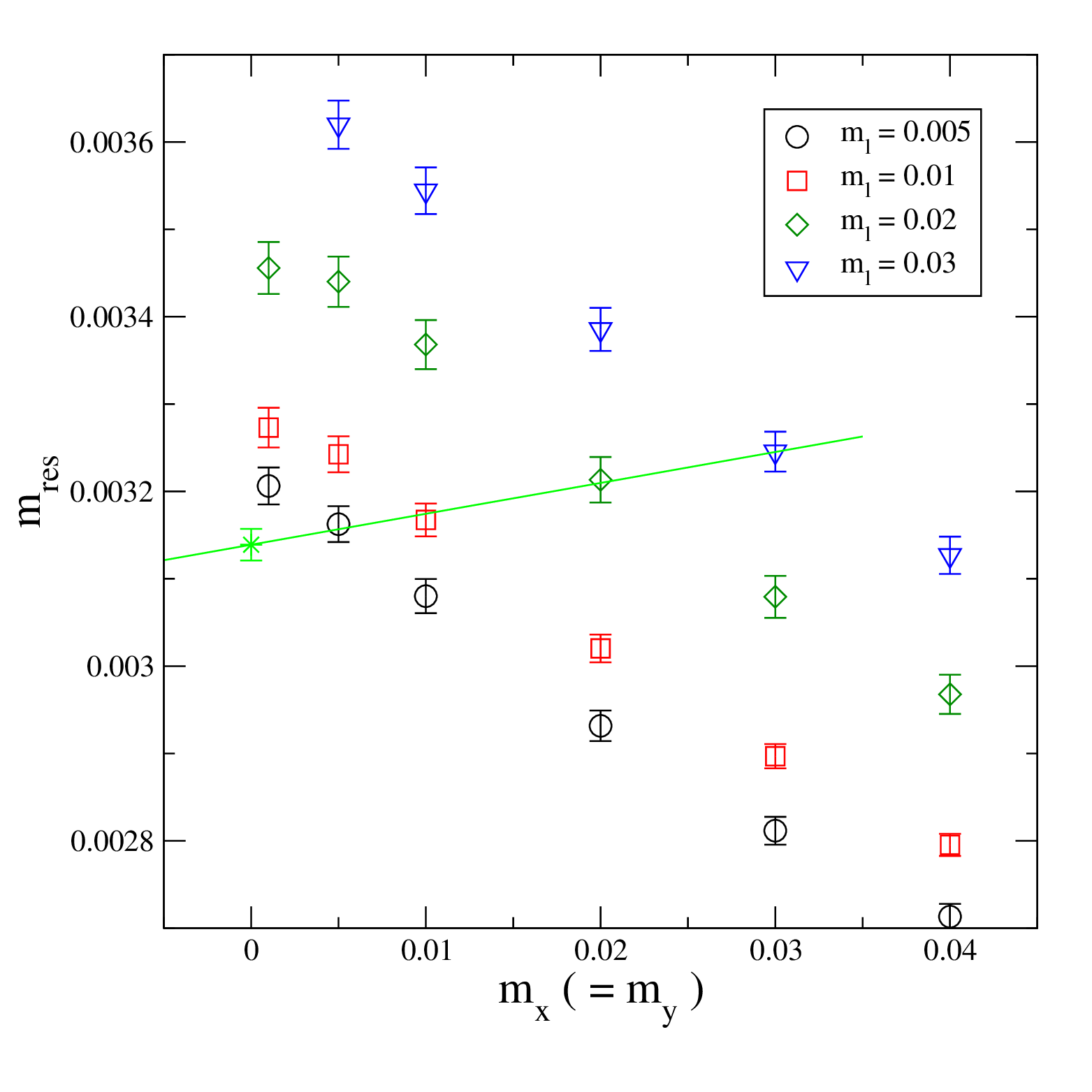}
\caption{The residual mass as a function of the valence quark mass. The solid line is a linear fit to the four unitary points with $m_l = 0.005, 0.01, 0.02$ and 0.03. The star is the extrapolation to $m_l = 0$, which we quote as the residual mass in our simulations.\label{fig:mres}}
\end{center}
\end{figure}

\begin{figure}[htbp]
\begin{center}
\includegraphics[width=0.8\textwidth]{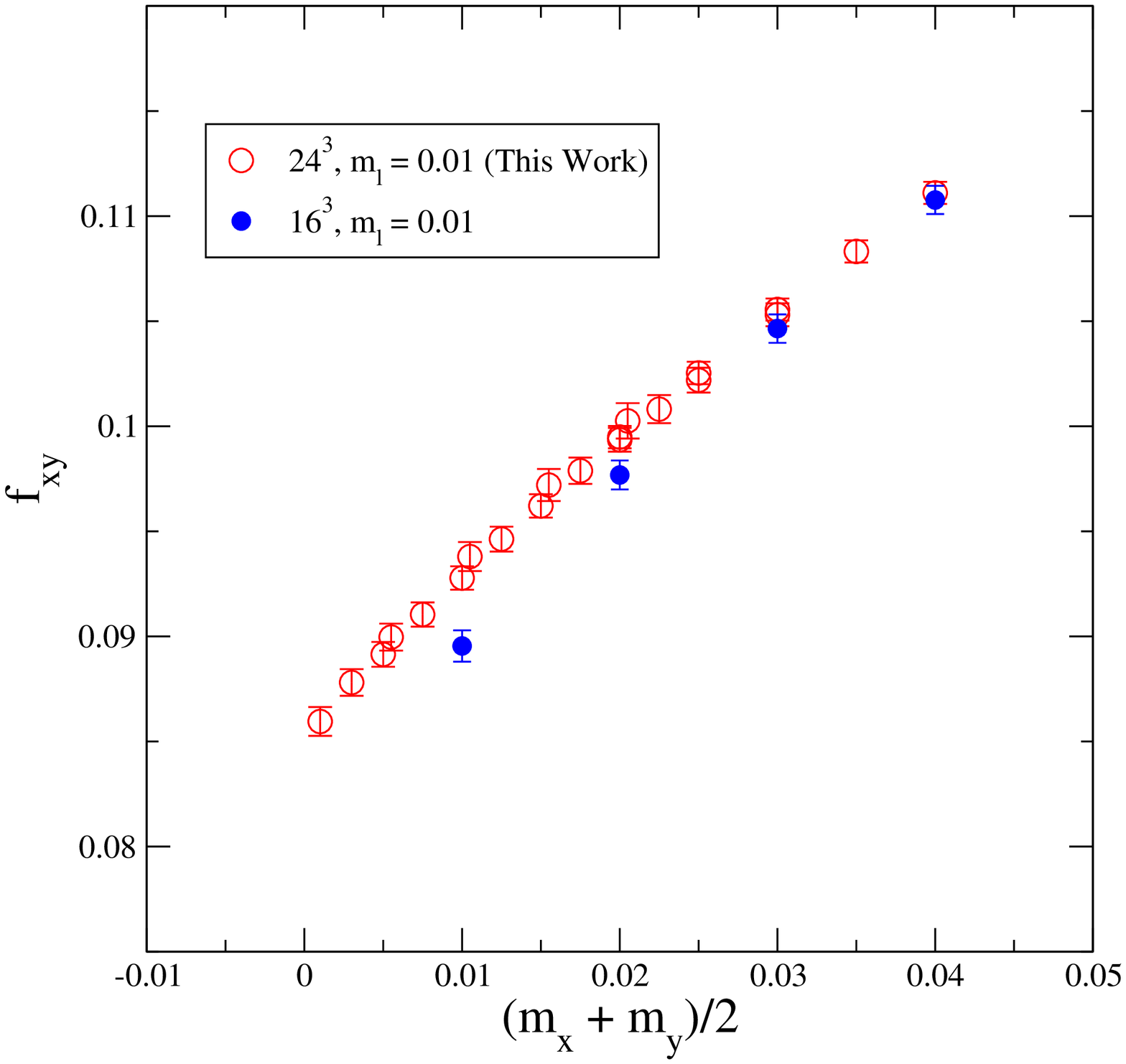}
\caption{Comparison of the lattice results for the pseudoscalar decay constants on the $16^3\times32$ and $24^3\times64$ lattices with $m_l = 0.01$. \label{fig:fpiFV}}
\end{center}
\end{figure}

%% file: fig/LatticeBaryon.fig
%
%


\begin{figure}[htbp]
\begin{center}
\includegraphics[angle=-90, width=0.6\textwidth]{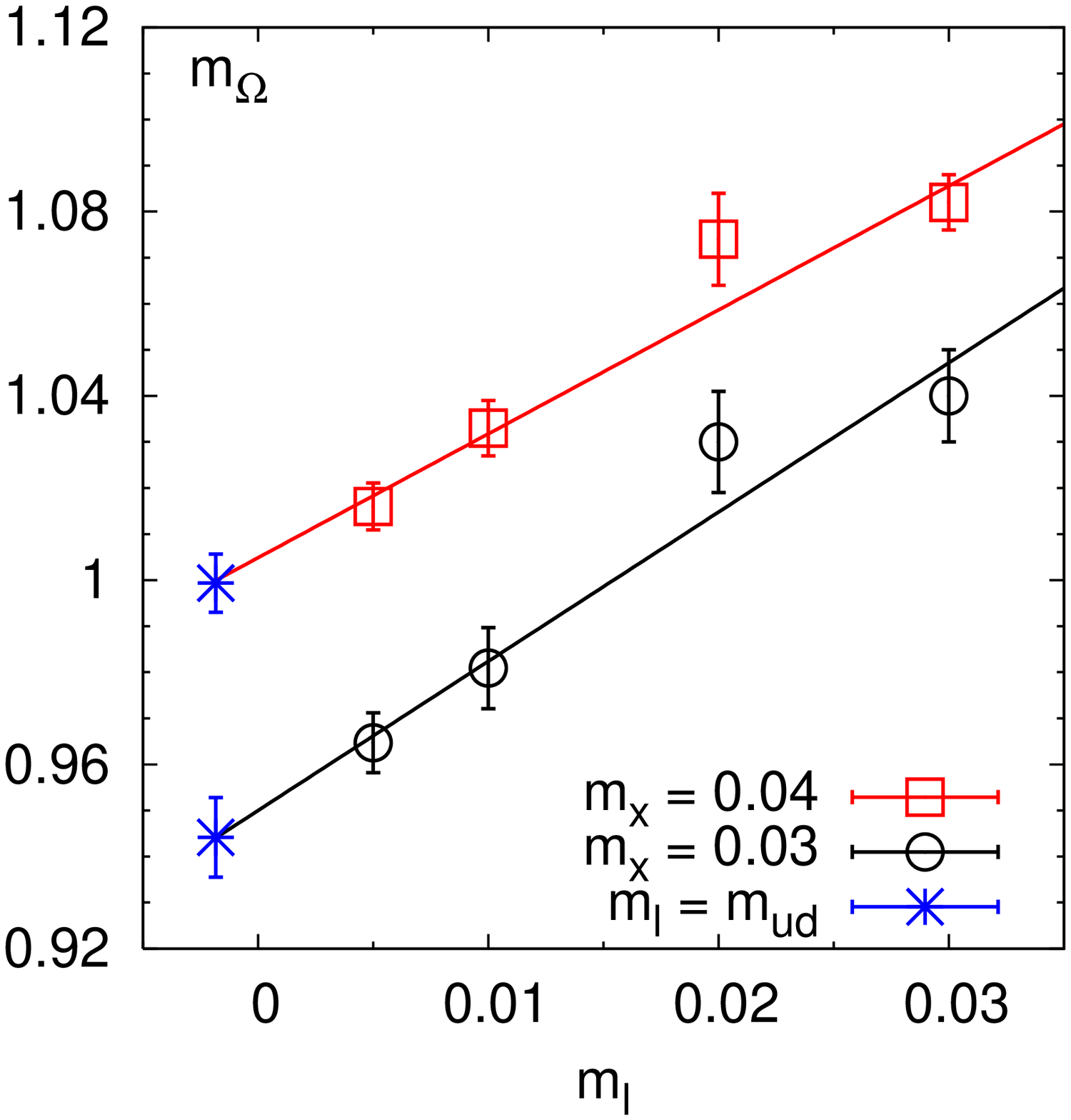} 
\caption{Sea quark mass dependence of the decuplet baryon masses for degenerate valence quarks $m_x=0.03$ \textit{(black circles)} and 0.04 \textit{(red squares)}. Linear fits are shown by solid lines. \textit{Blue asterisks} denote masses extrapolated to the physical value of the bare, light sea quark mass $m_{ud}$.  }
\label{fig:omega masses}
\end{center}
\end{figure}

\clearpage

%% file: fig/Su2.fig
%

\begin{figure}
\begin{center}
\begin{minipage}{.5\textwidth}
\begin{center}
\includegraphics[angle=-90, width=.9\textwidth]{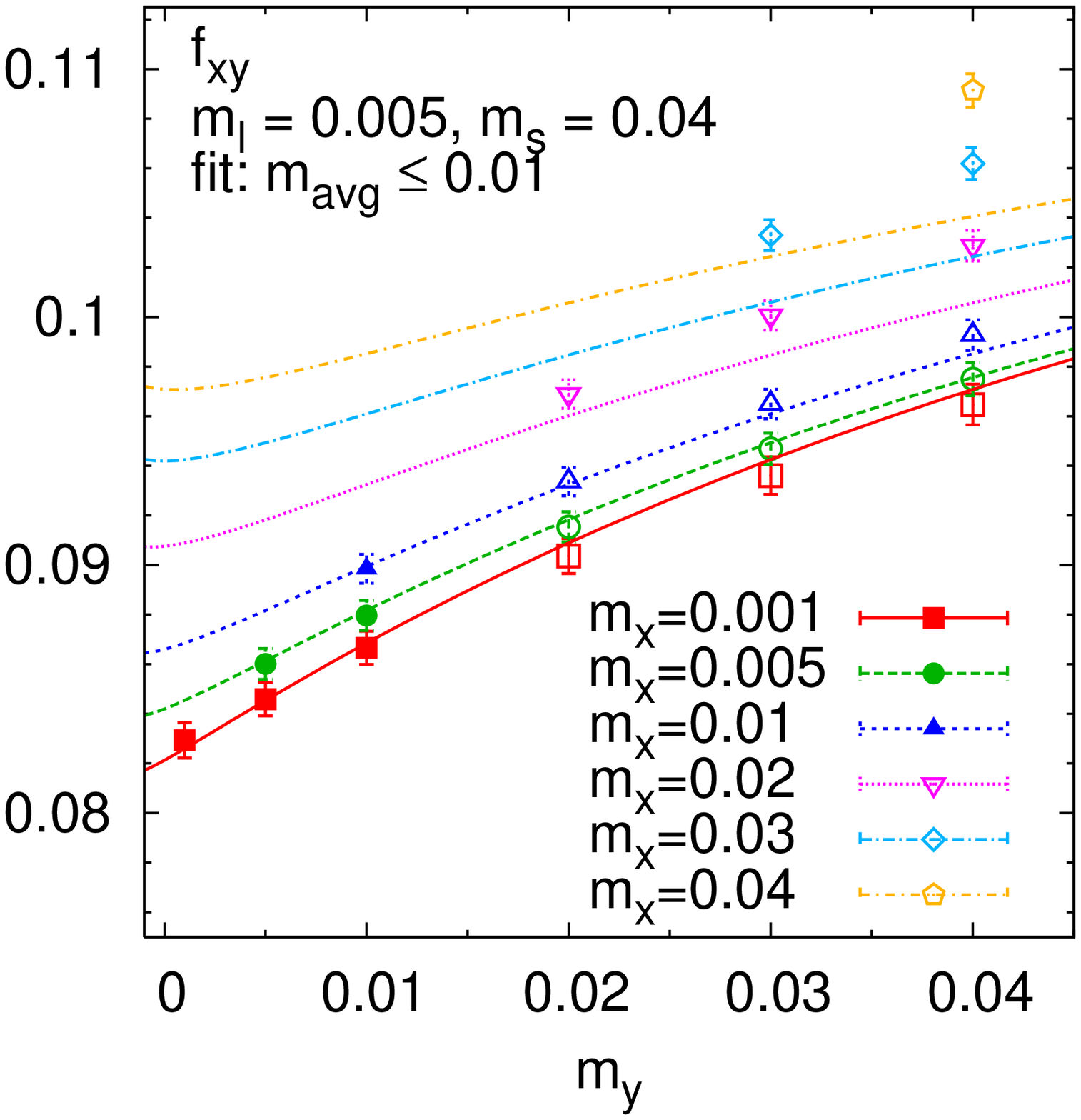}\\
\includegraphics[angle=-90, width=.9\textwidth]{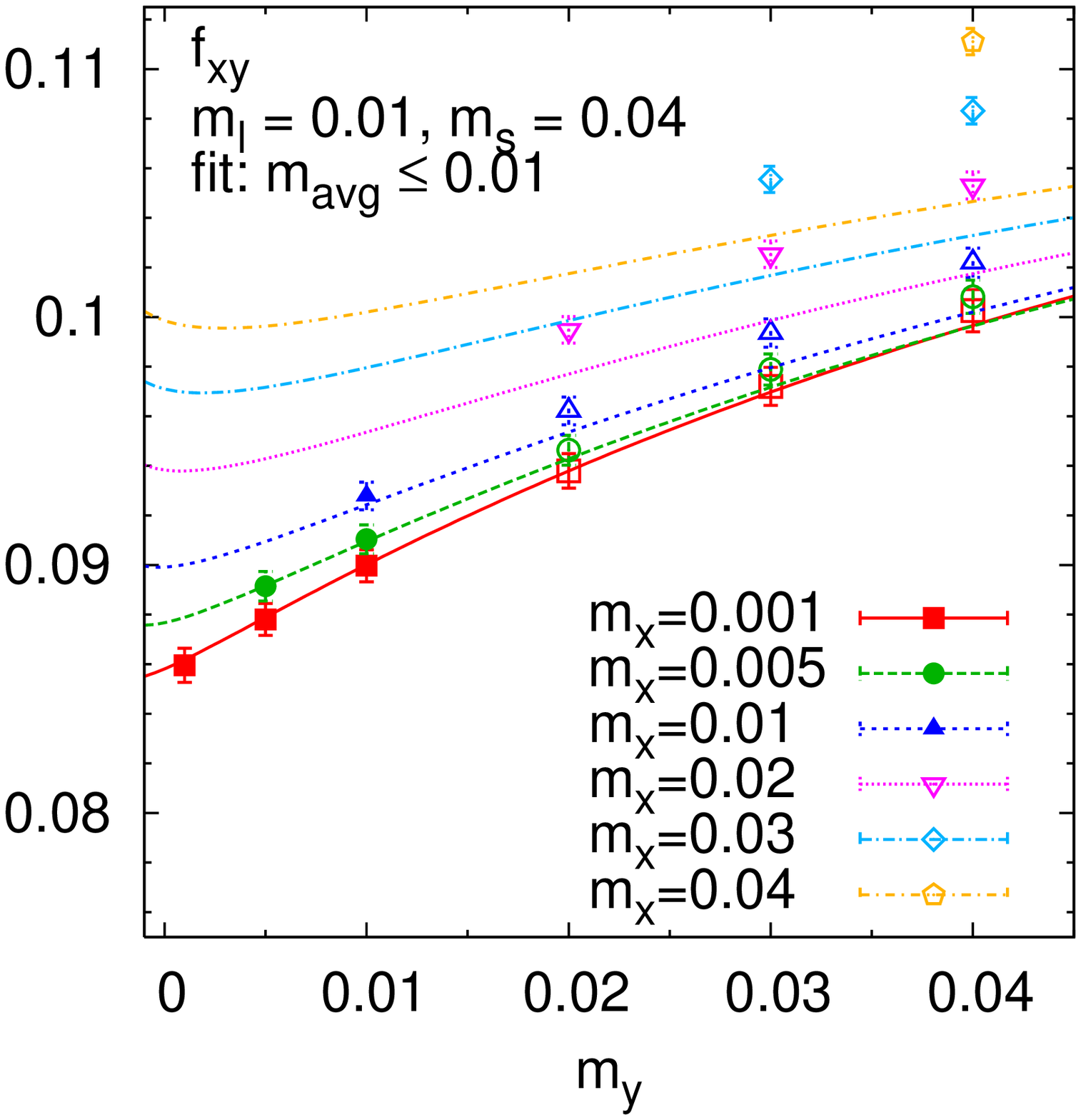}
\end{center}
\end{minipage}%
\begin{minipage}{.5\textwidth}
\begin{center}
\includegraphics[angle=-90, width=.87\textwidth]{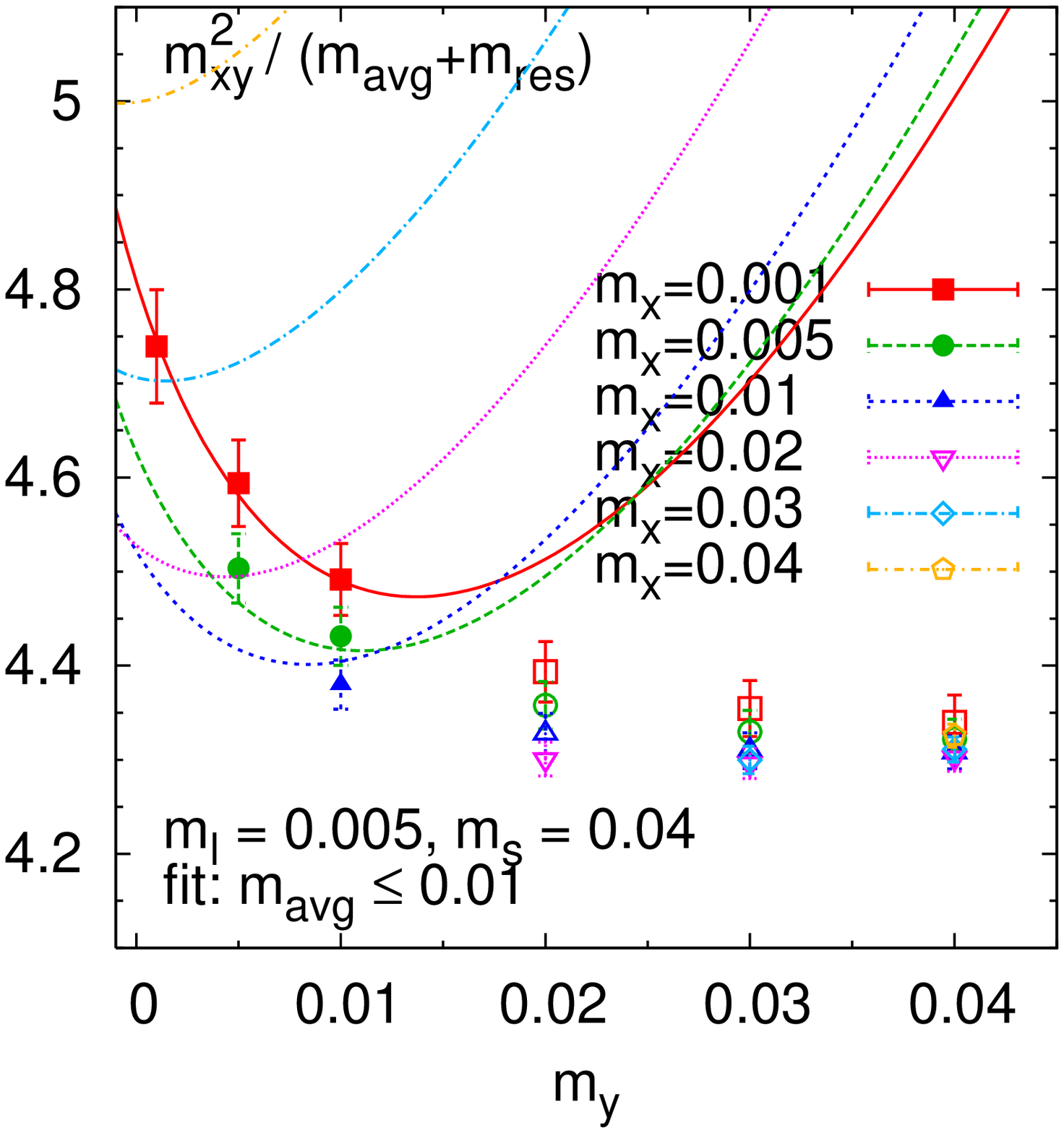}\\
\includegraphics[angle=-90, width=.87\textwidth]{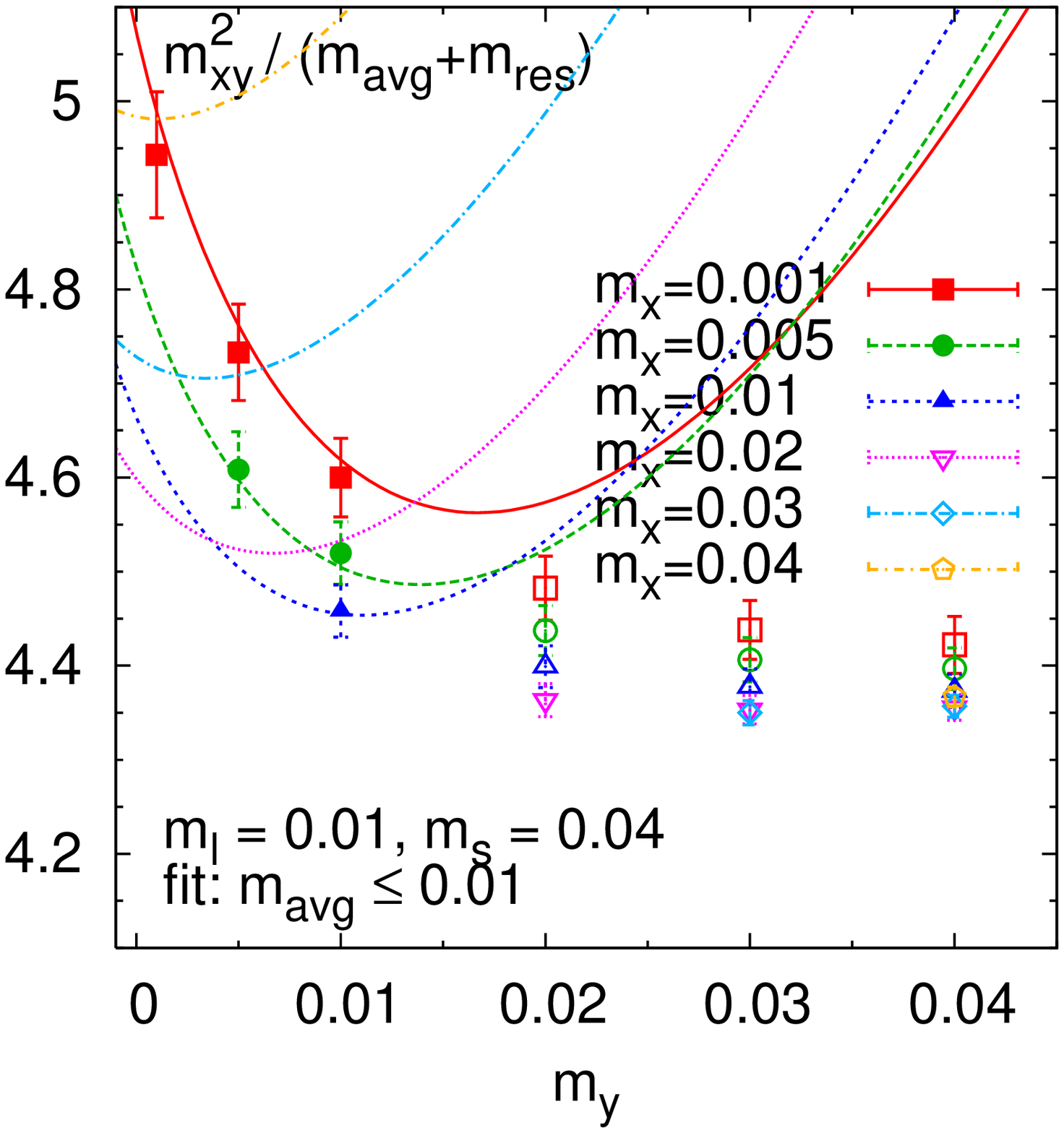}
\end{center}
\end{minipage}
\end{center}
\caption{\label{fig:SU2fits}Combined $\SU(2)\times\SU(2)$ fits for the meson decay constants \textit{(left panels)} and masses \textit{(right panels)} at two different values for the light sea quark mass, and with a cut on the valence masses $m_{\rm avg}\leq0.01$. Points marked by \textit{filled symbols} were included in the fit, while those with \textit{open symbols} were excluded.}
\end{figure}

\begin{figure}
\begin{center}
\hfill
\includegraphics[angle=-90, width=.45\textwidth]{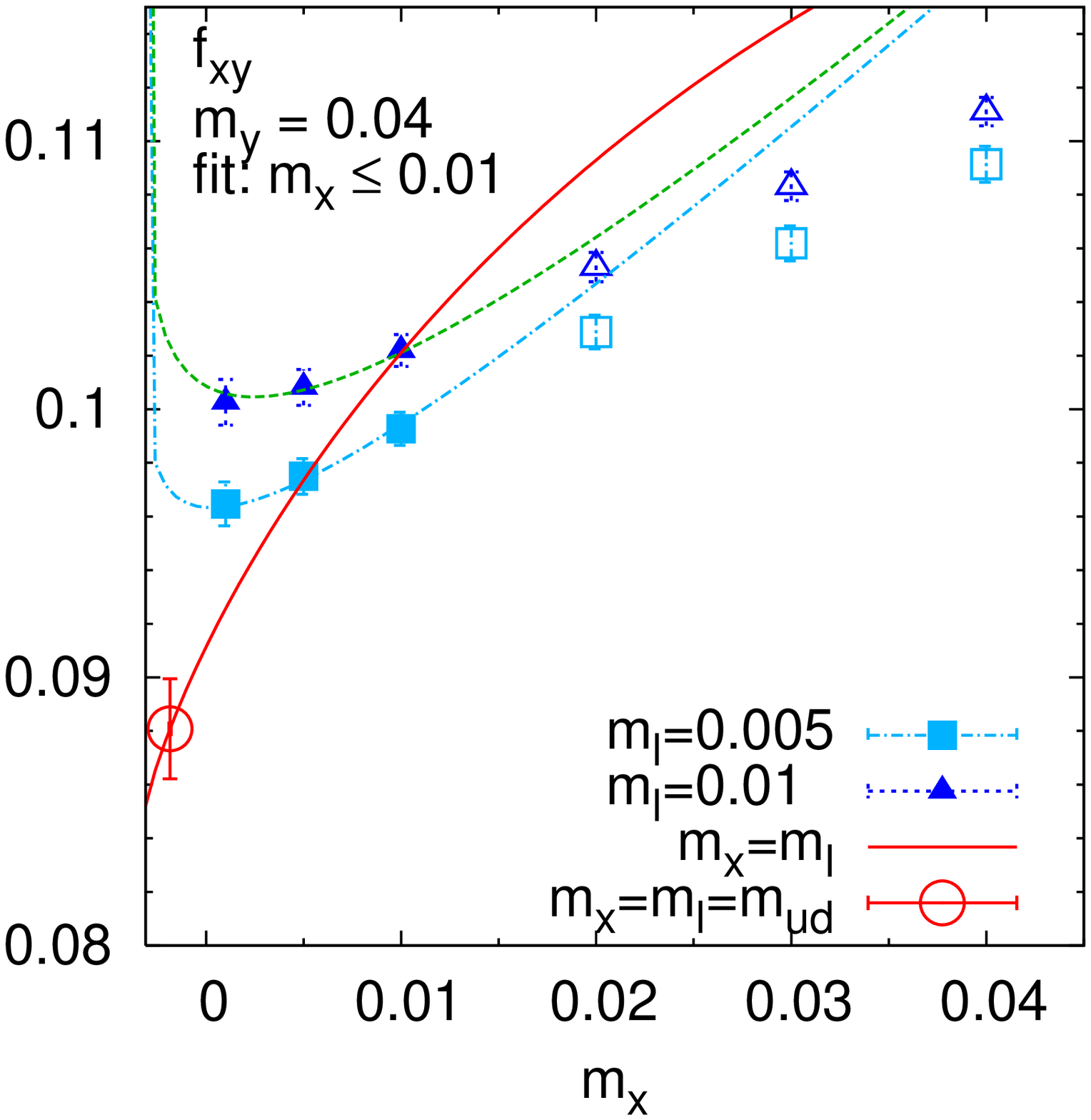}
\hfill
\includegraphics[angle=-90, width=.45\textwidth]{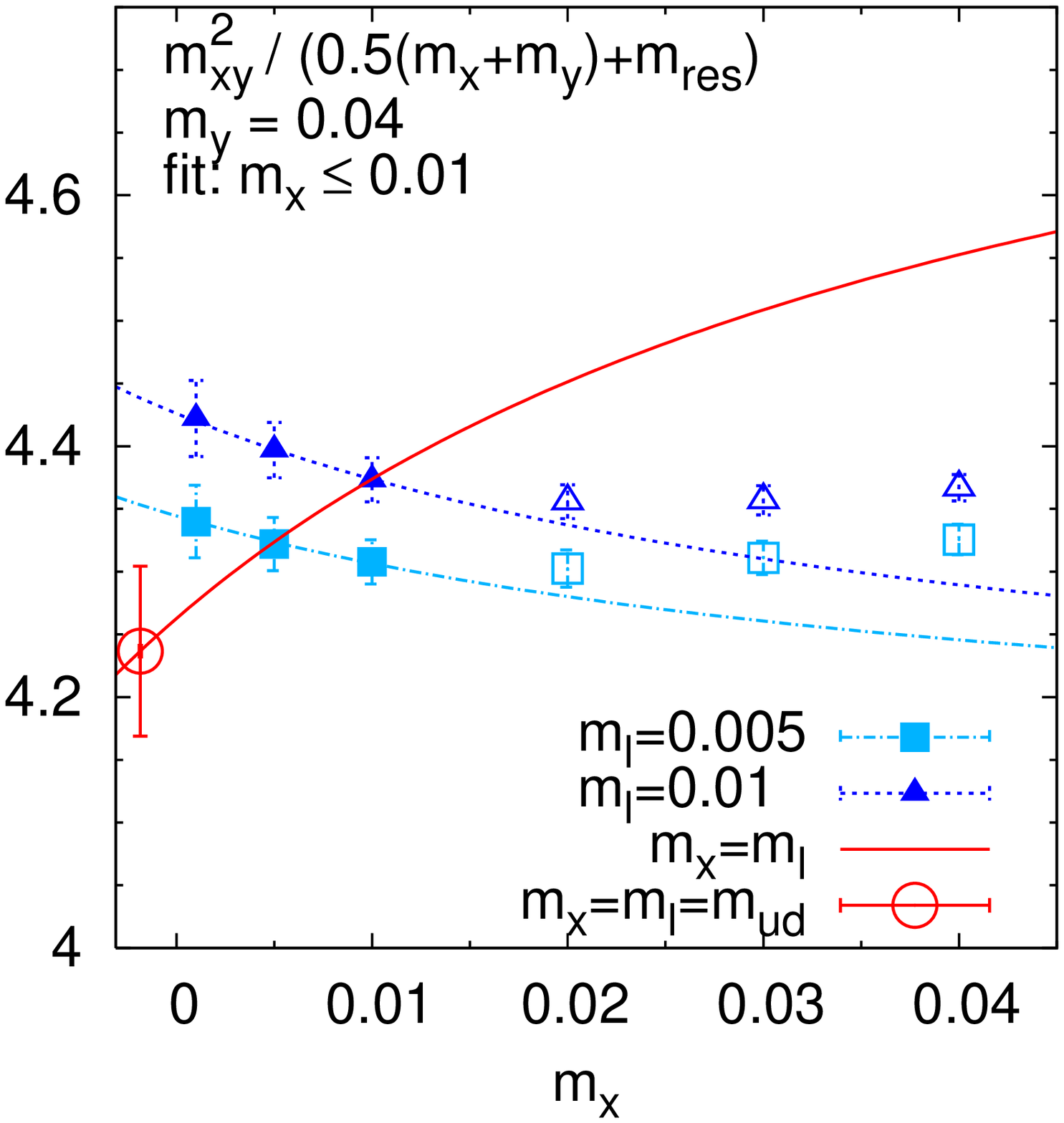}
\hfill
\end{center}
\caption{\label{fig:KSU2_ms0.04}$\SU(2)\times\SU(2)$ fits for the kaon sector. 
\textit{Left panel:} Kaon decay constant, \textit{right panel:} Kaon mass squared for $m_s=0.04$. 
Points with \textit{filled symbols} were included in the fit, while those with \textit{open symbols} were excluded.}
\end{figure}

\begin{figure}
\begin{center}
\hfill
\includegraphics[angle=-90, width=.45\textwidth]{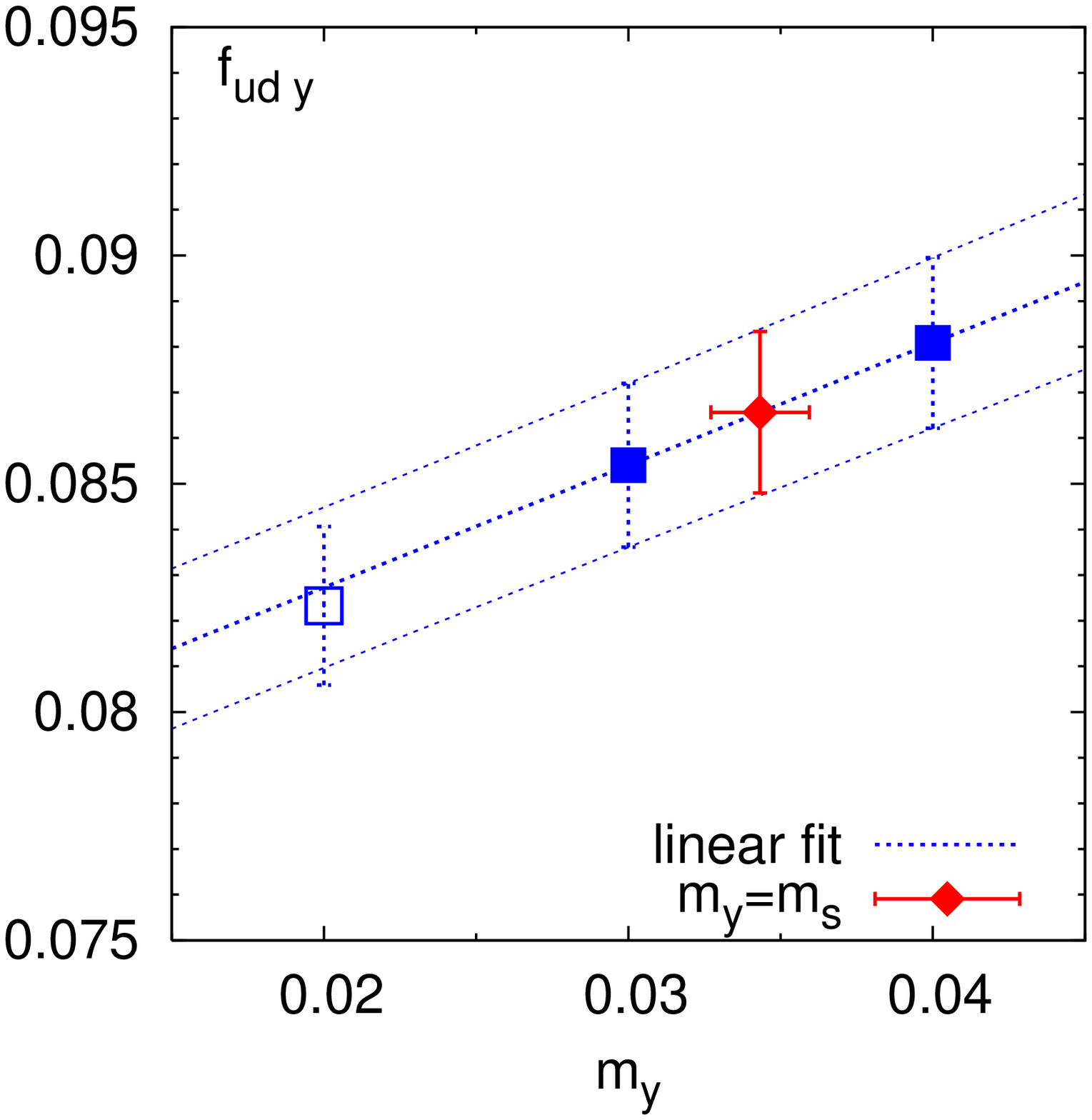}
\hfill
\includegraphics[angle=-90, width=.45\textwidth]{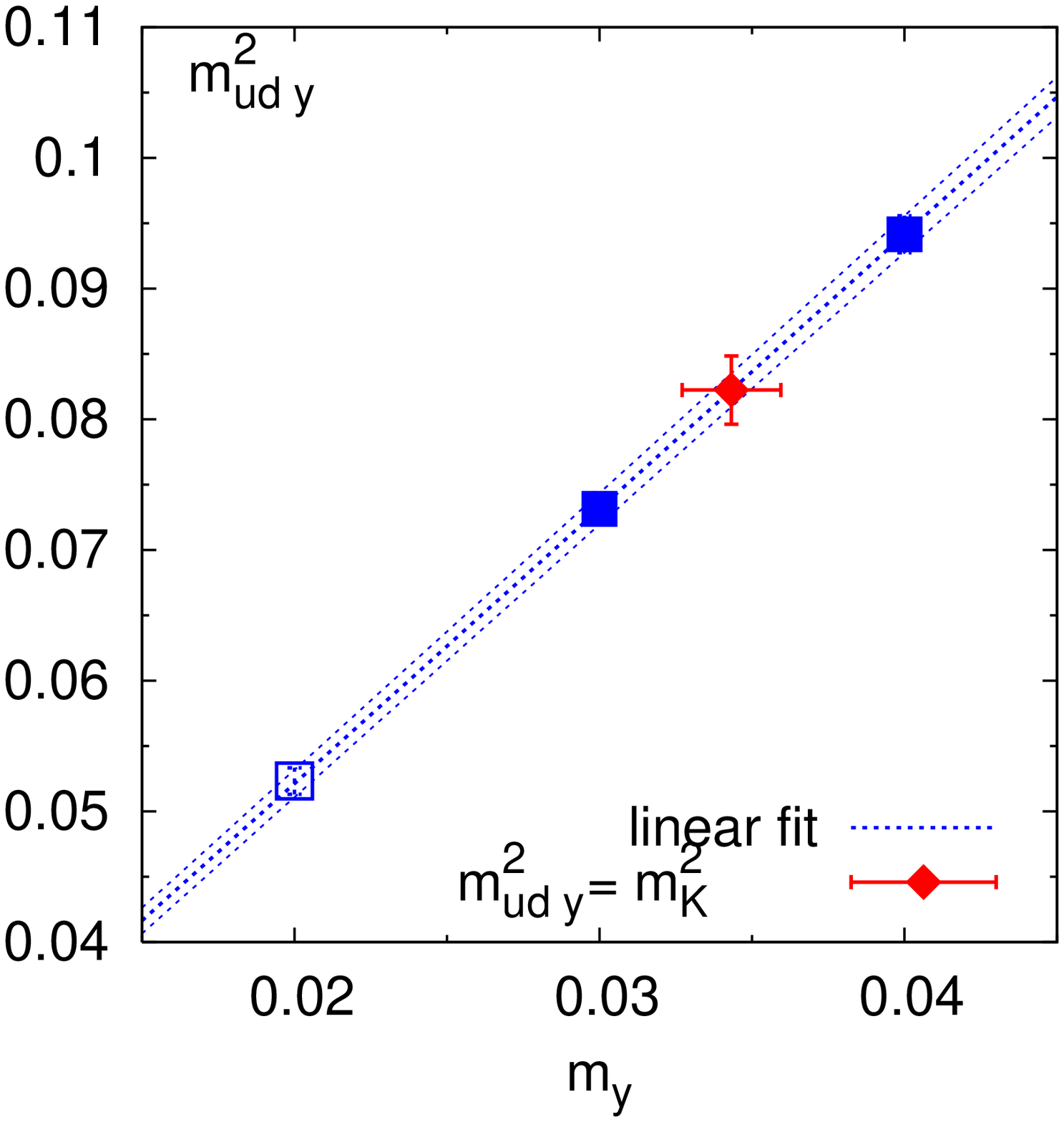}
\hfill
\end{center}
\caption{\label{fig:KSU2_interp} Interpolation to the physical strange quark mass (\textit{red diamond})
for $m_K^2$ and $f_K$. \textit{Open symbols} are not included in the interpolations. 
\textit{Left panel:} Kaon decay constant, \textit{right panel:} Squared kaon mass.
}
\end{figure}



\begin{figure}[ht]
\begin{center}
\includegraphics[angle=-90,width=.48\textwidth]{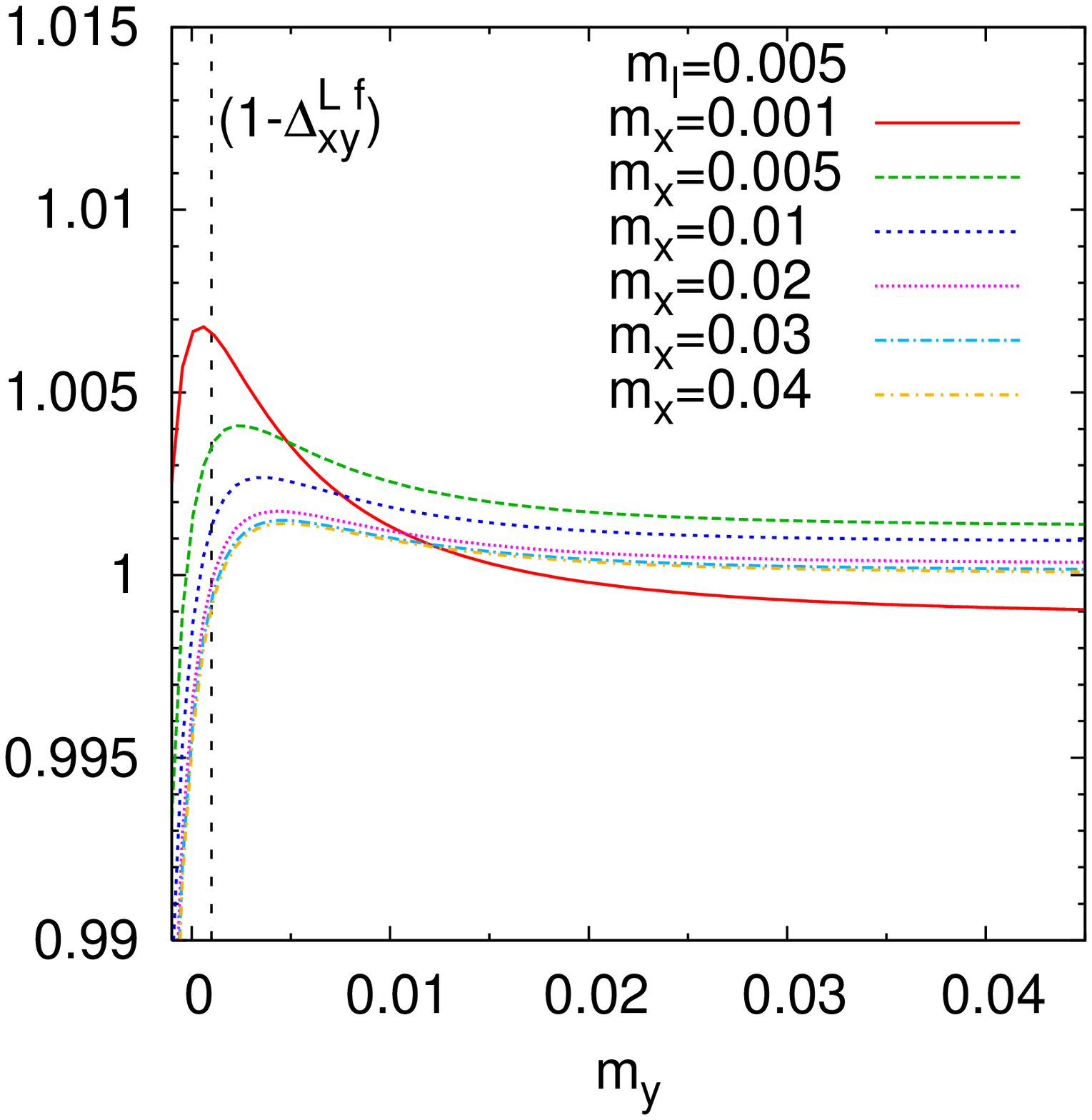}%
\includegraphics[angle=-90,width=.48\textwidth]{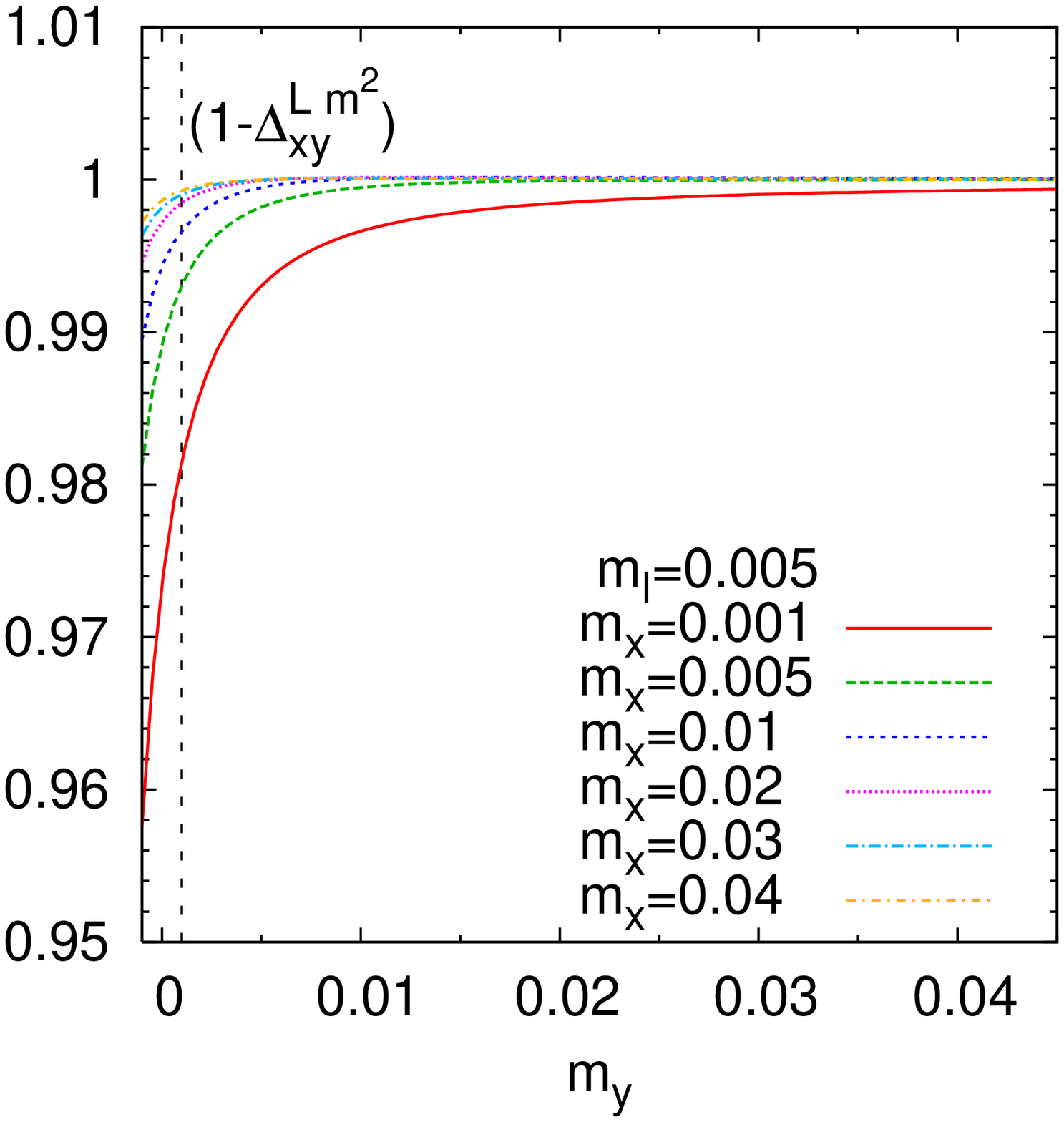}\\
\includegraphics[angle=-90,width=.48\textwidth]{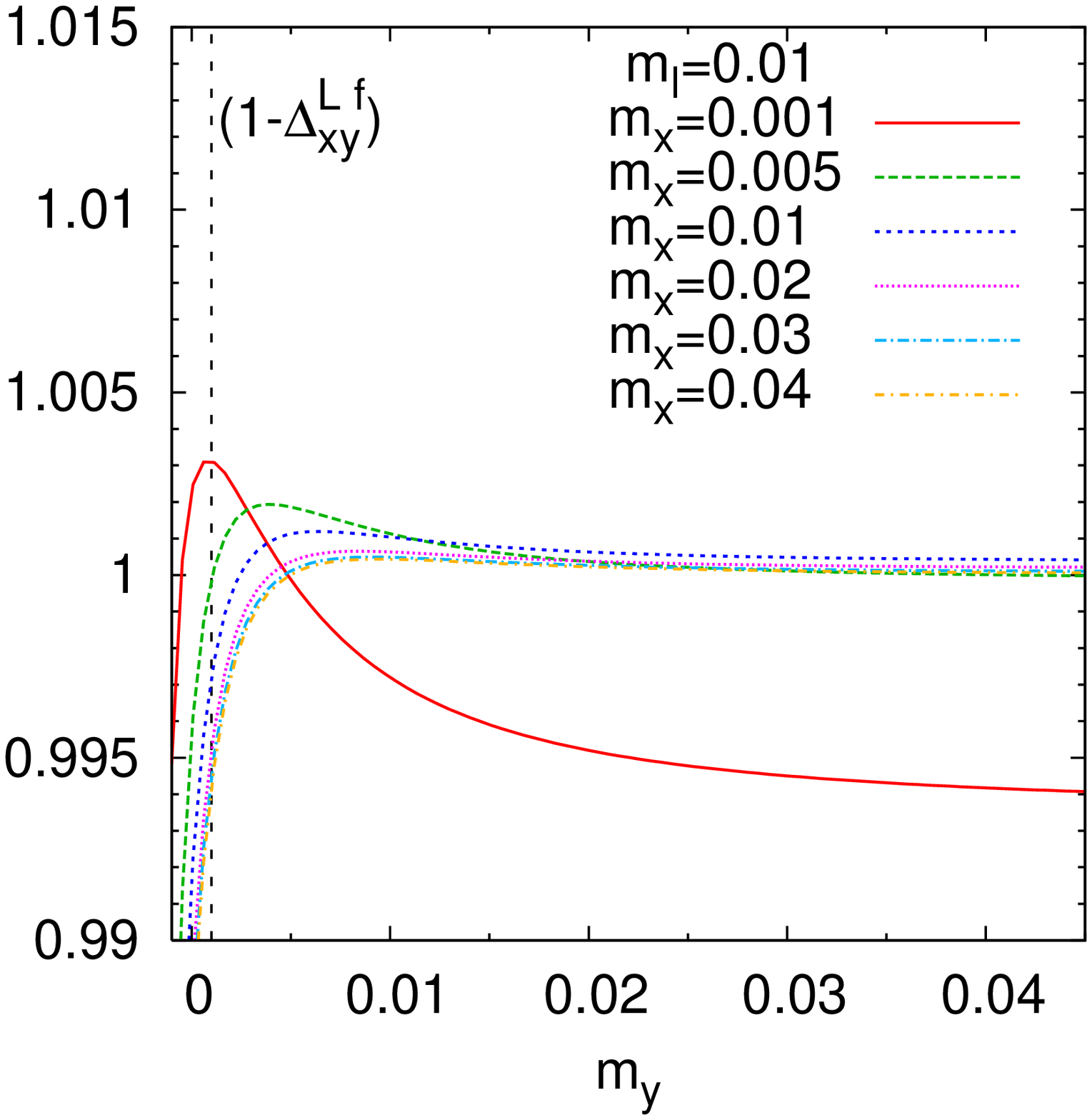}%
\includegraphics[angle=-90,width=.48\textwidth]{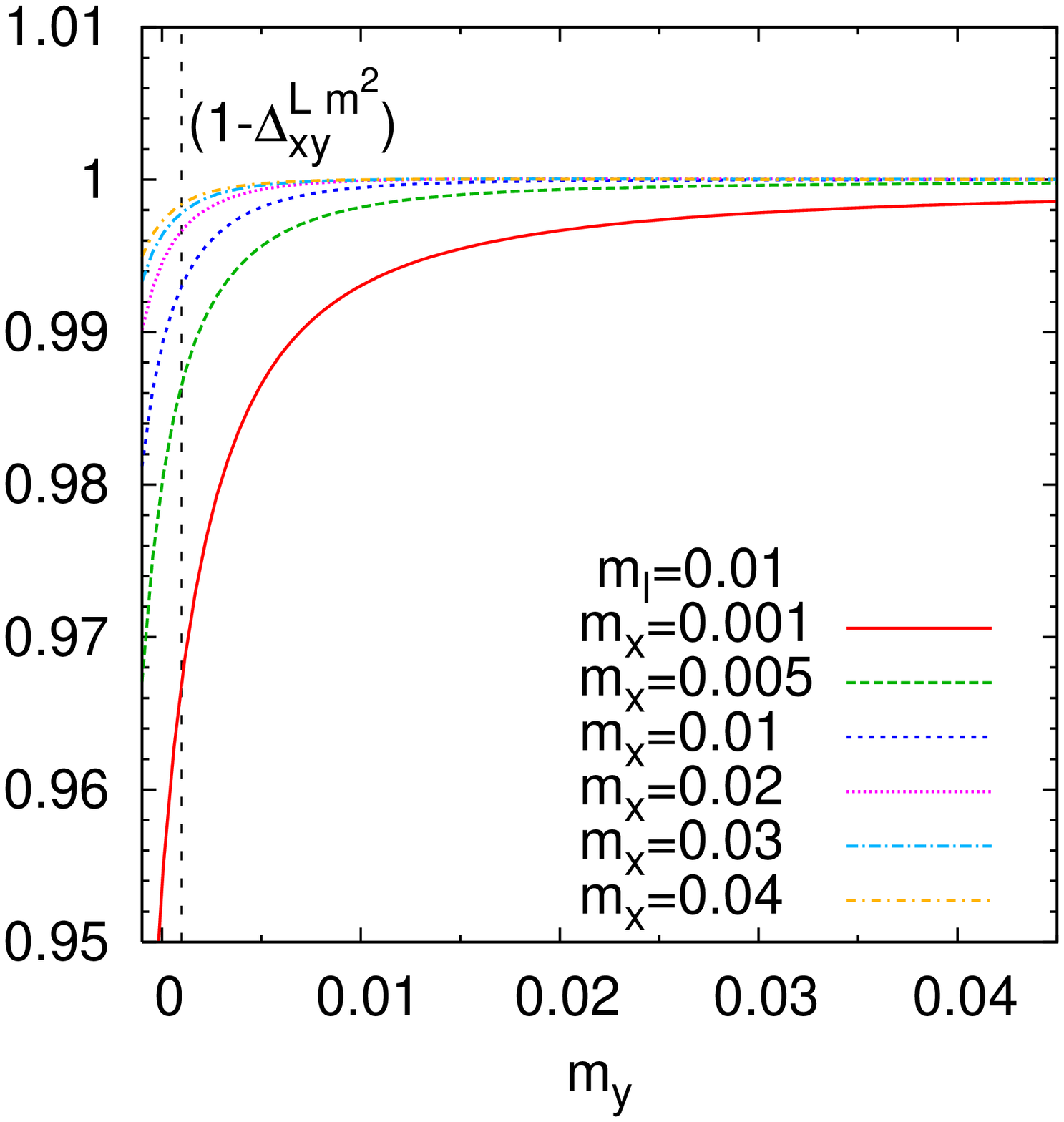}%
\end{center}
\caption{Finite volume correction factor obtained from the $\SU(2)$ fits for the decay constants \textit{(left panels)} and squared meson masses \textit{(right panels)} at sea quark masses $m_l=0.005$ (\textit{(top panels)} and 0.01 \textit{(bottom panels)}. The vertical dashed line denotes the lightest (valence) mass used in this analysis ($m_y=0.001$).}
\label{fig:Su2:fv}
\end{figure}


\begin{figure}[ht]
\begin{center}
\includegraphics[angle=-90, width=.48\textwidth]{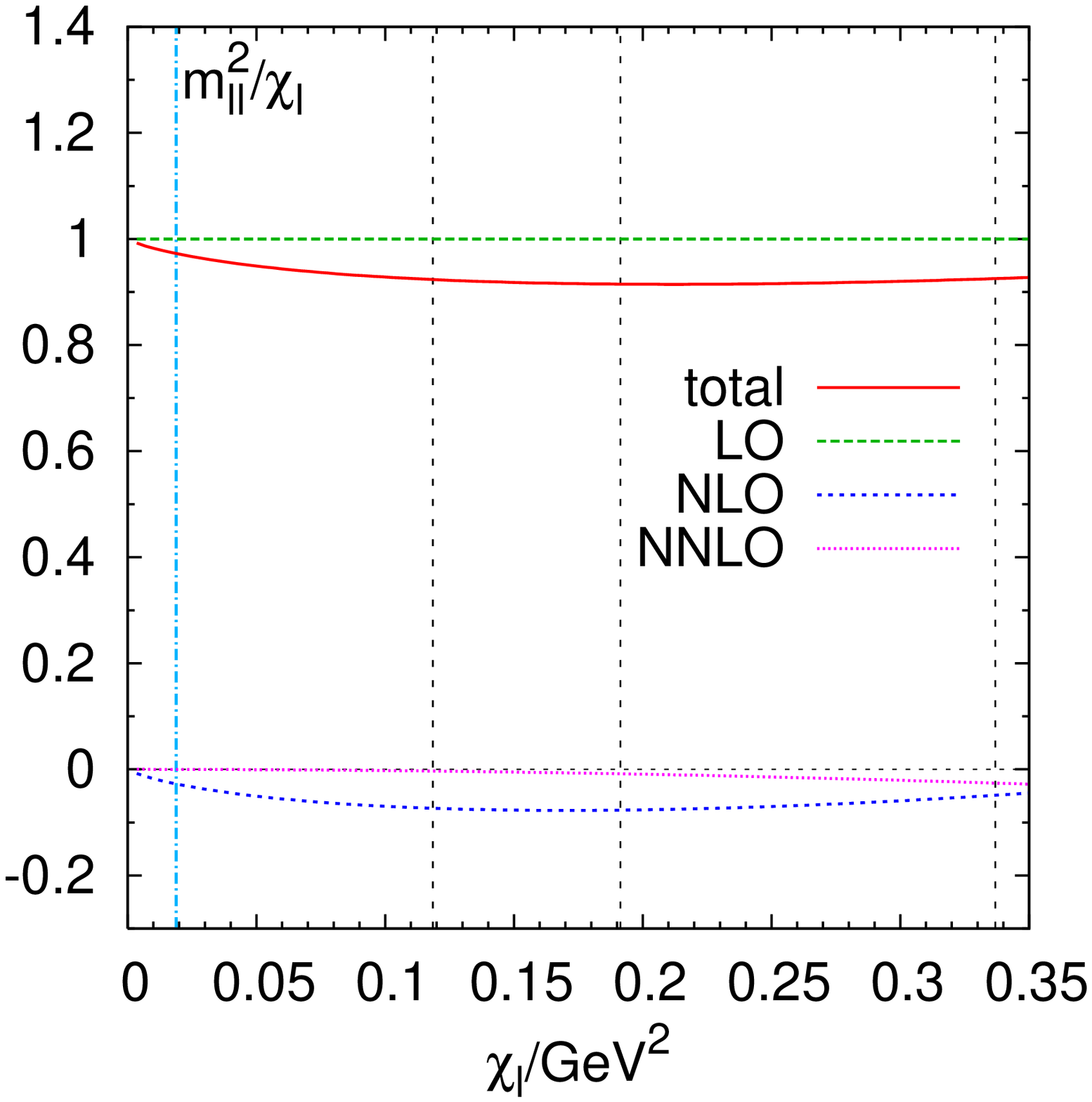}%
\includegraphics[angle=-90, width=.48\textwidth]{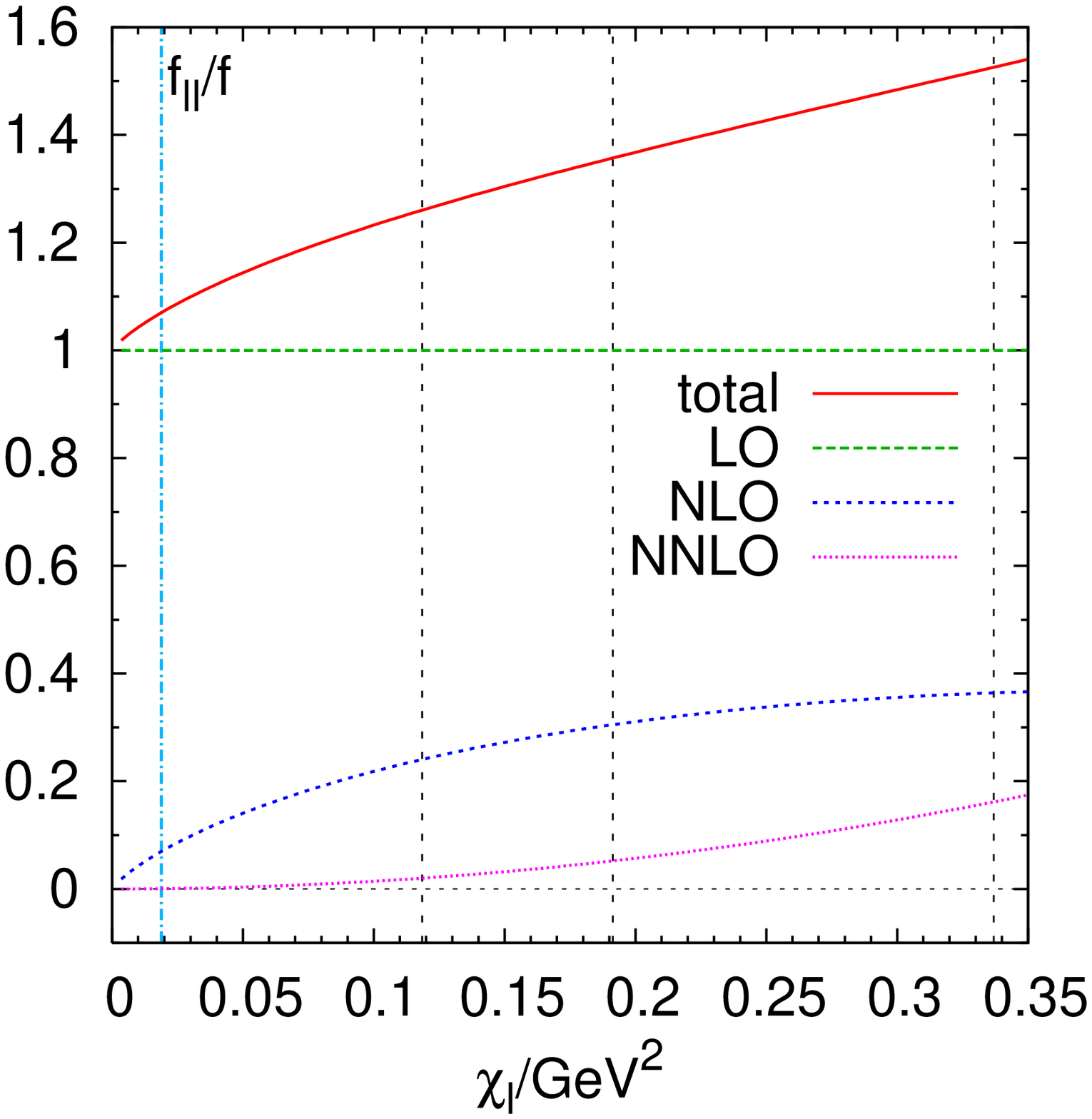} \\
\includegraphics[angle=-90, width=.48\textwidth]{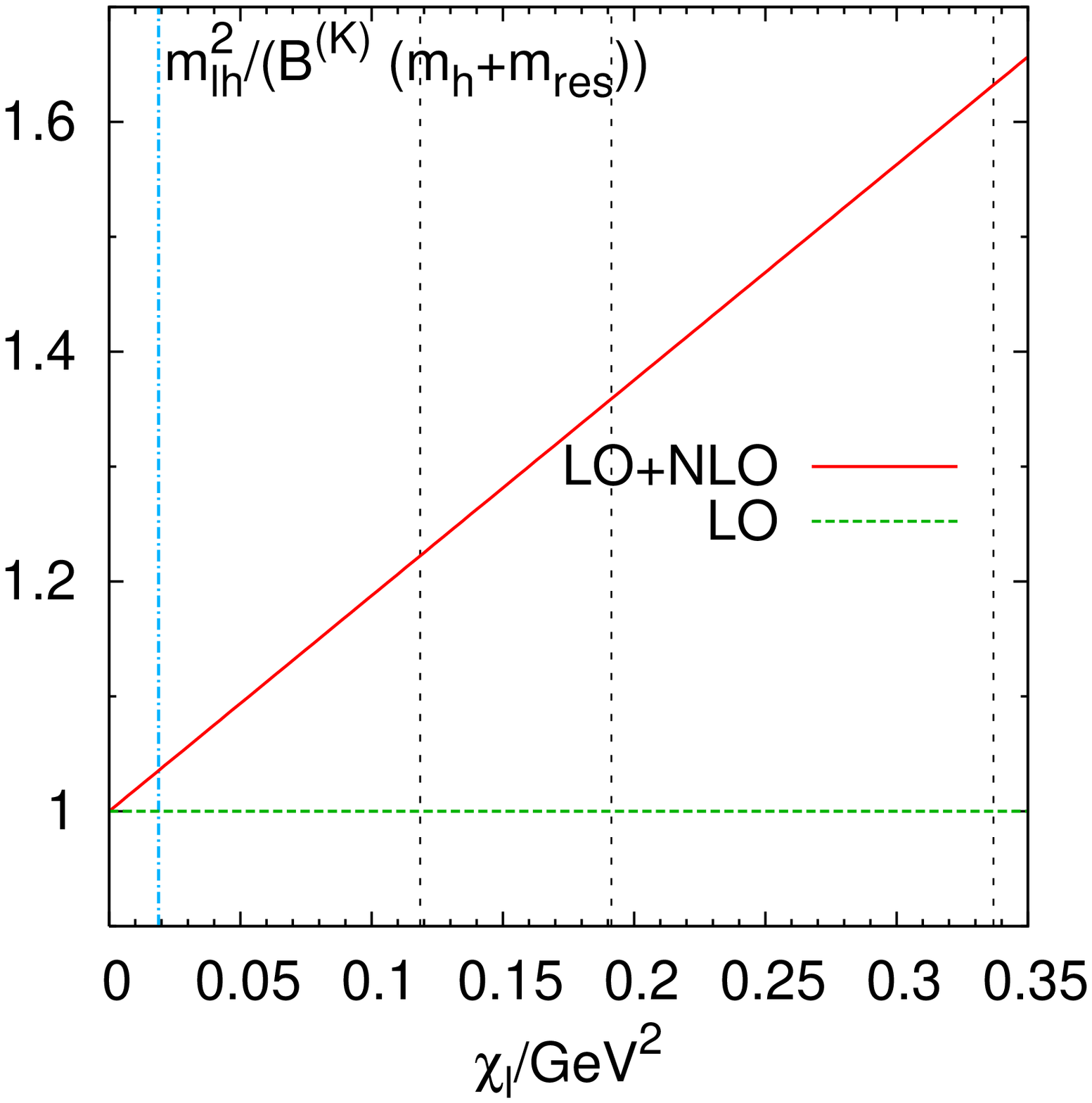}%
\includegraphics[angle=-90, width=.48\textwidth]{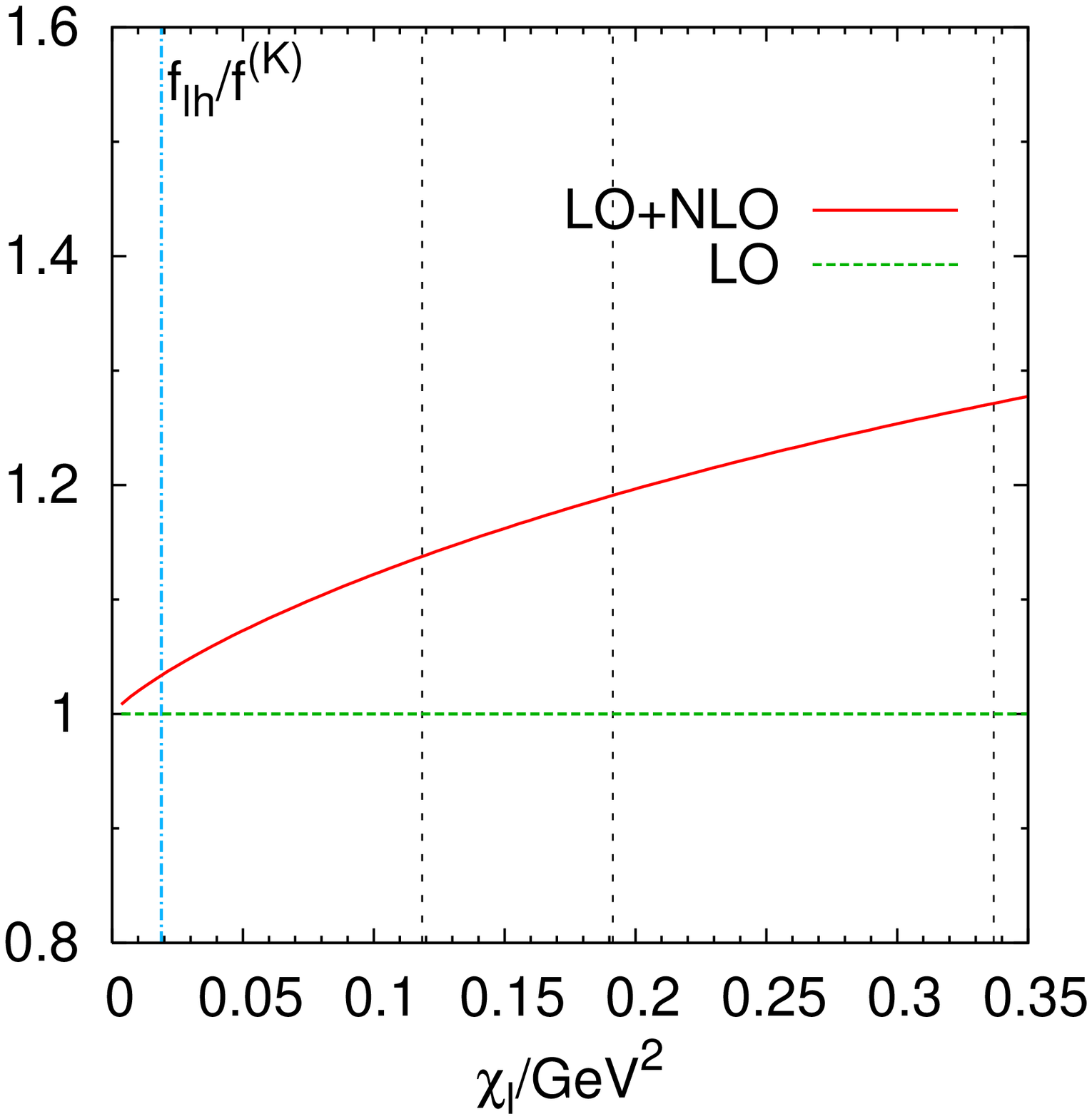}%
\end{center}
\caption{Comparison of the size of the LO, NLO, and NNLO contributions
in the $\SU(2)\times\SU(2)$ fits for unitary degenerate meson masses
squared $m_{ll}^2$ \textit{(upper left panel)} and decay constants
$f_{ll}$ \textit{(upper right panel)} as a function of the the light
sea quark mass parameter $\chi_l$ (for details see
Sec.~\ref{sec:su2:sysErr:chPT}). The LO-contribution is normalized
to one. \textit{Vertical dashed lines} indicate quark masses of
$m_l= m_{ud},\, 0.005$, 0.01, and 0.02 (from left to right).  The
lower panels give a comparison of the size of the LO and NLO
contributions in kaon ChPT for the kaon mass squared $m_{lh}^2$
\textit{(lower left panel)} and kaon decay constant $f_{lh}$
\textit{(lower right panel)} as a function of the the light sea quark
mass.}
\label{fit:Su2:nnloContr}
\end{figure}

%% file: fig/Su3.fig

\begin{figure}
\begin{center}
\begin{minipage}{.5\textwidth}
\begin{center}
\includegraphics[angle=-90, width=.9\textwidth]{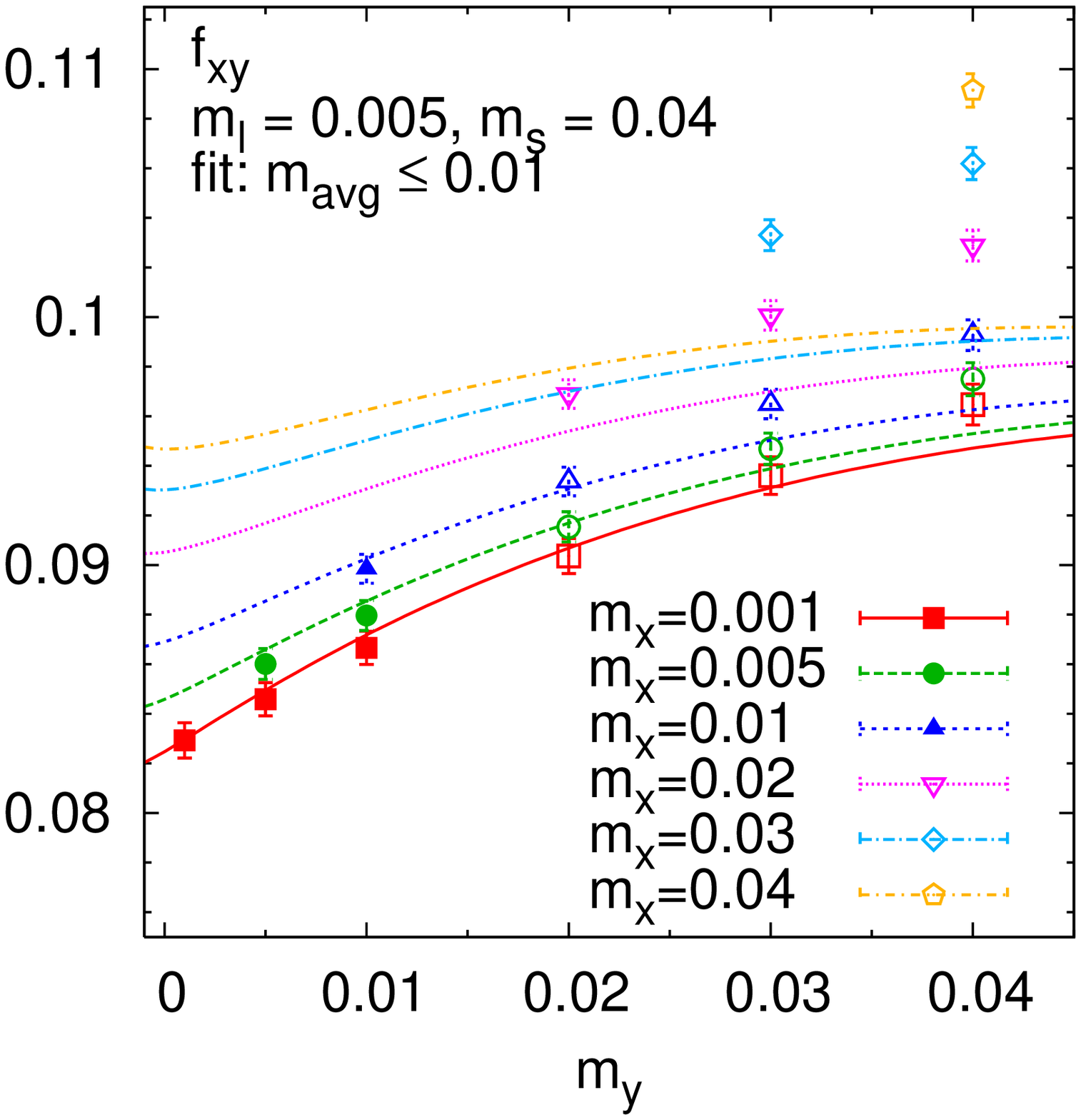}\\
\includegraphics[angle=-90, width=.9\textwidth]{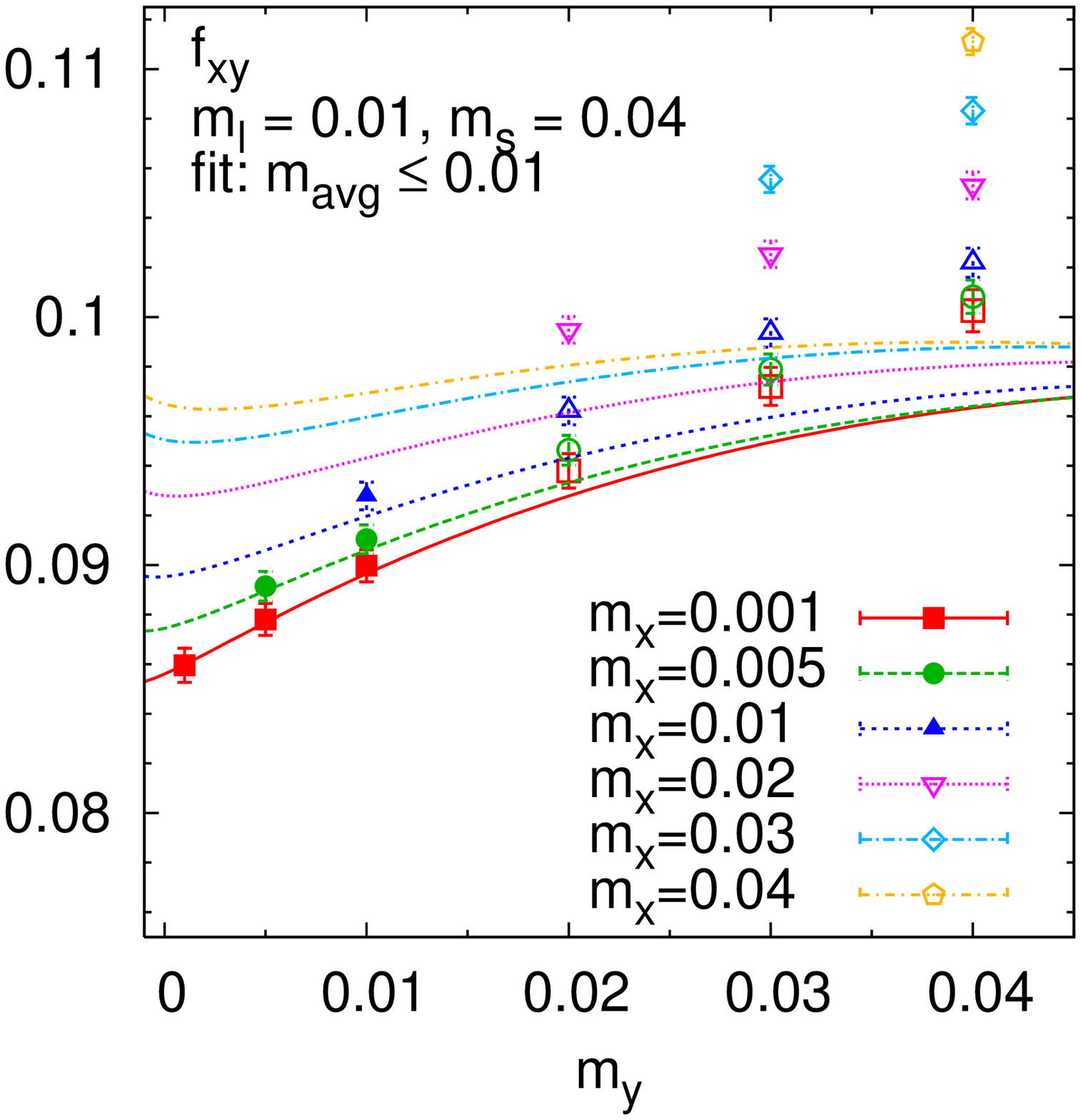}
\end{center}
\end{minipage}%
\begin{minipage}{.5\textwidth}
\begin{center}
\includegraphics[angle=-90, width=.87\textwidth]{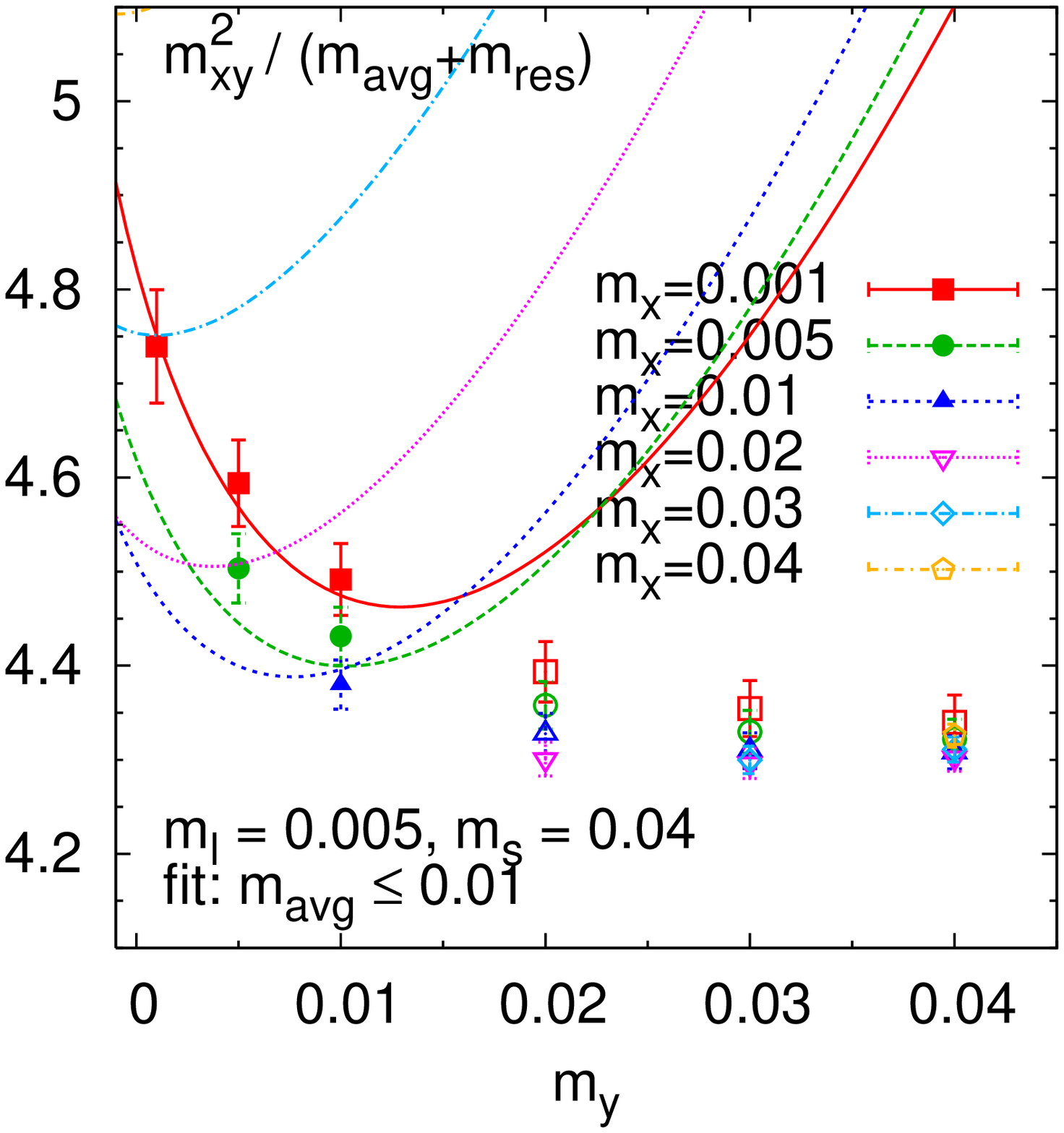}\\
\includegraphics[angle=-90, width=.87\textwidth]{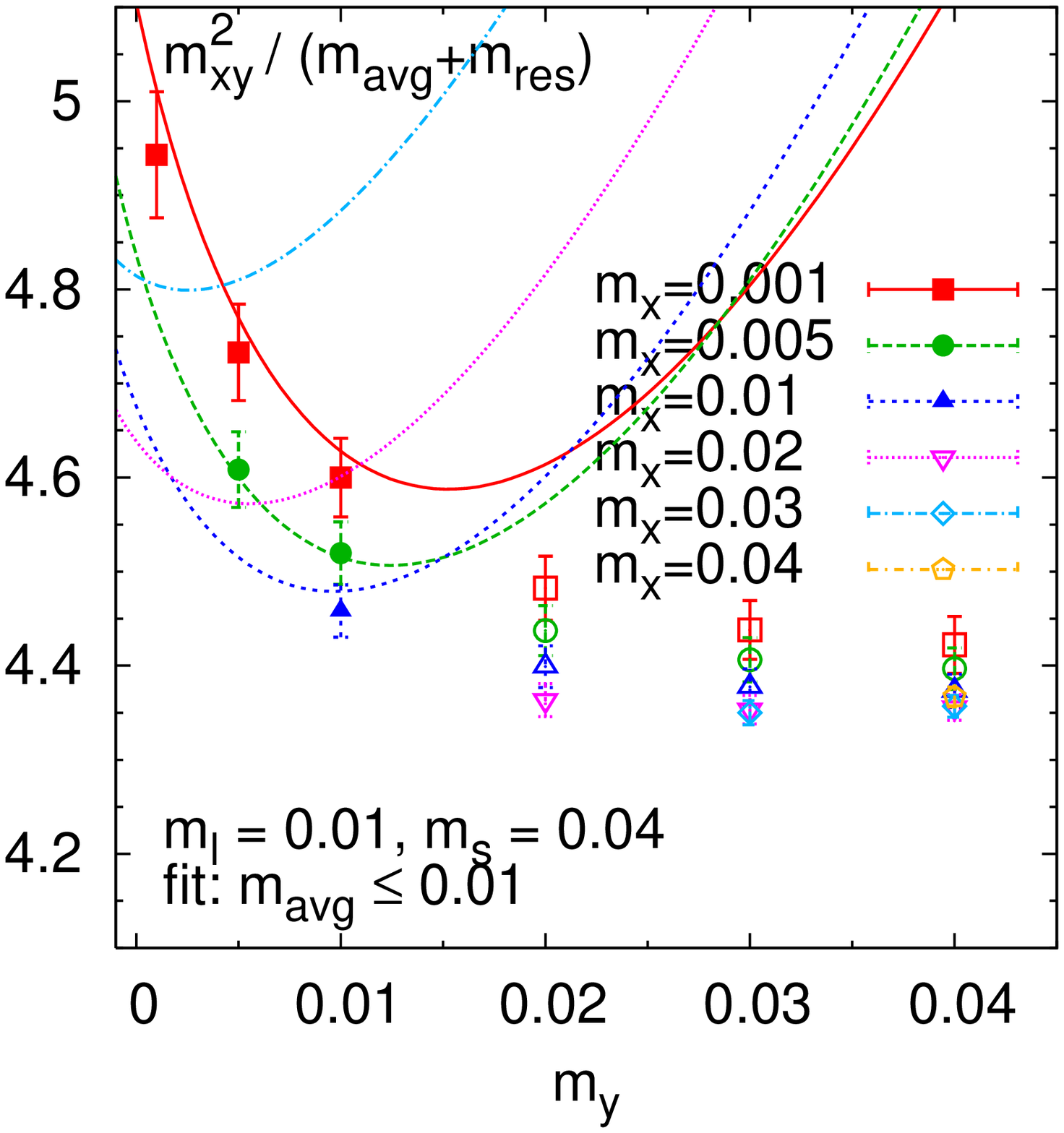}
\end{center}
\end{minipage}
\end{center}
\caption{\label{fig:SU3fits}Combined $\SU(3)\times \SU(3)$ fits for the meson decay constants \textit{(left panels)} and masses \textit{(right panels)} at two different values for the light sea quark mass, with valence mass cut $m_{\rm avg}\leq0.01$. Points marked by \textit{filled symbols} were included in the fit, while those with \textit{open symbols} were excluded.}
\end{figure}

\begin{figure}
\begin{center}
\begin{minipage}{.5\textwidth}
\begin{center}
\includegraphics[angle=-90, width=.9\textwidth]{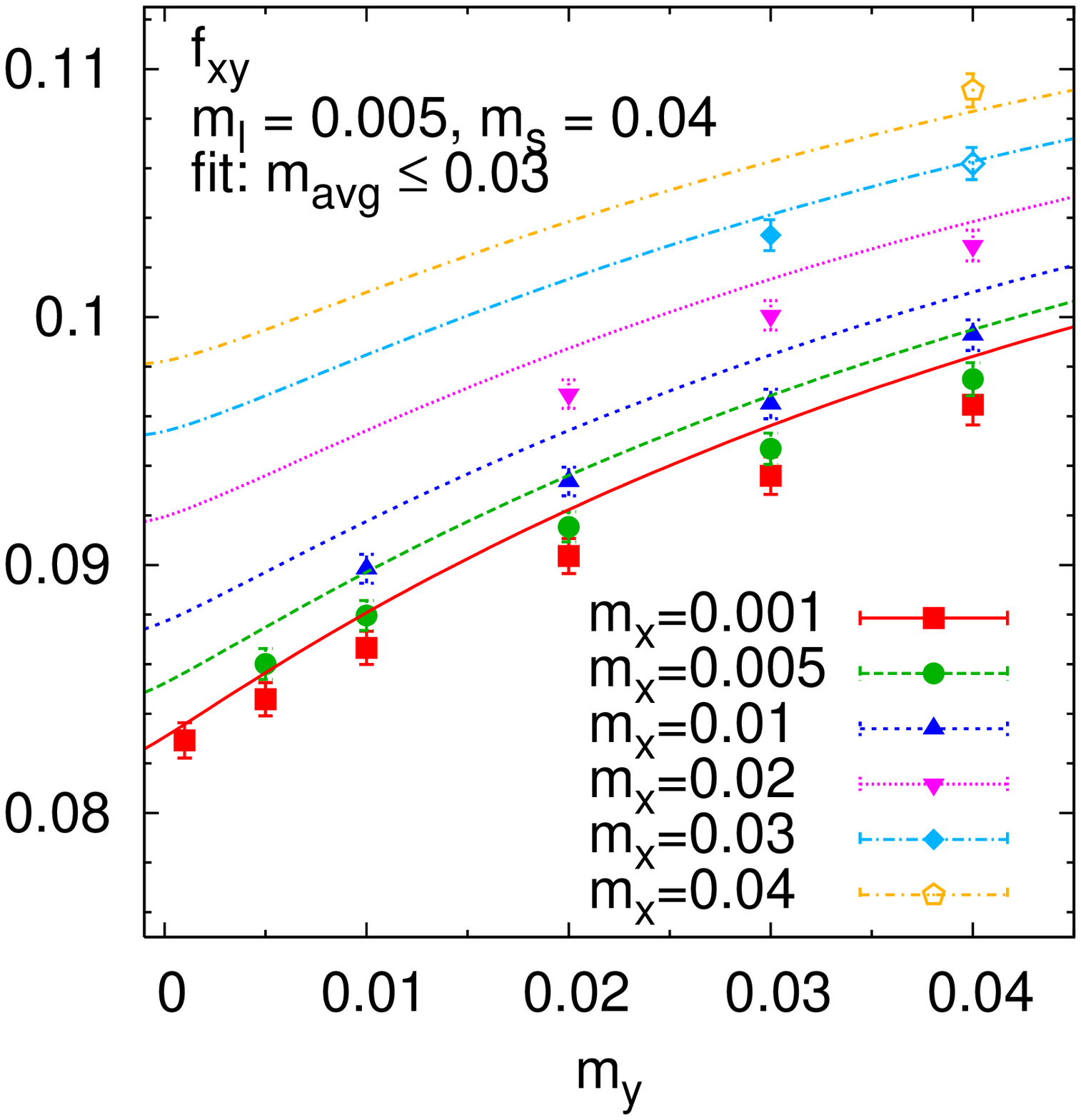}\\
\includegraphics[angle=-90, width=.9\textwidth]{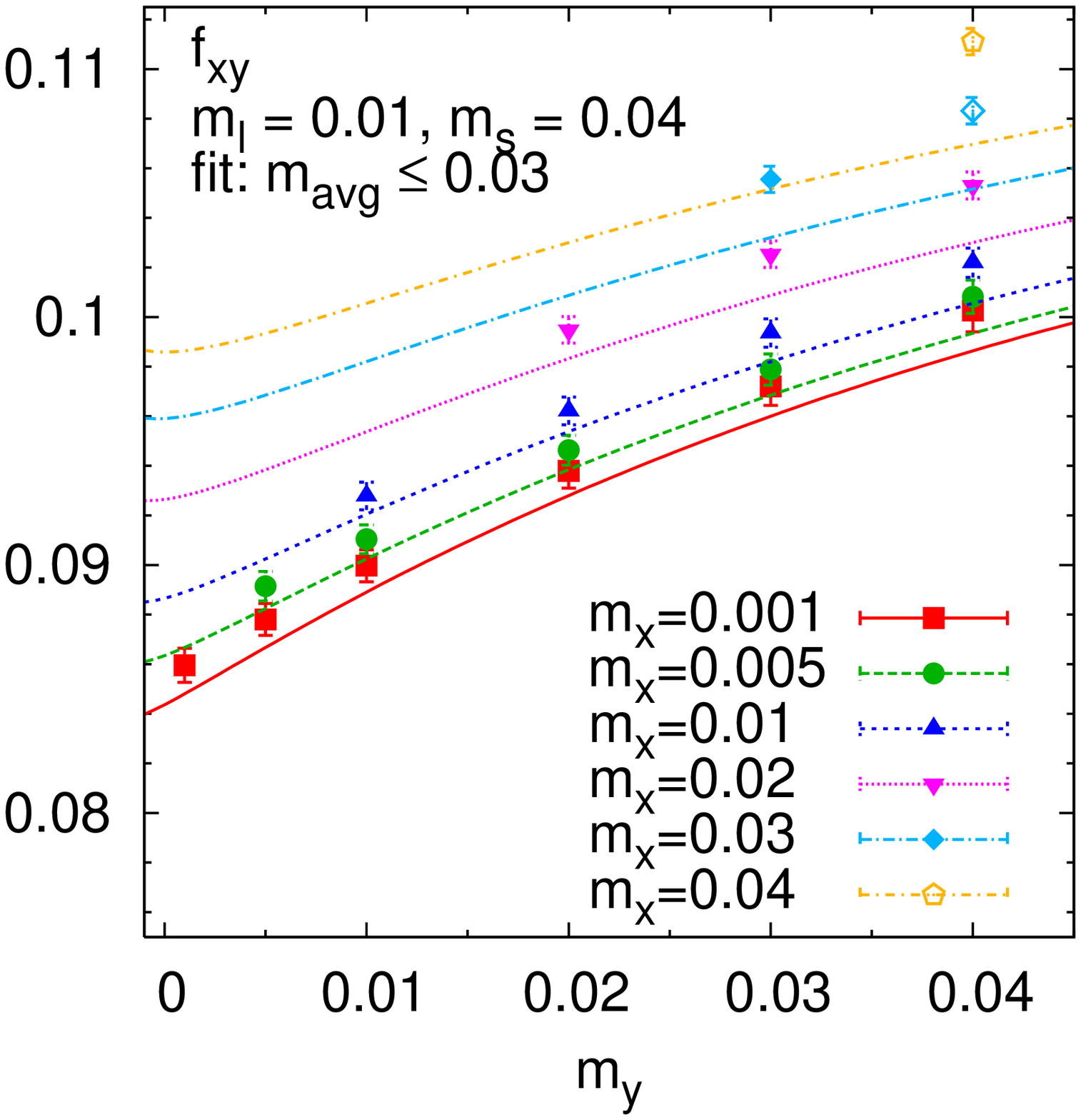}
\end{center}
\end{minipage}%
\begin{minipage}{.5\textwidth}
\begin{center}
\includegraphics[angle=-90, width=.87\textwidth]{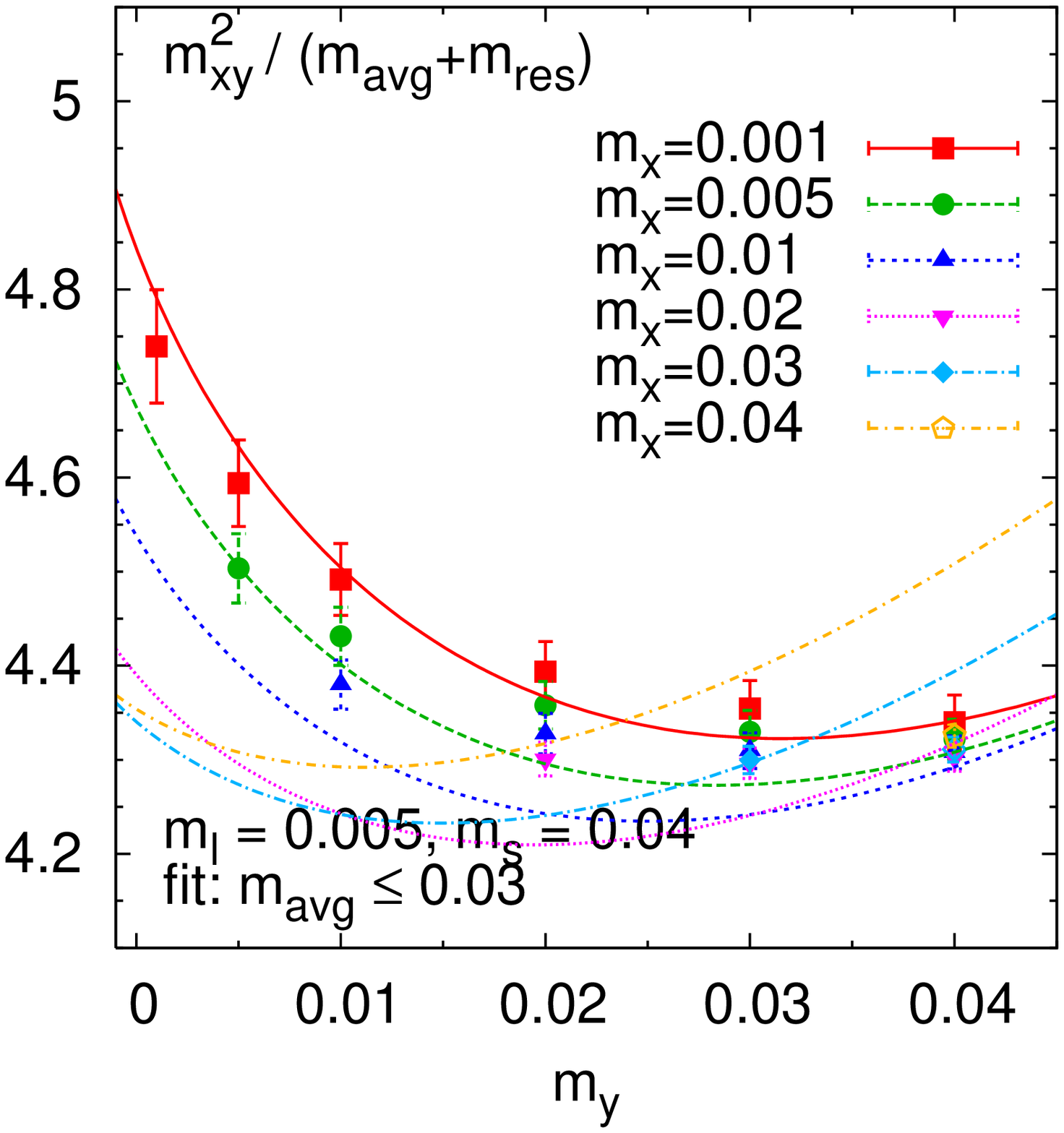}\\
\includegraphics[angle=-90, width=.87\textwidth]{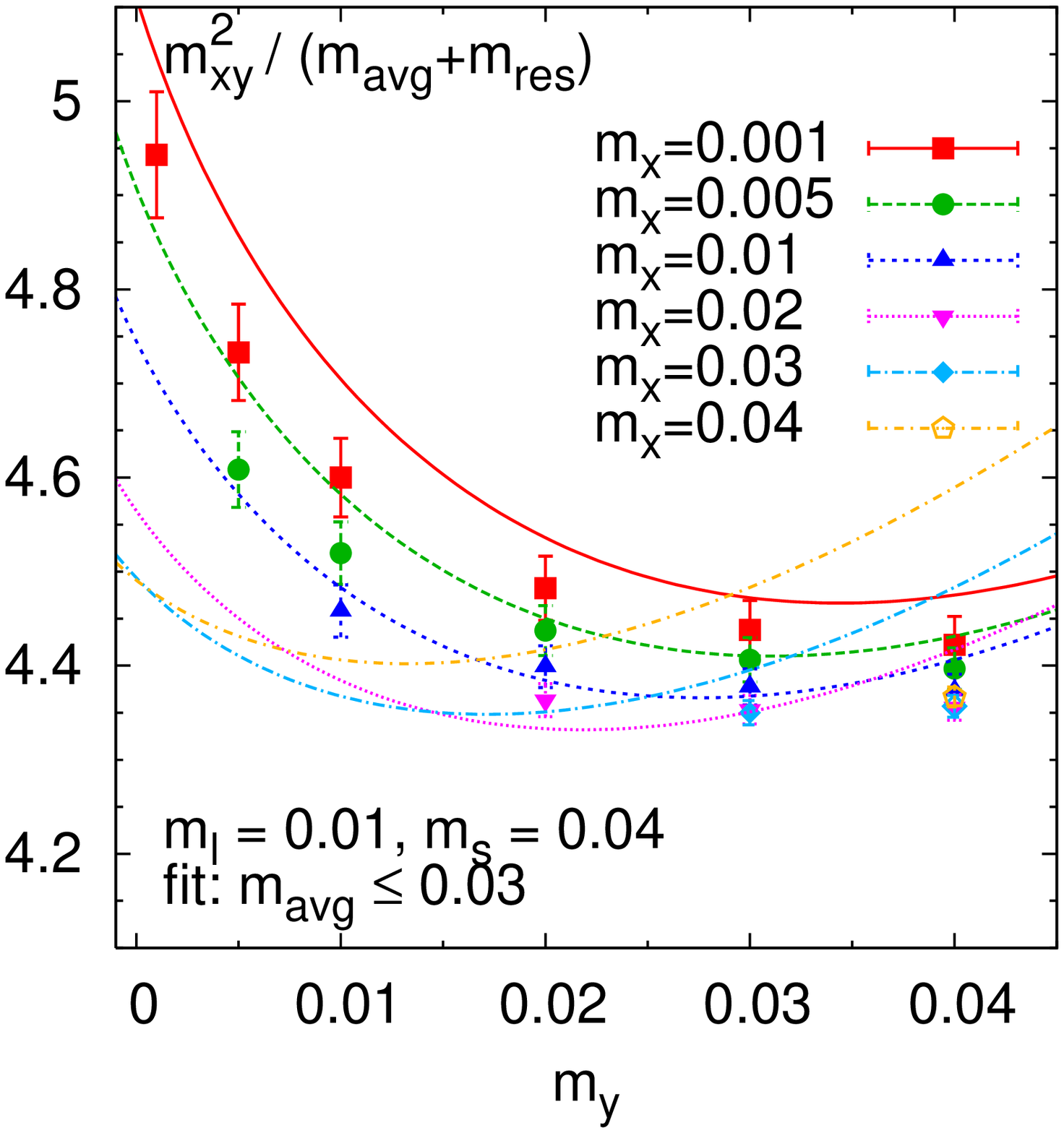}
\end{center}
\end{minipage}
\end{center}
\caption{\label{fig:SU3fits_bad}Combined $\SU(3)\times \SU(3)$ fits for the meson decay constants \textit{(left panels)} and masses \textit{(right panels)} at two different values for the light sea quark mass, with valence mass cut $m_{\rm avg}\leq0.03$. Points marked by \textit{filled symbols} were included in the fit, while those with \textit{open symbols} were excluded.}
\end{figure}

\clearpage
\begin{figure}[ht]
\begin{center}
\includegraphics[angle=-90, width=.48\textwidth]{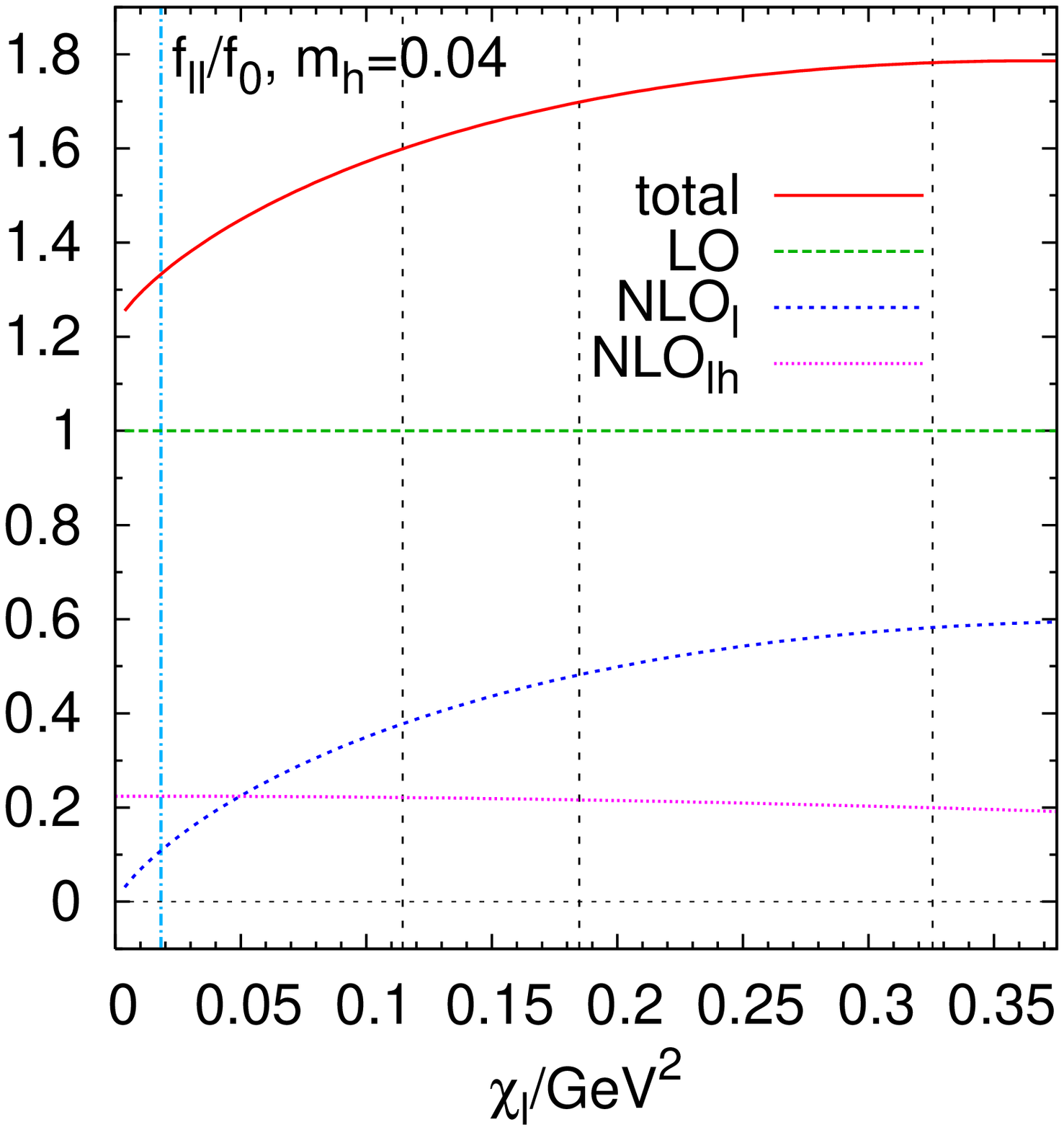}%
\includegraphics[angle=-90, width=.48\textwidth]{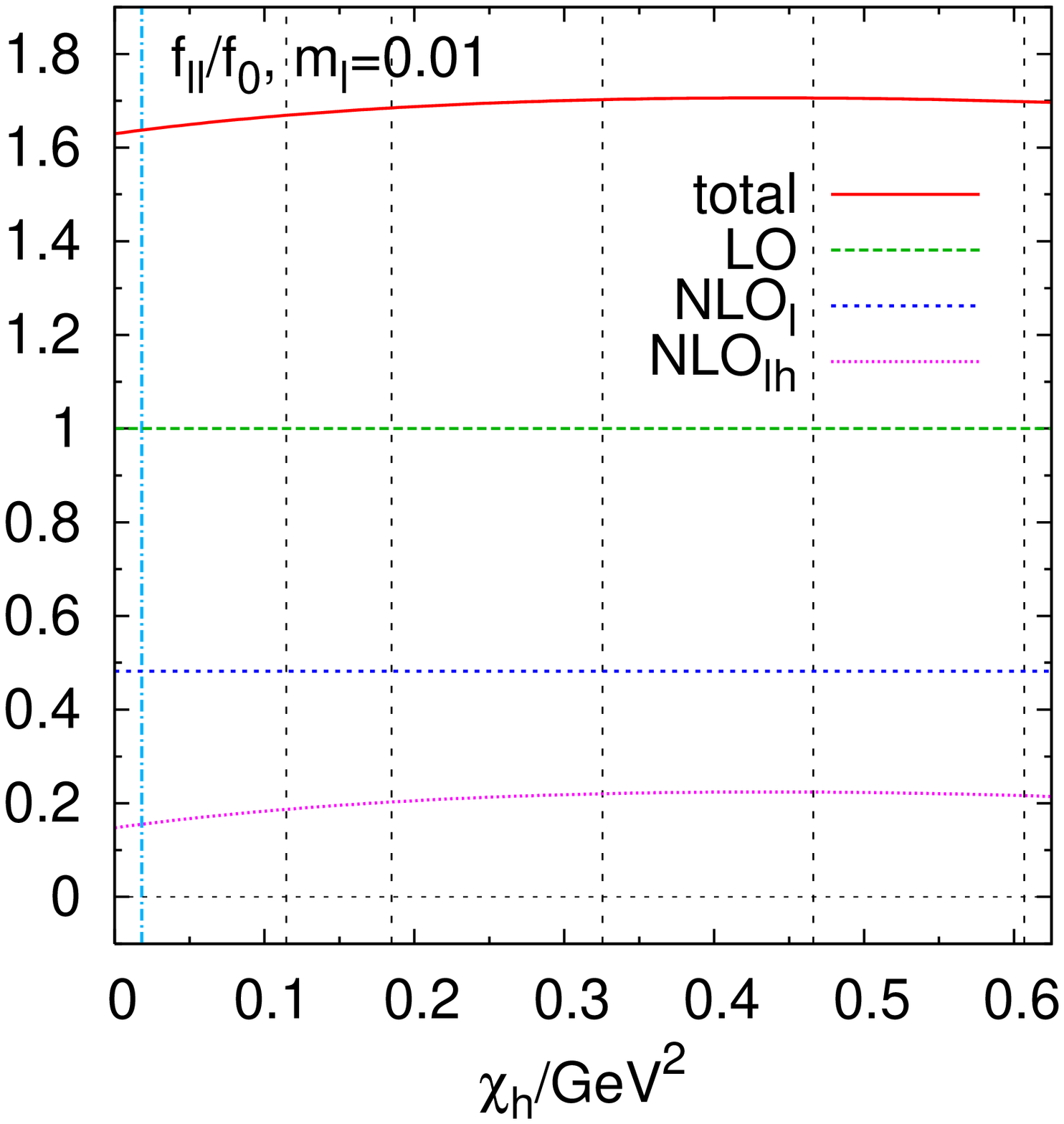}\\
\includegraphics[angle=-90, width=.48\textwidth]{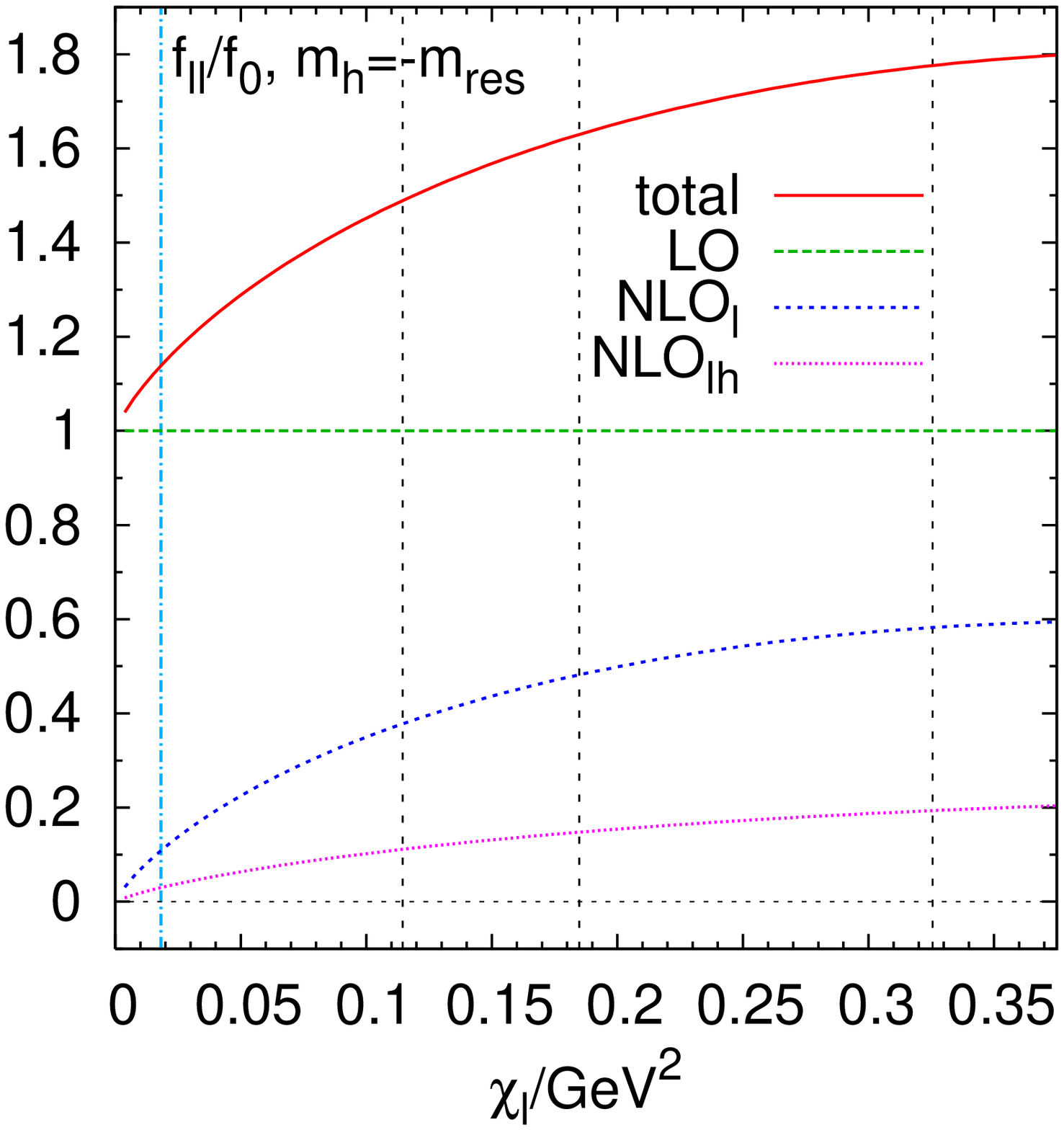}%
\includegraphics[angle=-90, width=.48\textwidth]{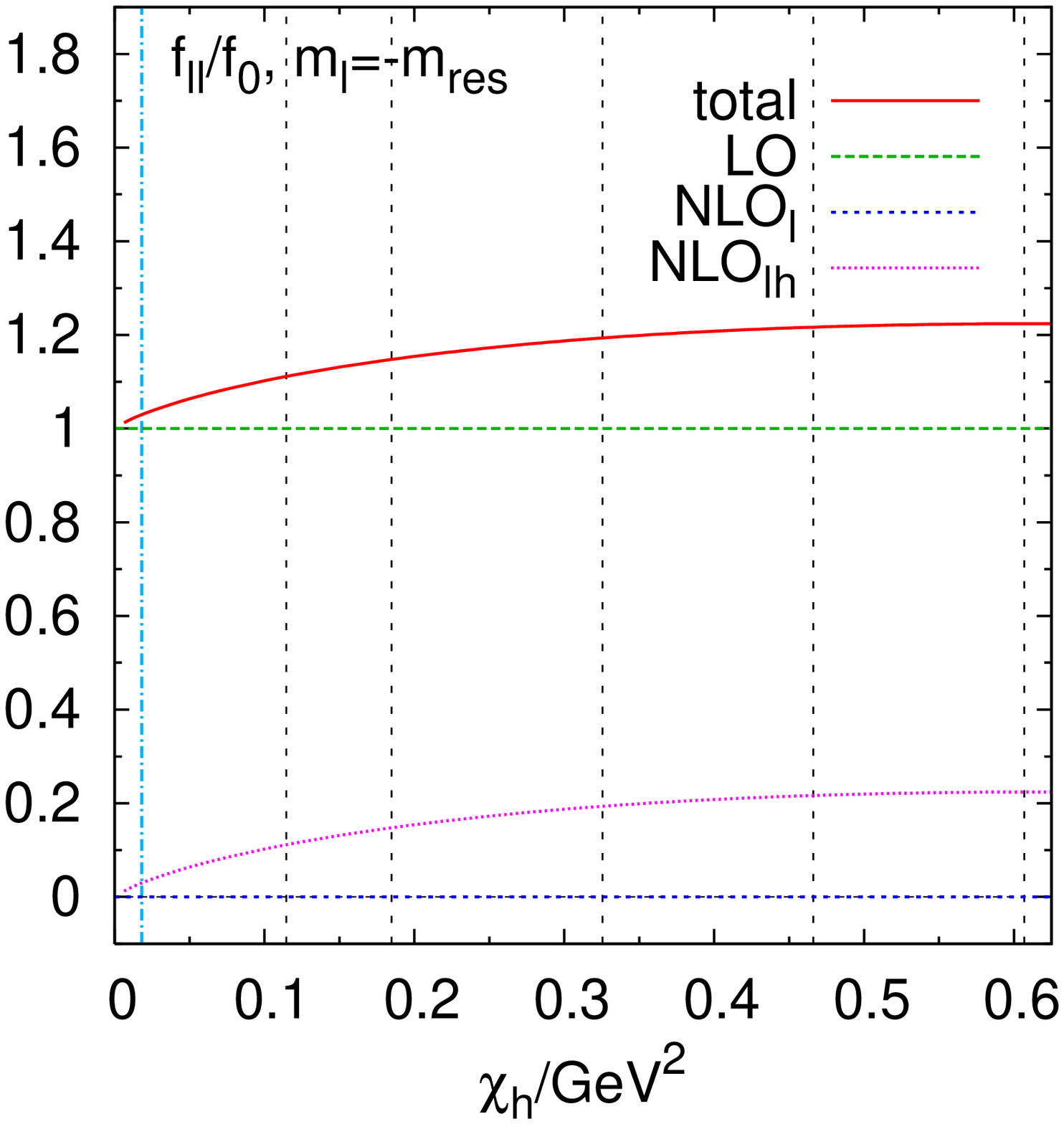}
\end{center}
\caption{Leading order (LO) and Next-to-Leading order (NLO) contributions in the unitary meson decay constant fit in $\SU(3)$ PQChPT as function of the light quark mass parameter \textit{(left panels)} and the heavy quark mass parameter \textit{(right panels)}. In the plots in the \textit{top panels} the other quark mass parameter was set to our highest value used in the fits ($m_l=0.01$ or $m_h=0.04$), while in those in the \textit{bottom panels} the remaining quark mass parameter was set to zero ($m_l$ or $m_h=-m_{\rm res}$). ${\rm NLO}_l$ denotes the NLO contribution proportional to $\log\chi_l$ and ${\rm NLO}_{lh}$ the NLO contribution proportional to $\log(\chi_l+\chi_h)/2$. The LO contribution is always normalized to one. \textit{Vertical dashed lines} indicate quark masses of $m_l=m_{ud},\, 0.005$, 0.01, 0.02, 0.03, and 0.04 (from left to right, left panels only show quark masses up to 0.02). }
\label{fig:su3:NLO}

\end{figure}

\clearpage
\begin{figure}[ht]
\begin{center}
\includegraphics[angle=-90, width=.9\textwidth]{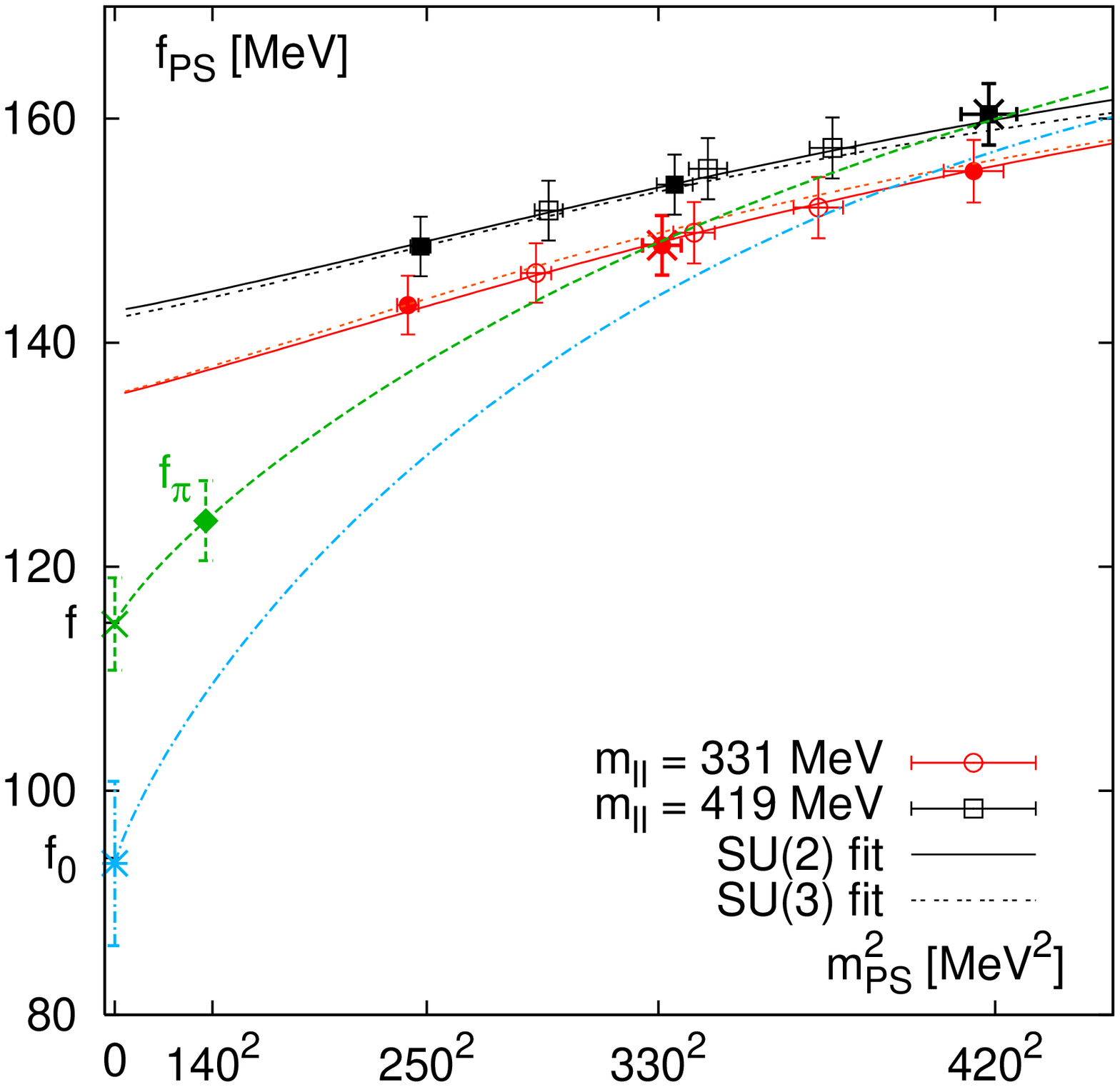}
\end{center}
\caption{ A plot of our measured values for the decay constant,
converted to physical units, versus the degenerate valence pseudoscalar
mass squared.  The data for degenerate quarks is denoted by filled
symbols and for non-degenerate quarks, open symbols are used.  The
graph shows that we see small effects for non-degenerate quarks.
The results of SU(2) and SU(3) partially quenched ChPT fits to our
data are shown, and both fits agree well with the data.  The unitary
SU(2) chiral extrapolation is also given, along with our value of
$\ftwo$, and two of our data points lie on this curve, as expected.
The value of $\ftwo$ differs by $\approx 30$\% from the
decay constant measured at $m_\pi = 420$ MeV.  We also plot the
SU(3) chiral limit curve, for which the horizontal axis is the
unitary meson mass for three degenerate mass quarks.  None of our
measured values must lie on this line.  The large difference between
$\fthree$ and our measurements is apparent, showing the poor convergence
of NLO SU(3) ChPT, with LECs as determined from our data.}
\label{fig:su2_su3_comp}
\end{figure}

%% file: fig/Bk.fig
%
%

%
%
\clearpage
\begin{figure}
\begin{center}
\includegraphics[angle=-90, width=.49\textwidth]{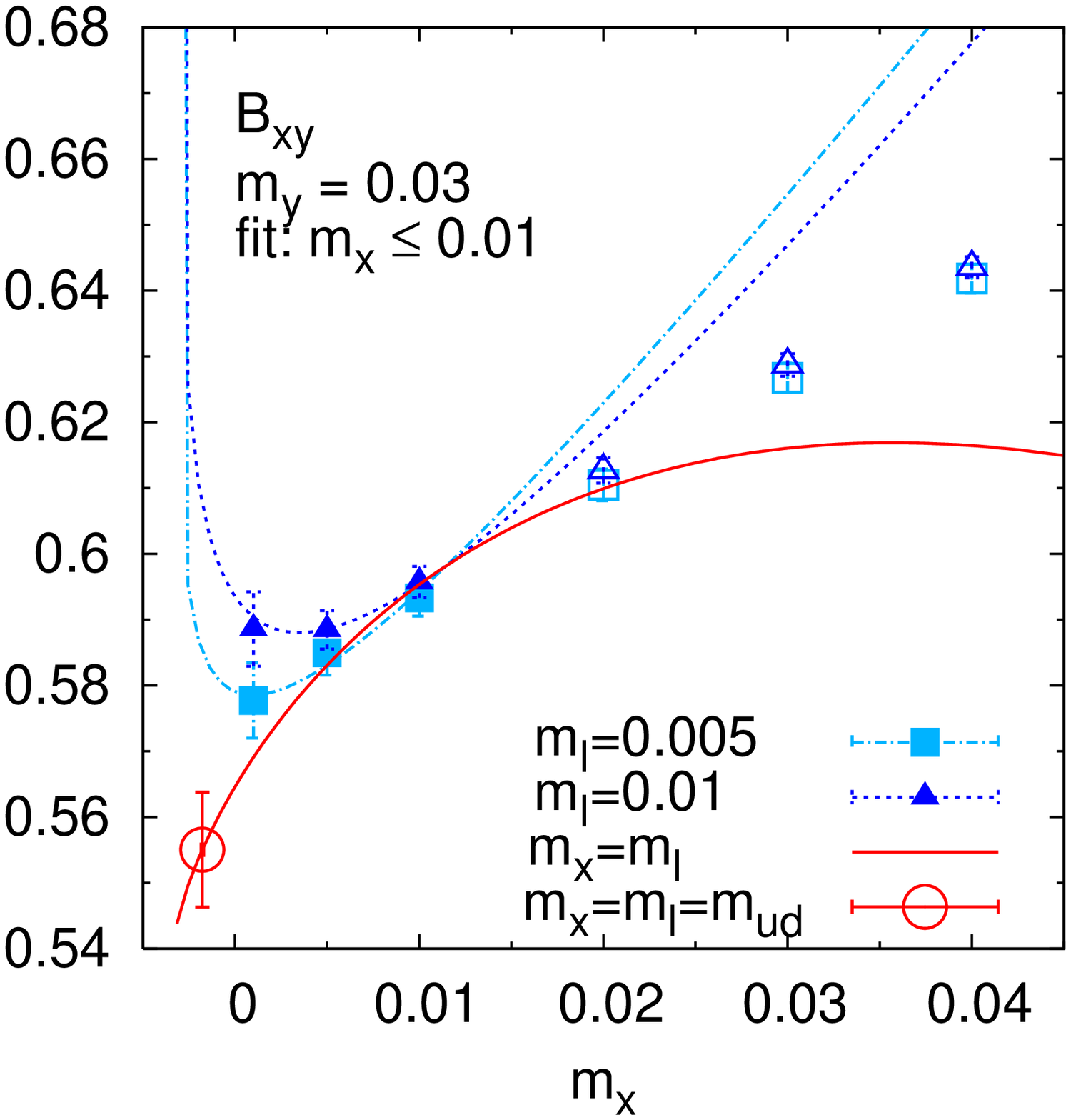}
\includegraphics[angle=-90, width=.49\textwidth]{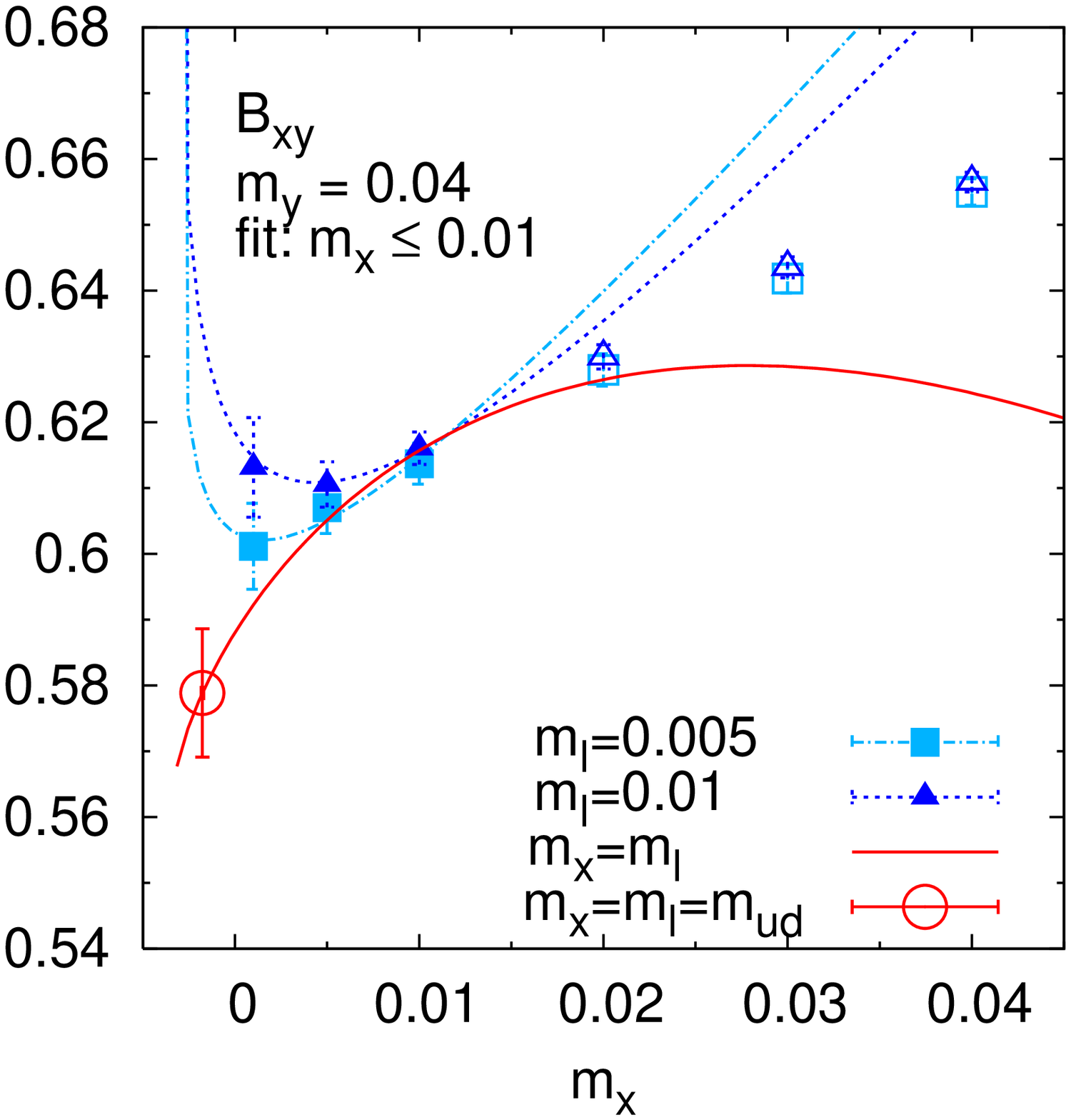}
\end{center}
\caption{Extrapolation of the pseudoscalar bag parameter $B_{xy}$ in the light quark mass $m_x$ to the physical point $m_{ud}$ for $m_y=0.03$ \textit{(left panel)} and $m_y=0.04$ \textit{(right panel)}. Points with \textit{filled symbols} are included in the fit while those marked by \textit{open symbols} are excluded.For fit parameters see Table \ref{table:bK:fitparamSU2}.}
\label{fig:bK:SU2lightExtr}
\end{figure}

\clearpage
\begin{figure}
\begin{center}
\includegraphics[angle=-90, width=.5\textwidth]{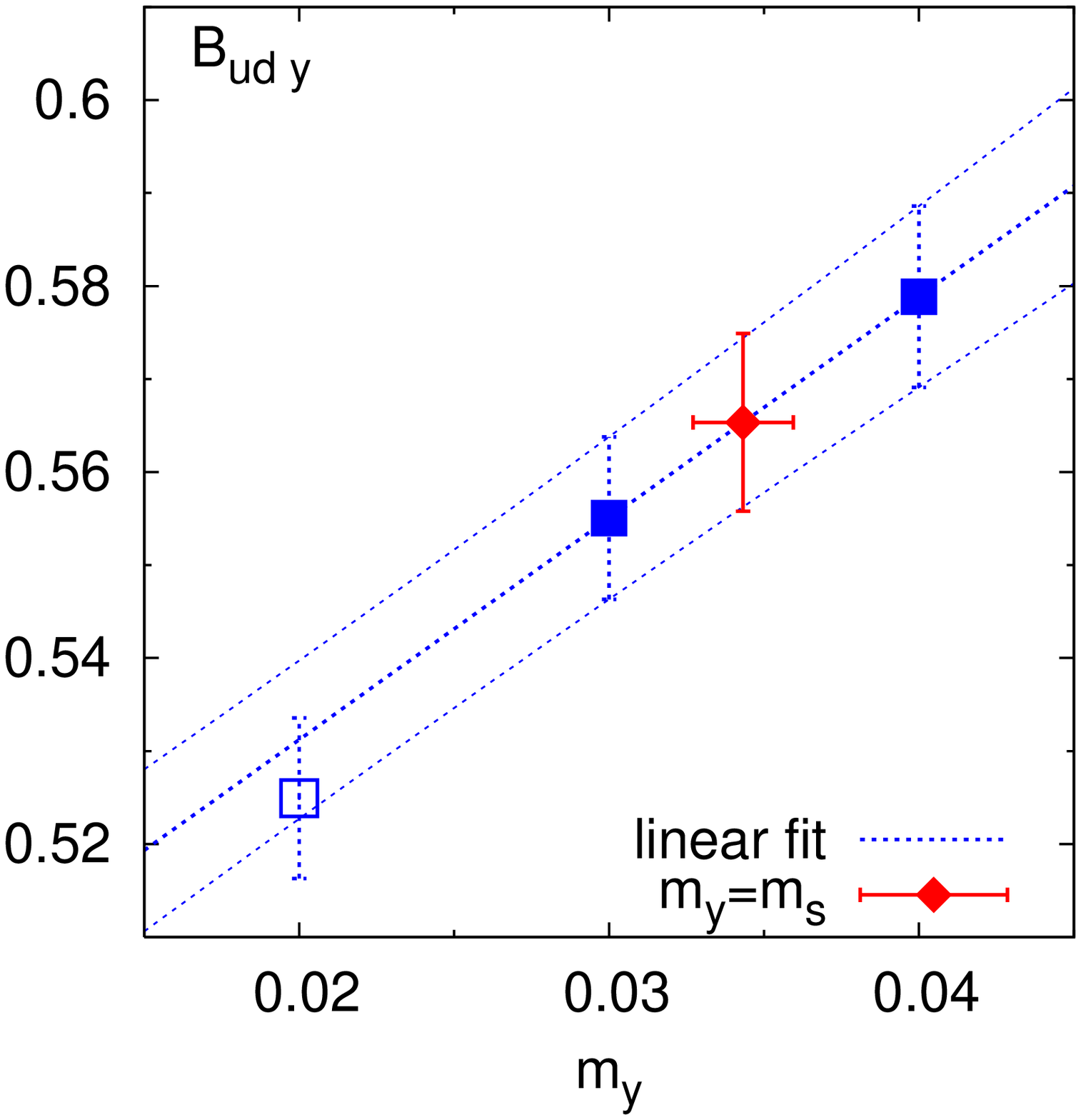}
\end{center}
\caption{Linear interpolation in $m_y$ of the pseudoscalar bag parameter $B_{ud\,y}$ to the physical strange quark mass $m_s$ (\textit{red diamond}). The \textit{open symbol} at $m_y=0.02$ was excluded from the interpolation.}
\label{fig:bK:SU2strInt}
\end{figure}

\clearpage
\begin{figure}
\begin{center}
\includegraphics[angle=-90, width=.48\textwidth]{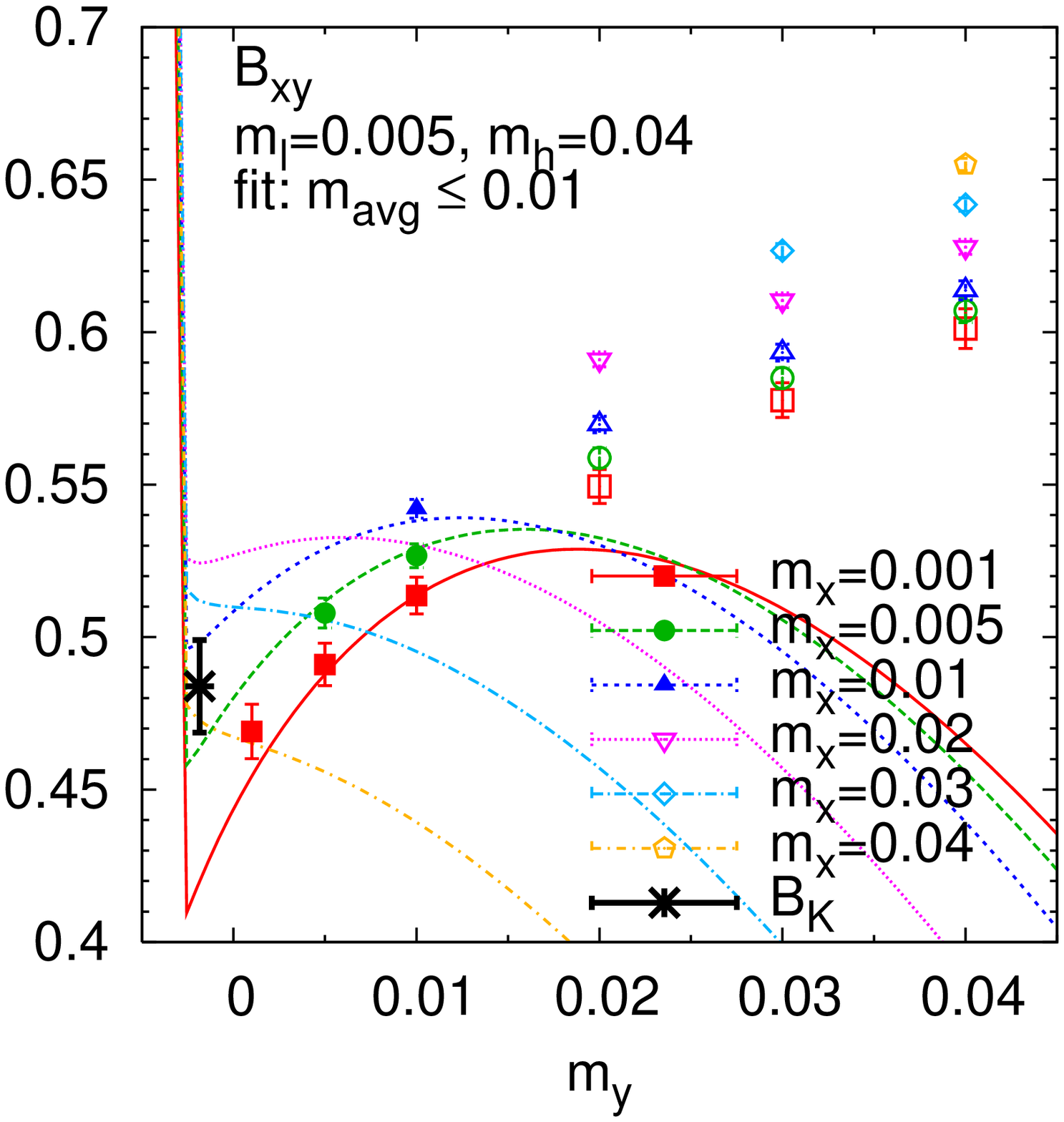}
\includegraphics[angle=-90, width=.48\textwidth]{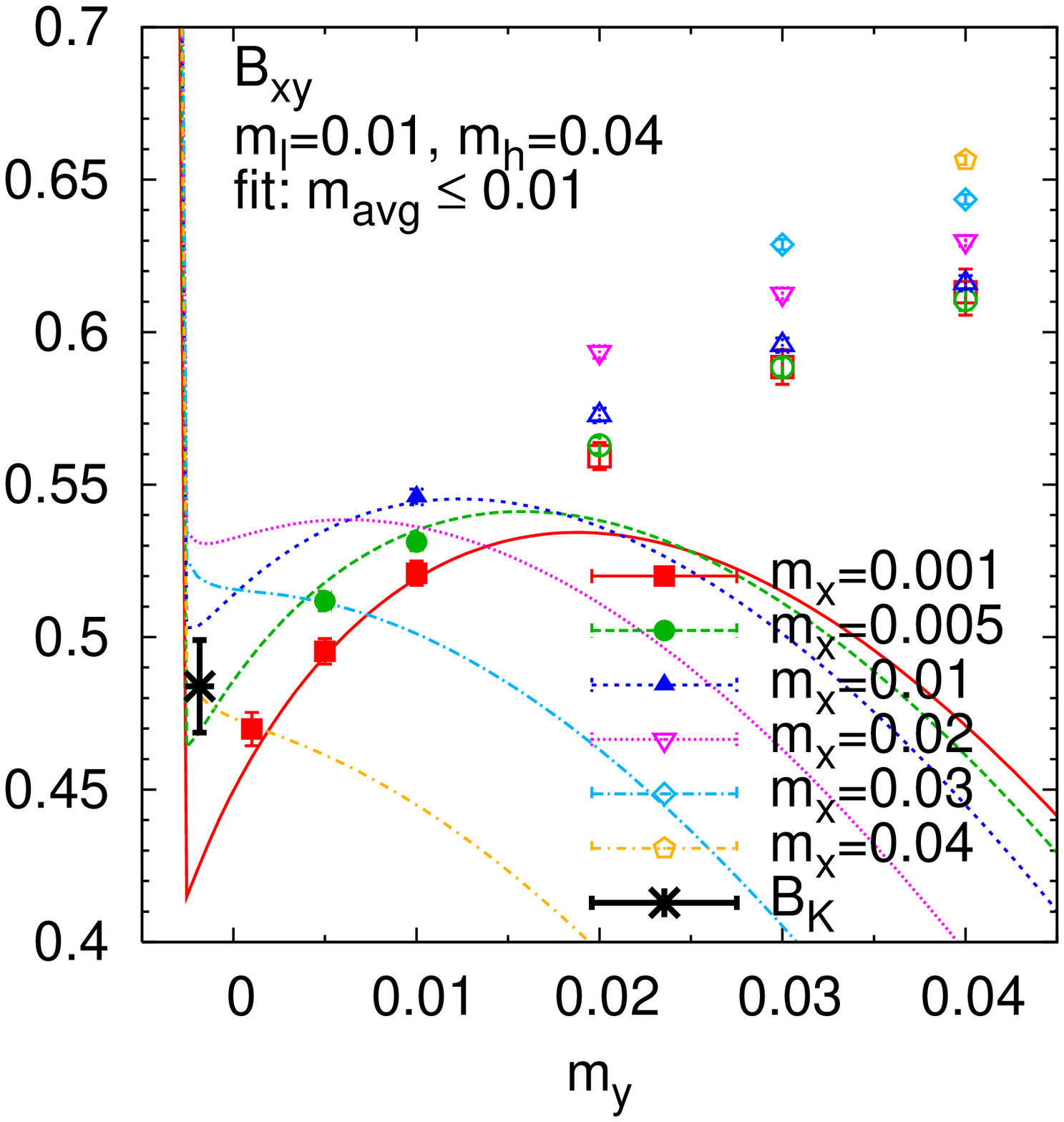}\\
\includegraphics[angle=-90, width=.48\textwidth]{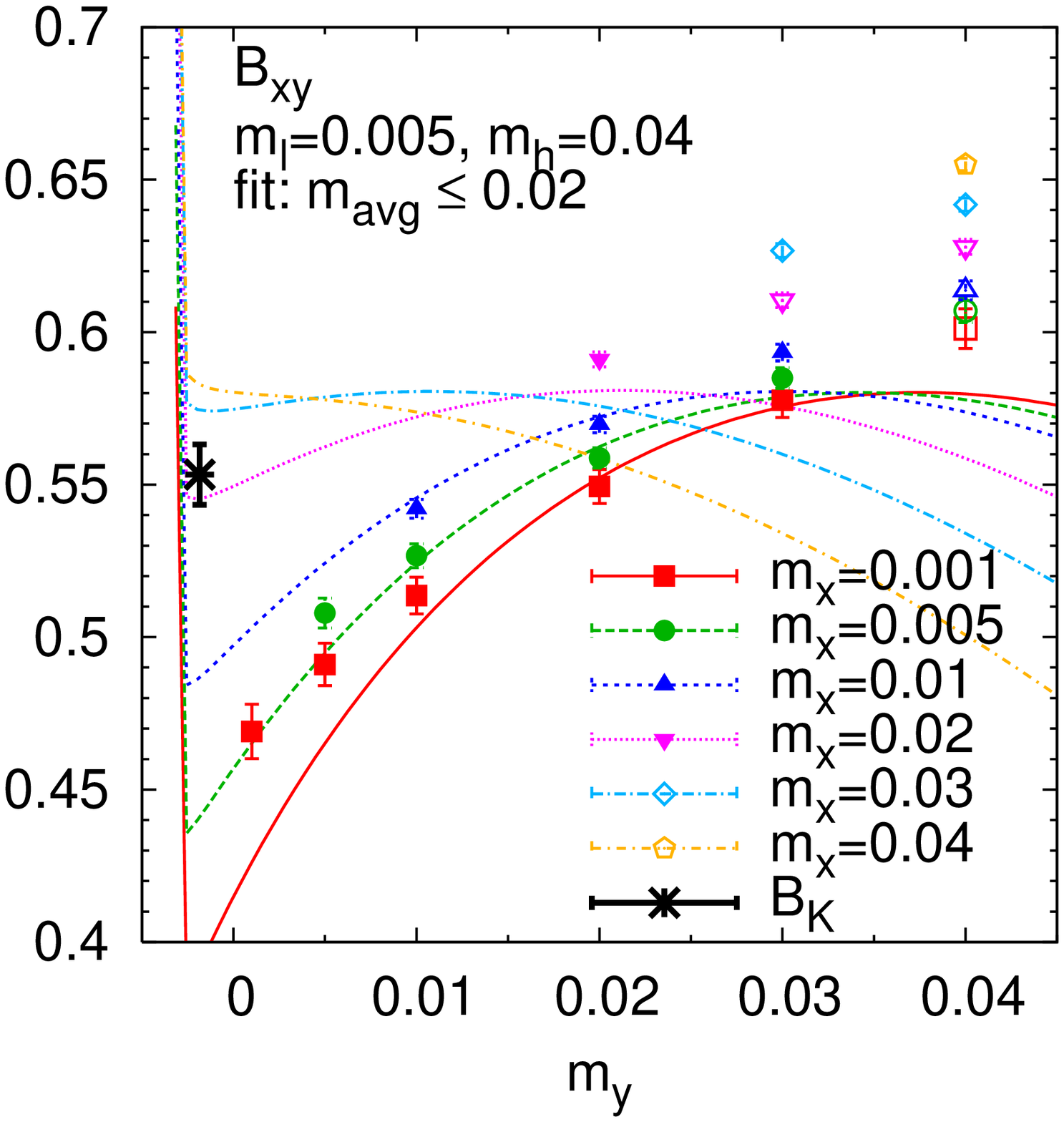}
\includegraphics[angle=-90, width=.48\textwidth]{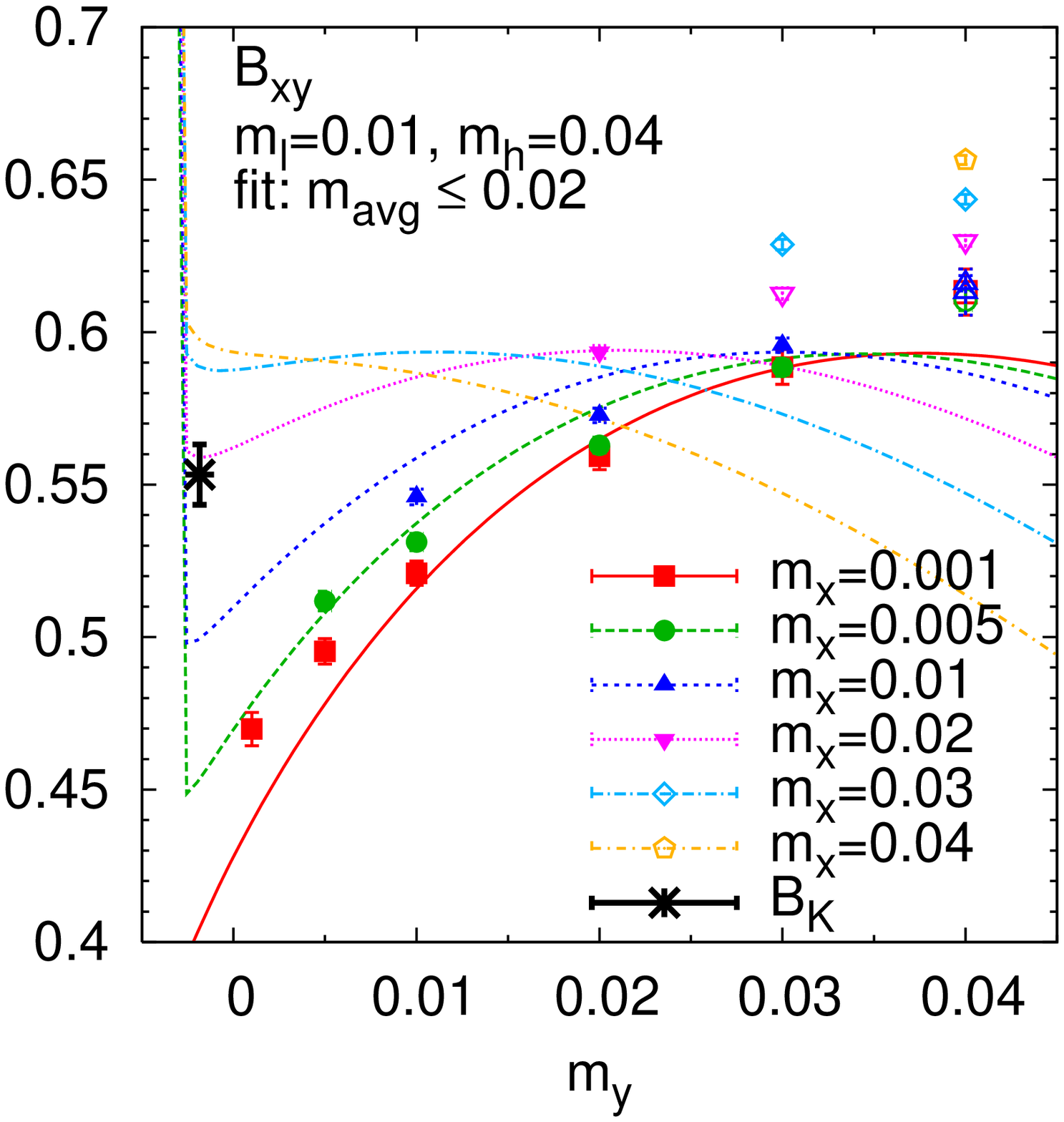}
\end{center}
\caption{Fitting the pseudoscalar bag parameter $B_{xy}$ to the $\SU(3)\times\SU(3)$ $\chi$PT formulae with a cut in the average valence mass value of 0.01 \textit{(top panels)} and 0.02 \textit{(bottom panels)}. Points marked by \textit{open symbols} are excluded from the fit. The \textit{left panels} are for light sea quark masses $m_l=0.005$, the \textit{right panels} for $m_l=0.01$.}
\label{fig:bk:SU3fits}
\end{figure}

\clearpage
\begin{figure}
\begin{center}
\includegraphics[angle=-90, width=.5\textwidth]{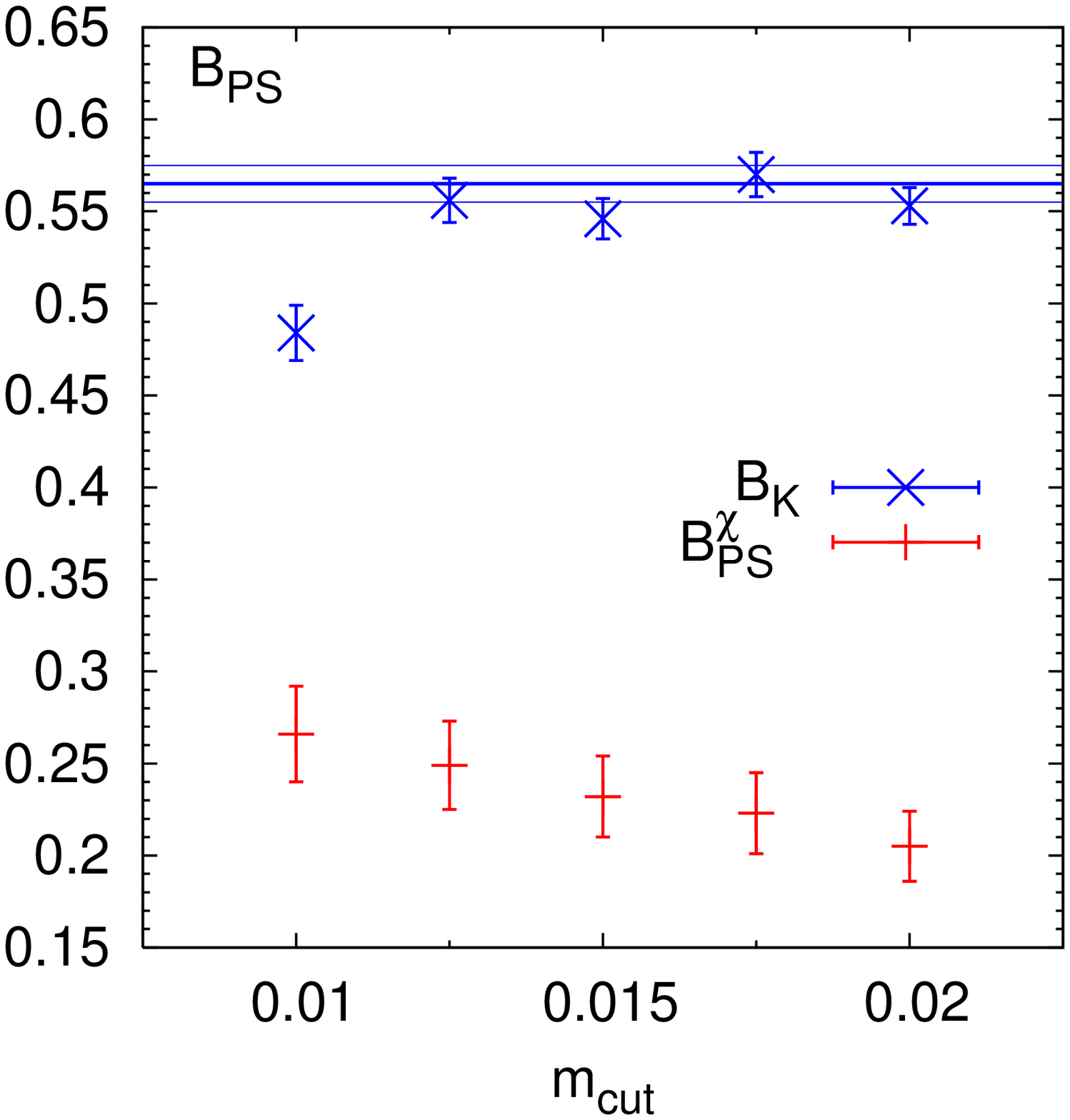}
\end{center}
\caption{Dependence of the fit parameter $\bagzero$ \textit{(red bars)} and the extrapolated physical value of $B_K$ \textit{(blue crosses)} on the average valence mass cut, when fitting to the $\SU(3)\times\SU(3)$ fit formula. The \textit{blue lines} indicate the result (with statistical error) for $B_K$ obtained from the $\SU(2)\times\SU(2)$ fit.}
\label{fig:bK:cutDepSU3}
\end{figure}

\clearpage
\begin{figure}
\begin{center}
\includegraphics[angle=-90, width=0.5\textwidth]{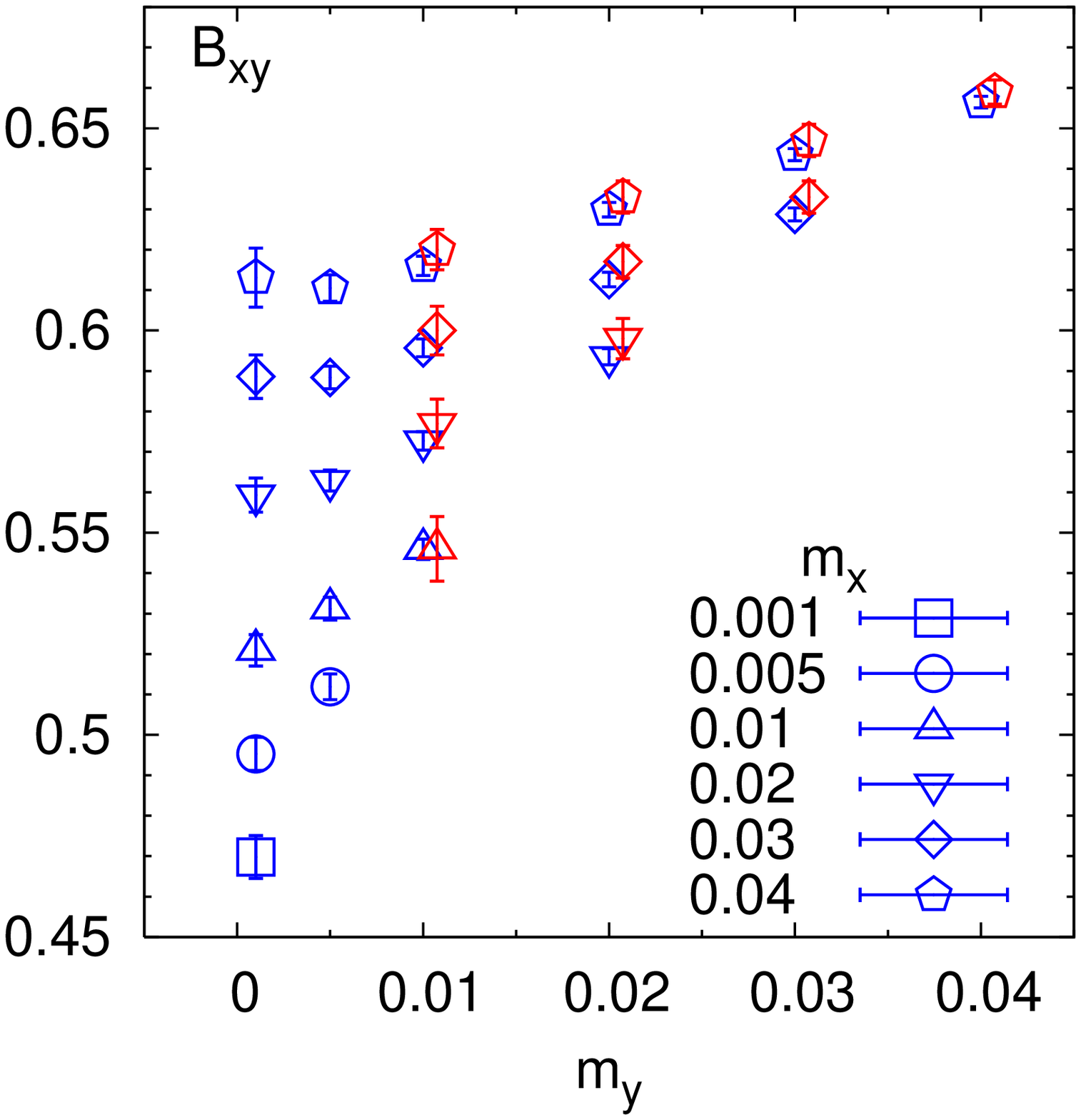}
\end{center}
\caption{Comparison of the pseudoscalar bag parameter $B_{xy}$ for valence quark masses $m_x$, $m_y$ measured on $16^3\times32$ \textit{(red symbols)} and $24^3\times64$ \textit{(blue symbols)} lattices at $m_l=0.01$, $m_h=0.04$. \textit{(Red symbols are slightly shifted to the right on this plot.)}}
\label{fig:bK:volume}
\end{figure}

%% file: fig/Vector.fig
%
%

\newpage

\begin{figure}
\begin{center}\begin{picture}(100,40)(-5,0)
\ArrowLine(0,30)(30,30)\ArrowLine(30,30)(60,30)
\ArrowLine(50,20)(70,10)\SetColor{Magenta}\Photon(30,30)(50,20){1.5}{5}
\GCirc(30,30){1.5}{0}\GCirc(50,20){1.5}{0}
\Text(74,10)[l]{\small$\rho,\,K^\ast$}
\Text(64,30)[l]{\small$\nu_\tau$} \Text(-4,30)[r]{\small$\tau$}
\end{picture}\end{center}
\caption{The diagram illustrating how $\tau$ decays can be used to
deduce $f_{\rho}$ and $f_{K^\ast}$.}
\label{fig:pda_feynman}
\end{figure}
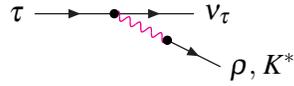

\clearpage
\begin{figure}\begin{center}
 \psfrag{xlabel}[t][b][0.8][0]{$m+m_{\textrm{res}}$}
 \psfrag{ylabel}[b][b][0.8][0]{$f_\rho^T/f_\rho$}
\hspace{4mm}\epsfig{scale=.5,angle=-90,file=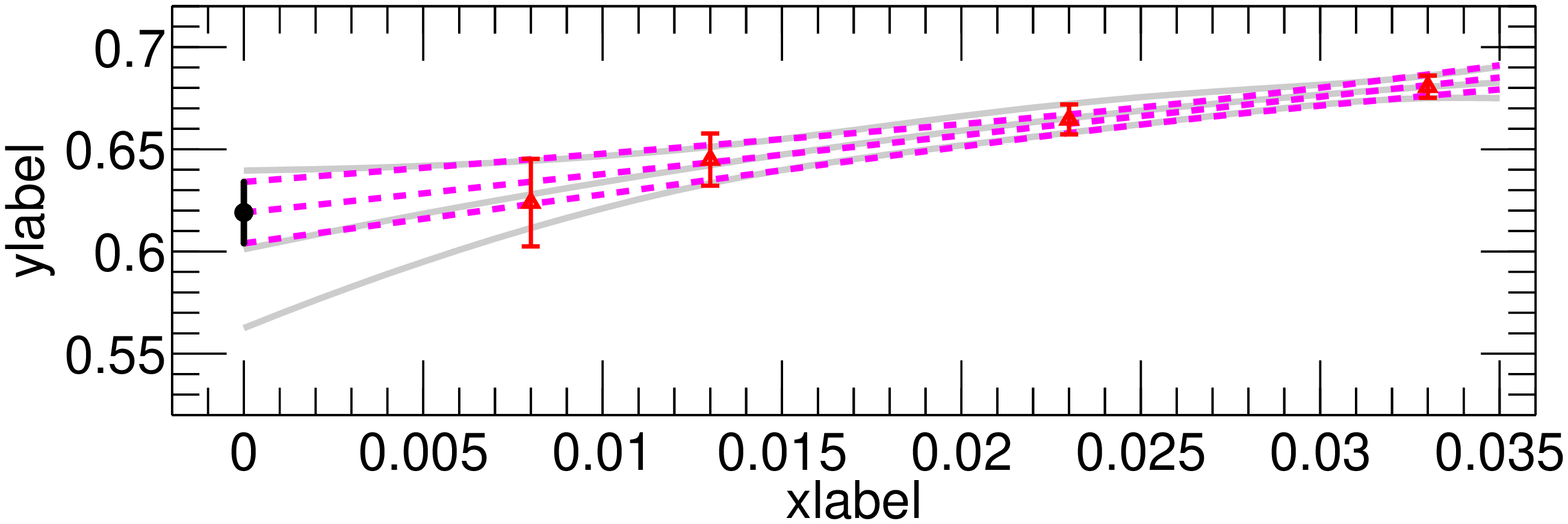}\\[4mm]
\quad \psfrag{ylabel}[b][c][0.75][0]{$f_{K^\ast}^T/f_{K^\ast}$}
\epsfig{scale=.5,angle=-90,file=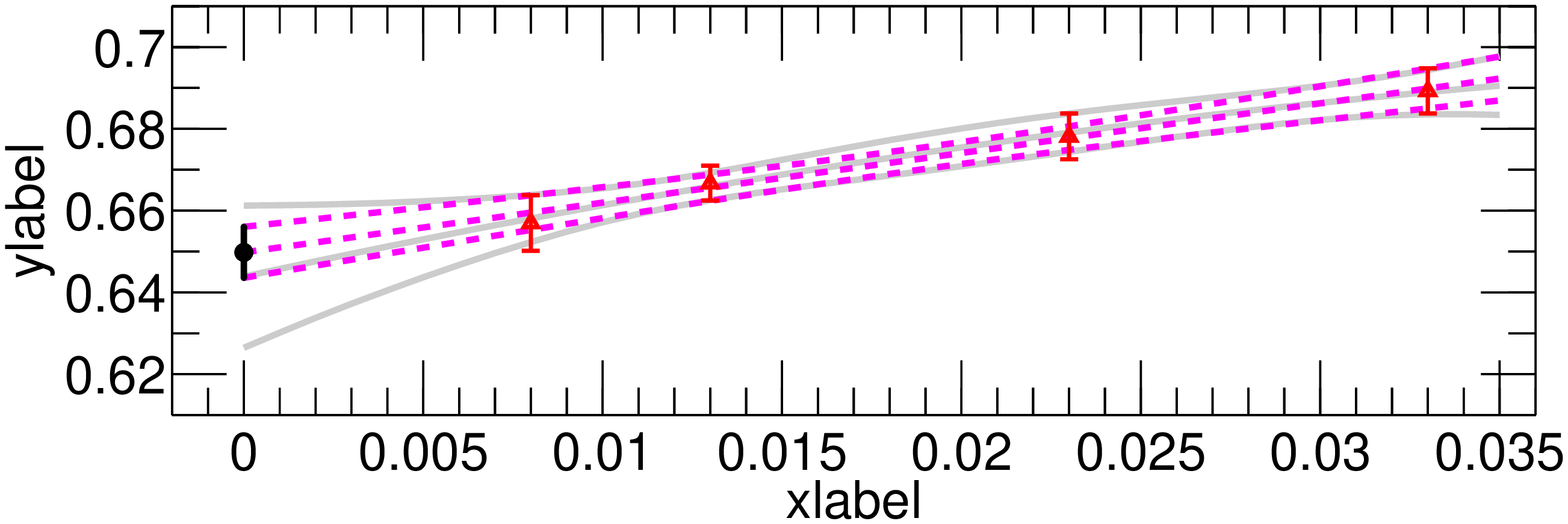}\\[4mm]
\quad \psfrag{ylabel}[b][b][0.75][0]{$f_\phi^T/f_\phi$}
\epsfig{scale=.5,angle=-90,file=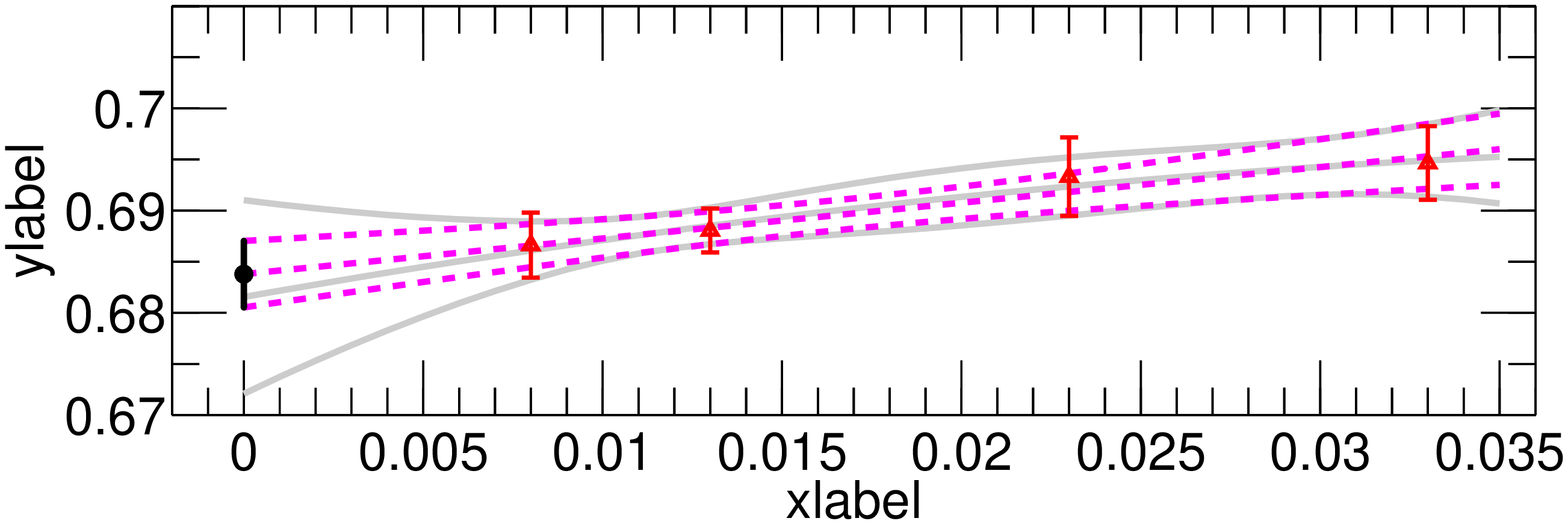}\\[4mm]
\end{center}\caption{Chiral extrapolations for $f_\rho^T/f_\rho$, $f_{K^\ast}^T/f_{K^\ast}$
and $f_\phi^T/f_\phi$ respectively. The broken red lines represent
a linear fit to the mass behaviour and the solid grey lines a
quadratic fit.\label{fig:pda_chiral}}\end{figure}